\documentclass[phd, titlesmallcaps, examinerscopy, copyrightpage, foronline]{mqthesis}

\usepackage{rotating}  
\usepackage{theorem}   
\usepackage{bm}  
\usepackage{amsmath,amsfonts,amssymb,mathabx}
\usepackage{booktabs}  
\usepackage{pdfsync}   
\usepackage{pdfpages}
\usepackage{float}
\usepackage{threeparttable}
\usepackage{hyperref}
\usepackage{natbib}
\usepackage{tcolorbox}
\usepackage{multirow}
\usepackage[LGR,T2A,T1]{fontenc}
\usepackage[utf8]{inputenc}
\usepackage[ukrainian,main=english]{babel}
\usepackage{float}

\makeatletter 
\disable@package@load{everypage}{}
\makeatother

\usepackage{silence}
\usepackage[pages=some]{background}

\backgroundsetup{
scale=1,
color=black,
opacity=1,
angle=0,
contents={%
  \includegraphics[trim=5.5cm 0 0 0, height=\paperheight]{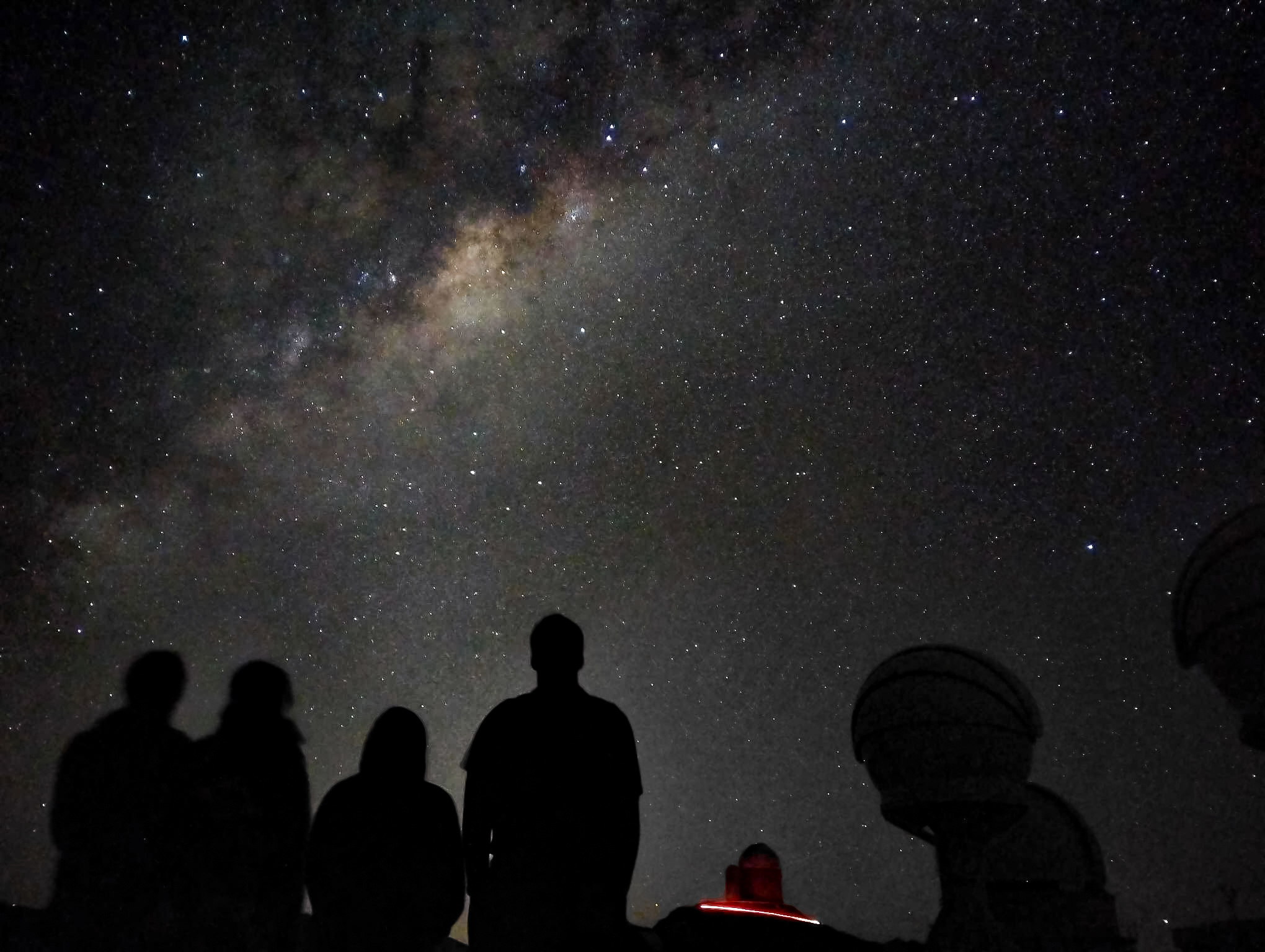} 
  }%
}

\ifpdf
    \pdfcatalog { /PageMode (/UseOutlines) /OpenAction (fitbh)  }
\fi

\begin{document}

\frontmatter

\title{Photospheric Chemical Depletion in Post-AGB/Post-RGB Binaries \\with Second-Generation Protoplanetary Discs} 

\ifthenelse{\boolean{foronline}}{
  \author{\href{mailto:maksym.mohorian@students.mq.edu.au}{Maksym Mohorian}} 
  \faculty{Faculty of Science and Engineering} 
  \department{School of Mathematical and Physical Sciences} 
}{
  \author{Maksym Mohorian} 
    \faculty{Faculty of Science and Engineering} 
  \department{School of Mathematical and Physical Sciences} 
}

 \submitdate{February 2025}


\clearpage
\thispagestyle{empty}
\null
\BgThispage
\vspace*{30ex}  
\begin{center}
\fontsize{30}{32}\selectfont
{\color{white} \textbf{Photospheric Chemical Depletion} \\\textbf{in Post-AGB/Post-RGB Binaries} \\\textbf{with Second-Generation} \\\textbf{Protoplanetary Discs}}
\end{center}
\vspace*{-30ex}
\clearpage

\titlepage

\chapter[Declaration]{Declaration of contributions to publications}

This thesis is an account of research undertaken between July 2021 and December 2024 at The School of Mathematical and Physical Sciences, Faculty of Science and Engineering, Macquarie University, Sydney NSW, Australia. This is a thesis by publication. Chapters \ref{chp:pap1}, \ref{chp:pap2}, and \ref{chp:pap3} contain the following three manuscripts, and my contribution and the contribution of co-authors are described here and also in each chapter:
\begin{itemize}
\item[$\bullet$] \textbf{M. Mohorian}, D. Kamath, M. Menon, P. Ventura, H. Van Winckel, D.~A. Garc{\'\i}a-Hern{\'a}ndez, and T. Masseron. \emph{The first measurements of carbon isotopic ratios in post-RGB stars: SZ Mon and DF Cyg}. Monthly Notices of the Royal Astronomical Society, 530:761, 2024, DOI: \url{https://doi.org/10.1093/mnras/stae791}

\textit{Author Contributions}\\
M.\,Mohorian performed the reduction and analysis of optical and near-IR spectra from HERMES/Mercator and the APOGEE survey, and was primarily responsible for interpretation of the results presented in this chapter. D.\,Kamath provided feedback and engaged in regular discussions throughout each stage of the research process. P.\,Ventura contributed the ATON models and provided feedback for the Discussion section. D.A.\,Garc{\'\i}a-Hern{\'a}ndez and H.\,Van\,Winckel are the principal investigators of the proposals that resulted in the data used in this study. M.\,Menon contributed to the development of E-iSpec -- the spectral analysis tool presented in this paper, while T.\,Masseron contributed ideas for E-iSpec development. The chapter was written by M.\,Mohorian, with all co-authors providing feedback and comments.

\item[$\bullet$] \textbf{M. Mohorian}, D. Kamath, M. Menon, A.~M. Amarsi, H. Van Winckel, C. Fava, and K. Andrych. \emph{Tracing Chemical Depletion in Evolved Binaries Hosting Second-Generation Transition Discs} (accepted for publication in Monthly Notices of the Royal Astronomical Society)

\textit{Author Contributions}\\
M.\,Mohorian performed the reduction and analysis of high-resolution optical spectra from HERMES/Mercator and UVES/VLT, and was primarily responsible for interpretation of the results presented in this chapter. D.\,Kamath provided feedback and engaged in regular discussions throughout each stage of the research process. H.\,Van\,Winckel and D.\,Kamath are the principal investigators of the proposals that resulted in the data used in this study. M.\,Menon and A.\,Amarsi contributed to the development of the E-iSpec and NLTE analysis methodology, respectively. C.\,Fava and K.\,Andrych contributed ideas and feedback for paper discussion. The chapter was written by M.\,Mohorian, with all co-authors providing feedback and comments.

\item[$\bullet$] \textbf{M. Mohorian}, D. Kamath, M. Menon, M. Jian, A.~M. Amarsi, H. Van Winckel. \emph{Abundance Analysis of Chemically Depleted Post-AGB/Post-RGB Binaries with Dust-Poor Discs} (to be submitted to Publications of the Astronomical Society of Australia)

\textit{Author Contributions}\\
M.\,Mohorian was primarily responsible for data reduction, analysis, and interpretation of the results presented in this chapter. D.\,Kamath provided feedback and engaged in regular discussions throughout each stage of the research process. M.\,Menon contributed to development of the \texttt{E-iSpec}. H.\,Van\,Winckel is the principal investigator of the proposals that resulted in the data used in this study. M.\,Jian and A.~M.\,Amarsi contributed to development of the NLTE analysis methodology. The chapter was written by M.\,Mohorian, with all co-authors providing feedback and comments.
\end{itemize}
\chapter{Acknowledgements}

\begin{tcolorbox}[colback=blue!10!white, boxrule=0mm]
    \textit{Acknowledgements for this thesis are written in English and Ukrainian (which is the native language of the author).}
\end{tcolorbox}

Completing this thesis has been a challenging yet deeply rewarding journey, one that would not have been possible without the support, guidance, and encouragement of many individuals. I would like to express my heartfelt gratitude to those who was always there for me.

First and foremost, I extend my deepest appreciation to Devika for her invaluable mentorship, patience, and encouragement throughout my research. Her insightful feedback and unwavering support have shaped this work and my growth as an astronomer. Additionally, I am grateful to other students of stellar research group at Macquarie University -- Meghna, Deepak, Amy, and Claudia -- for their stimulating discussions, technical assistance, and hospitality. Special thanks to my collaborators -- Hans, Paolo, Anish, and Toon -- for their contributions and for making the research environment both productive and enjoyable. Furthermore, I would like to thank Macquarie University for the financial support through iMQRES scholarship, which made this research possible. Also, I am grateful to Yakiv Volodymyrovych for mentoring my Undergraduate research -- you will never be able to see the results of your support, but your expertise and help were crucial for me. Last but not least, I am eternally grateful to Nataliia Al'bertivna for mentoring and supporting my first steps in astronomical research during my high school -- you sparked the flame that fuels my curiosity and ambition to this day.

To my family, who have been my foundation throughout this journey -- thank you for your endless patience, encouragement, and belief in me. To Katya, my partner and peer, -- your love and support have sustained me through the toughest moments, and I am forever grateful. To my mom -- you have always been strong for me and my siblings, and I am grateful for that. To my dad -- you will never see me finish my PhD, but I hope I made you proud. To Uncle Tolik and Aunt Ira -- thank you for being there for me and helping in the most difficult times. To my grandparents -- thank you for teaching me so many things. To Yulia (`uliublena naikrashcha super sister z shotlandiyi') -- thanks for making my life so much more fun, even when some co-op adventures were frustrating (like that octopus boss fight or achievement grinding in that short game about two detectives). To Sasha, Tima, and Mira -- thank you for giving me a different perspective on life and what is important. To Liza -- thanks for providing a sense of connection with home in this far away country. To Oksana Volodymyrivna and Dmytro Rostyslavovych -- thank you for believing in me.

A huge thank you to my friends, especially Pasha and Liva, for their unwavering encouragement. Your support has meant the world to me. Big thanks to Ilia for hospitality and support whenever I asked for help. Thanks to Nastia, Kostia, Alia, and Liosha for a three-year D\&D adventure distracting me from the worries of life on Saturday nights. Also, big thanks to Dania, Lina, Slava, Yura, and Vlada for ample investigations of Cthulhian eldritch horrors and other adventures on the weekends. Additionally, a thank you to Beth and my peers from Macquarie University -- Lachlan, Miguel, Gabbie, Oguzhan, Luis, Ana, Jack, and Kayla -- for hospitality, fruitful discussions, and friendly atmosphere. Furthermore, thanks to Kostia, Yehor, and Vania -- we grew apart over the last three years, but I am sure there are still some waffles and juice waiting for us to raid IC.

I am eternally indebted to the courageous people in the Armed Forces of Ukraine, whose bravery and resolution kept my family safe during most of my PhD candidature, thus allowing me to concentrate on my research.

Finally, thank you to everyone who has been part of this journey, whether through direct contributions to my research or by simply being there when I needed support. Personal thanks to Richard E. Cannon for capturing an incredible photo of our small exoplanetary quartet at the ESO Observing School 2024 (La Silla, Chile), a cropped version of which I used as the cover of this thesis, -- this picture will always remind me about the journey of my PhD. Final thanks for the AI helpers, who linked my messy mind palace with the outside world and helped to refine my most convoluted trains of thoughts.

\begin{center}
  * * *
\end{center}

\foreignlanguage{ukrainian}{
Моя аспірантура було складною, але дуже корисною подорожжю, яка була б неможливою без підтримки та допомоги багатьох людей. Я хотів би висловити свою сердечну подяку тим, хто допомагав мені на цьому шляху.

Перш за все, я висловлюю глибоку вдячність Девіці за її безцінне наставництво, терпіння та підтримку під час мого дослідження. Її проникливі відгуки та непохитна підтримка сформували цю роботу та мою еволюцію як вченого-астронома. Крім того, я вдячний іншим студентам зоряної дослідницької групи в Університеті Маккуорі -- Мегні, Діпаку, Емі та Клаудії -- за наукові дискусії, технічну допомогу та гостинність. Особлива подяка моїм колабораторам -- Гансу, Паоло, Анішу та Тоону -- за їхній внесок і за те, що зробили дослідницьке середовище продуктивним і приємним. Крім того, я хотів би подякувати Університету Маккуорі за фінансову підтримку через стипендію iMQRES, завдяки якій це дослідження стало можливим. Також я вдячний Якову Володимировичу за наставництво моїх бакалаврських досліджень -- нажаль, ви не зможете побачити результатів вашої підтримки, але ваш досвід і допомога були неоціненними на моєму шляху до аспірантури. І останнє, але не менш важливе: я назавжди вдячний Наталії Альбертівні за наставництво та підтримку моїх перших кроків в астрономічних дослідженнях під час мого навчання в старших класах -- ви розпалили полум'я, яке досі підживлює мою наснагу й амбіції.

Моїй родині, яка була моєю підтримкою протягом усієї цієї подорожі, дякую за ваше нескінченне терпіння та віру в мене. Каті, моїй нареченій та колезі, -- твоя любов і підтримка підтримували мене в найважчі моменти, і я тобі завжди вдячний. Моїй мамі -- ти завжди була сильною для мене та моїх братів і сестер, і я вдячний за це. Моєму батьку -- ти не зможеш побачити, як я отримаю докторську ступінь, але я сподіваюся, що ти мною пишаєшся. Дядьку Толіку і тітці Ірі -- дякую за те, що ви були поруч і допомагали у найтяжчі часи. Дідусю і бабусям -- дякую за те, що навчили мене стільком речам. Юлі (`улюбленій найкращій супер сістер з шотландії') -- дякую за те, що зробила моє життя набагато веселішим, навіть коли деякі кооперативні пригоди розлючували (як, наприклад, бій з босом-восьминогом чи грінд досягнень у короткій грі про двох детективів). Саші, Тімі та Мірі -- дякую за те, що ви дали мені інший погляд на життя, та на те, що є насправді важливим. Лізі -- дякую за відчуття зв'язку з домом у цій далекій країні. Оксані Володимирівні та Дмитру Ростиславовичу -- дякую за те, що ви прийняли мене у свою родину та вірили в мене.

Величезне спасибі моїм друзям, особливо Паші та Ліві, за їхню непохитну підтримку. Ваша підтримка означає для мене все. Велике спасибі Іллі за гостинність і підтримку, коли б я не попросив про допомогу. Спасибі Насті, Кості, Алі та Льоші за трирічну пригоду у D\&D, яка відволікала мене від життєвих турбот суботніми ночами. Також велика подяка Дані, Ліні, Славі, Юрі та Владі за розслідування ктулхіанських жахів та інші пригоди. Також дякую Бет та моїм колегам з Університету Маккуорі -- Лаклану, Мігелю, Геббі, Оузану, Ані, Джеку та Кейлі -- за гостинність, плідні дискусії та дружню атмосферу. Крім того, дякую Кості, Єгору та Вані -- ми віддалилися один від одного за останні роки, але я впевнений, що десь на нас ще чекають вафлі та сік для рейду до Цитаделі Крижаної Корони.

Я у вічному боргу перед мужніми людьми в Збройних Силах України, чия хоробрість і рішучість зберігали мою сім’ю у безпеці протягом більшої частини моєї аспірантури, що дозволило мені зосередитися на дослідженнях.

Нарешті, дякую всім, хто був частиною цієї подорожі, чи то через прямий внесок у мої дослідження, чи просто будучи поруч, коли я потребував підтримки. Особиста вдячність Річарду Е. Кеннону за неймовірну фотографію нашого маленького екзопланетного квартету в Школі спостережень Європейської південної обсерваторії 2024 (Ла-Сілья, Чилі), обрізану версію якої я використав як обкладинку цієї дисертації, —- це фото завжди нагадуватиме мені, який шлях я пройшов протягом моєї аспірантури. Наприкінці, висловлюю подяку ШІ-помічникам, які зв'язали мій безладний палац розуму із зовнішнім світом і допомогли сформулювати мої найзаплутаніші думки.
}
\chapter{List of Publications}
This thesis includes the articles published or prepared for publication over the course of my PhD candidature, presented in Chapters~\ref{chp:pap1}, \ref{chp:pap2}, and \ref{chp:pap3}. My contributions to each publication are listed in the respective chapters.

\begin{itemize}
\item[$\bullet$] \textbf{Mohorian M.}, Kamath D., Menon M., Ventura P., Van Winckel H., Garc\'{i}a-Hern\'{a}ndez D.\,A., Masseron T. \emph{The first measurements of carbon isotopic ratios in post-RGB stars: SZ Mon and DF Cyg.}  
        Monthly Notices of the Royal Astronomical Society \textbf{530}, 
        1 (2024)

\item[$\bullet$] \textbf{Mohorian M.}, Kamath D., Menon M., Amarsi A.~M., Van Winckel H., Fava C., Andrych K. \emph{Tracing Chemical Depletion in Evolved Binaries Hosting Second-Generation Transition Discs} (accepted for publication in Monthly Notices of the Royal Astronomical Society)

\item[$\bullet$] \textbf{Mohorian M.}, Kamath D., Menon M., Jian M., Amarsi A.~M., Van Winckel H. \emph{Abundance Analysis of Chemically Depleted Post-AGB/Post-RGB Binaries with Dust-Poor Discs} (to be submitted to Publications of the Astronomical Society of Australia)
\end{itemize}

In addition to the above, I contributed to the following papers during my PhD course:

\begin{itemize}
\item[$\bullet$] Kamath D., Van Winckel H., Ventura P., \textbf{Mohorian M.}, Hrivnak B.J., Dell'Agli F., Karakas A. \emph{Luminosities and Masses of Single Galactic Post-asymptotic Giant Branch Stars with Distances from Gaia EDR3: The Revelation of an s-process Diversity.}  
        The Astrophysical Journal \textbf{927}, 
        L13 (2022)

\item[$\bullet$] \textbf{Mohorian M.}, Bhatta G., Adhikari T.\,P., Dhital N., P{\'a}nis R., Dinesh A., Chaudhary S.\,C., Bachchan R.\,K., Stuchl{\'i}k Z. \emph{X-ray timing and spectral variability properties of blazars S5 0716+714, OJ 287, Mrk 501, and RBS 2070.}
        Monthly Notices of the Royal Astronomical Society, Volume 510, Issue 4, pp.5280-5301 (2022)

\item[$\bullet$] Dinesh A., Bhatta G., Adhikari T.\,P., \textbf{Mohorian M.}, Dhital N., Chaudhary S.\,C., P{\'a}nis R., G{\'o}ra D. \emph{Constraining X-Ray Variability of the Blazar 3C 273 Using XMM-Newton Observations over Two Decades.}
        The Astrophysical Journal, Volume 955, Issue 2, id.121, 13 pp. (2023)

\item[$\bullet$] Menon M., Kamath D., \textbf{Mohorian M.}, Van Winckel H., Ventura P. \emph{\textit{s}-Process Enriched Evolved Binaries in the Galaxy and the Magellanic Clouds.}  
        Publications of the Astronomical Society of Australia \textbf{41}, 
        e025 (2024)

\item[$\bullet$] Bhatta G., Chaudhary S.\,C., Dhital N., Adhikari T.\,P., \textbf{Mohorian M.}, P{\'a}nis R., Neupane R., Singh Maharjan Y. \emph{Probing X-ray Timing and Spectral Variability in the Blazar PKS 2155-304 Over a Decade of XMM-Newton Observations}  
        Monthly Notices of the Royal Astronomical Society, Volume 981, Issue 2, pp.118 (2025)

\item[$\bullet$] Menon, M., Kamath, D., \textbf{Mohorian, M.}, Van Winckel, H., and Ventura, P. \emph{J003643.94-723722.1: The First Post-AGB Star in the SMC with a High C/O Ratio and a Measured Pb Abundance} (to be submitted to Publications of the Astronomical Society of Australia)
\end{itemize}
\chapter{Abstract}

The origin and evolution of chemical elements in the Universe are governed not only by nucleosynthesis processes in stars, but also by mechanisms that alter observed photospheric compositions. Among these, chemical depletion (underabundance of refractory elements in stellar photospheres) presents a key puzzle in understanding the full chemical lifecycle. This PhD thesis explores the role of disc-binary interaction in shaping chemical abundances in evolved low- and intermediate-mass binary stars, focusing on systems that have undergone the red giant branch (RGB) or asymptotic giant branch (AGB) phase. In this thesis, we investigate binary systems containing post-asymptotic giant branch (post-AGB, $L_{\rm post-AGB}\,\gtrsim\,2\,500\,L_\odot$) and post-red giant branch (post-RGB, $L_{\rm post-RGB}\,\lesssim\,2\,500\,L_\odot$) binaries as key tracers of AGB/RGB nucleosynthesis. Although the effects of these interactions remain poorly understood, they are known to drive photospheric chemical depletion. This depletion closely resembles that observed in young planet-hosting stars with protoplanetary discs. Combined with other structural and dynamical similarities in disc properties, this suggests a potential link to second-generation planet formation in post-AGB/RGB binaries with circumbinary discs. Although direct imaging of such planets is not feasible, studying signatures such as photospheric depletion provides an indirect means of exploring their possible presence within these systems.

This thesis investigates the depletion mechanisms in post-AGB and post-RGB binaries by analysing high-resolution optical and near-infrared spectra across a diverse sample of full, transition, and dust-poor disc systems. By examining these disc types, we explore how disc structure influences photospheric chemical depletion and nucleosynthesis signatures in evolved binaries. Previous abundance studies of these binaries have often been piecemeal, typically assuming local thermodynamic equilibrium (LTE), which does not always hold for hot, metal-poor giants. This thesis presents the first homogeneous abundance analysis of post-AGB and post-RGB binaries, incorporating NLTE corrections and a multiwavelength approach to derive atmospheric parameters, elemental abundances, isotopic ratios, and depletion profiles. The multiwavelength approach, particularly near-infrared spectra, enables the accurate determination of CNO isotopic ratios through molecular bands such as CO, CN, and OH, as well as a comprehensive analysis of elemental abundances. Since existing spectral analysis codes are inadequate for post-AGB and post-RGB binaries, we developed E-iSpec, a specialised tool designed for evolved stars. E-iSpec was applied to conduct abundance analyses of post-AGB and post-RGB binaries in both the Galaxy and the Large Magellanic Cloud, facilitating a systematic investigation of depletion profiles across different metallicity environments.

We demonstrate that post-RGB binaries exhibit depletion profiles similar to those of post-AGB binaries, yet with a higher onset temperature of depletion. The first carbon isotopic ratios for post-RGB stars were measured, confirming consistency with single-star nucleosynthesis predictions. We further show that depletion efficiency is significantly enhanced in post-AGB and post-RGB binaries with transition discs, with depletion profiles resembling those observed in the ISM and being stronger than depletion profiles in young stars with transition discs. Finally, our analysis of dust-poor discs reveals a bimodal distribution in depletion onset temperature, allowing us to classify these systems into full-like and transition-like dust-poor discs. This suggests a possible evolutionary pathway from full and transition discs to dust-poor discs over time. These findings provide a systematic framework for understanding photospheric depletion in evolved binaries, offering indirect constraints on circumbinary disc evolution and second-generation planet formation. Future work integrating larger statistical samples and advanced modelling will further refine our understanding of these complex systems.

The thesis is structured as follows: In Chapter~\ref{chp:int}, we provide an overview of nucleosynthetic enrichment and chemical depletion in the Universe, and introduce photospheric chemical depletion in post-AGB/post-RGB binary systems with second-generation protoplanetary discs. In addition, we outline the motivation for the study and the gaps in understanding that this thesis aims to address. In Chapter~\ref{chp:mth}, we provide an overview of spectral analysis techniques and the spectroscopic data used in this thesis, as well as the development of the E-iSpec code. In Chapter~\ref{chp:pap1}, we present an abundance analysis of two dusty post-RGB binaries, SZ Mon and DF Cyg, and compare the derived abundances with predictions from the ATON evolutionary models. In Chapter~\ref{chp:pap2}, we investigate photospheric depletion in 12 post-AGB/post-RGB binaries with transition discs and compare the depletion efficiency in these targets with depletion efficiencies reported in the ISM and in the young planet-hosting stars. In Chapter~\ref{chp:pap3}, we explore the evolutionary status of these discs and their relation to full and transition discs by analysing 9 post-AGB/post-RGB binaries with dust-poor discs. In Chapter~\ref{chp:dsc}, we summarise the key findings of this thesis and outline future prospects in binary evolution, disc chemistry, and planet formation.

\tableofcontents
\listoffigures
\listoftables

\chapter{List of Acronyms}


\begin{list}{}{%
\setlength{\labelwidth}{24mm}
\setlength{\leftmargin}{35mm}}
\item[\textbf{2MASS}] 2-Micron All Sky Survey
\item[\textbf{AGB}] Asymptotic giant branch
\item[\textbf{APOGEE}] Apache Point Observatory Galactic Evolution Experiment
\item[\textbf{au}] Astronomical unit (1\,au\,=\,149\,597\,871\,km)
\item[\textbf{BBN}] Big Bang nucleosynthesis
\item[\textbf{ESO}] European Southern Observatory
\item[\textbf{EW}] Equivalent width
\item[\textbf{HERMES}] High Efficiency and Resolution Mercator Echelle Spectrograph
\item[\textbf{HR diagram}] Hertzsprung-Russell diagram
\item[\textbf{IR}] Infrared
\item[\textbf{IRAS}] Infrared Astronomical Satellite
\item[\textbf{LIM stars}] Low- and intermediate-mass stars
\item[\textbf{LMC}] Large Magellanic Cloud
\item[\textbf{LTE}] Local thermodynamic equilibrium
\item[\textbf{MARCS}] Model Atmospheres in Radiative and Convective Scheme
\item[\textbf{MS}] Main Sequence
\item[\textbf{NLTE}] Non-local thermodynamic equilibrium
\item[\textbf{PN}] Planetary nebula
\item[\textbf{Post-AGB}] Post-asymptotic giant branch
\item[\textbf{Post-RGB}] Post-red giant branch
\item[\textbf{$r$-process}] Rapid neutron capture process
\item[\textbf{RGB}] Red giant branch
\item[\textbf{S/N ratio}] Signal-to-noise ratio
\item[\textbf{$s$-process}] Slow neutron capture process
\item[\textbf{SED}] Spectral energy distribution
\item[\textbf{SMC}] Small Magellanic Cloud
\item[\textbf{UVES}] Ultraviolet and Visual Echelle Spectrograph
\item[\textbf{VALD}] Vienna Atomic Line Database
\item[\textbf{WD}] White dwarf
\item[\textbf{WISE}] Wide-field Infrared Survey Explorer
\item[\textbf{YSO}] Young stellar object
\end{list}

\mainmatter

\begin{savequote}[75mm]
\foreignlanguage{ukrainian}{``Мене часто запитують, чи було страшно перед польотом у космос, коли я сидів у космічному кораблі і чекав старту. Я кажу так — мені було страшно цікаво...''}
\qauthor{\foreignlanguage{ukrainian}{---Леонід Каденюк (1951-2018), \\перший космонавт незалежної України}}
``I’m often asked whether I was fearful before my spaceflight, sitting in the spacecraft and waiting for lift-off. I always reply: I was fearfully curious...''
\qauthor{---Leonid Kadenyuk (1951-2018), \\the first astronaut of independent Ukraine}
\end{savequote}

\chapter{Introduction}\label{chp:int}
\graphicspath{{ch_intro/figures/}} 

\clearpage

\textit{The goal of this thesis is to investigate the impact of binary interactions on element and isotope production in evolved binary systems with second-generation protoplanetary circumbinary discs, in particular, post-AGB binaries and their low-luminosity analogues, post-RGB binaries ($L_{\rm post-AGB}\,\gtrsim\,2\,500\,L_\odot$, $L_{\rm post-RGB}\,\lesssim\,2\,500\,L_\odot$). To achieve this, the optical and near-infrared spectra of post-AGB and post-RGB binary stars were analysed, providing insights into how binarity influences stellar nucleosynthesis and chemical evolution, as well as into the disc-binary interactions in these systems. While post-AGB and post-RGB single stars serve as excellent tracers of element production during the AGB and RGB stages, post-AGB and post-RGB binary systems are exceptional tracers of disc-binary interactions, exhibiting distinct signatures, including photospheric chemical depletion observed in the primary star. To establish a baseline for element and isotope production, we start this chapter with a brief overview of the nucleosynthetic processes occurring in stars and enriching the Universe with chemical elements (see Section~\ref{sec:intnuc}). We then examine the mechanisms that modify the chemical enrichment of the Universe, including chemical depletion in the interstellar medium (ISM) and photospheric chemical depletion in both young and evolved stars (see Section~\ref{sec:intdpl}). We specifically focus on photospheric chemical depletion in post-AGB/post-RGB binary systems with second-generation protoplanetary discs, highlighting existing gaps in our understanding of this process (see Section~\ref{sec:intpAR}). Finally, we conclude this chapter with the motivation and outline of this thesis (see Section~\ref{sec:intmtv}).}

\section{Origin of elements in the Universe}\label{sec:intnuc}
\begin{figure}[!ht]
    \centering
    \includegraphics[trim={0 15cm 0 1cm}, clip, width=.99\linewidth]{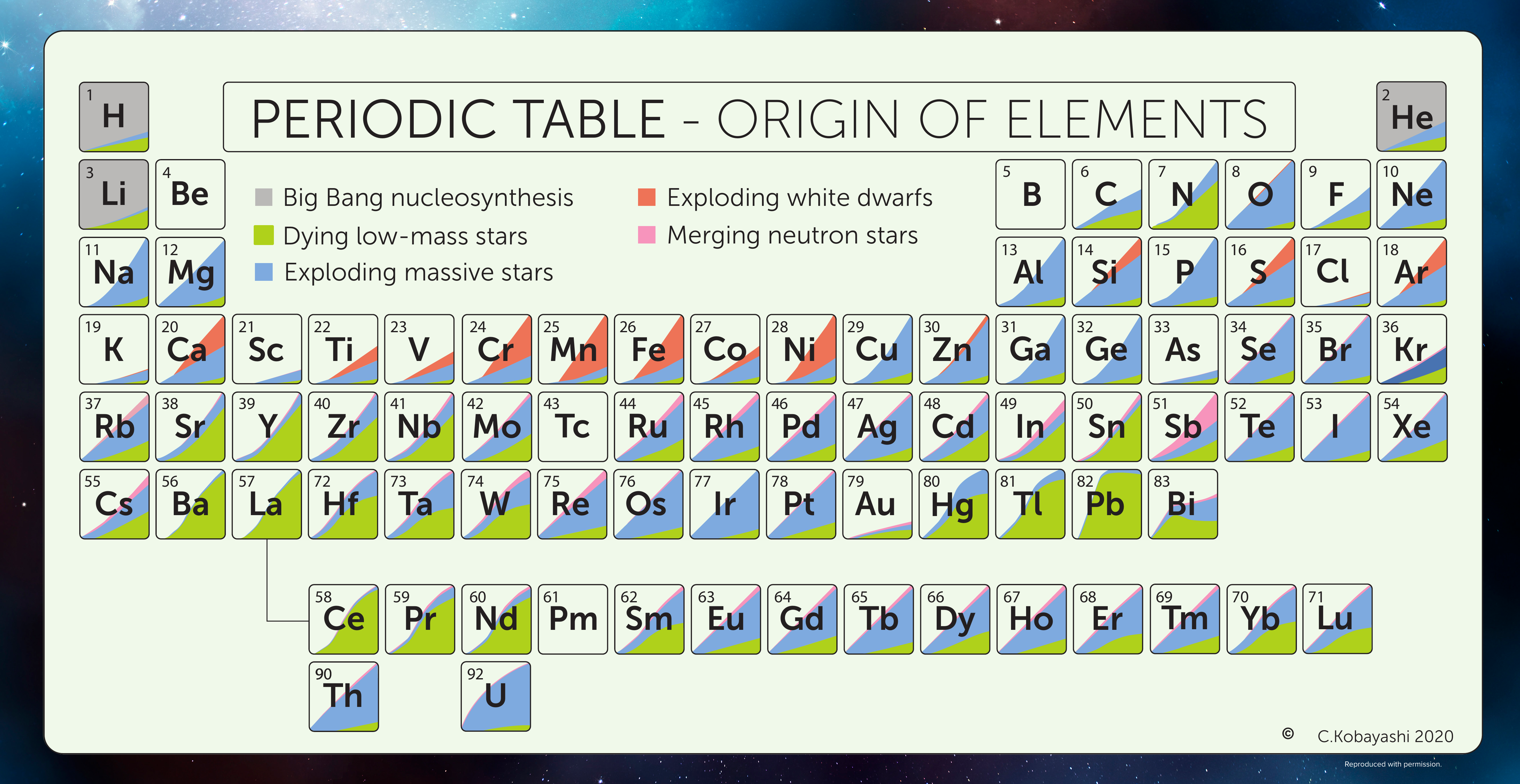}
    \caption[Origin of chemical elements in the Universe]{Origin of chemical elements in the Universe. The legend for the symbols and colours used is included within the plot. This figure was adapted from \href{https://astro3d.org.au/education-and-outreach/periodic-table-origin-of-elements/}{ASTRO 3D website} (ARC Centre of Excellence for All Sky Astrophysics in 3 Dimensions).}\label{fig:a3d_intro}
\end{figure}

Understanding the chemical enrichment of the Universe requires an examination of element and isotope production through nucleosynthesis. This forms the necessary baseline for studying how elemental abundances are altered by chemical depletion. In this section, we provide an overview of nucleosynthesis in the Universe (see Fig.~\ref{fig:a3d_intro}; adapted from \href{https://astro3d.org.au/education-and-outreach/periodic-table-origin-of-elements/}{ASTRO 3D website}). In Section~\ref{ssec:intnucbbn}, we briefly discuss Big Bang nucleosynthesis, which accounts for the formation of the lightest elements in the early Universe. In Section~\ref{ssec:intnuclim}, we focus on the final evolutionary stages of low- and intermediate-mass (LIM) stars, which are major contributors to the production of approximately half of the carbon and most of the \textit{s}-process elements. In Section~\ref{ssec:intnucmas}, we highlight the explosions of massive stars, specifically Types Ib, Ic, and II supernovae, enriching the Universe with \textit{r}-process elements. Finally, in Section~\ref{ssec:intnucwds}, we explore nucleosynthetic processes occurring in exploding white dwarfs (Type Ia supernovae) and neutron star mergers (kilonovae), contributing to the synthesis of Fe-peak and \textit{r}-process elements.

\subsection{Big Bang and primordial distribution of elements}\label{ssec:intnucbbn}
In this subsection, we briefly discuss the foundational assumptions and key stages of Big Bang nucleosynthesis, the process responsible for the formation of the Universe’s lightest elements during its first few minutes. For detailed reviews on Big Bang nucleosynthesis, see \citep{boesgaard1985BigBangNucleo, fields2006BigBangNucleo, pospelov2010BigBangNucleo, cyburt2016BigBangNucleo, grohs2022BigBangNucleo}.

Occurring within the rapidly cooling and expanding early Universe, Big Bang nucleosynthesis began once the temperature dropped below $\sim\,10^9$\,K, creating conditions for nuclear reactions \citep{fields2023BigBangNucleo}. These reactions unfolded in several key stages: 
\begin{itemize}
    \item \textbf{Neutrino decoupling}, during which neutrinos ceased to interact significantly with other particles, marking a critical shift in the energy distribution of the Universe;
    \item \textbf{Electron-positron annihilation and the freeze-out of the neutron-to-proton ratio}, which determined the relative abundance of neutrons and protons, setting the stage for the formation of hydrogen (H), helium (He) and other light nuclei;
    \item \textbf{The deuterium ($^2$H, D) bottleneck}, during which the synthesis of heavier nuclei was delayed until deuterium could form stably without being immediately photo-disintegrated by high-energy photons; and
    \item \textbf{The freeze-out of all nuclear reactions}, as the Universe expanded and cooled further, ending the production of new nuclei \citep{cyburt2016BigBangNucleo}.
\end{itemize}

These stages provided the baseline for the primordial chemical composition of the Universe.

After the first $\sim\,20$ minutes of the Universe, the fusion processes of Big Bang nucleosynthesis ceased. At this stage, the primordial composition of the Universe was firmly established, consisting predominantly of H ($\sim\,75$\% by mass) and He ($\sim\,25$\% by mass), with trace amounts of $^2$H, $^3$He, lithium (Li), and berillium (Be) \citep{coc2017BigBangNucleo}. Approximately 380\,000 years after the Big Bang, the Universe cooled enough for nuclei of H, He, and Li to capture free electrons, forming the first neutral atoms in a process known as recombination \citep{tanabashi2018ParticlePhysicsReview}. As a result, previously ionised plasma transitioned into a neutral gas, dramatically reducing the scattering of photons and allowing light to propagate freely through space. The photons released during this epoch formed the cosmic microwave background, a relic radiation that provides a direct observational window into the conditions of the early Universe \citep{grohs2022BigBangNucleo}. By studying the cosmic microwave background, we gain critical insight into the primordial density fluctuations, composition, and physical processes that shaped the evolution of the Universe.

Theoretical models of Big Bang nucleosynthesis provided a robust framework for predicting the abundances of H and He, which are in good agreement with observational data \citep[see, e.g.,][and references therein]{grohs2022BigBangNucleo}. However, a persistent discrepancy exists between Li abundances, predicted by Big Bang nucleosynthesis and observed in metal-poor (Pop II) stars in the Galaxy. The observed Li abundances are significantly lower than those predicted by Big Bang nucleosynthesis models, a phenomenon known as the `cosmological Li problem' \citep{boesgaard1985BigBangNucleo, hou2017LiProblem, tanabashi2018ParticlePhysicsReview}. This long-standing issue remains a key challenge in astrophysics, raising fundamental questions about the processes that determined the primordial abundances of light elements, and our broader understanding of nucleosynthesis and the early evolution of the Universe.

The process of Big Bang nucleosynthesis laid the foundation for the chemical composition of the Universe, creating the initial building blocks necessary for the formation of stars and galaxies. The first generation of stars (Pop III) subsequently produced heavier elements with atomic numbers $A\,\geq\,12$ via stellar nucleosynthesis \citep[see][and references therein]{sobral2015PopIIIStars}. These elements were released into the interstellar medium (ISM) through stellar winds or supernovae (SNe) explosions, enriching the star-forming regions (SFRs), where subsequent generations of stars (Pop II and Pop I) were formed. In Sections~\ref{ssec:intnuclim} and \ref{ssec:intnucmas}, we discuss these processes in greater detail.

\subsection{Contributions from dying LIM stars}\label{ssec:intnuclim}
\begin{figure}[!ht]
    \centering
    \includegraphics[width=.99\linewidth]{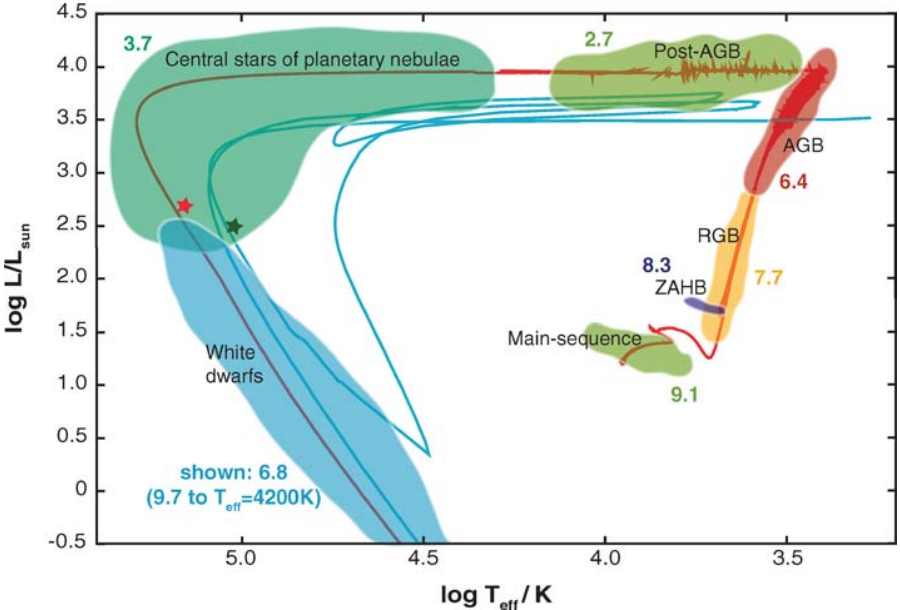}
    \caption[Evolutionary track from MS to WD evolution phase for a 2$M_\odot$ star of solar metallicity.]{Evolutionary track from MS to WD evolution phase for a 2$M_\odot$ star of solar metallicity. The numbered labels indicate the approximate logarithmic timescales of evolutionary phases for a 2$M_\odot$ star. We note that the post-RGB phase, which occurs only in binary systems, occupies a similar temperature range as the post-AGB phase, but at luminosities below the tip of the RGB ($\log\frac{L}{L_\odot}\,\lesssim\,3.4$; see Section~\ref{sec:intpAR}). This figure was adapted from \cite{herwig2005AGBEvolution}.}\label{fig:hrd_intro} 
\end{figure}

In this subsection, we present a brief overview of stellar nucleosynthesis \citep[an exhaustive review is presented in][]{nomoto2013StellarNucleo, karakas2014dawes}. In Fig.~\ref{fig:hrd_intro} \citep[adapted from][]{herwig2005AGBEvolution}, we show a schematic evolution of 1 $M_\odot$ star with the main evolutionary stages highlighted in different colours (the numbers besides evolution stages mark the approximate logarithmic duration of the stage).

\textbf{Evolution of LIM single stars}
 
As LIM stars ($0.8\,M_\odot\,\lesssim\,M\,\lesssim\,8\,M_\odot$) burn H into He in their cores during main sequence (MS) evolution through processes including $^2$H fusion, proton-proton (p-p) chain, or CNO cycle, inert He builds up in the core and H-burning reactions transition from the core to a H-burning shell surrounding an inert He core \citep{burbidge1957StellarNucleo, nomoto2013StellarNucleo}. This transition signals the end of the MS stage and the onset of red giant branch (RGB) evolution, characterised by significant structural changes. The He core contracts under gravitational forces, increasing its temperature $T$ and density $\rho$, while the outer layers expand and cool due to the thermal pressure generated by the intensifying H shell burning. These changes result in a dramatic increase in the stellar luminosity and radius, which are key markers of RGB evolution \citep[$L\,\sim\,100-2\,500\,L_\odot$, $R\,\sim\,1-200\,R_\odot$;][]{pols1998StellarNucleo, karakas2014dawes}.

During the RGB evolution, the convective envelope penetrates deeper into the stellar interior, reaching layers previously exposed to partial nuclear processing in the MS and early RGB. This process, known as the first dredge-up (FDU), brings nuclear-processed material to the stellar surface, altering the observed chemical composition of the star \citep{iben1984mixing, busso2007mixing, ventura2018mixing, ventura2022InternalProcesses}. The FDU enhances the surface abundances of isotopes, including $^{13}$C, $^{14}$N, $^{17}$O, and $^{23}$Na, while depleting $^{12}$C, $^{16}$O, and $^{18}$O. Beyond the FDU, additional non-convective mixing processes (e.g., thermohaline mixing, driven by molecular weight inversions, and magnetic-buoyancy-induced mixing) further modify the surface isotopic ratios \citep{sneden1986ExtraMixing, tautvaisiene2013ExtraMixing, drazdauskas2016ExtraMixing, charbonnel2020ExtraMixing}. In particular, the carbon isotopic ratio $^{12}$C/$^{13}$C was proven to be a reliable proxy of the efficiency of mixing processes on the RGB \citep{shetrone2019FDU}.

At the tip of the RGB, the stellar interior achieves temperatures, which are sufficient to start He burning in the core, prompting the Horizontal Branch (HB) stage \citep{sackmann1980mixing, siess2002mixing, karakas2014dawes}. Following core He flash or quiescent He ignition, HB stars, including red clump (RC) stars, consist of three major layers: i) He-burning cores, ii) H-burning shells, and iii) inert H envelope. HB stars decrease in luminosity and shift blueward on the HR diagram at approximately constant luminosity \citep{karttunen2007HBandRC, bertolami2016tracks}. HB stars with initial masses $M\,\gtrsim\,4-5\,M_\odot$ experience an inward extension of deep convective envelope, marking the onset of the second dredge-up. The second dredge-up results in significant changes to the stellar surface composition, characterised by an enhancement of $^{14}$N and $^{23}$Na and a depletion of $^{12}$C, $^{13}$C, $^{15}$N, and $^{16}$O \citep{kwok2000SDU, karakas2014dawes}.

\textbf{Shell He burning and asymptotic giant branch}

The asymptotic giant branch (AGB) stage is characterised by the presence of two active nuclear burning shells: a He-burning shell and a H-burning shell, both surrounding a degenerate C-O core \citep{vassiliadis1993AGB, karakas2014dawes}. The He shell undergoes periodic instabilities, known as thermal pulses, which trigger episodic convective mixing events, termed the third dredge-up. This process brings to the stellar surface carbon (C), nitrogen (N), fluorine (F), and elements produced by neutron-capture nucleosynthetic processes wherein the atom has enough time to experience $\beta$-decay between the consecutive captures \citep[slow neutron-capture process or \textit{s}-process;][]{gallino1998sProcess, goriely2000sProcess, kwok2000SDU, busso2001sProcess, lugaro2003sProcess, straniero2006sProcess, straniero2014sProcess, rees2024TDU}, including yttrium (Y), zirconium (Zr), barium (Ba), lanthanum (La), and lead (Pb). In more massive AGB stars with $3\,M_\odot\,\lesssim\,M\,\lesssim\,8\,M_\odot$ \citep[depending on initial metallicity and numerics of the stellar model;][]{kamath2023models}, the base of the convective envelope reaches temperatures sufficient to sustain the Hot Bottom Burning process \citep{garciahernandez2013HBB, ventura2015HBB}. This proton-capture nucleosynthesis process (see Section~\ref{ssec:intnucwds}) leads to the production of Li and the conversion of $^{12}$C into $^{13}$C and $^{14}$N, further diversifying the surface chemical composition of AGB stars.

Super-AGB stars, occupying the upper mass range of LIM stars with $5\,M_\odot\,\lesssim\,M\,\lesssim\,10\,M_\odot$ \citep[depending on metallicity and model numerics;][]{doherty2015SAGB, doherty2017SAGB}, represent a transitional population between traditional AGB stars and more massive stellar progenitors \citep{garciaberro1994SAGB, siess2009ThermohalineSAGB, siess2010SAGB, farmer2015SAGB}. These stars are distinguished by their more intense thermal pulses and highly efficient third dredge-up episodes in comparison with AGB stars. Super-AGB stars either evolve into O-Ne white dwarfs (WDs) or undergo electron-capture SNe, depending on the interplay between their initial mass and the extent of envelope stripping \citep{doherty2017SAGB}. Observational studies reveal that super-AGB stars exhibit a notable overabundance of heavy \textit{s}-process elements, including Rb, coupled with a comparatively low enhancement of light \textit{s}-process elements, such as Zr \citep{garciahernandez2006sProcessSAGB, garciahernandez2007sProcessSAGB}.

As AGB single stars evolve, mass-loss processes reduce the stellar envelope, eventually making it insufficient to sustain further AGB evolution. Once the envelope is reduced enough, these stars move across the HR diagram blueward from AGB at approximately constant luminosity \citep{kamath2023models}. Post-asymptotic giant branch (post-AGB) single stars offer critical insights into AGB/RGB evolution, as the surface chemistry of post-AGB single stars remains relatively unchanged after leaving the giant branch \citep{vanwinckel2003Review, desmedt2014LeadMCs, desmedt2016LeadMW, kamath2019depletionLMC}. Post-AGB stars contract at constant luminosity until reaching sufficient effective temperature to ionise the shed layers ($T_{\rm eff}\,\sim\,30\,000$\,K), marking the onset of the final evolutionary stage in LIM stars -- transformation into a WD surrounded by a PN \citep{kwok2000SDU}. The constant luminosities of post-AGB stars enable linking observed abundances directly to stellar evolution models using core mass-luminosity relations \citep{bertolami2016tracks, kamath2023models}.

\textbf{Late stages of evolution in LIM binary systems}

The late stages of evolution in LIM binaries are shaped by complex interactions that drive unique chemical and structural outcomes, including termination of AGB/RGB evolution or extrinsic chemical enrichment. In binary systems, the evolution during and after MS stage is profoundly influenced by binary interactions, including mass transfer, common envelopes, and tidal forces \citep{demarco2017CommonEnvelope, izzard2018CommonEnvelopeCBD, oomen2020MESAdepletion, chen2024BinaryInteractions}. During the AGB or RGB phases, one of the binary components may overfill its Roche lobe, leading to the shedding of its outer envelope \citep{paczynski1971RLOF, vanwinckel2009}. This mass transfer exposes deeper, nuclear-processed layers, truncating the traditional giant branch evolution and significantly altering the star’s surface composition. Such interactions lead to inability of the primary component to sustain further AGB or RGB evolution, resulting in the formation of post-AGB binaries and their low-luminosity analogues -- post-red giant branch (post-RGB) binaries. In LIM binary systems, the disc-binary interactions enhance the transfer of chemically processed material to the companion star, resulting in distinct evolutionary paths that differ from single star evolution. These evolutionary outcomes include the formation of chemically peculiar stars, including S stars, Ba stars, carbon-enhanced metal-poor (CEMP) stars, O/B-type subdwarf stars, and post-AGB/post-RGB binaries (which are the main focus of this thesis).

In the following paragraphs, we provide a brief overview of the main types of chemically peculiar binaries starting with S stars. S stars in binary systems are cool AGB giants transitioning from the M spectral type to the C spectral type \citep[0.5<C/O<1;][]{iben1983Sstars}, characterised by the presence of titanium oxide (TiO) and zirconium oxide (ZrO) molecular bands in their spectra. The origin of \textit{s}-process enhancement in S stars can be tracked by the surface abundances of technetium (Tc) and niobium (Nb) \citep{karinkuzhi2018Sstars, shetye2019Sstars, shetye2021Sstars}: Tc-rich and Nb-poor S stars are enriched intrinsically via AGB nucleosynthesis, whereas Tc-poor and Nb-rich S stars are enriched extrinsically via mass transfer from the companion. Moreover, a subclass of S stars with mild Tc enhancement (bitrinsic S stars) are enriched via combination of internal nucleosynthesis and external pollution.

\textbf{Chemically peculiar evolved binaries}

Ba stars are a subset of binary systems in which a mass transfer episode from an evolved AGB primary enriches an MS secondary with \textit{s}-process elements, including barium (Ba) \citep{mcclure1980BaStars}. Ba stars are observed when the primary star already evolved into a dim C-O WD, whereas the secondary star became a bright RGB star of spectral types G-K. The observed secondary retains the \textit{s}-process enrichment long after mass transfer ended, showing strong absorption lines of singly ionised Ba (\ion{Ba}{ii}). Observational studies \citep{mcclure1980BaStars, jorissen1998BaStars, jorissen2019BaStars, escorza2019BaStars, escorza2023BaStars} confirm that all Ba stars are found in binary systems.

CEMP stars represent a diverse population of chemically peculiar stars with varying origins tied to binary evolution. They are classified into several subtypes, including CEMP-s, CEMP-no, CEMP-r, and CEMP-rs \citep{spite2013CEMPstars, hansen2015CEMPstars, sharma2018CEMPstars, caffau2019CEMPstars}. CEMP-s stars exhibit strong enhancements of \textit{s}-process elements due to mass transfer from an AGB companion, whereas CEMP-no stars show little to no such enhancements, suggesting internal processes as the primary source of their chemical composition. CEMP-r stars are enhanced primarily by the \textit{r}-process, often linked to the pollution of birth cloud from compact binary mergers (see Section~\ref{ssec:intnucwds}), whereas CEMP-rs stars exhibit a mixed contribution from both \textit{s}- and \textit{r}-processes, suggesting enrichment via the intermediate neutron-capture process \citep[\textit{i}-process;][]{dardelet2015iProcess, choplin2021iProcess}.

Hot O/B-type subdwarf stars are LIM stars with He-burning cores and thin H outer layers \citep{heber2016HotsdOsdBstars, byrne2021HotsdOsdBstars, lynasgray2021HotsdOsdBstars, blomberg2024HotsdOsdBstars}. The merger of two He WDs can explain the formation of single hot O/B-type subdwarf stars (see Section~\ref{ssec:intnucwds}), whereas stable Roche lobe overflow or common envelope ejection can explain the formation of binary O/B-type subdwarf stars \citep[see][and references therein]{li2024sdOsdBstars}. The formation of hot subdwarfs depends on the He ignition in the core. In binary systems, this can occur when the donor star begins mass transfer during the RGB evolution \citep[RGB channel;][]{wu2018HotsdOsdBstars}. Observational evidence suggests that the RGB channel accounts for most O/B-type subdwarf stars. However, while this channel typically produces He-deficient subdwarfs, it cannot explain the observed population of He-rich hot subdwarfs \citep{heber2016HotsdOsdBstars}. We note that O/B-type subdwarf stars exhibit a weak depletion of $\alpha$-elements (by $\sim$\,--0.5\,dex), including oxygen (O), magnesium (Mg), and silicon (Si), despite the theoretically predicted enrichment of these elements during the core He burning \citep{heber2009sdB, naslim2010sdOsdBabundances}. Conversely, Fe-peak elements (e.g., iron Fe and nickel Ni) and \textit{s}-process elements (e.g., Sr and Ba) are enhanced, reflecting their synthesis during the AGB phase and subsequent dilution due to envelope loss \citep{han2002HotsdOsdBstars}.

Post-AGB/post-RGB binaries are systems where primary component overfills its Roche lobe during AGB/RGB evolution, leading to formation of a circumbinary disc \citep[CBD; ][]{vanwinckel2003Review}. As the system evolves, the disc accretes matter onto the central binary and enhances the stellar surfaces with elements not involved in dust evolution within the disc. In Section~\ref{ssec:intpARevo}, we provide a more detailed discussion of post-AGB/post-RGB binary stars, which are the centrepieces of this thesis.

In summary, LIM stars contribute significantly to the chemical enrichment of the Universe, particularly in C, N, F, and \textit{s}-process elements through nucleosynthesis, mixing processes, and binary interactions. In the next subsections, we will discuss the enrichment of the Universe through more violent explosive processes occurring in massive stars, exploding white dwarfs, and merging neutron stars.

\subsection{Contributions from exploding massive stars}\label{ssec:intnucmas}
In this section, we discuss the crucial role of massive stars in driving galactic chemical evolution, focusing on their contributions to $\alpha$- and \textit{r}-process nucleosynthesis. At the end of their lives, massive stars with $M\,\gtrsim8\,M_\odot$ \citep[depending on metallicity and model numerics;][]{karakas2014dawes} undergo a complex evolution culminating in SN explosions. These explosions create the extreme conditions necessary for \textit{r}-process nucleosynthesis, including high temperatures ($T\,>\,10^9$\,K) and high densities of neutrons \citep[$n_n\,>\,10^{20}$ cm$^{-3}$;][]{vescovi2022rProcess}. Through the energetic explosions, supernovae inject into the ISM synthesised metals, including $\alpha$-elements (O, Mg, Si, sulphur S, calcium Ca) and \textit{r}-process elements with peaks around $A\,=\,82$ (selenium Se, bromine Br, and krypton Kr), $A\,=\,130$ (tellurium Te, iodine I, and xenon Xe), and $A\,=\,196$ \citep[osmium Os, iridium Ir, and platinum Pt;][]{kobayashi2020OriginOfElements}.

Core-collapse supernovae (CCSNe), including Types II, Ib, and Ic, occur due to the gravitational collapse of massive stars (with $E\,\sim\,10^{51}$ erg). This collapse produces O, Si, and Si-burning elements, including Fe and Ni \citep{werner2008rProcess, bohringer2010rProcess, boccioli2024CCSNe}. This distinct yield pattern allows to distinguish the enrichment from CCSNe and Type Ia SNe (see Section~\ref{ssec:intnucwds}). \citet{woosley1978pProcess} demonstrated that stellar interiors may undergo complete reprocessing as the explosive shock wave propagates through the star. The O/Ne-rich layers in CCSNe were shown to be the optimal environment for re-forming proton-rich nuclei from $^{74}$Se to $^{196}$Hg through a sequence of photodisintegration reactions \citep[$\gamma$-process;][]{roberti2024gammaProcess}.

In an extreme subclass of CCSNe, hypernovae (HNe), the explosions of rapidly-rotating massive stars are highly energetic \citep[$E\,>\,10^{52}$ erg;][]{grimmett2020Hypernovae}, which results in a different pattern of enhancement of heavy elements, primarily including Fe, cobalt (Co), copper (Cu), zinc (Zn), gallium (Ga), and germanium (Ge). \citet{nakamura2001HNe} and \citet{umeda2005HNe} demonstrated that raising the explosion energy by an order of magnitude above the standard value for SNe ($10^{51}$ erg) expands the Si-burning regions in mass coordinates. This expansion increases the mass ratio of complete-to-incomplete Si-burning products, which aligns well with the observed trend of declining [(Cr, Mn)/Fe]\footnote{$\text{[X/Y]}\,=\,\log\left(\dfrac{N_X}{N_Y}\right) - \log\left(\dfrac{N_X}{N_Y}\right)_\odot$, where $N_X$ and $N_Y$ are number abundances of elements X and Y, respectively, scaled to the solar values (see Section~\ref{ssec:mththratm}).} and rising [(Co, Zn)/Fe] ratios in stars with lower [Fe/H] \citep{kobayashi2020OriginOfElements}. Furthermore, HNe explosions may be enhanced by aspherical effects, including rotation and magnetic fields \citep{mosta2014HNe, obergaulinger2017HNe}. HNe explosions involving bipolar jets were shown to produce Fe-peak elements and certain intermediate-mass elements (including C, scandium Sc, and titanium Ti) in ratios consistent with those observed in CEMP stars \citep{tominaga2009HNe}. Additionally, these jets are considered a potential site for \textit{r}-process nucleosynthesis \citep{winteler2012rProcessJets, halevi2018rProcessJets}.

Overall, massive stars provide ideal conditions for the nucleosynthesis of elements from O to U, subsequently enriching the Universe through CCSNe and HNe explosions.

\subsection{Contributions from exploding white dwarfs and merging neutron stars}\label{ssec:intnucwds}
In this subsection, we outline the explosions of WDs in binary systems \citep[Type Ia SNe;][]{mazzali2007TypeIaSNe}. In such systems, a C-O WD accretes matter from its companion, leading to a runaway thermonuclear reaction once the WD approaches the Chandrasekhar limit. Type Ia SNe produce most Cr, Mn, Fe, and Ni, as well as significant amounts of even-$Z$ $\alpha$-elements, including Si, S, and Ca \citep{hillebrandt2000TypeIaSNe, kobayashi2020OriginOfElements, aouad2024TypeIaSNe}. Beyond the production of Fe-peak elements, exploding C-O WDs synthesise isotopes from $^{74}$Se to $^{196}$Hg through proton-capture process  \citep[\textit{p}-process;][]{lambert1992pNuclei, zhou2023pProcess}. The \textit{p}-process is particularly effective in creating proton-rich isotopes that are otherwise rare in the Universe, adding further complexity to the chemical yields of Type Ia supernovae.

While SNe are crucial for producing Fe-peak and \textit{p}-process elements, kilonovae (luminous transients following the merger of two neutron stars or a neutron star and a black hole) produce \textit{r}-process elements, including Ru, In, Sb, and lanthanides \citep{kobayashi2020OriginOfElements, curtis2023Kilonovae}. Half a century ago, kilonovae were proposed as potential sites for \textit{r}-process nucleosynthesis \citep{lattimer1974Kilonovae}, and this hypothesis was confirmed with the detection of the gravitational wave event GW170817 \citep{abbott2017Kilonovae}, linked to the transient AT 2017gfo \citep{smartt2017Kilonovae, valenti2017Kilonovae}. The spectra of AT 2017gfo revealed near-infrared (near-IR) emissions from heavy \textit{r}-process elements, including lanthanides, in the dynamical ejecta, alongside optical emissions from outflows around black hole discs \citep{metzger2014Kilonovae, pian2017Kilonovae, tanaka2017Kilonovae}.

Current GCE models are largely consistent with the observed abundances of most elements and isotopes in the Universe. However, more computationally-demanding chemo-hydrodynamical simulations are required to make these models more realistic. Additionally, while GCE models incorporate binary evolution to some extent (primarily through Type Ia SNe and kilonovae), the effects of LIM star binary interactions on AGB/RGB nucleosynthesis remain poorly understood \citep{izzard2006PopSynth, kobayashi2020OriginOfElements, izzard2023DiscBinaryInteractionAccretion, osborn2024PopSynth}. These binary interactions not only influence nucleosynthesis but also produce unique chemical signatures that can deviate from the expected trends of single-star evolution (e.g., enhancement of \textit{s}-process or $\alpha$- elements; see Section~\ref{sec:intnuc}), including photospheric chemical depletion, which we will discuss in detail in Section~\ref{sec:intdpl}.

\section{Chemical depletion in the Universe}\label{sec:intdpl}
To understand how the chemical composition of the Universe evolves beyond its initial enrichment through nucleosynthesis (see Section~\ref{sec:intnuc}), it is crucial to examine the chemical depletion process, which preferentially binds chemical elements in the dust particles, modifying the observed abundances of the gas phase. The chemical depletion phenomenon is particularly relevant in post-AGB/post-RGB binary systems surrounded by second-generation protoplanetary discs, where gas-dust separation within the circumbinary environment can imprint distinct abundance patterns on the stellar photospheres (see Section~\ref{sec:intpAR}). Investigating photospheric depletion in post-AGB/post-RGB binaries provides key insights into the role of disc-binary interaction in shaping chemical evolution.

In this section, we discuss the process of gas condensation into dust in an environment, which depends on multiple parameters, including the condensation properties of the gas mixture (individual abundances of chemical elements) and the physical conditions of the environment (e.g., temperature, density, and the pressure). In Section~\ref{ssec:intdpldef}, we examine the mechanisms driving gas condensation into dust. In Sections~\ref{ssec:intdplism} and \ref{ssec:intdplstr}, we outline the resulting depletion patterns observed in various astrophysical environments, including the ISM and the stellar photospheres, respectively. In Table~\ref{tab:dpluni_intro}, we compare the efficiency of the chemical depletion in various sites across the Universe.

\subsection{Process of chemical depletion}\label{ssec:intdpldef}

In this subsection, we examine the mechanism and parameters governing chemical depletion, with a particular focus on the role of volatility and condensation temperatures in shaping elemental distributions in astrophysical environments.

Volatility is a fundamental concept in geochemistry and astrochemistry, referring to the temperature at which elements condense from a gas with solar composition. By definition, volatile elements (volatiles), including C, N, O, S, and Zn, primarily accumulate in gaseous reservoirs, forming atmospheres. However, during the early stages of planet formation, these elements can also preferentially bond with iron and concentrate in core-forming metallic compounds. In contrast, refractory elements (refractories), including Al, Ti, Fe, and \textit{s}-process elements, condense into dust and are resistant to re-entering the gas phase. This distinction helps interpret chemical and isotopic differences among meteorites and planetary bodies, tracing processes in the solar nebula and planetary accretion.

Volatility in a chemical equilibrium is quantified by the 50\% condensation temperature ($T_{\rm cond,\,50\%}$ or $T_{\rm cond}$) that indicates the temperature at which half of an element exists in the gas phase, whereas another half exists in the dust phase. The abundances of major dust phases (abundant dust species, including silicates and carbides) for different C/O ratios at various temperatures are well established from meteorite analyses, chemical affinities, and phase equilibrium studies \citep{lodders2003CondensationTemperatures}. Calculations for minor dust phases (dust species present in trace amounts, including oxides of \textit{s}-process elements) incorporate activity coefficients to account for non-ideal solid solutions \citep{kornacki1986Condensation, wood2019CondensationTemperatures}. In chemical equilibrium calculations for major and minor dust phases, chemical elements can be classified as volatile or refractory depending on the individual temperatures, at which the elements condense into dust \citep{lodders2003CondensationTemperatures, wood2019CondensationTemperatures}. In this thesis, to study photospheric chemical depletion (see Section~\ref{ssec:intdplstr}), we adopt a demarcating temperature of 1200\,K to distinguish between volatiles and refractories, with Mn ($T_{\rm cond}$ = 1123\,K) as the least volatile element and Cr ($T_{\rm cond}$ = 1291\,K) as the least refractory element.

\begin{sidewaysfigure}[ph!]
    \centering
    \includegraphics[width=0.8\textwidth]{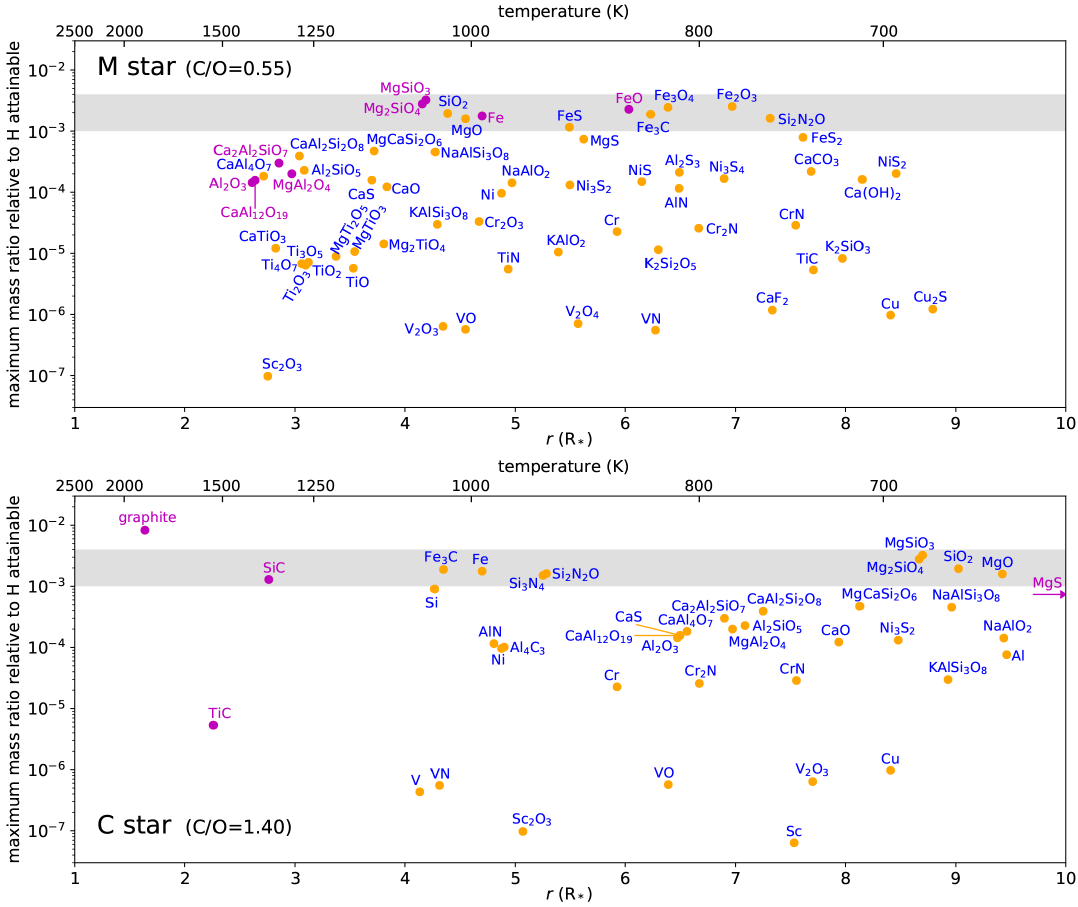}
    \caption[Main condensates expected to form within the 1–10 $R_\ast$ range in the atmospheres of M-type and C-type AGB stars]{Main condensates expected to form within the 1–10 $R_\ast$ range in the atmospheres of M-type (\textit{upper panel}) and C-type (\textit{lower panel}) AGB stars. The condensates are sorted based on the radius from the star's centre (bottom x-axis) and corresponding condensation temperature (top x-axis). Observed condensates in AGB envelopes are highlighted in magenta. The gray horizontal band represents the dust-to-gas mass ratio range estimated by \citet{ramstedt2008AGBDust2GasRatio} for AGB star envelopes. This figure was adapted from \citep{agundez2020AGBDepletionModelling}.}\label{fig:mjrcnd_intro}
\end{sidewaysfigure}

The 50\% condensation temperatures are calculated for all elements in a solar-mixture gas existing in chemical equilibrium at a total pressure of 10$^{-4}$ bar. Major condensates in the mixture strongly depend on the C/O ratio of the gas mixture, leading to entirely different main compounds for the same element in O-rich and C-rich environments. A notable example is Al: in O-rich environments, the first and primary condensate of Al is corundum Al$_2$O$3$ (condensing at $T\,\sim\,1400$\,K), whereas in C-rich environments, the first and primary condensate is nitride AlN (condensing at $T\,\sim\,1000$\,K). However, \citet{agundez2020AGBDepletionModelling} showed that a solar-mixture-at-chemical-equilibrium approach (with C/O$_\odot\,\sim\,0.6$) is fairly suitable for most O-rich AGB envelopes. In the upper panel of Fig.~\ref{fig:mjrcnd_intro} \citep[adapted from][]{agundez2020AGBDepletionModelling}, the major condensates in O-rich environments are shown, including corundum Al$_2$O$_3$, hibonite (Ca,Ce)(Al,Ti,Mg)$_{12}$O$_{19}$, melilite (Ca,Na)$_2$(Al,Mg,Fe)[(Al,Si)SiO$_7$], spinel MgAl$_2$O$_4$, and silicates (olivines and pyroxenes). In the lower panel of Fig.~\ref{fig:mjrcnd_intro} \citep[adapted from][]{agundez2020AGBDepletionModelling}, the major condensates in C-rich environments are shown, including graphite, carbides (TiC and SiC), and alloys (V, Si, Fe).

\begin{table}[!ht]
    \centering
    \caption{Chemical depletion in the Galaxy (see Section~\ref{sec:intdpl}).\\} \label{tab:dpluni_intro}
    \begin{tabular}{|c|c|c|c|}\hline
         \textbf{Site} & \textbf{[V/R] (dex)} & \textbf{Note} & \textbf{Ref.} \\\hline
         ISM & [0.75:1.75] & [Zn/Fe] & a, b \\\hline
         YSO & [0.0:0.5] & [O/Fe] & c, d, e, f \\\hline
         Sun-like & [0.00:0.05] & \begin{tabular}{c} [Zn/Fe] inferred from \\ GCE-corrected trend \end{tabular} & g \\\hline
         AGB & -- & \begin{tabular}{c} Dust species observed \\ in the extended envelope \end{tabular} & h \\\hline
         Post-AGB & [0.0:2.5] & [Zn/Fe] & i, j, k \\\hline 
         WD & [--1.0:0.0] & [O/Fe], planet engulfment & l \\\hline 
    \end{tabular}\\
    \textbf{References:} $^a$\citet{decia2016ISMdepletion}, $^b$\citet{konstantopoulou2022ISMdepletion}, $^c$\citet{sturenburg1993lambdaBooAbundances}, $^d$\citet{andrievsky2002lambdaBooAbundances}, $^e$\citet{heiter2002lambdaBooAbundances}, $^f$\citet{kama2015DiscDepletionLinkinYSOs}, $^g$\citet{yun2024SolarDepletion}, $^h$\citet{agundez2020AGBDepletionModelling}, $^i$\citet{gorlova2012BD+46442}, $^j$\citet{kluska2022GalacticBinaries}, $^k$\citet{mohorian2025TransitionDiscs}, $^l$\citet{lebourdais2024WDDepletion}.\\
\end{table}

\subsection{Observed chemical depletion in the ISM}\label{ssec:intdplism}
All condensed astronomical objects, from stars to comets, asteroids, and cosmic dust grains, originate from the ISM, underscoring the importance of the chemical composition of the ISM. The ISM comprises matter in the form of gas molecules and solid dust grains, accounting for approximately 99\% and 1\% of the ISM mass, respectively (Witt 2001; Jenkins 2009). Within dense molecular clouds (regions characterised by extremely low temperatures $\sim5-10$\,K and gas densities $\sim10^4$\,cm$^{-3}$, where star formation begins) most interstellar grains have sub-micron sizes \citep{greenberg1974SubmicronGrains, wardthompson2015StarFormation}. These dust grains have a refractory core composed of silicates or carbonaceous material (see Section~\ref{ssec:intdpldef}), which is coated with ice mantles primarily consisting of water H$_2$O, along with other volatile compounds, including carbon monoxide CO, carbon dioxide CO$_2$, ammonia NH$_3$, and methanol CH$_3$OH \citep[see, e.g.,][and references therein]{draine2003ISMdust}. These compounds were detected using near-IR observational techniques \citep{boogert2015IcyUniverse, mcclure2023DenseMolecularClouds}. In contrast to the dust grains, the gas-phase components include nearly 300 interstellar molecules discovered through rotational emission spectroscopy in the millimeter and submillimeter wavelengths \citep{mcguire2022InterstellarCircumstellarMolecules}.

In the ISM, there are two channels of gas condensation into dust: gas phase reactions and grain surface reactions. Gas phase reactions involve the direct condensation into dust, whereas the grain surface reactions are two-fold: i) gas-phase molecules freeze onto grain surfaces and undergo chemical reactions, and ii) the modified molecules return to the gas phase as more complex species, enabling further transformations and reactions. Moreover, the formation of molecules and grains is strongly influenced by the physical conditions of the environment. A notable example is H$_2$O, which can form through distinct pathways depending on the environment: in diffuse clouds, it forms via reactions between OH and H; while in dense clouds, it originates from the reaction of OH with H$_2$ \citep{cuppen2007IceFormationISM, dulieu2010IceFormationISM}. In contrast to the diffuse environments, in dense molecular clouds, a significant fraction of O is locked into icy mantles on dust grains or incorporated into complex molecules, including CO, H$_2$O, and carbonyl sulphide (OCS), effectively removing it from the gas phase and altering the chemical balance of the region \citep{martinezbachs2023GasPhaseGrainSurfaceReactions}.

\begin{figure}[!ht]
    \centering
    \includegraphics[width=.99\linewidth]{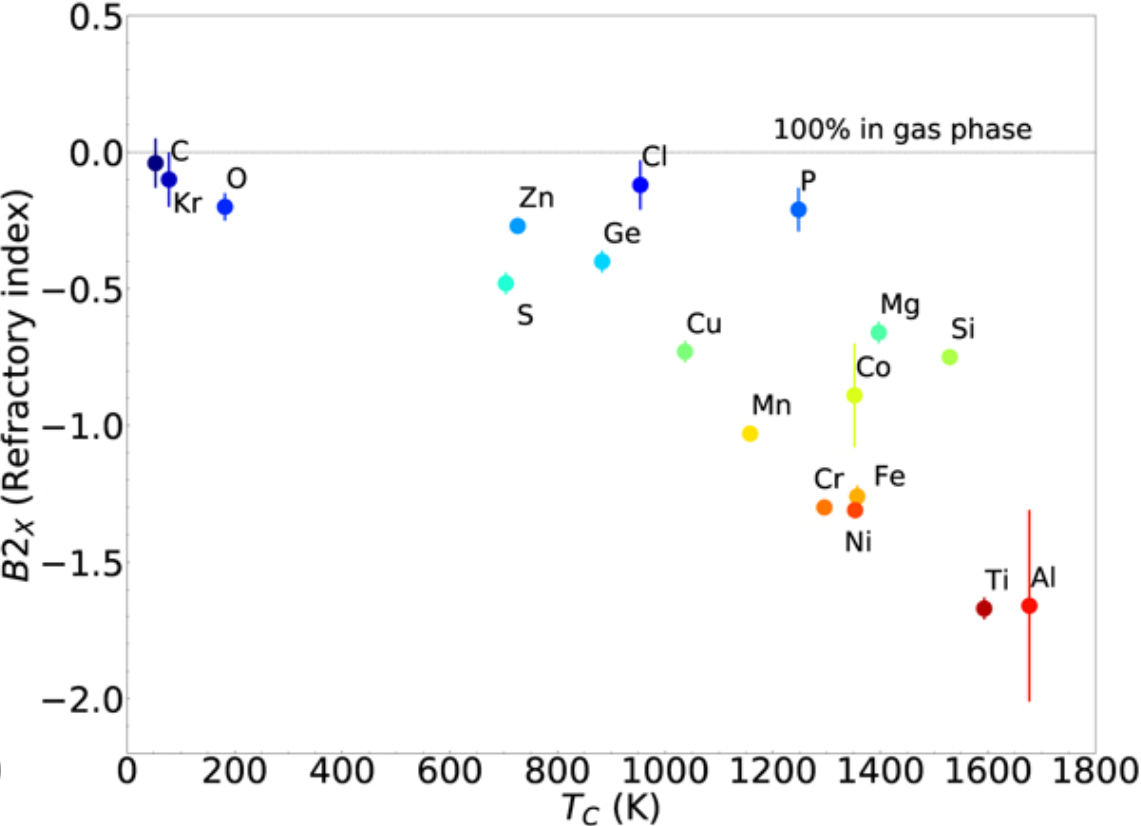}
    \caption[Chemical depletion in the ISM]{The slopes of the linear depletion profiles (refractory index) as a function of $T_{\rm cond}$ (see Section~\ref{ssec:intdplism}). The chemical depletion profile in the ISM is given by $\delta X = A_{2X} + B_{2X}\times$[Zn/Fe], where $A_{2X}$ and $B_{2X}$ are the fitting coefficients, $\delta X$ is the depletion measure derived from [X/Zn] by subtracting the nucleosynthetic baseline from the GCE models. The condensation temperatures $T_{\rm cond}$ are adopted from \cite{lodders2003CondensationTemperatures}. This figure was adapted from \cite{konstantopoulou2022ISMdepletion}.}\label{fig:ism_intro}
\end{figure}

Multiple observational studies of the Galactic ISM demonstrated a correlation between the efficiency of dust depletion and the condensation temperature along sightlines to Galactic stars \citep{gondhalekar1985ISMdepletion, keenan1986ISMdepletion, sembach1994ISMdepletion, jenkins2009ISMdepletion}. More recent studies of the ISM composition included sightlines to damped Lyman-$\alpha$ absorbers (DLAs), where metallicity is lower than in the Galaxy \citep{pettini1994DLAdepletion, decia2016ISMdepletion, decia2021ISMdepletion}. \citet{konstantopoulou2022ISMdepletion} compiled all available literature on metal column densities across diverse ISM environments, including the Galaxy, the Magellanic Clouds, and DLAs around gamma-ray bursts (GRBs) and quasi-stellar objects (QSOs). For a broad range of elements from C to Fe, the chemical depletion of the ISM was found to be well described by linear fits \citep[see Fig.~\ref{fig:ism_intro}; adapted from][]{konstantopoulou2022ISMdepletion}, with small deviations attributed to nucleosynthesis rather than dust depletion (for S, Mg, Ti, Cr, and Mn). The similarity of dust depletion patterns across diverse ISM environments highlight the significance of dust grain growth in the ISM, suggesting that dust formation processes in a galaxy are largely independent of star-formation history.

\subsection{Observed chemical depletion in the stars}\label{ssec:intdplstr}
Unlike chemical depletion in the ISM, photospheric depletion in stars is linked to the presence of a circumstellar disc accreting matter onto the star. The accreted matter dominates over the original stellar surface, altering the observed chemical composition of the star. The composition of accreted matter is defined by the distribution and dynamics of gas and dust within the disc, including dust processes, including dust growth, dust settling, inward drift, and trapping by pressure bumps \citep{drazkowska2023DustPlanetFormation, oberg2023ProtoplanetaryDiscChemistry, vandermarel2023TransitionDiscFormation}. These dust processes predominantly affect refractory elements, which constitute the majority of dust in the circumstellar disc. The trapping of refractory dust in the disc and the accretion of volatile-rich gas onto the star is indirectly confirmed by multiple chemical analysis studies of young and old stars \citep{herczeg2002VolatileAccretion, deruyter2006discs, mcclure2019CarbonDepletionTTau, oomen2019depletion, oomen2020MESAdepletion}.

\textbf{Depletion in young stars and its connection to planet formation}

A small fraction of young Pop I stars with spectral types B9-F3 ($<\,200$ discovered stars in the Galaxy, including candidates) exhibit significant depletion in $\alpha$- and Fe-peak elements, while retaining near-solar abundances of volatile elements, including C, N, O, and S \citep[$\lambda$ B\"{o}o phenomenon;][]{paunzen2004lambdaBooStars, venn1990lambdaBooStars, waters1992GasDustFractionation}. \citet{turcotte1993lambdaBooStars, turcotte2002AccretionDomination} showed that meridional circulation homogenises the surfaces of the $\lambda$ B\"{o}o stars within $\sim\,10^6$ years after accretion episodes, highlighting the transient nature of $\lambda$ B\"{o}o phenomenon.

\begin{figure}[!ht]
    \centering
    \includegraphics[width=.99\linewidth]{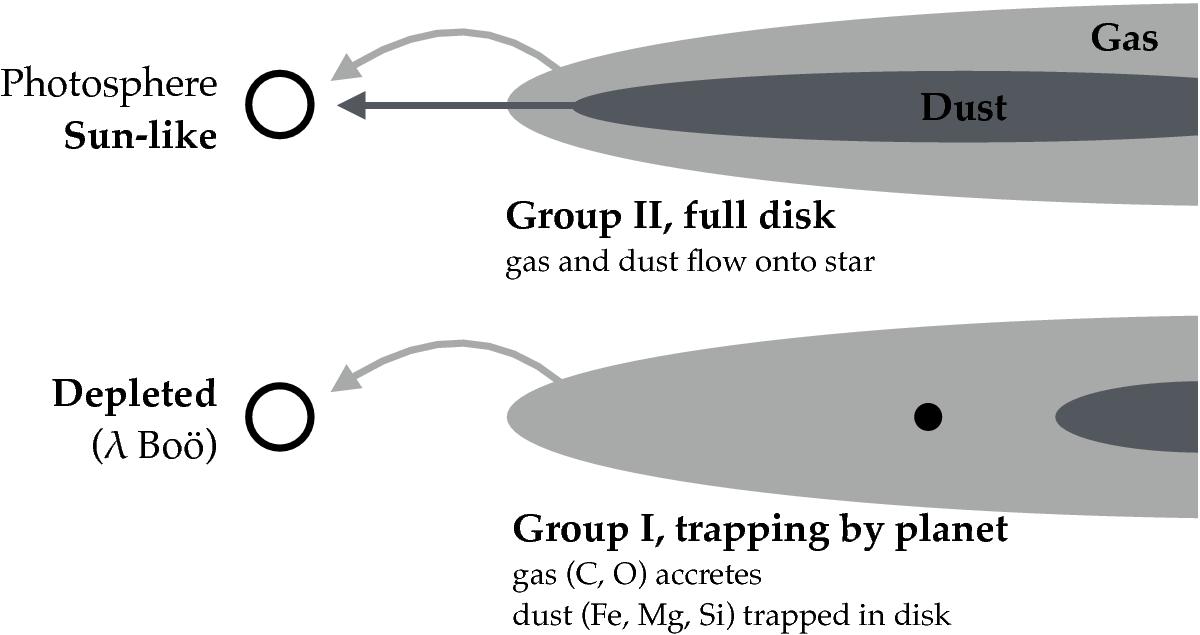}
    \caption[Depletion in YSOs]{The structure of the full (group II, top) and transition (group I, bottom) disc around YSOs. The photospheric depletion (observed as $\lambda$ B\"{o}o phenomenon) results from fractionation of dust and gas and consequent accretion of the matter from the disc to the star (shown with arrows). Illustration is not to scale. This figure was adapted from \cite{kama2015DiscDepletionLinkinYSOs}.}\label{fig:yso_intro}
\end{figure}

$\lambda$ B\"{o}o phenomenon is linked to the dust distribution processes occurring in circumstellar discs, including effective fractionation of gas and dust. To explain this fractionation, several mechanisms were proposed, including remnants of star formation \citep{holweger1993lambdaBooStars}, gas from dense ISM regions \citep{kamp2002lambdaBooStars}, matter ablated from hot Jupiters \citep{jura2015lambdaBoo}, and planet formation \citep{waters1992GasDustFractionation, kama2015DiscDepletionLinkinYSOs, jermyn2018Depletion}. Observations from VLTI and ALMA confirm the presence of planets embedded in the protoplanetary or debris discs\footnote{Protoplanetary discs have inner dust cavities detected in their spectral energy distributions (SEDs), indicative of potential planet formation. In contrast, debris discs have low IR dust excess in their SEDs and are considered gas-poor with dust coagulating into planetesimals \citep{wyatt2008DebrisDiscEvolution, oberg2023ProtoplanetaryDiscChemistry}.} of $\lambda$ B\"{o}o stars \citep[see, e.g.,][]{matter2016VLTIdepletion, fedele2017ALMAdepletion, cugno2019VLTdepletion, toci2020ALMAdepletion}. In Fig.~\ref{fig:yso_intro} \citep[adapted from][]{kama2015DiscDepletionLinkinYSOs}, we show a sketch of the depletion process in young stars with different types of circumstellar discs.

\begin{figure}[!ht]
    \centering
    \includegraphics[width=.99\linewidth]{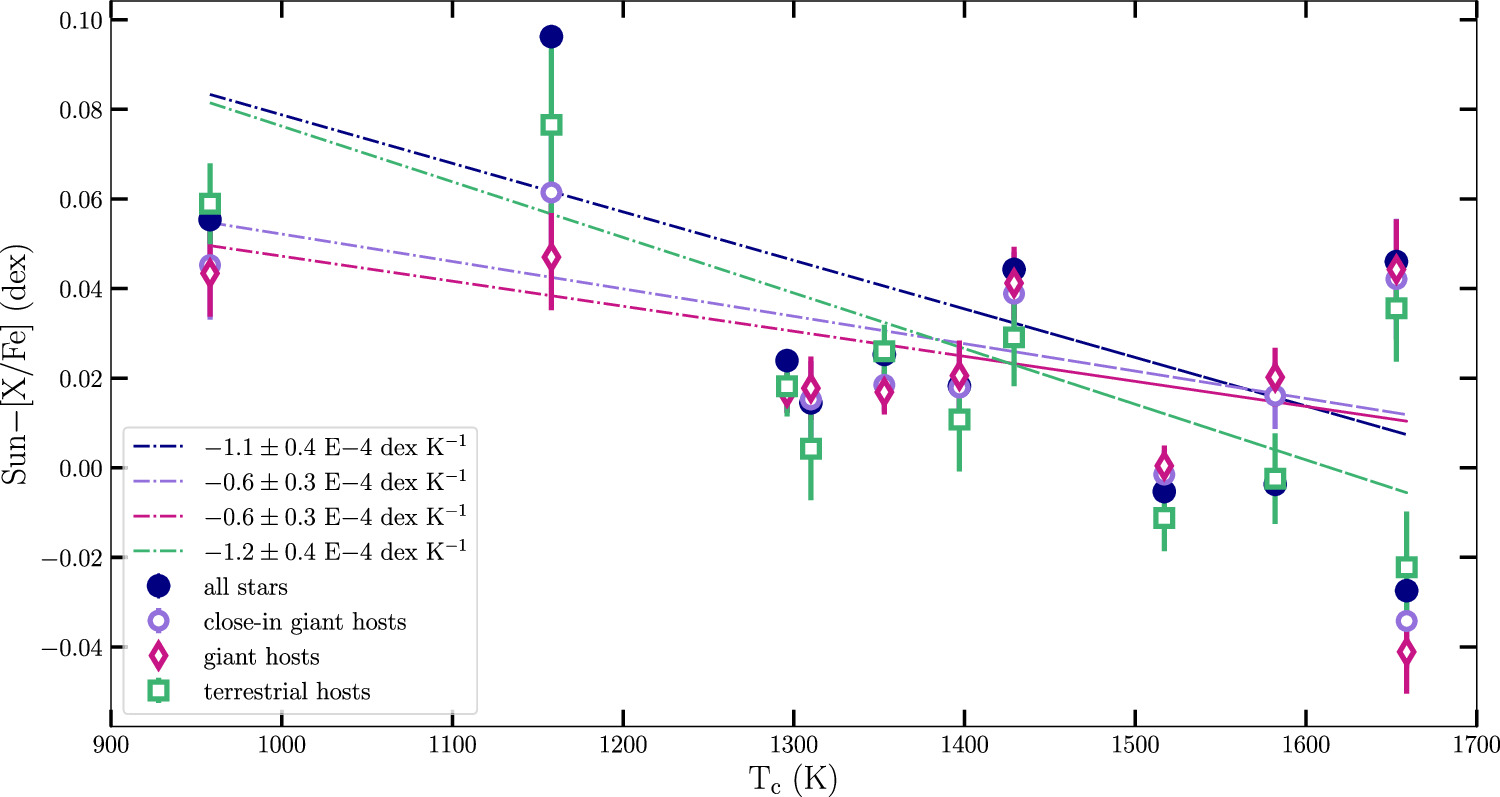}
    \caption[Photospheric depletion in Sun-like stars]{Photospheric depletion in Sun-like stars (navy filled circles), close-in giant planet hosts (purple unfilled circles), giant planet hosts (pink unfilled diamonds), and terrestrial/small planet hosts (green squares). The linear fits of the depletion profiles are coloured accordingly. The condensation temperatures $T_{\rm C}\equiv T_{\rm cond}$ are adopted from \cite{lodders2003CondensationTemperatures}. For Sun-like stars, the similarity in slopes is within the error margins suggesting that the photospheric depletion in Sun-like stars does not strongly depend on the planet types and sizes. This figure was adapted from \cite{rampalli2024SolarDepletion}.}\label{fig:sol_intro}
\end{figure}

Comparative abundance studies of the Sun and Sun-like stars revealed depletion trends with condensation temperature, which are similar to $\lambda$ B\"{o}o phenomenon \citep{melendez2009SolarDepletionPlanetFormation, booth2020DepletionSolarTwins, yun2024SolarDepletion, rampalli2024SolarDepletion, sun2024SunLikeDepletion}. \citet{melendez2009SolarDepletionPlanetFormation} analysed 11 solar twin stars and found that the Sun exhibits relative depletion in refractory elements with T$_{\rm cond}\,>\,1200$\,K (e.g., Mg, Al, Si). Chambers (2010) estimated that refractory deficit in the Sun corresponds to approximately 4 $M_\oplus$ of terrestrial material. Furthermore, elemental abundances in the bulk silicate Earth (BSE) showed a negative correlation with condensation temperature when compared to volatile-rich meteorites like CI chondrites \citep{mezger2020SolarVolatileAccretion}. These findings motivated extensive investigations into chemical abundance differences between planet-hosting stars (PHSs) and non-planet-hosting stars (NPHSs), using abundance trends as a function of condensation temperature \citep[see, e.g.,][]{ramirez2009SolarTwinDepletion, gonzalezhernandez2011SolarTwinDepletion, gonzalezhernandez2013SolarTwinDepletion, adibekyan2014SolarTwinDepletion, nissen2015SolarTwinDepletion, bedell2018SolarTwinDepletion, liu2020DustPlanetFormation, nibauer2021SolarTwinDepletion, tautvaisiene2022PlanetFormationDepletion}. In Fig.~\ref{fig:sol_intro} \citep[adapted from][]{rampalli2024SolarDepletion}, we show the depletion profile in planet-hosting solar-like stars.

A unique avenue for exploring photospheric depletion involves young and solar-like binaries where one component (A or B) hosts circumcompanion planets. This setup allows investigation of relative elemental abundances of the binary ($\Delta\text{[X/H]}_{A-B}\,=\,\log\left(\frac{N_X}{N_H}\right)_A-\log\left(\frac{N_X}{N_H}\right)_B$) formed from the same material. However, binaries with circumcompanion planets may display opposing trends due to planet engulfment, which enhances refractory abundances in the engulfing star. Studies suggest that up to 20-35\% of Sun-like stars may ingest planets \citep{spina2021PlanetEngulfment, liu2021PlanetEngulfment, behmard2023PlanetEngulfment, spina2024PlanetEngulfment, teske2024PlanetEngulfment, liu2024PlanetEngulfment}, though disentangling photospheric depletion from planet engulfment remains challenging. A notable example is the depletion trend in 16 Cyg (a binary system comprising two Sun-like stars), which was initially attributed to the rocky core of 16 Cyg Bb, with 16 Cyg B showing a depletion in refractories \citep{tuccimaia2014PlanetEngulfment}. However, \citet{maia2019PlanetEngulfment} found 16 Cyg A to be unusually rich in Li and Be, suggesting the trend is better explained by 16 Cyg A having engulfed 2.5-3 M$_\oplus$ of terrestrial material.


\textbf{Depletion in evolved stars}

As mentioned in Section~\ref{sec:intnuc}, binary interactions lead to unique chemical processes, which are significantly altering the surface composition of evolved stars from AGB/RGB stars to WDs (i.e. matter accretion results in chemical enrichment of the surface of one component with products of AGB/RGB nucleosynthesis from another component). Moreover, if one component overfills its Roche lobe during AGB/RGB evolution, the resulting mass loss leads to the formation of a post-AGB/post-RGB binary and a circumbinary disc \citep[CBD;][]{vanwinckel2003Review}. The interaction of this disc with the central binary results in the observed photospheric chemical depletion of post-AGB/post-RGB binaries \citep{vanwinckel1992depletion, deruyter2005discs, deruyter2006discs, oomen2019depletion}, which shows similar patterns as in young planet-hosting stars, though $\sim\,100$ times stronger (for more details, see Section~\ref{sec:intpAR}).

\section[Post-AGB/post-RGB binary systems]{Post-AGB/post-RGB binary systems and their photospheric depletion}\label{sec:intpAR}
In this section, we introduce post-AGB/post-RGB binary systems and how their disc-binary interactions result in the photospheric chemical depletion of the primary star. In Section~\ref{ssec:intpARdis}, we discuss the parameters and methods that led to the discovery of these binaries. In Section~\ref{ssec:intpARevo}, we detail the evolution and structure of post-AGB/post-RGB binary systems. In Section~\ref{ssec:intpARcbd}, we introduce the second-generation CBD which forms from the matter ejected when the star evolves off the AGB/RGB. In Section~\ref{ssec:intpARjet}, we discuss the main channels of disc-binary interaction in post-AGB/post-RGB binary stars. In Section~\ref{ssec:intpARdpl}, we detail the photospheric chemical depletion of post-AGB/post-RGB binaries. In Table~\ref{tab:pagbpar_intro}, we show the ranges for the key stellar and orbital parameters of post-AGB/post-RGB binaries \citep{vanwinckel2003Review, vanaarle2011PAGBsInLMC, vanwinckel2012SMCLMCPost-AGB, gezer2015WISERVTau, kluska2022GalacticBinaries}.

\begin{table}[!ht]
    \centering
    \caption[Ranges of observed parameters for post-AGB/post-RGB binaries (see Section~\ref{sec:intpAR})]{Ranges of stellar parameters for post-AGB/post-RGB binaries (see Section~\ref{sec:intpAR}).\\} \label{tab:pagbpar_intro}
    \begin{tabular}{|c|c|c|}\hline
         \textbf{Parameter} & \textbf{Min} & \textbf{Max} \\\hline
         \multicolumn{3}{|c|}{\textit{Stellar parameters}} \\\hline
         $L$ ($L_\odot$) & 100 & 25\,000 \\\hline
         $T_{\rm eff}$ (K) & 4\,000 & 12\,000 \\
         $\log g$ (dex) & 0.0 & 2.5 \\
         $[$Fe/H$]$ (dex) & -5.0 & 0.0 \\\hline
         {[Zn/Ti]} (dex) & 0.0 & 3.5 \\\hline
         $P_{\rm puls}$ (d) & 10 & 80 \\\hline
         \multicolumn{3}{|c|}{\textit{IR colours}} \\\hline
         $H-K$ (mag) & 0 & 1.7 \\
         $W_1-W_3$ (mag) & 0 & 7 \\\hline
         \multicolumn{3}{|c|}{\textit{Orbital parameters}} \\\hline
         $P_{\rm orb}$ (d) & 100 & 2500 \\
         $e$ & 0 & 0.6 \\\hline
         \multicolumn{3}{|c|}{\textit{Disc type breakdown}} \\\hline
         Full (\%) & \multicolumn{2}{|c|}{$\sim$77} \\
         Transition (\%) & \multicolumn{2}{|c|}{$\sim$12} \\
         Dust-poor (\%) & \multicolumn{2}{|c|}{$\sim$11} \\\hline
    \end{tabular}\\
    \textbf{Note:} The parameter limits are defined based on the following studies: \citet{vanwinckel2003Review, vanaarle2011PAGBsInLMC, vanwinckel2012SMCLMCPost-AGB, gezer2015WISERVTau, kluska2022GalacticBinaries}.\\
\end{table}

\subsection{Discovery of post-AGB/post-RGB binary stars}\label{ssec:intpARdis}
The discovery of post-AGB binaries marked a significant milestone in the study of evolved LIM stars. These systems were first identified in the Galaxy through pioneering works by \citet{kwok1987PostAGBDiscovery, vanwinckel1995ExtremelyDepletedPostAGB, waelkens1991depletion, pollard1995UMon, waelkens1996RedRectangle}. Subsequent studies expanded the known sample of post-AGB binaries to the Magellanic Clouds \citep{vanaarle2011PAGBsInLMC, vanwinckel2012SMCLMCPost-AGB, kamath2014SMC, kama2015DiscDepletionLinkinYSOs}, allowing to investigate post-AGB/post-RGB evolution in different metallicity environments. More recently, the less luminous analogues of post-AGB binaries -- post-RGB binaries -- were discovered in the Galaxy and the Magellanic Clouds \citep[$L_{\rm post-RGB}\,\lesssim\,2500\,L_\odot$;][]{kamath2016PostRGBDiscovery, kamath2022GalacticSingles}, further enriching our understanding of binary evolution of evolved stars.

\begin{figure}[!ht]
    \centering
    \includegraphics[width=.99\linewidth]{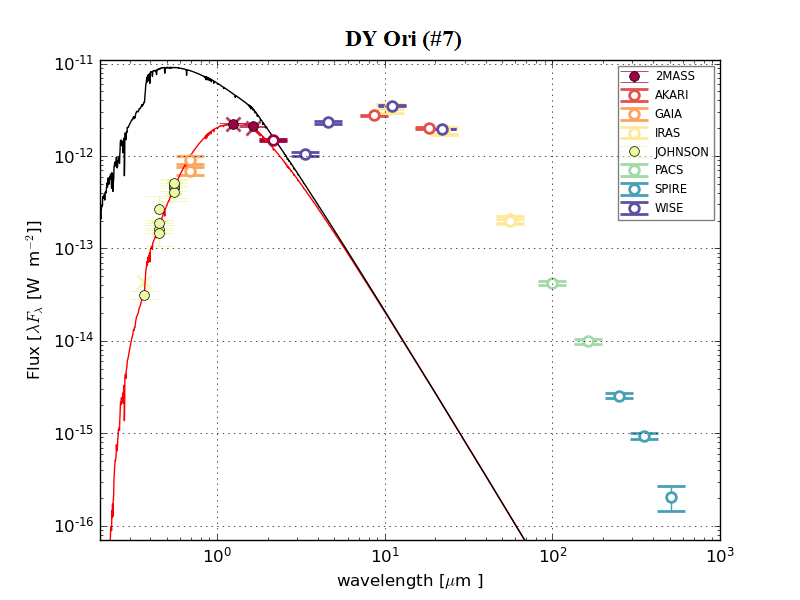}
    \caption[SED of the transition disc target DY Ori]{SED of the transition disc target DY Ori. The red line represents the fitted Kurucz model atmosphere, the black line shows the scaled, de-reddened Kurucz model atmosphere. The integration of de-reddened model yields the SED luminosity (see Section~\ref{ssec:obslum_paper1}). A legend within the plot explains the symbols and colours used. This figure was adapted from \citet{mohorian2025TransitionDiscs}.}\label{fig:SED_intro}
\end{figure}

The features of post-AGB/post-RGB binaries, which enabled the discovery, were their SEDs, typically containing a photospheric contribution from the luminous primary star ($L_{\rm post-AGB}\,\gtrsim\,2500\,L_\odot$) and distinct dust emission in the IR range \citep[see Fig.~\ref{fig:SED_intro}; adapted from][]{mohorian2025TransitionDiscs}. The near-IR excess of post-AGB/post-RGB binaries begins at wavelengths shorter than 2 $\mu$m and peaks around 10 $\mu$m, whereas its long-wavelength tail follows the Rayleigh–Jeans slope up to sub-millimetre wavelengths. The dust emission feature in the SED of post-AGB/post-RGB binaries originates in a stable, compact CBD \citep{hillen2017DiscInterferometry, kluska2018IRAS08}, with the broad near-IR excess originating in circumstellar dust close to the sublimation radius of the binary system \citep[$T_{\rm sublim}\,\sim\,1500$\,K;][]{kluska2019DiscSurvey}. Therefore, SEDs with broad near-IR excess are classified as disc-type SEDs.

Following photometric observations, long-term spectroscopic monitoring allowed to determine the orbits of $\sim\,50$ discovered post-AGB/post-RGB binaries in the Galaxy \citep{oomen2018OrbitalParameters, bodi2019RVTauVars}. Over 70\% of Galactic post-AGB/post-RGB binaries exhibit significant non-zero eccentricities, despite the relatively small orbital separations, which are insufficient to accommodate a fully developed AGB star \citep{vanwinckel2003Review, vanwinckel2007VLTIsurvey, vanwinckel2009}. Moreover, orbital periods of post-AGB/post-RGB binaries range from 100 to 3000 days, which contrasts with the predictions from binary evolution theories (see Section~\ref{ssec:intpARevo}).

\subsection{Evolution and structure of post-AGB/post-RGB binaries}\label{ssec:intpARevo}
As mentioned in Section~\ref{ssec:intnuclim}, post-AGB and post-RGB binary stars are systems in which the primary component is a LIM star that evolved off the AGB/RGB, but lacks the required temperature ($T\,\sim\,30\,000$\,K) to ionise its circumstellar material and to form a PN \citep[typically lasting $\,\sim\,10^4-10^5$ years;][]{vanwinckel2003Review, bertolami2016tracks}. We note that the post-AGB evolutionary track crosses the instability strip in the HR, which results in a significant fraction of post-AGB sample pulsating as Type II Cepheids \citep[W Vir and RV Tau variables;][]{giridhar1998RVTauVars, giridhar2005rvtau, maas2007t2cep, gezer2015WISERVTau, kiss2017RVTau, bodi2019RVTauVars}.

The transition of the primary star off the AGB/RGB in a binary system is primarily driven by significant mass-loss episodes, often triggered by binary interactions. Mass transfer models suggest that stars filling their Roche lobes undergo unstable mass transfer, which can lead to the formation of a common envelope. This process results in the inward spiralling of the system, ultimately causing short-period orbits \citep{hjellming1987MassTransferInBinaries, izzard2023DiscBinaryInteractionAccretion}. In contrast, systems that experience minimal interaction are expected to extend their orbits due to mass loss via isotropic winds from the AGB star \citep{nie2012PostRGBBinaryNebula}. These two evolutionary pathways should produce a bimodal period distribution, with short-period systems ($<\,100$ days) forming via common envelope channel and wide-orbit systems ($>\,3000$ days) that did not undergo interaction. However, the observed period distribution of post-AGB/post-RGB binaries \citep{vanwinckel2003Review, vanwinckel2009, manick2017RVTauBinarity, oomen2018OrbitalParameters} deviates from standard predictions of population synthesis models \citep{izzard2010WDkicksBaStars, nie2012PostRGBBinaryNebula}. The discrepancies in the observed period distribution suggest that additional physical mechanisms must be considered, including high accretion rates from wind-Roche-lobe overflow \citep{abate2013WindRocheLobeOverflow}, efficient angular momentum loss due to mass loss \citep{chen2018WindMassLoss, saladino2018WindMassLoss}, and/or efficient common-envelope ejection \citep{soker2015CommonEnvelope}.

\begin{figure}[!ht]
    \centering
    \includegraphics[width=.99\linewidth]{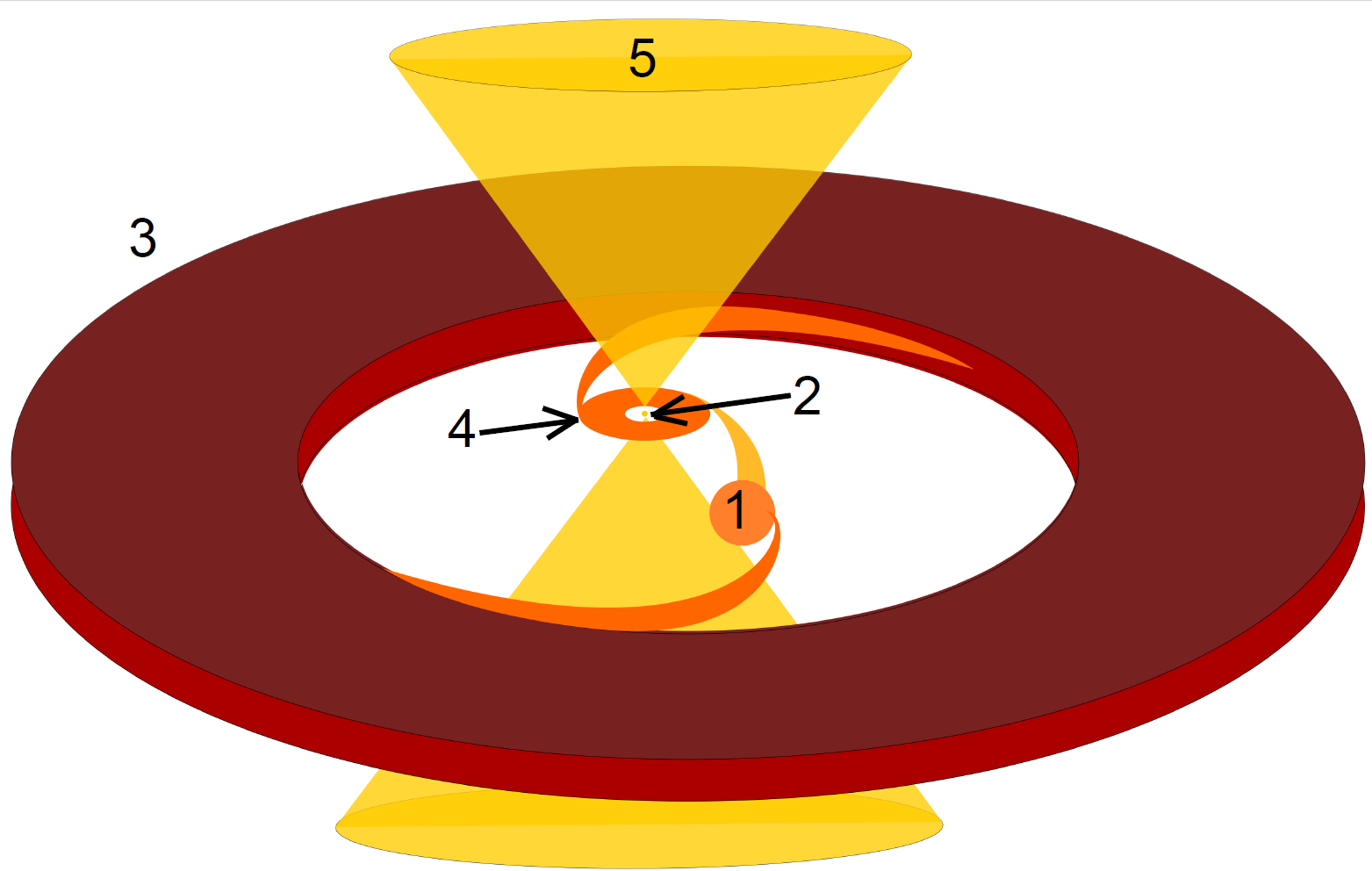}
    \caption[Illustration depicting the key structural components of post-AGB binaries]{Illustration depicting the key structural components of post-AGB binaries: 1) post-AGB primary star, 2) MS secondary star, 3) the circumbinary disc, and 4) the circumcompanion accretion disc that generates 5) the jet. This figure was adapted from \cite{bollen2022Jets}.}\label{fig:str_intro}
\end{figure}

Typically, post-AGB/post-RGB binary systems comprise the following components:
\begin{itemize}
    \item a post-AGB/post-RGB primary star with $L\,\sim\,10^3-10^5\,L_\odot$ \citep{kamath2023models, mohorian2024EiSpec}, 
    \item an MS secondary companion with  $L\,\sim\,L_\odot$ and $M\,\sim\,1.09\,\pm\,0.62\,M_\odot$ \citep{oomen2018OrbitalParameters}, 
    \item a CBD of gas and dust (see Section~\ref{ssec:intpARcbd}), 
    \item a circumcompanion disc launching a bipolar high-velocity outflow (jet; see Section~\ref{ssec:intpARjet}), 
    \item an extended circumstellar gas structure \citep{bujarrabal2015KeplerianRotation, gallardocava2021PostAGBOutflows, gallardocava202389HerNebula}.
\end{itemize}
In Fig.~\ref{fig:str_intro} \citep[adapted from][]{bollen2022Jets}, we show a schematic sketch of the typical structure of post-AGB/post-RGB binary systems.

Overall, post-AGB/post-RGB binaries are ideal tracers for investigating the impact of binarity on element and isotope production. The spectra of post-AGB/post-RGB binaries, dominated by atomic transitions, enable precise analysis of both photospheres (surface elemental abundances and isotopic ratios) and circumstellar environments (jet modelling). The formation of CBDs due to binarity and subsequent disc-binary interactions further emphasise the significance of post-AGB/post-RGB binaries (see Section~\ref{ssec:intpARcbd}).

\subsection{Second-generation protoplanetary discs around post-AGB and post-RGB binary stars}\label{ssec:intpARcbd}
As the component of a binary star overfills its Roche lobe during the AGB/RGB evolution, the ejected matter may form a CBD, consisting of gas and dust \citep{nie2012PostRGBBinaryNebula}. The inner rim of gas in CBD is located beyond the distance at which the gravitational influence of the binary system truncates the gas distribution ($\lesssim\,3-5$ AU), whereas the inner rim of dust in CBD is located at the sublimation radius \citep[$\sim\,5-30$ AU;][]{corporaal2023DiscParameters, gallardocava202389HerNebula}. High-resolution interferometric observations in mm and sub-mm range with ALMA revealed the CBDs have masses $\sim\,0.01-0.1\,M_\odot$ and experience the Keplerian rotation based on the $^{12}$CO position–velocity maps  \citep{bujarrabal2015KeplerianRotation, gallardocava202389HerNebula}. Recent studies using advanced direct imaging with extreme adaptive optics confirmed the complex structure of CBDs around post-AGB/post-RGB binaries \citep{ertel2019Imaging, andrych2023Polarimetry, andrych2024IRAS08}. High-angular resolution imaging in the $V$-, $I$-, and $H$-bands with VLT/SPHERE revealed that CBDs around post-AGB/post-RGB binaries display complex asymmetries (including rings, gaps, and arc-like features) potentially caused by disc winds, outflows, or shadows from inner disc structures. Similar structural asymmetries were also observed in protoplanetary discs \citep[PPDs;][]{fukagawa2010MIRimagingHerbigAeStars, garufi2016PPDwithSPHERE}, suggesting similarities in the morphology and dynamics of CBDs around post-AGB/post-RGB binaries and PPDs around young stellar objects (YSOs). 

\begin{figure}[!ht]
    \centering
    \includegraphics[width=1.\linewidth]{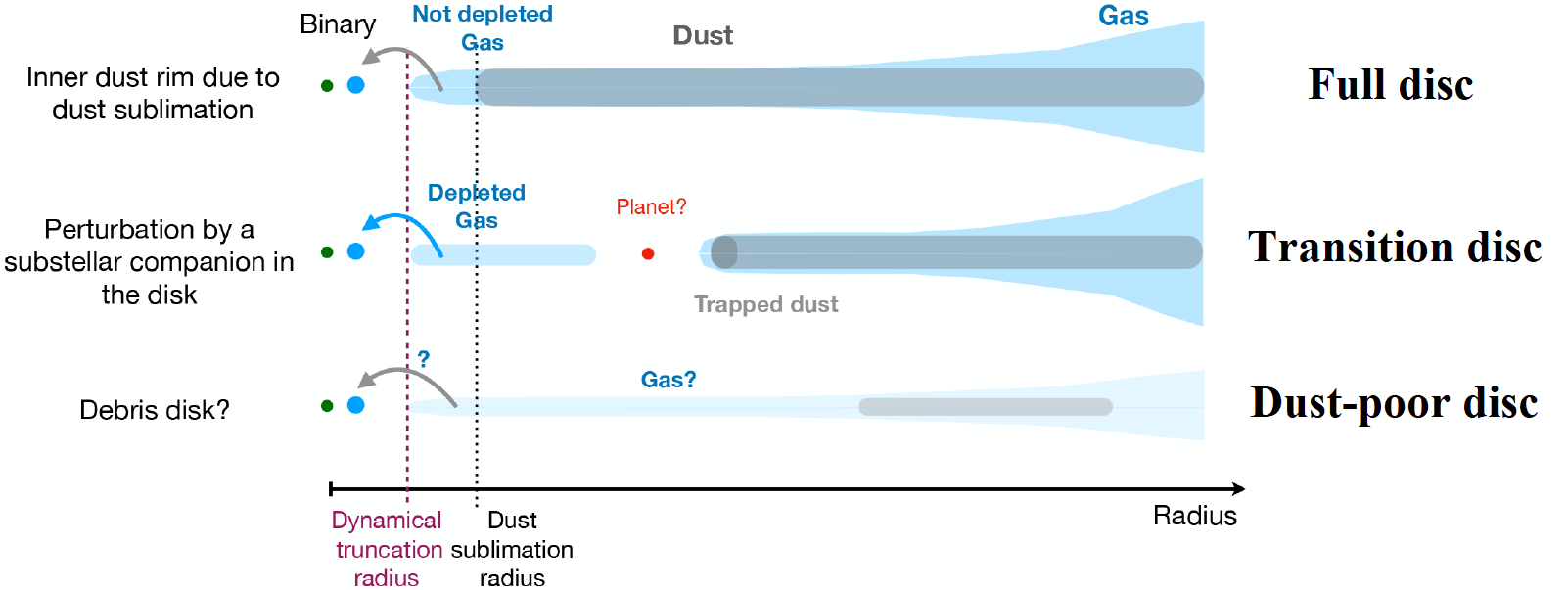}
    \caption[Schematic depiction of the disc types hosted by post-AGB binaries]{Schematic depiction of the disc types hosted by post-AGB binaries. Illustration is not to scale. This figure was adapted from \cite{kluska2022GalacticBinaries}.}\label{fig:cbd_intro} 
\end{figure}

Based on IR colours 2MASS $H-K$ and WISE $W_1-W_3$, post-AGB/post-RGB binaries with CBDs can be categorised into three distinct types \citep{kluska2022GalacticBinaries}: full disc targets, transition disc targets, and dust-poor disc targets (see Fig.~\ref{fig:cbd_intro}). Full disc targets exhibit broad IR excesses, indicating a continuous distribution of dust in the disc ($H-K\,\gtrsim\,0.3^m$, $2.3^m\,\lesssim\,W_1-W_3\,\lesssim\,4.5^m$). Transition disc targets display an SED with a lack of near-IR excess, indicating a cavity in the inner disc ($H-K\,\lesssim\,0.3^m$, $W_1-W_3\,\gtrsim\,2.3^m$). Dust-poor disc targets have significantly weaker IR excesses, hinting at their low dust content ($H-K\,\lesssim\,0.3^m$, $W_1-W_3\,\lesssim\,2.3^m$). We note that in edge-on systems the disc veils the binary, which results in increased IR colours ($H-K\,\gtrsim\,0.3^m$, $W_1-W_3\,\gtrsim\,4.5^m$). In Fig.~\ref{fig:cbd_intro} \citep[adapted from][]{kluska2022GalacticBinaries}, we show the schematic models of full, transition, and dust-poor disc targets from post-AGB/post-RGB binary sample. Furthermore, high-resolution interferometric IR observations with VLTI confirmed that the radius of dust inner rim in post-AGB/post-RGB binaries with full discs is close to the radius of dust sublimation \citep[typically, at $\sim\,1500$\,K;][]{kluska2019DiscSurvey, corporaal2023DiscParameters}. Similarly, optical and near-IR long-baseline interferometric studies of YSOs with full discs (Group II PPDs) revealed that the radius of dust inner rim aligns with the dust sublimation radius, though at higher temperature \citep[typically, at $\sim\,1800$\,K;][]{tannirkulam2008SublimationTemperaturePPD, lazareff2017HerbigDiscs}.

Despite the observed evidences for Keplerian rotation and dust grain growth \citep{gielen2008SPITZERsurvey, gielen2011silicates, sahai2011DustGrains, scicluna2020GrainGrowth}, the evolution of CBDs around post-AGB/post-RGB binaries remains poorly understood. To further explore the dynamics and evolution of CBDs, the models of PPDs are often adapted, providing valuable insights into substructures and chemical processes in CBDs (see Section~\ref{ssec:intpARcbd}). In Table~\ref{tab:ppdcbd_intro}, we compare the most relevant parameters of PPDs around young stars and CBDs around post-AGB/post-RGB binaries.

The interaction of CBDs with post-AGB/post-RGB binaries results in distinct processes, including the jet formation (see Section~\ref{ssec:intpARjet}) and photospheric chemical depletion (see Section~\ref{ssec:intpARdpl}). Similarly, the interaction of PPDs with YSOs (T Tau and Herbig Ae/Be stars) leads to photospheric chemical depletion, commonly attributed to planet formation (see Section~\ref{ssec:intdplstr}), highlighting the importance of studying the surface composition of post-AGB/post-RGB binaries.

\begin{sidewaystable}[ph!]
    \centering
    \caption[Comparison of the properties of protoplanetary discs around YSOs and circumbinary discs around post-AGB binaries]{Comparison of the properties of protoplanetary discs around young stars and circumbinary discs around post-AGB binaries (see Section~\ref{sec:intpAR}). This table was adapted from \href{https://fys.kuleuven.be/ster/pub}{Corporaal 2023 (PhD thesis)} and \href{https://figshare.mq.edu.au/theses/categories/Astronomical_sciences/30085}{Andrych 2025 (PhD thesis)}.\\} \label{tab:ppdcbd_intro}
    \begin{tabular}{|c|c|c|}\hline
         \textbf{Property} & \textbf{PPD} & \textbf{CBD} \\\hline
         Disc mass ($M_\odot$) & $10^{-5}-10^{-1}$ & $\sim10^{-2}$ \\\hline
         Disc lifetime & $\sim1$ Myr & $\sim0.1$ Myr \\\hline
         Velocity distribution & Keplerian & Keplerian \\\hline
         Vertical dust settling & Yes & Yes \\\hline
         Disc inner rim & Puffed-up and rounded & Puffed-up and rounded \\\hline
         Surface dust grains & \begin{tabular}{c} Dust aggregates, \\ >1 $\mu$m size \end{tabular} & \begin{tabular}{c} Porous dust aggregates, \\ $\sim$1 $\mu$m size \end{tabular} \\\hline
         Radial structure & \begin{tabular}{c} Full and transition discs, \\ substructures and asymmetries \end{tabular} & \begin{tabular}{c} Full and transition discs, \\ substructures and asymmetries \end{tabular} \\\hline
         Warps and disc tearing & Yes & Maybe \\\hline
         Disc flaring & Yes & Yes \\\hline
         \begin{tabular}{c} Size-luminosity \\ relation \end{tabular} & \begin{tabular}{c} $R_{\rm sublim}\propto L^{1/2}$, \\ $T_{\rm sublim}\sim 1500-1800$ K \end{tabular} & \begin{tabular}{c} $R_{\rm sublim}\propto L^{1/2}$, \\ $T_{\rm sublim}\sim 1200-1300$ K \end{tabular} \\\hline
    \end{tabular}\\
\end{sidewaystable}

\subsection{Formation and modelling of jets in post-AGB/post-RGB binaries}\label{ssec:intpARjet}
Disc-binary interactions in post-AGB/post-RGB binaries lead to distinct processes, including the formation of non-relativistic high-velocity ($\sim\,100-400$\,km/s) bipolar outflows \citep[jets;][]{gorlova2012BD+46442}. As mentioned in Section~\ref{ssec:intpARdis}, the long-term high-resolution spectroscopic monitoring with HERMES \citep[High-Efficiency and high-Resolution Mercator Echelle Spectrograph;][]{vanwinckel2009, raskin2011hermes} in stalled at Mercator telescope (see Section~\ref{ssec:mthdather}) enabled the discovery and investigation of jets in post-AGB/post-RGB binaries. The analysis of variable H$\alpha$ line profiles in dynamical spectra revealed that these jets are wide ($>\,30^\circ$) with an angle-dependent density profile \citep{gorlova2012BD+46442, gorlova2015IRAS19jet, bollen2022Jets, deprins2024Jets}, resembling those observed in PNe and YSOs \citep{lopez1993PNeJets, ray2021YSOjets, clairmont2022PNjet, lora2023PNjet}. Moreover, jet velocities in post-AGB/post-RGB binaries are consistent with the escape velocity of MS stars, confirming the evolutionary nature of the secondary component \citep{bollen2017BD+46442Jet, bollen2019JetModelling, bollen2021JetParameters, bollen2022Jets}.

\begin{figure}[!ht]
    \centering
    \includegraphics[width=.99\linewidth]{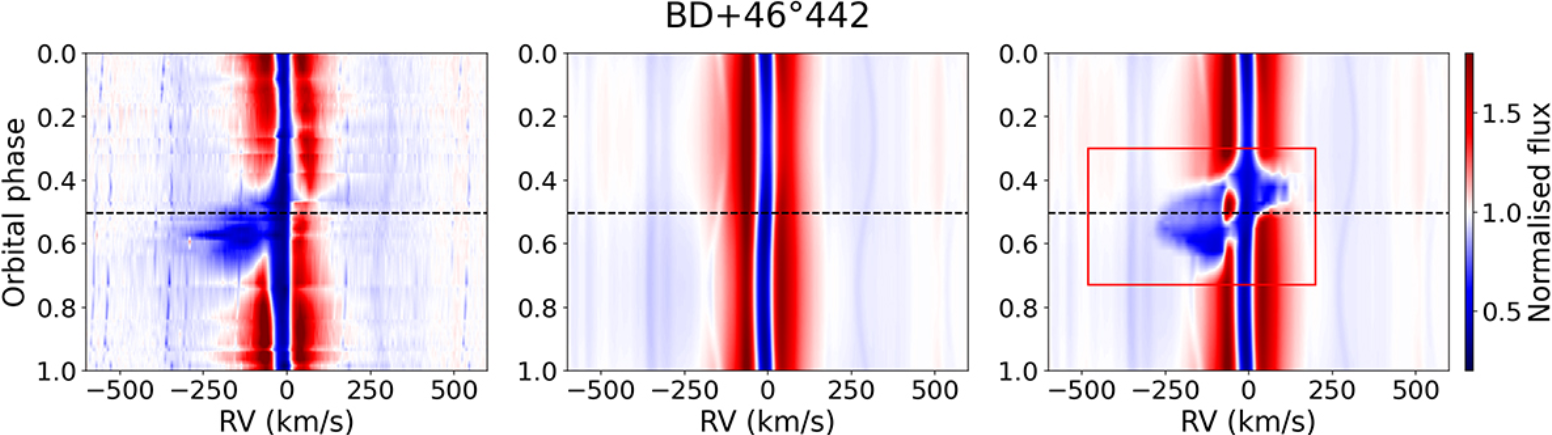}
    \caption[Dynamic H$\alpha$ line spectra of BD+46$^\circ$442]{Dynamic H$\alpha$ line spectra of BD+46$^\circ$442. \textit{Left panel}: observed time-series. \textit{Middle panel}: zero-iteration spectrum. \textit{Right panel}: best-fit disc wind model. The black dashed line indicates superior conjunction, and the red box highlights the area used for $\chi_\nu^2$ calculations. This figure was adapted from \cite{deprins2024Jets}.}\label{fig:jet_intro}
\end{figure}

To better connect these jets with the underlying physics, \citet{verhamme2024DiscWindModelling} employed magnetohydrodynamic disc wind models \citep{jacqueminide2019DiscWindModelling}, adopting the extended disc wind paradigm, which is commonly used as the main mechanism forming jets in YSOs \citep{ferreira2006TTauJetLaunching, moscadelli2022MHDDiscWind, launhardt2023TTauDiscWind}. \citet{deprins2024Jets} advanced the jet formation modelling by assuming self-similar magnetohydrodynamic disc wind models and fitting dynamical spectra of observationally diverse sample of 5 post-AGB/post-RGB stars. In Fig.~\ref{fig:jet_intro} \citep[adapted from][]{deprins2024Jets}, we show an example of jet modelling for BD+46 442 (\textit{left panel}: observations, \textit{middle panel}: model without rotation, \textit{right panel}: model with rotation). Current jet models provide valuable insights but can be further refined by incorporating magnetothermal models that account for irradiation from the luminous post-AGB primary, enabling more realistic predictions for wind rotation and accretion rates.

\subsection{Photospheric chemical depletion in post-AGB/post-RGB binaries}\label{ssec:intpARdpl}
\begin{figure}[!ht]
    \centering
    \includegraphics[width=.99\linewidth]{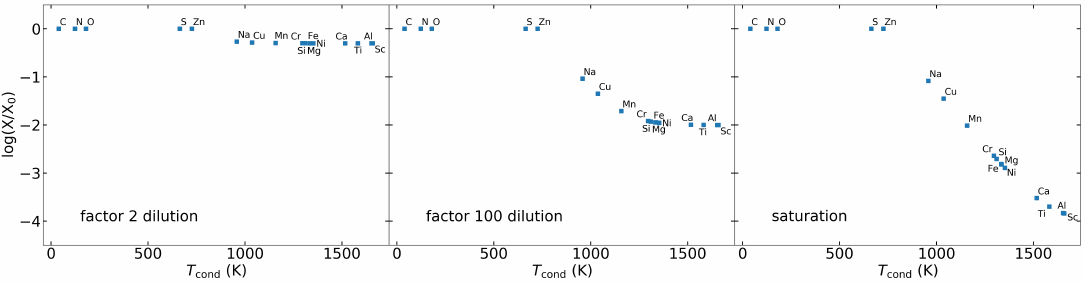}
    \caption[The modelled chemical composition of a gas mixture after diluting a solar-composition gas (stellar surface) with refractory-poor gas (accreted disc matter)]{The modelled chemical composition of a gas mixture after diluting a solar-composition gas (stellar surface) with refractory-poor gas (accreted disc matter). The left, middle, and right panels display the result of a 2-fold (not depleted), 100-fold (plateau), and 10\,000-fold dilution (saturation), respectively. The position of the break in depletion profile between the less depleted volatile elements and significantly depleted refractory elements depends on the chemical mixture of re-accreted gas. This figure was adapted from \cite{oomen2019depletion}.}\label{fig:dil_intro}
\end{figure}

As mentioned in Section~\ref{ssec:intpARcbd}, interactions between post-AGB/post-RGB binaries and their CBDs play a crucial role in modifying the surface composition of the central stars \citep{gielen2009Depletion, gezer2015WISERVTau, kamath2019depletionLMC}. While the exact mechanism leading to photospheric depletion remains uncertain, it is hypothesised to involve gas-dust fractionation in the CBD, followed by selective re-accretion of volatile-rich gas onto the surface of the primary star \citep{waters1992GasDustFractionation, oomen2020MESAdepletion}. In this scenario, volatile-rich gas is re-accreted onto the binary, whereas refractory-rich dust is expected to settle in the disc mid-plane \citep{mosta2019ReaccretionInnerRim, munoz2019ReaccretionInnerRim}. In Fig.~\ref{fig:dil_intro} \citep[adapted from][]{oomen2019depletion}, we show the modelled impact of dilution on surface abundances of post-AGB primary star. The dilution factor is a proxy of depletion efficiency and represents mass ratio of volatile-rich re-accreted matter to a solar-composition gas (\textit{left panel:} dilution with ratio 2:1, \textit{middle panel:} dilution with ratio 100:1, \textit{right panel:} saturation, i.e. dilution with ratio 10\,000:1).

Observations of photospheric chemical depletion in post-AGB/post-RGB binaries in the Galaxy and the Magellanic Clouds \citep{giridhar1998RVTauVars, maas2002RUCenSXCen, deruyter2005discs, giridhar2005rvtau, deruyter2006discs, maas2007t2cep, vanwinckel2012AFCrt, rao2014RVTauAbundances, desmedt2012j004441, desmedt2014LeadMCs, desmedt2015LMC2sEnrichedPAGBs, desmedt2016LeadMW} revealed that their relative abundances [X/H] behave differently for volatile and refractory elements (see Section~\ref{ssec:mththratm}). The relative abundances of volatile elements remain comparable to each other, whereas the relative abundances of refractory elements decrease significantly (for most refractory elements, the underabundance can reach 4\,dex). This creates a distinct break in the depletion profiles of post-AGB/post-RGB binaries (see Fig.~\ref{fig:dil_intro}). The turn-off temperature $T_{\rm turn-off}$, marking the temperature of the break, varies widely among Galactic systems \citep[800–1500\,K;][]{kluska2022GalacticBinaries}. We note that CNO elements (C, N, O) are typically excluded from depletion studies because their abundances are heavily influenced by nucleosynthesis and mixing during the AGB/RGB phases \citep[see Section~\ref{ssec:intnuclim}][]{mohorian2024EiSpec, menon2024EvolvedBinaries}. While mixing processes on AGB/RGB significantly alter C and N abundances, O abundance is expected to remain relatively unchanged in low-mass ($M\,<\,2\,M_\odot$) AGB stars and RGB stars \citep{ventura2008aton3, karakas2014dawes, ventura2020CNOinAGB, kobayashi2020OriginOfElements, kamath2023models}.

\textbf{Quantification of depletion efficiency}

Depletion efficiency is typically quantified using the volatile-to-refractory abundance ratios, including [Zn/Ti]\footnote{$\text{[X/Y]}\,=\,\log\left(\dfrac{N_X}{N_Y}\right) - \log\left(\dfrac{N_X}{N_Y}\right)_\odot$, where $N_X$ and $N_Y$ are number abundances of elements X and Y, respectively, scaled to the solar values (see Section~\ref{ssec:mththratm}).}, [S/Ti], or [Zn/Fe] \citep{gezer2015WISERVTau, oomen2019depletion}. Based on depletion strength, post-AGB/post-RGB binaries can be classified as follows:
\begin{itemize}
    \item Mildly depleted ([Zn/Ti]\,<\,0.5\,dex),
    \item Moderately depleted (0.5\,<\,[Zn/Ti]\,<\,1.5\,dex), or 
    \item Strongly depleted ([Zn/Ti]\,>\,1.5\,dex).
\end{itemize}
Mild depletion is typically observed in post-AGB/post-RGB systems with full discs, while strong depletion tends to be displayed by transition disc targets, although this separation is not definitive \citep{kluska2022GalacticBinaries}. Additionally, depletion profiles for $T_{\rm cond}\,>\,T_{\rm turn-off}$ can exhibit either a linear trend (``saturated'') or a two-segment fit with a plateau-like part at higher $T_{\rm cond}$ \citep[``plateau'';][]{waelkens1991depletion, oomen2019depletion, oomen2020MESAdepletion}. Stellar evolution models using MESA suggest that photospheric depletion profiles, including turn-off breaks and plateaus, result from the dilution of re-accreted volatile-rich gas with the original composition of the stellar surface \citep{oomen2019depletion, oomen2020MESAdepletion}.

Notably, the Galactic subsample of post-AGB/post-RGB binaries \citep[85 Galactic targets;][]{kluska2022GalacticBinaries} contains 6 binaries, which show an extremely strong depletion ([Fe/H]\,$\lesssim\,-3.0$\,dex, [Zn/Ti]\,$\gtrsim$\,2.0\,dex). This occurs despite their SEDs, luminosities, and turn-off temperatures being typical for the whole Galactic post-AGB/post-RGB subsample \citep[$T_{\rm turn-off}\,\lesssim\,1\,000$\,K;][]{kluska2022GalacticBinaries}. Among these 6 extremely depleted systems, 3 full disc targets are surrounded by carbonaceous dust, whereas the other 3 with dust-poor discs show little to no dust excess. However, the limited availability of IR observations for post-AGB/post-RGB binaries highlights the need for additional data, particularly from facilities like MIRI/JWST, CRIRES/VLT, and NIRPS/3.6m telescope, to better understand the link between depletion efficiency and carbonaceous dust.

Finally, we note that the photospheric depletion patterns observed in post-AGB/post-RGB binaries exhibit similarities to those observed in young stars displaying $\lambda$ B\"{o}o phenomenon (see Section~\ref{ssec:intdplstr}).  In $\lambda$ B\"{o}o stars, photospheric depletion is commonly attributed to processes associated with planet formation. However, the refractory elements in post-AGB/post-RGB photospheres are depleted $\sim\,100$ times stronger than those in YSO photospheres. This prominent difference suggests more efficient fractionation of gas and dust in CBDs around post-AGB/post-RGB binaries or that photospheric depletion in post-AGB/post-RGB binaries likely arises from a combination of processes in addition to/instead of planet formation (including photoevaporation, grain growth, and dead zones). Further investigation into this phenomenon provides a valuable opportunity to constrain key parameters of disc-binary interactions, including accretion rates, spatial origins of accreted matter in the CBD, and chemical composition of the accreted matter.

\section{Motivation and outline of this thesis}\label{sec:intmtv}
The goal of this thesis is to investigate the impact of binarity on element production in evolved stars, using post-AGB/post-RGB binaries as unique tracers of AGB/RGB nucleosynthesis. The post-AGB/post-RGB binaries exhibit photospheric chemical depletion, a phenomenon directly linked to the interaction between the central binary system and the surrounding circumbinary disc. Notably, chemical depletion is a well-established pattern in young stars with PPDs, suggesting a strong connection between planet formation and disc evolution. Furthermore, the striking similarity between PPDs around young stars and CBDs around post-AGB/post-RGB binaries led to the compelling hypothesis of the second-generation protoplanetary nature of CBDs around a subclass of post-AGB/post-RGB binaries. Although second-generation planets remain beyond the reach of direct imaging, studying disc-binary interactions provides a rare and powerful indirect probe of the potential presence of these planets. By examining the chemical and dynamical signatures imprinted on post-AGB/post-RGB binaries by their CBDs, this thesis aims to establish a statistically robust representation of the depletion process in evolved stellar environments and to gain critical insights into the interplay between stellar evolution, binary interactions, and second-generation planet formation.

In this thesis, we analyse high-resolution optical and near-IR spectra of post-AGB and post-RGB binaries to derive precise elemental abundances and isotopic ratios. Near-IR spectra are essential for the derivation of CNO isotopic ratios, since these ratios can only be derived from near-IR molecular bands (e.g., CO, CN, and OH). Additionally, multi-wavelength approach broadens the range of chemical elements that can be analysed, enhancing the overall accuracy of abundance derivation. However, current tools for spectral analysis face notable limitations, since existing model atmosphere grids do not fully encompass the typical atmospheric parameters of post-AGB/post-RGB binaries. In addition, standard radiative transfer calculations rely on the local thermodynamic equilibrium (LTE) assumption, which is not valid for post-AGB/post-RGB binaries. Non-local thermodynamic equilibrium (non-LTE, NLTE) corrections are particularly crucial for our target sample, as departures from LTE can lead to abundance differences of up to 0.5\,dex for individual spectral lines. To overcome these challenges and investigate disc-binary interactions in post-AGB/post-RGB systems, we developed E-iSpec, a spectral analysis tool tailored for evolved stars. In this thesis, we used E-iSpec for detailed abundance analyses of post-AGB/post-RGB binaries in the Galaxy and in the Large Magellanic Cloud, incorporating NLTE corrections consistently across a representative set of elements from C to Ba.

In Chapter~\ref{chp:mth}, we provide an overview of current spectral analysis techniques and outline the spectroscopic data used in this thesis. Additionally, we present the theoretical framework that underpins spectral analysis. We then describe E-iSpec, a semi-automated spectral analysis code developed and used in this thesis to investigate disc-binary interactions driving photospheric depletion in post-AGB/post-RGB binaries. Finally, we conclude the chapter by discussing the limitations of E-iSpec and its applications in collaborative research projects.

In Chapter~\ref{chp:pap1}, we present a detailed abundance analysis of two dusty post-RGB binaries, SZ Mon and DF Cyg, using multi-wavelength spectroscopic data from HERMES/Mercator (optical) and APOGEE (near-infrared; Apache Point Observatory Galactic Evolution Experiment; see Section~\ref{ssec:mthdatapo}). We examined the photospheric depletion patterns in post-RGB binaries and compared them with those observed in post-AGB binaries. In addition, we derived the first photospheric $^{12}$C/$^{13}$C isotopic ratios of post-RGB binary stars and compared their CNO abundances with theoretical predictions from ATON single-star evolutionary models, demonstrating that the observed abundances of the studied stars primarily reflect nucleosynthetic processes that occurred before the RGB phase was prematurely terminated by binary interaction.

In Chapter~\ref{chp:pap2}, we investigate photospheric depletion in 12 post-AGB/post-RGB binary systems with transition discs using high-resolution optical spectra from HERMES/Mercator and UVES/VLT (Ultraviolet and Visual Echelle Spectrograph installed at Very Large Telescope; see Section~\ref{ssec:mthdatuvs}). We perform a detailed abundance analysis, applying 1D LTE methods and incorporating 1D NLTE corrections to individual abundances of a representative set of chemical elements from C to Fe. Our findings revealed that depletion efficiency, measured by [S/Ti] abundance ratio, is significantly higher in transition disc targets compared to the general post-AGB/post-RGB binary population. We also examined correlations between depletion patterns and binary system parameters and compared the derived abundance trends in our sample to those observed in the ISM and in young stars with transition discs. This chapter provides valuable insights into the mechanisms driving chemical depletion in CBDs around evolved stars.

In Chapter~\ref{chp:pap3}, we extend the abundance analysis to 9 post-AGB/post-RGB binaries with dust-poor discs, based on the methodology and results presented in Chapters~\ref{chp:pap1} and \ref{chp:pap2}. Given that this subsample includes both weakly and strongly depleted post-AGB/post-RGB binaries, we investigated the evolutionary status of dust-poor discs and their potential connection to full and transition discs.

Finally, in Chapter~\ref{chp:dsc}, we summarise the key findings of this thesis and explore potential future research directions, with a particular focus on binary evolution, disc chemistry, and second-generation planet formation.

\begin{savequote}[75mm]
\foreignlanguage{ukrainian}{
``Він, швидко поробивши човни,\\
На синє море поспускав,\\
Троянців насажавши повні,\\
І куди очі почухрав...''}
\qauthor{\foreignlanguage{ukrainian}{---Іван Котляревський, ``Енеїда'' (1798)}}
``He quickly built some boats of timber,\\
Then launched them in the quiet sea\\
And filling them with muscle limber\\
He hit the foam where eyes could see...''
\qauthor{---Ivan Kotliarevsky, ``Eneida'' (1798)}
\end{savequote}

\chapter{Data, observations, and methodology}\label{chp:mth}
\graphicspath{{ch_methods/figures/}}

\clearpage

\textit{In this thesis, we investigate the impact of binarity on element production in evolved stars, focusing on post-AGB/post-RGB binaries with second-generation protoplanetary discs as tracers of AGB/RGB nucleosynthesis and the role of disc-binary interactions in photospheric chemical depletion. By examining the chemical signatures in these binaries, we aim to uncover insights into stellar evolution, binary interactions, and potential hints of second-generation planet formation in evolved stellar environments. To investigate photospheric chemical depletion in post-AGB/post-RGB binaries, we use high-resolution optical spectra complemented by near-infrared spectra. In this chapter, we begin by presenting the spectroscopic data used in this thesis (see Section~\ref{sec:mthdat}). We then outline the theoretical framework underpinning spectral analysis and the key assumptions of radiative transfer theory (see Section~\ref{sec:mththr}). Finally, we describe E-iSpec, a semi-automated spectral analysis code developed and utilised in this thesis to investigate disc-binary interactions driving photospheric depletion in post-AGB and post-RGB binaries (see Section~\ref{sec:mtheis}).}

\section{Spectroscopic data used for photospheric depletion study}\label{sec:mthdat}
Stellar spectral analysis has advanced significantly in recent years, driven by the growing availability of high-resolution spectroscopic data from both ground-based and space-based observatories. Ground-based surveys, such as \textit{RAVE-6} \citep{steinmetz2020RAVE}, \textit{WEAVE} \citep{jin2024WEAVE}, \textit{DESI} \citep{levi2019DESI}, \textit{GALAH DR3} \citep{buder2021GALAH}, \textit{APOGEE-2} \citep{majewski2017APOGEE}, and \textit{LAMOST-II} \citep{zong2020LAMOST}, provide extensive multiwavelength datasets that enable detailed studies of stellar and galactic properties. Additionally, space missions, such as \textit{JWST} \citep{gardner2006JWST}, \textit{Gaia} \citep{creevey2023Gaia}, \textit{TESS} \citep{ricker2015TESS}, and \textit{Euclid} \citep{serrano2024Euclid} complement these efforts by offering precise, atmosphere-free multiwavelength observations. Combining these high-resolution spectroscopic datasets enables detailed multiwavelength chemical studies, providing insights into stellar and exoplanetary atmospheres, binary interactions, and the broader chemical evolution of the Universe.

In this thesis, we investigate the chemical composition of post-AGB/post-RGB binaries in the Galaxy and in the Large Magellanic Cloud (2 full disc targets, 12 transition disc targets, and 9 dust-poor disc targets; see Tables~\ref{tab:varpro_paper1}, \ref{tab:tarlst_paper2}, and \ref{tab:sample_paper3}) using high-resolution optical and near-IR spectra obtained from three major spectroscopic facilities: HERMES/Mercator Telescope \citep[][see Section~\ref{ssec:mthdather}]{raskin2011hermes}, UVES/VLT \citep[][see Section~\ref{ssec:mthdatuvs}]{dekker2000UVES}, and the APOGEE survey \citep[][see Section~\ref{ssec:mthdatapo}]{jonsson2020DR16APOGEE}. To investigate the impact of binarity on the element/isotope production in post-AGB/post-RGB binaries, we derived precise atmospheric parameters, elemental abundances, and isotopic ratios from these spectra using E-iSpec (see Section~\ref{sec:mtheis}).

\begin{figure}[!ht]
    \centering
    \includegraphics[width=.99\linewidth]{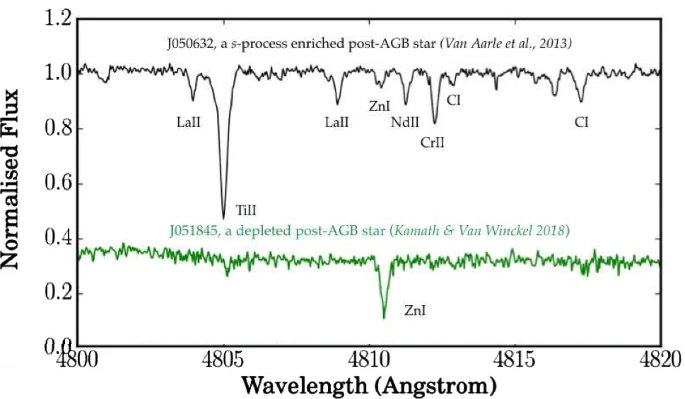}
    \caption[UVES/VLT spectra of J050632 (single) and J051845 (binary) showcasing chemical diversity observed in post-AGB stars]{UVES/VLT spectra of J050632 and J051845 showcasing chemical diversity observed in post-AGB stars. This figure was adapted from \citet{kamath2020Review}.}\label{fig:spc_methods}
\end{figure}

In Fig.~\ref{fig:spc_methods}, we illustrate the importance of high resolution (UVES/VLT data) for studying atomic lines in spectra of post-AGB/post-RGB stars. These examples illustrate the need for high spectral resolution ($R\,\gtrsim\,50\,000$) and broad wavelength coverage: optical spectra (UVES/VLT and HERMES/Mercator) are essential for resolving weak atomic lines of depleted refractory elements, while near-infrared spectra (APOGEE) are crucial for accessing molecular bands (including isotopologues of CO, CN, and OH) that allow the derivation of CNO isotopic ratios (see Section~\ref{sssec:abuiso_paper1}).

\subsection{HERMES/Mercator}\label{ssec:mthdather}
\begin{figure}[!ht]
    \centering 
    \includegraphics[height=5.5cm]{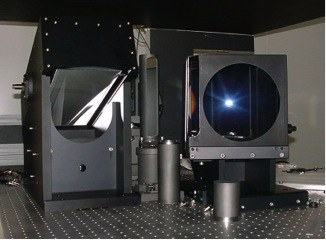}
    \includegraphics[height=5.5cm]{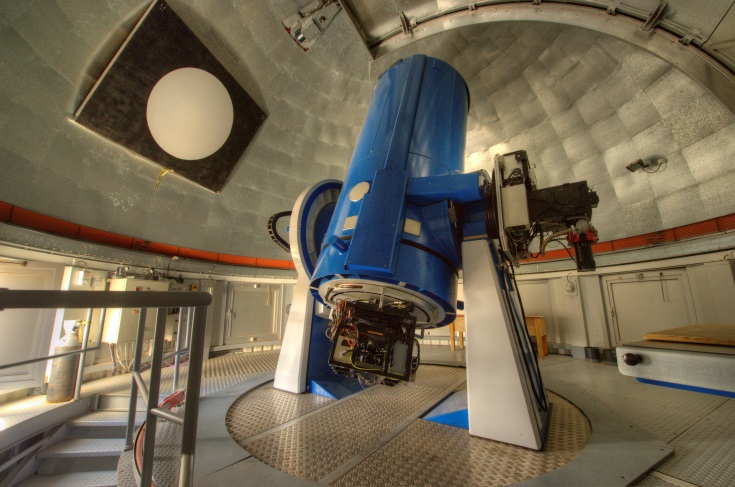}
    \caption{HERMES (\textit{left panel}) and Mercator (\textit{right panel}). For more details, see Section~\ref{ssec:mthdather}. Image courtesy of KU Leuven.}\label{fig:her_methods}
\end{figure}

HERMES is mounted on the 1.2 m Mercator Telescope at the Roque de los Muchachos Observatory in La Palma, Spain. HERMES is a high-resolution spectrograph (R$\sim$85\,000) operating in the optical wavelength range (377–900 nm). HERMES is widely used for asteroseismology and binarity research, but the sensitivity and high resolution of this instrument also make HERMES well-suited for detailed abundance studies of optically bright post-AGB/post-RGB binaries. Furthermore, HERMES plays a key role in the ongoing long-term monitoring program of Galactic post-AGB/post-RGB binaries in the Northern Hemisphere, which began in June 2009 \citep{vanwinckel2018Binaries}. In Fig.~\ref{fig:her_methods} (image courtesy of KU Leuven), we show HERMES instrument and Mercator telescope.

\subsection{UVES/VLT}\label{ssec:mthdatuvs}
\begin{figure}[!ht]
    \centering 
    \includegraphics[height=5.5cm]{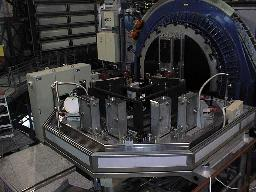}
    \includegraphics[height=5.5cm]{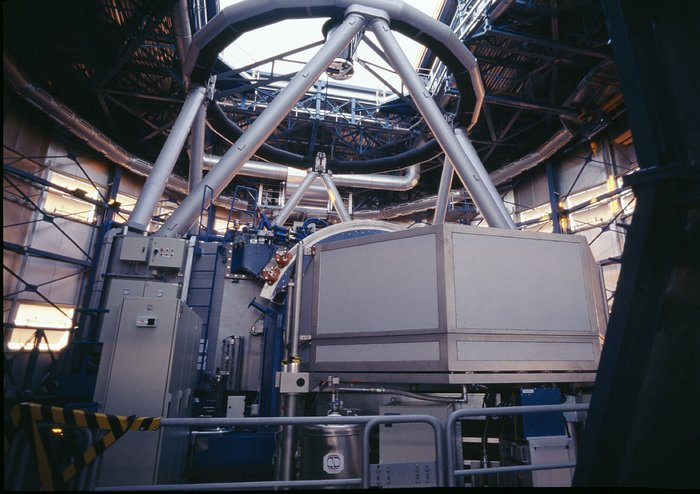}
    \caption{UVES (\textit{left panel}) and UT2/Kueyen (\textit{right panel}). For more details, see Section~\ref{ssec:mthdatuvs}. Image courtesy of ESO.}\label{fig:uvs_methods}
\end{figure}

UVES is located at the Nasmyth B focus of UT2/Kueyen, one of the unit telescopes of the Very Large Telescope (VLT) at the Paranal Observatory in Chile, part of the European Southern Observatory (ESO). UVES is a high-resolution echelle spectrograph (R$\sim$40\,000–110\,000) covering a wide spectral range from the near-ultraviolet to the red optical ($\sim$300–1100 nm). The large collecting area of the VLT further enhances the utility of UVES for faint targets (up to 18$^m$ in Blue arm and up to 19$^m$ in Red arm; see Table~\ref{tab:spcfac_methods}). The resolution, sensitivity, and stability of the UVES/VLT enable precise abundance analysis of a broad set of chemical elements, which is critical for studying the detailed chemical properties of post-AGB/post-RGB binaries. In Fig.~\ref{fig:uvs_methods} (image courtesy of ESO), we show the UVES instrument and the UT2/Kueyen telescope.

\begin{table}[!ht]
    \centering
    \caption{Key parameters of the spectroscopic facilities used in this thesis (see Section~\ref{sec:mthdat}).\\} \label{tab:spcfac_methods}
    \begin{tabular}{|c|ccc|}\hline
         \textbf{Parameter~\textbackslash~Facility} & \textbf{HERMES/Mercator} & \textbf{UVES/VLT} & \textbf{APOGEE} \\\hline
         Spectral resolution ($\frac{\lambda}{\Delta\lambda}$) & 85\,000 & \begin{tabular}{c}80\,000 (Blue)\\110\,000 (Red)\end{tabular} & 22\,500 \\
         Wavelength range ($\mu$m)  & 0.377-0.900 & \begin{tabular}{c}0.300-0.500 (Blue)\\0.420-1.100 (Red)\end{tabular} & 1.51-1.70 \\
         Diameter (m) & 1.2 & 8.2 & 2.5 \\
         \begin{tabular}{c}Limiting magnitude\\(mag)\end{tabular} & 13.4 (V) & \begin{tabular}{c}17-18 (V; Blue)\\18-19 (V; Red)\end{tabular} & 13.8 (H) \\\hline
    \end{tabular}\\
\end{table}

\subsection{APOGEE survey}\label{ssec:mthdatapo}
\begin{figure}[!ht]
    \centering 
    \includegraphics[height=5.5cm]{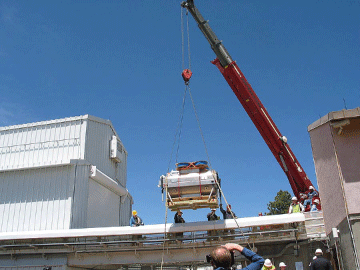}
    \includegraphics[height=5.5cm]{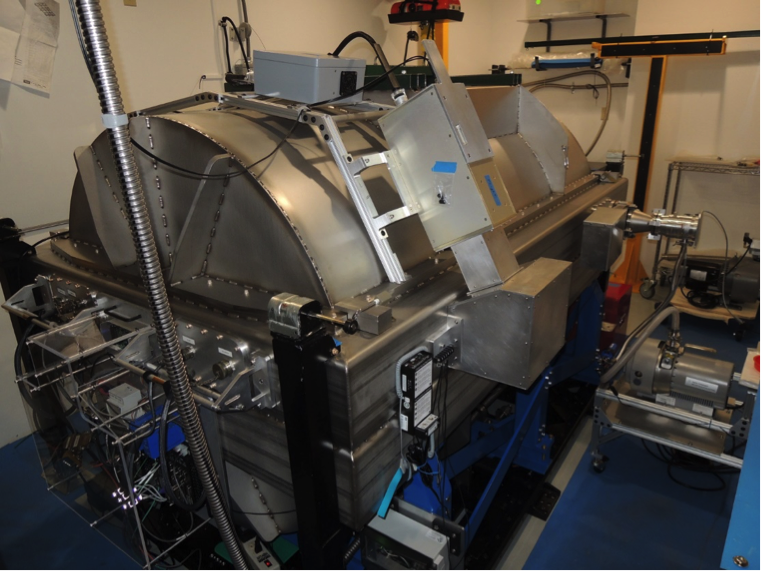}
    \caption{APOGEE-N (\textit{left panel}) and APOGEE-S (\textit{right panel}). For more details, see Section~\ref{ssec:mthdatapo}. Image courtesy of SDSS.}\label{fig:apo_methods}
\end{figure}

The APOGEE survey, part of the Sloan Digital Sky Survey (SDSS), provides high-resolution (R$\sim$22,500) near-infrared spectra (1.5–1.7 $\mu$m) of stars in the Galaxy and its satellites, such as the Magellanic Clouds, the Sagittarius Dwarf, Fornax Dwarf, and Gaia Sausage/Enceladus (GSE) system \citep{hasselquist2021APOGEEgalaxies}. The near-infrared wavelength coverage of APOGEE enables the analysis of not only atomic lines but also molecular bands, including CO, CN, OH, and C$_2$, underscoring APOGEE's significance as a crucial data source for dusty post-AGB/post-RGB binaries. In Appendix~\ref{tabA:allapo_paper1}, we provide the full sample of post-AGB/post-RGB binaries observed with APOGEE. In Fig.~\ref{fig:apo_methods} (image courtesy of SDSS), we show APOGEE-N and APOGEE-S instruments. The joint analysis of APOGEE near-IR spectra with data from optical facilities enables derivation of CNO isotopic ratios, as well as abundances of chemical elements with small number of features ($\lesssim\,5$ spectral lines) in the optical range (see Section~\ref{ssec:mththrrtc} for more details on the derivation process).

\section{Theoretical framework underpinning stellar spectral analysis}\label{sec:mththr}
In this section, we provide a brief outline of relevant fundamentals of stellar spectral analysis. For a more detailed discussion, see \citet{bohmvitense1992StellarAstrophysics}, \citet{gray2005StellarPhotospheres}, and \citet{leblanc2010StellarAstrophysics}, among others. In Section~\ref{ssec:mththrequ}, we present the basic equations of stellar structure, underlying stellar model atmospheres. In Section~\ref{ssec:mththrrtc}, we outline the key assumptions and equations of radiative transfer theory that form the basis of modern spectral analysis. In Section~\ref{ssec:mththrmda}, we discuss key approximations of radiative transfer in stellar model atmospheres. In Section~\ref{ssec:mththropa}, we highlight the opacity sources in stellar atmospheres, leading to formation of spectral absorption lines, which we analyse in this thesis. In Section~\ref{ssec:mththrews}, we explore the equivalent widths, which are used as a primary observational parameter of stellar spectral analysis. In Section~\ref{ssec:mththrcog}, we present the basics of curve of growth method, which forms the foundation of abundance analysis. In Section~\ref{ssec:mththratm}, we detail the atmospheric parameters and their convergence conditions.

\subsection{Basic equations of stellar structure}\label{ssec:mththrequ}
The stellar structure equations describe the internal properties and dynamics of stars, providing a theoretical framework to understand stellar evolution. These equations are based on the following basic concepts:
\begin{itemize}
    \item Hydrostatic equilibrium: the inward gravitational force is balanced by the outward pressure gradient at every point within the star, given by $\frac{dP_g(r)}{dr} = -\frac{GM(r)\rho(r)}{r^2} = -g(r)\rho(r)$, where $P_g$ is the gas pressure, $M$ is the mass of stellar interior, $\rho$ is the mass density, and $g$ is the gravitational acceleration.
    \item Mass conservation: the mass is distributed continuously within the star, given by $\frac{dM(r)}{dr} = 4\pi r^2\rho(r)$.
    \item Energy conservation: the rate of energy production per unit volume depends on the density, temperature, and the specific nuclear reactions occurring, given by $\frac{dL(r)}{dr} = 4\pi r^2\rho(r)\varepsilon(r)$, where $L$ is the luminosity and $\varepsilon$ is the emission coefficient (emissivity).
    \item Radiative transport: the energy moves outward from stellar core to the surface, either through radiation, convection, or conduction, given by $\frac{dT(r)}{dr} = -\frac{3}{4ac}\frac{\bar{\kappa}\rho(r)}{T^3(r)}\frac{L(r)}{4\pi r^2}$, where $T$ is the temperature, $\bar{\kappa}$ is the mean opacity (see Section~\ref{ssec:mththropa}), $a$ and $c$ are the Stefan-Boltzmann constant and the speed of light, respectively. Alternatively, a convective transport is given by $\frac{dT(r)}{dr} = \left(1-\frac{1}{\gamma}\right)\frac{T(r)}{P(r)}\frac{dP(r)}{dr}$, where $\gamma$ is the adiabatic constant.
    \item Equation of state, given by $P(r)=\frac{k\rho(r)T(r)}{\mu m_H}$, where $k$ is the Boltzmann constant, $\mu$ is the molecular weight, and $m_H$ is the mass of H atom.
    \item Boundary conditions (for $r\rightarrow0$: $M(r)\rightarrow0$ and $L(r)\rightarrow0$; for $r\rightarrow R_\ast$: $T(r)\rightarrow0$, $P(r)\rightarrow0$, $\rho(r)\rightarrow0$.
\end{itemize}

These equations and conditions offer a basic framework for stellar structure, which are further refined to create more realistic models of stellar atmospheres.

\subsection{Radiative transfer calculations}\label{ssec:mththrrtc}
Radiative transfer (RT) calculations are crucial for understanding how energy and radiation propagate through stellar atmospheres. In this thesis, we use spectral analysis tools based on RT calculations to interpret stellar spectra and to derive atmospheric parameters, elemental abundances, and isotopic ratios.

The propagation process is governed by the RT equation, which accounts for the absorption, emission, and scattering of radiation as it travels through a stellar atmosphere. The general RT equation describes how intensity of radiation $I_\lambda\,=\,\frac{dE}{d\omega d\sigma d\lambda}$ (the amount of energy per second $E$ going to the opening of the cone with solid angle $d\omega$ through the area $d\sigma$ per wavelength band $d\lambda$) at a given wavelength $\lambda$ changes in the outer layer of the stellar atmosphere along a path through the stellar medium given by:
\begin{equation}\label{equ:rte0}
    \dfrac{dI_\lambda}{ds} = \varepsilon_\lambda - \alpha_\lambda I_\lambda,
\end{equation}
where $s$ is the spatial coordinate along the ray, $\varepsilon_\lambda$ is the emission coefficient, and $\alpha_\lambda$ is the absorption coefficient. For simplicity, this equation is often rewritten in terms of the source function $S_\lambda=\frac{\varepsilon_\lambda}{\alpha_\lambda}$ given by:
\begin{equation}\label{equ:rtegen}
    \dfrac{dI_\lambda}{\alpha_\lambda ds} = \dfrac{dI_\lambda}{d\tau_{\lambda s}} = S_\lambda-I_\lambda,
\end{equation}
where $\tau_{\lambda s}$ is the optical depth along $s$ for the outgoing beam of light.

In the simplest case of LTE, $S_\lambda$ in each atmospheric layer of the star equals the Planck function, given by
\begin{equation}\label{equ:planck}
    B_\lambda\,(T)\,=\,\frac{2hc^2}{\lambda^5\left[\exp\left(hc/\lambda kT\right)-1\right]},
\end{equation}
which only depends on the local temperature $T$. However, the conditions inside real stellar atmospheres often depart from LTE, requiring more sophisticated treatments that account for reduced collisional interactions in photospheres of stars with low $\log g$ and/or low [Fe/H] (see Section~\ref{ssec:mththrmda}).

\subsection{Numerical approximations of model atmospheres}\label{ssec:mththrmda}
Analytical solutions to the general RT equation (see Eq.~\ref{equ:rtegen}) do not exist due to the complexity of real stellar atmospheres. Instead, numerical methods are employed, assuming either plane-parallel or spherically-symmetric approximation. Plane-parallel approximation is suitable for stars with thin atmospheres (in relation to the stellar radius), treating the layers as horizontal slabs. Introducing perpendicular optical depth $d\tau_\lambda = -\cos\theta d\tau_{\lambda s}$ (sign change is due to opposite directions of stellar radius and optical depth) results in the RT equation for the plane-parallel approximation of stellar atmosphere given by:
\begin{equation}\label{equ:rteppa}
    \cos\theta\dfrac{dI_\lambda (\tau_\lambda,\,\theta)}{d\tau_\lambda} = I_\lambda(\tau_\lambda,\,\theta)-S_\lambda.
\end{equation}

In contrast, spherically-symmetric approximation is suitable for extended atmospheres of giant stars or stars with strong winds. The RT equation for spherically-symmetric atmospheres is given by:
\begin{equation}\label{equ:genssa}
    \cos\theta\dfrac{\partial I_{\lambda\,\theta}}{\partial r}\,-\,\dfrac{\sin{\theta}}{r}\cdot\dfrac{\partial I_{\lambda\,\theta}}{\partial\theta} =\varepsilon_\lambda\,-\,\alpha_\lambda I_{\lambda\,\theta}.
\end{equation}

Discretisation of the spatial and wavelength domains is fundamental to numerical solutions. The RT equation can be solved over small segments of a ray path, balancing local accuracy with computational efficiency (short characteristics). Alternatively, the RT equation can be integrated over the entire ray, offering better accuracy at a higher computational cost (long characteristics). Iterative methods refine the solution, particularly in optically thick regions where radiation and matter are tightly coupled. A basic technique is Lambda iteration, but it suffers from slow convergence in high-opacity cases. Modern alternatives, including accelerated Lambda iteration (ALI) and operator splitting, dramatically improve efficiency of RT calculations for a wide range of opacities. Advancements in RT calculations enabled precise determination of the relationship between elemental abundances and observed line profiles (curve of growth method; see Section~\ref{ssec:mththrcog}).

\textbf{3D and $\langle$3D$\rangle$ model atmospheres}

Recent advancements in spectral analysis include the incorporation of 3D hydrodynamical model atmospheres or their time-averaged analogues, $\langle$3D$\rangle$ models, which significantly improved the accuracy of abundance determinations. Traditional 1D static models, which have been the cornerstone of stellar spectral analysis for decades, assume that stellar atmospheres are stratified into concentric layers with uniform conditions at each layer. 1D models simplify radiative transfer calculations and are computationally efficient, making them widely applicable. However, 1D models are limited by their inability to account for horizontal inhomogeneities and time-dependent processes, such as convection and turbulence, that play a critical role in shaping the spectra of many stars. Hence, the microturbulent velocity $\xi_t$ is introduced in 1D models to indirectly account for line broadening caused by convection and turbulence.

To overcome these limitations, 3D hydrodynamical models have been developed, providing a more realistic representation of stellar atmospheres. 3D models simulate the complex interplay between convective motions, turbulence, and temperature fluctuations across the stellar surface, capturing the dynamic nature of these environments. These processes are the most prominent for stars with extended or metal-poor atmospheres \citep[see, e.g.,][and references therein]{amarsi2020NLTEgalah, masseron20213DNLTE}. $\langle$3D$\rangle$ models offer a compromise between the simplicity of 1D models and the computational intensity of full 3D simulations. By averaging the spatial and temporal variations of 3D models, $\langle$3D$\rangle$ models retain essential physical effects, such as temperature gradients and convective inhomogeneities, while being less computationally heavy for abundance analyses \citep{magic2013Stagger, zhou2023Stagger}.

\textbf{Assumption of thermodynamical equilibrium}

In addition to the structural improvements introduced by 3D and $\langle$3D$\rangle$ models, advances in accounting for departures from LTE further enhanced the accuracy of abundance determinations. In stellar interiors, the LTE assumptions are known to hold, defined as:
\begin{itemize}
    \item Maxwell velocity distribution, which describes the number of particles $N$ of mass $m$ with a velocity amplitude between $v$ and $v + dv$, given by $\frac{N(v)dv}{N_{\rm total}} = \left(\frac{m}{2\pi kT}\right)^\frac{3}{2}\cdot4\pi v^2e^{-\frac{mv^2}{2kT}}dv$, where $N_{\rm total}$ is the total number of particles, $k$ is the Boltzmann constant, and $T$ is the local temperature,
    \item Boltzmann excitation equilibrium, which describes the number of atoms or molecules occupying a specific excited energy state, given by $\frac{N_{i,s}}{N_i}=\frac{g_{i,s}}{U_i}\cdot e^{-\frac{\chi_{i,s}}{kT}}$, where $N_{i,s}$ is the number density of ionisation state $i$ in excitation level $s$, $N_i$ is the number density of ionisation state $i$ in all excitation levels, $g_{i,s}$ is the statistical weight of ionisation state $i$ in excitation level $s$, $U_r$ is the partition function ($U_r=\sum_s g_{i,s}\cdot e^{-\frac{\chi_{i,s}}{kT}}$), $\chi_{i,s}$ is the excitation energy of the ionisation state $i$ in excitation level $s$ ($\chi_{i,s}=E_{i,s}-E_{i,0}$),
    \item Saha ionisation equilibrium, which describes the ratio between ions of subsequent ionisation levels, given by $\frac{N_{i+1}}{N_i}=\frac{0.665}{P_e}\cdot\frac{U_{i+1}}{U_i}\cdot T^\frac{5}{2}\cdot10^{-\frac{5040}{T}\cdot\chi}$, where $P_e$ is the electron pressure ($P_e=N_e kT$),
    \item individual Planck source function in each layer of stellar atmosphere (see Eq.~\ref{equ:planck}).
\end{itemize}

However, in collision-poor atmospheres of hot metal-poor giants, radiation field dominates over collisions, altering the distribution of level populations. Hence, derivation of atmospheric parameters and elemental abundances in NLTE requires the simultaneous solution of the RT equation (see Eq.~\ref{equ:rtegen}) and the kinetic (statistical) equilibrium equations \citep[instead of Boltzmann distribution;][]{amarsi2016NLTEiron, amarsi2020NLTEgalah} given by:
\begin{equation}\label{equ:rtenlt}
    \sum_j(n_i P_{ij}-n_j P_{ji})=0,
\end{equation}
where $n_i$ and $n_j$ are the populations of levels $i$ and $j$, and $P_{ij}$ includes radiative and collisional transition probabilities. NLTE modelling is computationally demanding, as it involves iterative solutions that ensure consistency between the radiation field, level populations, and temperature structure. Solving the radiative transfer equations in a kinetic equilibrium framework provides grids of departure coefficients $\beta_i = \frac{n_i^{\rm NLTE}}{n_i^{\rm LTE}}$ for level populations $n_i$. Using the combination of departure coefficients, observed equivalent widths, and derived LTE atmospheric parameters, the relative NLTE correction $\Delta_i = [{\rm X/H}]_i^{\rm NLTE} - [{\rm X/H}]_i^{\rm LTE}$ can be calculated for each individual spectral line $i$.

Despite the recent advancements in spectral analysis tools, the application of 3D and $\langle$3D$\rangle$ models remains computationally heavy, limiting their widespread adoption, unlike the presence of NLTE calculations. Furthermore, many current tools employing 3D/$\langle$3D$\rangle$ models and NLTE corrections are optimised for MS stars or a narrow range of stellar types, leaving a gap in their applicability to more complex systems, including dusty and chemically peculiar post-AGB/post-RGB binaries (see Section~\ref{sec:intpAR}). This highlights the importance of developing analysis methods suited to these stars, making full use of modern spectroscopic datasets (see Section~\ref{sec:mthdat}).

\subsection{Origins of opacity}\label{ssec:mththropa}
Opacity ($\kappa_\lambda(r) = -\frac{1}{\rho(r) I_\lambda(r)}\frac{dI_\lambda(r)}{dr}$) quantifies how strongly radiation is absorbed or scattered by atoms (primarily, H$^-$, He, and Fe) in stellar atmospheres. Accurately modelling opacity requires detailed knowledge of atomic and molecular physics, including transition probabilities, energy levels, and collisional cross-sections. For complex atmospheres of cool stars with molecular absorption bands ($T_{\rm eff}\lesssim4\,000$) or highly ionised hot stars ($T_{\rm eff}\gtrsim10\,000$), opacity becomes wavelength-dependent and computationally challenging.

In stellar atmospheres, opacity is influenced by several key processes:
\begin{itemize}
    \item Electron (Thomson) scattering: A dominant source of opacity in high-temperature, low-density atmospheres, where free electrons scatter radiation with a wavelength-independent cross-section, given by $\sigma_{el}\,=\,\frac{8\pi}{3}\left(\frac{e^2}{mc^2}\right)^2$.
    \item Free-free transitions: Occur when an electron moves in a hyperbolic orbit around an ion and absorbs a photon, increasing its energy.
    \item Bound-bound and bound-free transitions: Bound-bound transitions correspond to electron transitions between discrete energy levels, producing absorption and emission lines in stellar spectra (e.g., the Balmer series in the Sun or molecular bands in cool stars). Bound-free transitions occur during photoionization, where an electron absorbs a photon and escapes from an atom or ion, contributing to the continuum opacity.
    \item Conduction (conductive opacity): A diffusive heat transport mechanism, analogous to radiative transfer, but with electrons as the primary energy carriers instead of photons.
\end{itemize}

Opacity governs the absorption and scattering of photons, shaping the line formation. Variations in opacity as a function of wavelength influence the shape and equivalent width of spectral lines (see Section~\ref{ssec:mththrews}). Accurate treatment of opacity is essential for the analysis of realistic model atmospheres (see Section~\ref{ssec:mththrmda}), providing critical information about atmospheric parameters and elemental abundances (see Section~\ref{ssec:mththratm}).

\subsection{Concept of equivalent width}\label{ssec:mththrews}
\begin{figure}[!ht]
    \centering
    \includegraphics[width=.75\linewidth]{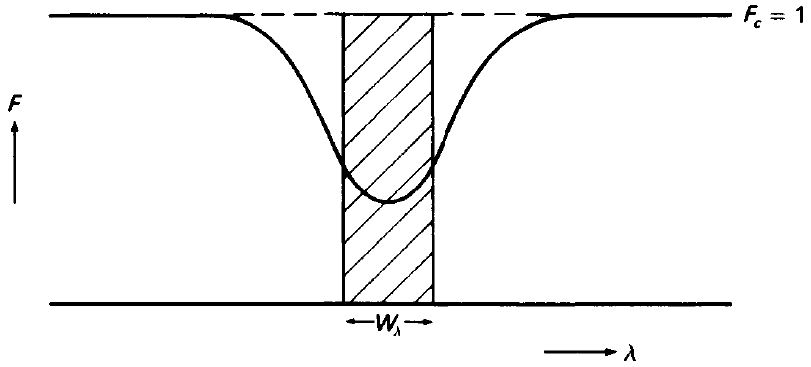}
    \caption[Equivalent width of a normalised absorption line]{Equivalent width of a normalised absorption line. The hatched rectangle has a width $W_\lambda\,=\,EW$, defined such that hatched area is equal to the area of the absorption line. This figure was adapted from \citet{bohmvitense1992StellarAstrophysics}}\label{fig:ews_methods}
\end{figure}

The formation and strength of spectral lines in stellar atmospheres is governed by the opacity, which dictates the transport of radiation through the stellar atmosphere (see Section~\ref{ssec:mththropa}), directly influencing the depth and shape of spectral lines. Equivalent width ($EW$) provides a quantifiable measure of absorption features, serving as a critical diagnostic tool for derivation of atmospheric parameters and elemental abundances given by:
\begin{equation}
    EW = \int_{line}R_\lambda d\lambda = \int_{line}\left(1-\dfrac{F_\lambda}{F_{cont}}\right)d\lambda,
\end{equation}
where $R_\lambda$ is the line depth at a given wavelength, $F_\lambda$ is the flux at a given wavelength, and $F_{cont}$ is the continuum flux used to normalise $F_\lambda$. In Fig.~\ref{fig:ews_methods} \citep[adapted from][]{bohmvitense1992StellarAstrophysics}, we show an example of an absorption line profile with a rectangle of width $W_\lambda\,=\,EW$ marking region of the same area. We note that interpretation of EWs is inherently tied to assumptions about temperature stratification, pressure broadening, and radiative transfer (see Section~\ref{ssec:mththrcog}).

\subsection{Curve of growth method}\label{ssec:mththrcog}
\begin{figure}[!ht]
    \centering
    \includegraphics[width=.75\linewidth]{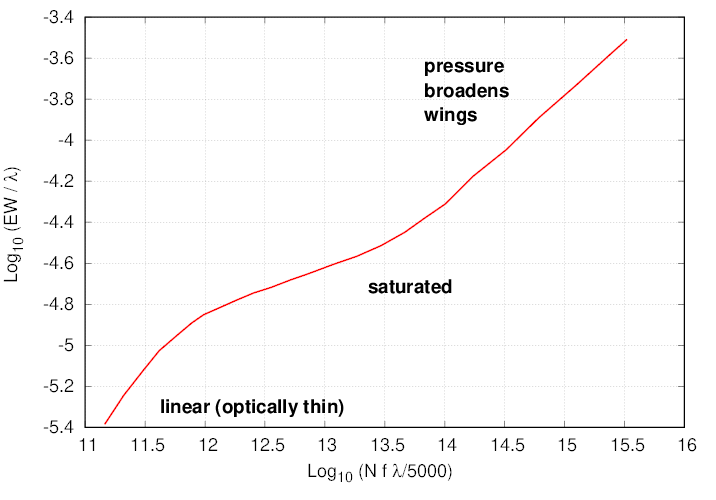}
    \caption[A typical curve of growth showing dependence of the observed reduced EW ($rEW$) on the number abundance $N$ of an ion]{A typical curve of growth showing dependence of the observed reduced EW ($rEW$) on the number abundance $N$ of an ion. Three marked parts Equivalent width of a normalised absorption line. The hatched rectangle has a width $W_\lambda$, defined such that hatched area is equal to the area of the absorption line. This figure was adapted from \citet{carroll1996CoG}.}\label{fig:cog_methods}
\end{figure}

In stellar spectroscopy, curves of growth (CoG) are functions that relate the reduced EW of an atomic spectral line to the number of absorbing atoms of ion i. In Fig.~\ref{fig:cog_methods} \citep[adapted from][]{carroll1996CoG}, we show a typical curve of growth, which has three distinct parts:
\begin{itemize}
    \item Linear part, where $rEW\,\sim\,N_i$ (few absorbers, small optical depth, weak lines).
    \item Saturated part, where $rEW\,\sim\,\log N_i$ (spectral line has a saturated core and slowly changing wings, $\tau\,\gtrsim\,5$).
    \item Damping part, where $rEW\,\sim\,\sqrt{N_i}$ (Lorentzian wings due to collisional broadening dominating Doppler wings).
\end{itemize}

For weak spectral lines ($\tau\,\ll\,1$) forming in a cold homogeneous atmosphere layer, the linear relationship between the observed EW and derived abundance for a given ion is defined as:
\begin{equation}\label{equ:mthcog}
    \log(rEW)\,=\,\log\left(\dfrac{\pi e^2}{m_e c^2}\right) + \log N_i + \log(gf\lambda),
\end{equation}
where
\begin{itemize}
    \item $rEW\,=\,\frac{EW}{\lambda}$ is the reduced equivalent width,
    \item $\log\left(\dfrac{\pi e^2}{m_e c^2}\right)$ is the term containing physical constants $e$, $m_e$, and $c$ (elemental charge, electron mass, and speed of light in vacuum, respectively),
    \item $N_i$ is the number abundance (column density) of ion $i$, and
    \item $gf$ is the transition probability.
\end{itemize}

\textbf{Empirical curves of growth}

Since the column density $N_i$ is not directly observable, empirical CoGs instead represent $\log(rEW)$ as a function of $\log(gf\lambda)$. To construct an empirical CoG, we adopt the Milne-Eddington approach, which assumes LTE, linearity of source function $S_\lambda$ with optical depth $\tau$, and independence of absorption in the line and in the continuum from optical depth $\tau$. Under these conditions, the empirical CoG is given by:

\begin{equation}
    \log(rEW)\,=\,\log\left(\dfrac{\pi e^2}{m_e c^2}\cdot\dfrac{C_0}{U_r}\cdot\dfrac{N_{i+1}}{N_i}\cdot N_H\right) + \log A(X) + \log(gf\lambda) - \dfrac{5040}{T}\cdot\chi - \log\kappa_\lambda,
\end{equation}
where
\begin{itemize}
    \item $C_0$ is the parameter describing the flux contrast between line and continuum,
    \item $U_r$ is the partition function,
    \item $\frac{N_{i+1}}{N_i}$ is the term from Saha equation (see Section~\ref{ssec:mththrmda}),
    \item $N_H$ is the number abundance of H,
    \item $A(X)$ is the abundance of element X ($A(X) = \log\varepsilon(X) = \log\frac{N_X}{N_H}$),
    \item $gf$ is the oscillator strength,
    \item $T$ is the local temperature in the atmospheric layer,
    \item $\chi$ is the excitation potential of the lower energy level of the line, and $\kappa_\lambda$ is the opacity (see Section~\ref{ssec:mththropa}).
\end{itemize}

Determining elemental abundances relies on extensive input data, primarily supplied by model atmospheres, which include local parameters of atmosphere layers (including temperature $T$, pressure $P$, and optical depth $\tau$) as well as spectral line lists that provide essential atomic properties (including rest wavelengths $\lambda_0$, lower excitation potentials $\chi$, and oscillator strengths $\log gf$). However, many spectral lines have poorly constrained $\log gf$ values, introducing substantial uncertainties. Even minor variations in these values can lead to significant changes in the derived abundances. Consequently, the impact of $\log gf$ uncertainties must be carefully considered in the derivation of atmospheric parameters, elemental abundances, and isotopic ratios.

The CoG method is implemented in atmospheric parameter and abundance analysis in two distinct ways: EW method and synthetic spectral fitting (SSF) technique. Using EW method, EWs of spectral lines are directly related to the line strengths $\log gf$ and can be used to infer the elemental abundance. Alternatively, SSF technique involves generating theoretical (synthetic) spectra from stellar atmosphere models and fitting these spectra to the observed profile of spectral line. Unlike EW method, SSF technique allows for a more detailed analysis of complex combinations of multiple spectral features (blends). However, SSF technique requires an additional parameter ($v_{\rm broad}$) and is generally computationally heavier than EW method. In this thesis, we derive atmospheric parameters, elemental abundances, and isotopic ratios using: i) EW method for unblended atomic lines in optical and near-IR ranges and ii) SSF technique for molecular bands in near-IR range.

\subsection{Outputs of spectral analysis}\label{ssec:mththratm}
\textbf{Atmospheric parameters}

In this thesis, we use EW method and SSF technique to investigate the impact of binarity on element/isotope production in post-AGB/post-RGB binaries. To perform the abundance analysis of high-resolution optical and near-IR spectra, we first need to derive the following atmospheric parameters:
\begin{itemize}
    \item Effective temperature $T_{\rm eff}$ is the temperature at the optical depth $\tau=2/3$ in a grey atmosphere ($\alpha_\lambda\,=\,\alpha$, $\tau_\lambda\,=\,\tau$) with a linear depth dependence of the source function. This parameter converges when excitation equilibrium is enforced (derived [Fe/H] abundances are independent of excitation potential of spectral lines).
    \item Surface gravity $\log g$ connects stellar physical parameters (including $M(r)$ and $\rho(r)$) to the local parameters governing the propagation of radiation (including $\kappa(r)$ and $\varepsilon(r)$). This parameter converges when ionisation equilibrium is enforced (abundances derived from \ion{Fe}{i} and \ion{Fe}{ii} lines match).
    \item {[Fe/H]} abundance is commonly used as a proxy for the global metallicity of the star due to ample number of \ion{Fe}{i} and \ion{Fe}{ii} lines in optical spectra of AFGKM stars. This parameter converges when $T_{\rm eff}$, $\log g$, and $\xi_t$ are converged.
    \item Microturbulent velocity $\xi_t$ implicitly represents microturbulence in 1D model atmospheres, and is given by $P_t\,=\,\rho\xi_t^2$. This parameter is obsolete in 3D model atmospheres, as the microturbulence in 3D models is accounted for explicitly. This parameter converges when line intensity balance is enforced (derived [Fe/H] abundances are independent of EW$_\lambda$ of spectral lines).
    \item Macroturbulent velocity $v_{\rm macro}$, rotational velocity $v\sin i$, and resolution $R$ are parameters of SSF technique, but not of EW method. As these three parameters are degenerate, a universal broadening velocity $v_{\rm broad}$ allows to simulate the impact of all three parameters on the synthetic spectra.
\end{itemize}

\textbf{Elemental abundances}

Once atmospheric parameters are determined through spectral analysis of Fe lines, Eq.~\ref{equ:mthcog} is used for precise derivation of elemental abundances and isotopic ratios using EW method or SSF technique. However, to place elemental abundances within a broader astrophysical context, these abundances must be referenced against a well-defined abundance scale. The absolute scale of number abundances $N_X$ for element X is rarely used in stellar spectral analysis. One of the common abundance scales involves scaling to H abundance given by:
\begin{equation}\label{equ:abueps}
    A(X)\,=\,\log\varepsilon(X)\,=\,\log\left(\dfrac{N_X}{N_H}\right)+12,
\end{equation}
assuming $\log\varepsilon(H)\,=\,12$. Another commonly used abundance scale is relative to the Sun and is given by:
\begin{equation}\label{equ:abux_h}
    {\rm [X/H]}\,=\,A(X) - A_\odot(X)\,=\,\log\left(\dfrac{N_X}{N_H}\right) - \log\left(\dfrac{N_X}{N_H}\right)_\odot.
\end{equation}

Finally, elemental abundances may be also expressed relative to Fe given by:
\begin{equation}\label{equ:abuxfe}
    {\rm [X/Fe]}\,=\,{\rm [X/H]}-{\rm [Fe/H]}.
\end{equation}

In this thesis, we primarily use the [X/H] scale for elemental abundances of post-AGB/post-RGB binaries.

\section{E-iSpec: a solution for spectral analysis of evolved stars}\label{sec:mtheis}
As discussed in Section~\ref{sec:mththr}, spectral analysis is a cornerstone of stellar astrophysics, providing insights into the physical and chemical properties of stars. E-iSpec, a modified version of the widely used iSpec, is tailored to address the specific challenges associated with analysing the spectra of evolved stars, such as post-AGB/post-RGB binaries. In Fig.~\ref{fig:eis_methods}, we present a schematic structure of E-iSpec. In Appendices~\ref{tabA:tstsmp_paper1}, \ref{tabA:tstabs_paper1}, and \ref{tabA:tststp_paper1}, we provide the testing steps we invoked to ensure precise and reliable results of abundance analysis. In Section~\ref{ssec:mtheistls}, we discuss the modern spectral analysis tools and their limitations, which lead to the development of E-iSpec. In Section~\ref{ssec:mtheisovr}, we overview the analysis of stellar spectra in E-iSpec. In Section~\ref{ssec:mtheislim}, we present the current limitations of E-iSpec. In Section~\ref{ssec:mtheisapp}, we discuss the examples of E-iSpec applied to scientific cases beyond the scope of this thesis.

\subsection{Challenges in modern spectral analysis of evolved stars}\label{ssec:mtheistls}
The analysis of multiwavelength spectra is crucial for understanding processes happening in stellar and circumstellar environments of post-AGB/post-RGB binaries. Different wavelength regimes probe distinct physical processes and regions, from the hot, ionised stellar atmospheres visible in UV and optical wavelengths to the cooler, dusty circumstellar material visible in IR and radio spectra. By disentangling overlapping contributions from photospheric, chromospheric, and circumstellar emissions, multiwavelength studies provide a more complete picture of stellar systems. This approach is particularly important for evolved stars, where interactions between stellar winds, accretion flows, and surrounding discs or envelopes produce features across a broad spectral range.

The analysis of the modern spectral datasets requires robust tools. The most widely used radiative transfer codes include MOOG \citep{sneden1973moog, sneden2012Moog}, SYNTHE \citep{kurucz1993SYNTHE, sbordone2004SYNTHE}, SPECTRUM \citep{gray1994spectrum}, Turbospectrum \citep{alvarez1998turbospectrum, plez2012Turbospectrum, gerber2023Turbospectrum}, SYNSPEC \citep{hubeny2011SYNSPEC, hubeny2017SYNSPEC, hubeny2021SYNSPEC}, SME/pySME \citep{valenti1996SME, piskunov2017SME, wehrhahn2023pySME}, and Korg \citep{wheeler2023Korg}. Over the past decade, these tools continue to be refined to meet the demands of increasingly complex and high-quality spectral data.

In this thesis, we analyse the high-precision spectral data from HERMES/Mercator, UVES/VLT, and APOGEE survey (see Section~\ref{sec:mthdat}). The APOGEE survey has an in-built spectral tool to derive atmospheric parameters and elemental abundances effectively for many stellar populations, called APOGEE Stellar Parameter and Chemical Abundances Pipeline (ASPCAP). However, the applicability of ASPCAP to dusty giants, including post-AGB/post-RGB binaries \citep{schultheis2020APOGEE-AGB}, is limited due to relying on the MARCS grid of model atmospheres. This grid lacks coverage for key parameter ranges relevant to post-AGB/post-RGB binaries, such as $T_{\rm eff}$\,>\,6000\,K, $\log g$\,<\,1\,dex, and [Fe/H]\,<\,--2\,dex (see Table~\ref{tab:pagbpar_intro}), which hinders the effective analysis of these systems. This coverage problem is also common for other popular wrappers of radiative transfer codes, such as pySME \citep{wehrhahn2023pySME} or Brussels Automatic Code for Characterising High-accuracy Spectra \citep[BACCHUS;][]{masseron2016BACCHUS}, which offer flexible methods for abundance analysis.

\textbf{iSpec and E-iSpec}

iSpec is a spectral analysis tool, which addresses these challenges by integrating multiple radiative transfer codes, model atmospheres, and line lists. However, iSpec also has drawbacks, including outdated infrared line lists, reliance on older solar composition standards, and limited methods for abundance error analysis. Furthermore, the use of 1D LTE models in iSpec restricts iSpec's application to the complex atmospheric processes of post-AGB/post-RGB binaries, which are predominantly hot, metal-poor giants.

To address these gaps, we developed E-iSpec as a semi-automatic version of iSpec, incorporating updates, such as a revised infrared line list, modern solar composition standards, and a framework for abundance error analysis. In Section~\ref{ssec:mtheisovr}, we provide further details on the structure and implementation of E-iSpec, highlighting its advances for studying post-AGB/post-RGB binaries.

\subsection{Spectral analysis using E-iSpec}\label{ssec:mtheisovr}
In this subsection, we outline the key steps of spectral analysis implemented in E-iSpec, such as: i) initial spectra preparation, ii) derivation of atmospheric parameters, elemental abundances, and isotopic ratios, and iii) applying NLTE corrections to the derived abundances. For more details on E-iSpec, see Section~\ref{ssec:saneis_paper1}.

The initial step in spectral analysis in E-iSpec involves preparing the observed spectra for accurate interpretation. Continuum normalisation is performed to isolate the intrinsic spectral features by correcting for the overall shape of the spectrum caused by instrumental response, interstellar extinction, and atmospheric effects. This step ensures that absorption and emission features can be analysed against a flat continuum baseline. Radial velocity correction is applied to align the observed spectra of the star with the rest frame, compensating for Doppler shifts caused by the stellar motion relative to the observer. Additionally, we remove the cosmic ray artifacts, which manifest as sharp, spurious spikes in the spectra. These steps ensure that the processed spectra accurately represent the intrinsic properties of the star, setting the foundation for precise derivation of atmospheric parameters and elemental abundances.

\textbf{Derivation of atmospheric parameters and elemental abundances}

Using the spectrum prepared for analysis, we derive the atmospheric parameters of the star (effective temperature $T_{\rm eff}$, surface gravity $\log g$, metallicity [Fe/H], and microturbulent velocity $\xi_t$). E-iSpec determines the atmospheric parameters through 1D LTE analysis of Fe lines, leveraging the fact that Fe offers numerous lines across a broad wavelength range in the parameter range of our sample (see Table~\ref{tab:pagbpar_intro}). The excitation equilibrium is achieved by ensuring that the derived Fe abundances are independent of the excitation potential of the lines (fixing $T_{\rm eff}$). Similarly, ionisation equilibrium is enforced by equating the abundances derived from \ion{Fe}{i} and \ion{Fe}{ii} lines (fixing $\log g$). Finally, the dependence of Fe abundances on line strength is minimised (fixing $\xi_t$). These iterative steps ensure a self-consistent solution for all atmospheric parameters ($T_{\rm eff}$, $\log g$, [Fe/H] and $\xi_t$).

With atmospheric parameters fixed, elemental abundances are derived using the EW method. This step relies on radiative transfer calculations based on 1D MARCS/ATLAS model atmospheres. For each element, the spectral lines are carefully selected to avoid blending. Elemental abundances of elements from C to Pb, as well as $^{12}$C/$^{13}$C isotopic ratios, are derived to investigate nucleosynthesis and mixing processes in post-AGB/post-RGB binaries, where processes such as third dredge-up, hot bottom burning, and binary interactions significantly modify the original chemical composition (see Chapter~\ref{chp:int}).

Since the 1D LTE assumption is used in E-iSpec, to account for departures from LTE, NLTE corrections are adopted from different software (Balder or pySME). We note that NLTE corrections for our sample of post-AGB/post-RGB binaries are the most significant for elements with strong lines influenced by radiative pumping, including Na, Mg, and Ba, where corrections can exceed 0.5\,dex. Incorporating NLTE corrections minimises the differences between the abundances of various ionisations of chemical elements (e.g., \ion{Cr}{i} and \ion{Cr}{II}), improving the reliability of depletion studies in post-AGB/post-RGB binaries (see Chapters~\ref{chp:pap2} and \ref{chp:pap3}).

\begin{figure}[!ht]
    \centering
    \includegraphics[width=.99\linewidth]{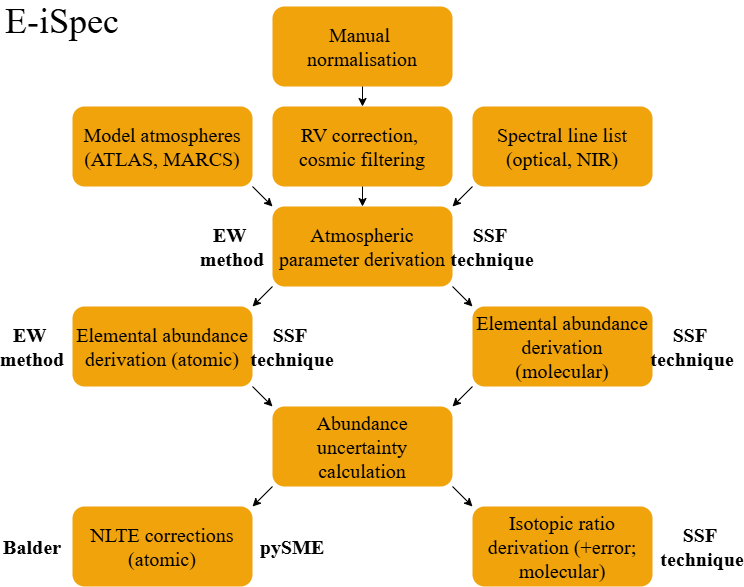}
    \caption[Schematic structure of E-iSpec]{Schematic structure of E-iSpec. The text inside orange boxes marks the steps of our chemical analysis, while the bolded text outside marks the available methods for the corresponding steps (EW method or SSF technique) and the software used for calculating the individual NLTE corrections (Balder or pySME). We note that we adopted solar abundances from \citet{asplund2021solar}.}\label{fig:eis_methods}
\end{figure}

\subsection{Limitations of E-iSpec}\label{ssec:mtheislim}
In this subsection, we outline the limitations of E-iSpec, assess their relevance to our scientific objectives, and highlight areas where refinements may be necessary. While some of these constraints do not significantly impact the specific goals of our study, others could influence the broader application of E-iSpec.

E-iSpec operates with 1D model atmospheres and does not include implementations for 3D or $\langle$3D$\rangle$ models. Although 3D models are more important for giants with low-density photospheres, radiative transfer studies of late-type stars \citep{amarsi2016NLTEiron, nordlander2017NLTEaluminium, norris20193DNLTEcorrections, bergemann2019NLTEMn} show that the average differences between 3D and 1D models for metal-poor giants are typically $\lesssim$0.1\,dex, which is significantly smaller than the typical differences between NLTE and LTE models for metal-poor giants \citep[up to 0.5\,dex;][]{amarsi2020NLTEgalah}. Given our current uncertainty thresholds ($\sim$0.1\,dex), we conclude that while 3D modelling could enhance precision in some cases, the absence of such functionality in E-iSpec does not significantly affect the abundance analysis of post-AGB/post-RGB binaries.

The error analysis in E-iSpec is currently based on bootstrapping, which tends to yield conservative estimates of uncertainties. A more precise approach, such as diagonalising the error covariance matrix, could provide tighter constraints on error estimates. However, the abundance uncertainties in our sample are generally below 0.2\,dex, which is sufficient for studying chemical depletion profiles. While refining the error analysis framework would improve the accuracy of abundance determinations for a broader sample of stars (e.g., precision of elemental abundances of MS stars is $\sim0.05$\,dex in current surveys, such as GALAH DR3 and APOGEE-2), the currently implemented error analysis in E-iSpec enables precise investigation of photospheric chemical depletion in post-AGB/post-RGB binaries.

Additionally, in the current version of E-iSpec, the atmospheric parameters and elemental abundances are derived using ATLAS model atmospheres (since MARCS grid is limited in the range of metal-poor giants), while NLTE corrections are calculated using MARCS model atmospheres. This hybrid approach introduces some inconsistencies, since the two grids are not fully interchangeable. However, these inconsistencies are minor ($\lesssim$0.1\,dex) and are considered of secondary importance for our study. Future work could explore the absolute effects of this assumption on individual spectral lines, which would provide a more rigorous evaluation of its impact.

E-iSpec is currently optimised to derive atmospheric parameters using Fe-line analysis, requiring at least few unblended atomic spectral lines of \ion{Fe}{i} and \ion{Fe}{ii}. This limits the application of E-iSpec to stars with effective temperatures 4000\,K$\lesssim T_{\rm eff}\lesssim$9000\,K. Additionally, E-iSpec is designed to derive elemental abundances and isotopic ratios for elements from C to Pb. While this functionality is sufficient for addressing our scientific goals, expanding E-iSpec’s capabilities would increase its versatility. Potential enhancements include streamlining derivation of atmospheric parameters using features, such as the Ca triplet or Balmer jump, and extending abundance analysis to lighter elements, such as He and Li. These advancements would broaden the scope of E-iSpec and enable its application to a wider range of stellar environments.

\subsection{Applications of E-iSpec to broader scientific investigations}\label{ssec:mtheisapp}
In this thesis, we use E-iSpec to explore the impact of binarity on stellar nucleosynthesis in chemically depleted post-AGB/post-RGB binaries (see Chapters~\ref{chp:pap1}, \ref{chp:pap2}, and \ref{chp:pap3}). Additionally, E-iSpec has been applied to studies investigating the origins of \textit{s}-process enrichment in post-AGB/post-RGB single stars and binaries \citep[][Menon et al., in prep.]{menon2024EvolvedBinaries}, highlighting the versatility of E-iSpec in different contexts.

\begin{figure}[!ht]
    \centering
    \includegraphics[width=.75\linewidth]{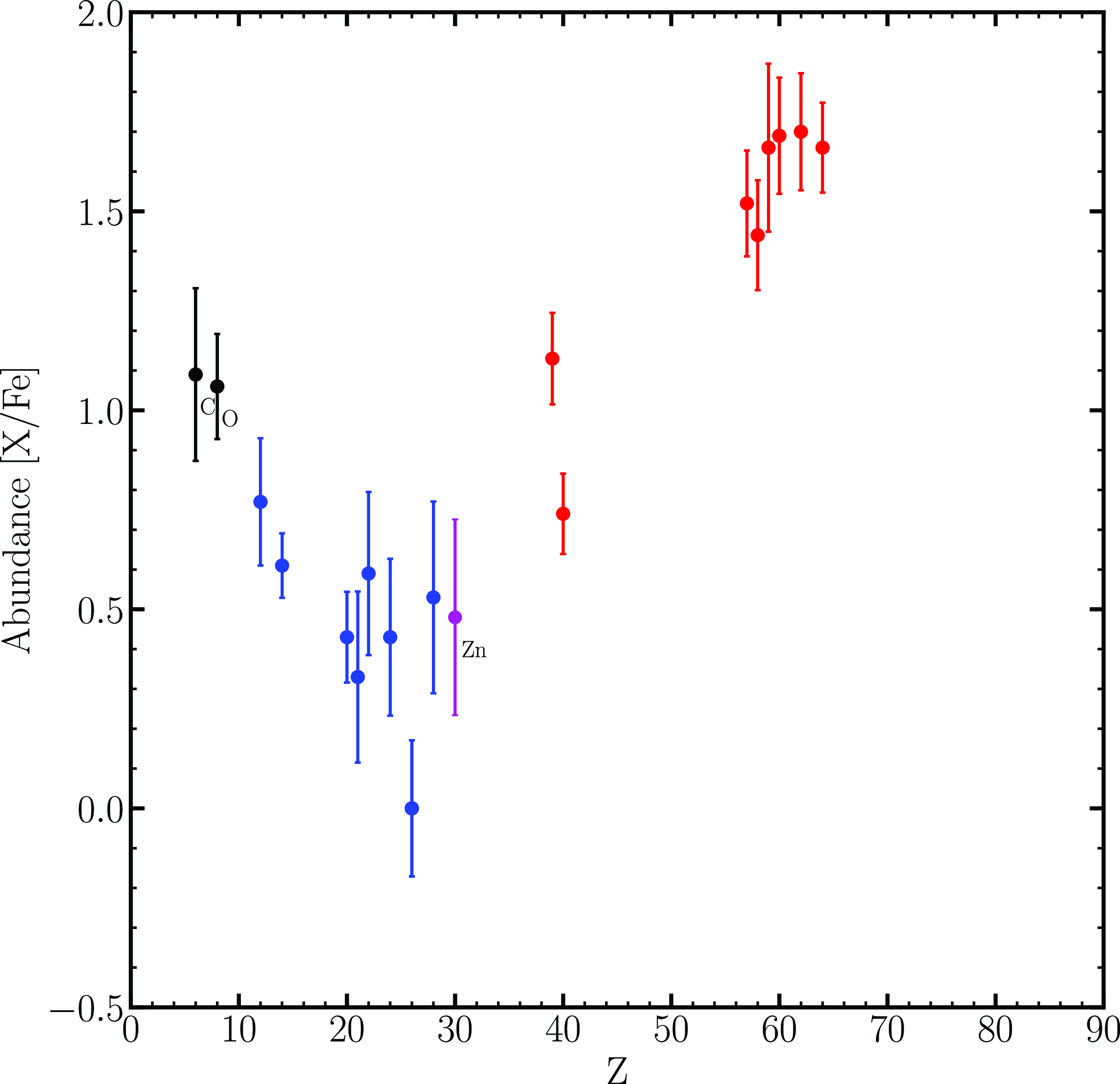}
    \caption[Spectroscopically derived abundances of J005107]{Spectroscopically derived abundances of J005107. Some elements are labelled for reference. The black colour data points represent CNO elements, the blue represents Fe-peak elements, the magenta represents Zn used as a proxy for volatile elements, and the red represents \textit{s}-process elements. This figure was adapted from \cite{menon2024EvolvedBinaries}.}\label{fig:mnn_methods}
\end{figure}

\citet{menon2024EvolvedBinaries} conducted an abundance analysis of post-AGB/post-RGB binaries both in the Milky Way and the Magellanic Clouds, revealing an enhancement in carbon and \textit{s}-process elements, contrary to the expected photospheric depletion. By combining derived atmospheric parameters and elemental abundances of post-AGB/post-RGB singles and binaries with predictions from ATON evolutionary models for AGB/RGB single stars, \citet{menon2024EvolvedBinaries} concluded that intrinsic enrichment is the most likely explanation for the observed chemical anomalies in \textit{s}-process enhanced post-AGB/post-RGB binaries. These results emphasise the need for further research into disc chemistry to better understand the role of disc-binary interactions in shaping the chemical properties of post-AGB/post-RGB binaries. In Fig.~\ref{fig:mnn_methods}, we present the elemental abundances of an \textit{s}-process enhanced post-AGB binary, 2MASS J00510723-7341334, as derived using E-iSpec \citep[adopted from][]{menon2024EvolvedBinaries}.

Menon et al. (in prep.) further investigated an \textit{s}-process and carbon-enriched post-AGB single star in the SMC (2MASS J00364393-7237223) using high-resolution optical spectra from UVES/VLT. The spectral analysis conducted with E-iSpec revealed an exceptionally high C/O ratio (C/O $\sim$ 12) and a significant enhancement of \textit{s}-process elements ([s/Fe] $\sim$ 2 dex), including the first precise measurement of Pb abundance ([Pb/Fe] $\sim$ 3 dex). The derived elemental abundances were compared with predictions from advanced stellar evolutionary models such as ATON, FRUITY, and MONASH. These findings challenge existing nucleosynthesis models and provide new insights into how mass-loss rates and mixing processes influence element production in evolved stars.

These applications of E-iSpec demonstrate its wide-ranging capabilities and efficiency in addressing complex spectral analysis challenges, especially in cases of unusual atmospheric parameters and atypical abundance patterns found in evolved stars.

\begin{savequote}[75mm]
\foreignlanguage{ukrainian}{
``Розширювати свої знання можна тільки тоді, коли дивишся прямо в очі власному незнанню...''}
\qauthor{\foreignlanguage{ukrainian}{---Костянтин Ушинський (1823-1871), \\український педагог}}
``You can only expand your knowledge when you look straight in the eyes of your own ignorance...''
\qauthor{---K. D. Ushinsky (1823-1871), \\Ukrainian pedagogist}
\end{savequote}

\chapter{Photospheric depletion in post-RGB binaries with full discs}\label{chp:pap1}
\graphicspath{{ch_paper1/figures_paper1/}} 

\clearpage

\textit{This chapter was originally published as:}\\
\textbf{The first measurements of carbon isotopic ratios in post-RGB stars: SZ Mon and DF Cyg}\\
M. Mohorian, D. Kamath, M. Menon, P. Ventura, H. Van Winckel, D.~A. Garc{\'\i}a-Hern{\'a}ndez, T. Masseron\\
\textit{Monthly Notices of the Royal Astronomical Society}, 2024, Volume  530, Issue 1, pp.761-782\\
\\
\textit{In this chapter, we present an abundance analysis of the dusty post-RGB binaries SZ Mon and DF Cyg using optical (HERMES/Mercator) and near-infrared (APOGEE) spectroscopy. We investigated the photospheric depletion patterns of these post-RGB binaries and compared them with depletion patterns of post-AGB binaries. Furthermore, we derived the first photospheric $^{12}$C/$^{13}$C ratios for post-RGB binaries. By examining CNO abundances of post-RGB binaries against ATON model predictions for RGB single stars, we showed that the observed CNO chemistry in studied post-RGB binaries mainly reflects RGB nucleosynthesis before binary interaction terminated their RGB stages.}\\
\\
\textbf{\textit{Author Contributions}}\\
M.\,Mohorian performed the reduction and analysis of optical and near-IR spectra from HERMES/Mercator and the APOGEE survey, and was primarily responsible for interpretation of the results presented in this chapter. D.\,Kamath provided feedback and engaged in regular discussions throughout each stage of the research process. P.\,Ventura contributed the ATON models and provided feedback for the Discussion section. D.A.\,Garc{\'\i}a-Hern{\'a}ndez and H.\,Van\,Winckel are the principal investigators of the proposals that resulted in the data used in this study. M.\,Menon contributed to the development of E-iSpec -- the spectral analysis tool presented in this paper, while T.\,Masseron contributed ideas for E-iSpec development. The chapter was written by M.\,Mohorian, with all co-authors providing feedback and comments.

\begin{tcolorbox}[colback=blue!10!white, boxrule=0mm]
    \section*{Original paper abstract}
    
    Dusty post-red giant branch (post-RGB) stars are low- and intermediate-mass stars where the RGB evolution was prematurely terminated by a poorly understood binary interaction. These binary stars are considered to be low-luminosity analogues of post-asymptotic giant branch (post-AGB) binary stars. In this study, we investigated the chemical composition of two dusty post-RGB binary stars, SZ~Mon and DF~Cyg, using multi-wavelength spectroscopic data from HERMES/Mercator (optical) and the APOGEE survey (near-infrared). Owing to challenges posed by existing spectral analysis tools for the study of evolved stars with complex atmospheres, we developed E-iSpec: a dedicated spectral analysis tool for evolved stars, to consistently determine atmospheric parameters, elemental abundances, and carbon isotopic ratios. Our abundance analysis revealed that observed depletion patterns and estimated depletion efficiencies resemble those found in post-AGB binary stars. However, the onset of chemical depletion in post-RGB targets occurs at higher condensation temperatures ($T_{\rm turn-off,~post-RGB}\approx1\,400$~K), than in most post-AGB stars ($T_{\rm turn-off,~post-AGB}\approx1\,100$~K). Additionally, our study resulted in the first estimates of carbon isotopic ratios for post-RGB stars ($^{12}$C/$^{13}$C$_{\rm SZ~Mon}=8\pm4$, $^{12}$C/$^{13}$C$_{\rm DF~Cyg}=12\pm3$). We found that the observationally derived CNO abundances and the carbon isotopic ratios of our post-RGB binary targets are in good agreement with theoretical predictions from the ATON single star evolutionary models involving first dredge-up and moderately-deep extra mixing. This agreement emphasises that in post-RGB binary targets, the observed CNO abundances reflect the chemical composition expected from single star nucleosynthesis (i.e., convective and non-convective mixing processes) occurring during the RGB phase before it is terminated.
\end{tcolorbox}    

\section{Introduction}\label{sec:int_paper1}

Low- and intermediate-mass (LIM) stars, which have masses roughly between 0.9 to 8~M$_\odot$, account for $\sim95\%$ of the aging stellar population and produce $\sim90\%$ of the ejected material in terms of silicates and carbonaceous dust \citep{sloan2008dust}. This makes LIM stars key contributors to the chemical enrichment of the Universe \citep{kobayashi2020OriginOfElements}.

For single LIM stars, during their giant branch phases, i.e., the red giant branch (RGB) phase and the asymptotic giant branch (AGB) phase, the chemical elements synthesised through nuclear fusion in stellar interiors are brought to the surface through convective (and non-convective) mixing processes \citep{busso2007mixing, karakas2014dawes, ventura2015HBB, ventura2018mixing}. However, the physical mechanisms that govern the  nucleosynthesis of elements and elemental isotopes, different mixing processes and mass loss, remain poorly understood \citep{ventura2022InternalProcesses}.

Based on past theoretical studies \citep[see][and references therein]{karakas2014dawes}, it is widely accepted that the first significant alteration in the photospheric chemical composition of LIM stars occurs during the RGB phase, primarily due to convective-driven mixing processes, commonly referred to as first dredge-ups (FDUs). The observational studies of single RGB stars \citep{carbon1982FDU, pilachowski1986FDU, kraft1994FDU, shetrone2019FDU} further evidenced that the FDU leads to an enrichment of surface abundances of $^4$He, $^{13}$C, $^{14}$N, $^{17}$O, and $^{23}$Na, while causing a depletion in surface abundances of $^7$Li, $^{12}$C, $^{16}$O, and $^{18}$O. Overall, the mixing processes on RGB lead to a substantial decrease in the $^{12}$C/$^{13}$C, $^{14}$N/$^{15}$N, and $^{16}$O/$^{17}$O isotopic ratios and slight increase in $^{16}$O/$^{18}$O isotopic ratio \citep{abia2017CNOisotopesAGB, mccormick2023RGBExtraMixing}. Moreover, subsequent observational studies have revealed peculiar abundance patterns of elements, such as He, Li, C, N, O, and Na, on the RGB, where the FDU alone fails to account for the observed changes in these elements \citep{sneden1986ExtraMixing, gilroy1989ExtraMixing, smiljanic2009ExtraMixing, tautvaisiene2013ExtraMixing, drazdauskas2016ExtraMixing, charbonnel2020ExtraMixing}. To address the observationally derived abundances from RGB stars, theoretical models have incorporated non-convective `extra' mixing processes, such as rotation-induced mixing, thermohaline mixing, meridional circulation, shear mixing, and various hydrodynamic and magnetic mixing processes \citep[e.g.,][and references therein]{palacios2003ExMixMod, charbonnel2007ExMixMod, lagarde2012ExMixMod, karakas2014dawes}.

In this work, we aim to trace the evolutionary processes and nucleosynthesis that occur during the RGB phase using dusty post-red giant branch (post-RGB) stars \citep{kamath2016PostRGBDiscovery}. Dusty post-RGB stars are low-luminosity \citep[with luminosities below the tip of the RGB, $L_{\rm RGB\ tip}\approx2\,500 L_\odot$;][]{kamath2014SMC, kamath2022GalacticSingles} analogues of binary post-asymptotic giant branch (post-AGB) stars, where their RGB phase was likely terminated by binary interaction. Post-RGB stars therefore occupy a lower luminosity region in the Hertzsprung-Russell (HR) diagram (compared to post-AGB stars), situated bluewards of the RGB, with the presence of circumstellar dust attributed to the mass loss process induced by binary interaction \cite{kamath2019depletionLMC}.

Previously, several studies \citep[e.g.,][]{desmedt2012j004441, desmedt2015LMC2sEnrichedPAGBs, kamath2019depletionLMC, dellagli2023silicates} extensively explored low-mass stars in the post-AGB phase as valuable tracers of evolution, nucleosynthesis, mass loss through thermal pulses, and mixing processes that occur during the AGB phase. More recently, the study by \cite{kamath2023models} showed that the luminosity and the surface carbon abundance of a post-AGB star serve as the most valuable indicators, unveiling the preceding evolution and nucleosynthetic history of the star. \cite{kamath2023models} derived the masses of AGB progenitors for 31 single post-AGB stars by comparing their observed chemical composition with the predictions from ATON stellar evolutionary models. This highlights the wealth of using post-AGB stars as tracers of AGB nucleosynthesis. In this study, we extend the concept of using post-AGB stars as tracers of nucleosynthesis by leveraging observations of their low-luminosity analogues -- dusty post-RGB stars.
 
The dusty post-RGB stars display properties resembling post-AGB binary stars. For example, they have similar atmospheric parameters such as effective temperatures ($T_{\rm eff}$) ranging from 3\,500 to 8\,500 K, surface gravities ($\log g$) ranging from 0 to 2.5 dex, and metallicities ranging from --5 to 0 dex (depending on their host galaxy). Comparable to post-AGB binary stars, the majority of the post-RGB stars exhibit a `disc-type' spectral energy distribution (SED), characterized by a near-infrared (NIR) excess, indicative of the presence of a circumbinary disc \citep[][and references therein]{kluska2022GalacticBinaries}. Furthermore, a subset of post-RGB stars also demonstrates Type II Cepheid variability \citep{kamath2014SMC, kamath2015LMC, kamath2016PostRGBDiscovery, manick2019RVTauDFCyg}.

Post-RGB stars and binary post-AGB stars also share a common characteristic in their photospheric chemistry known as `chemical depletion'. This chemical anomaly was first identified through various studies focusing on post-AGB stars \citep[such as][]{waelkens1991depletion, vanwinckel1992depletion, venn2014depletion, kamath2019depletionLMC, oomen2019depletion, kluska2022GalacticBinaries}, and then confirmed for post-RGB stars by piecemeal studies \citep[e.g.,][]{giridhar2005rvtau, maas2007t2cep, manick2019RVTauDFCyg, gezer2019GKCarGZNor}.

Chemical depletion is characterised by a photospheric abundance pattern resembling the patterns observed in the interstellar gas phase \citep[][and references therein]{konstantopoulou2022ISMdepletion}. In this pattern, refractory elements like Al, Ti, Fe, Y, and La are typically underabundant, whereas volatile elements such as Zn and S tend to maintain their initial abundance levels, similar to those observed in the interstellar medium. Consequently, these stars exhibit extrinsically metal-poor characteristics with [Fe/H] ranging from --5.0 to --0.5 \citep{vanwinckel1992depletion, oomen2019depletion}. However, it is important to note that the abundances of C, N, and O deviate from the pattern of other volatile elements. This deviation arises from likely alterations in CNO abundances due to dredge-ups occurring during previous RGB and AGB evolutionary phases.

In this study, we targeted two post-RGB stars that are known to show photospheric chemical depletion \citep{maas2007t2cep, giridhar2005rvtau}. We used high-resolution optical and mid-resolution NIR spectra to derive accurate elemental abundances and carbon isotopic ratios. Our goal is to investigate whether the combined knowledge of the luminosity together with the surface carbon and nitrogen (i.e., CNO abundances, the C/O and $^{12}$C/$^{13}$C ratios) could serve as valuable tracers of the nucleosynthetic history that occurs during the RGB phase before transitioning to the post-RGB phase. This study also let us investigate whether binary interactions cause any drastic effect on the mixing processes that are known to alter the surface chemical composition.

In Section~\ref{sec:tar_paper1}, we provide an overview of the targets under investigation. In Section~\ref{sec:obs_paper1}, we introduce the data and observations. In Section~\ref{sec:san_paper1}, we describe E-iSpec, a dedicated spectral analysis tool developed for this study to derive elemental abundances and isotopic ratios in evolved stars, and present our research findings. In Section~\ref{sec:mod_paper1}, we compare observationally derived abundances and carbon isotopic ratios with theoretical predictions from the ATON evolutionary models \citep{ventura2008aton3} of single RGB stars that have similar atmospheric parameters and luminosities as our binary post-RGB targets. Finally, in Section~\ref{sec:con_paper1}, we delve into the potential of using observationally derived atmospheric parameters, elemental abundances, and isotopic ratios to explore the impact of binary interactions on RGB nucleosynthesis and mixing processes.

\section{Target sample}\label{sec:tar_paper1}
The target sample for this study comprised of two binary post-RGB stars: SZ~Mon and DF~Cyg (see Table~\ref{tab:varpro_paper1}), which were derived from a larger sample of spectroscopically verified evolved candidates in the Galaxy and the Magellanic Clouds (see Appendix~\ref{app:tar_paper1} for more details on the target selection). We note that both targets are Type II Cepheid variables \citep{maas2007t2cep, giridhar2005rvtau}. We present target details below.

\begin{table}[!ht]
    \centering
    \caption{Variability properties and luminosity estimates (see Section~\ref{ssec:obslum_paper1}) obtained in this study for our post-RGB binary pulsating variables, SZ Mon and DF Cyg.\\} \label{tab:varpro_paper1}
    \begin{tabular}{|l|cc|}\hline
         \textbf{Parameter} & \textbf{SZ Mon} & \textbf{DF Cyg} \\\hline
         RVb phenomenon & no$^{a}$ & yes$^{b}$ \\
         $P_\textrm{puls}$ (days) & 16.34$^{c}$ & 24.91$^{d}$ \\
         $P_\textrm{orb}$ (days) & -- & 780$^{d}$\\\hline
         L$_{\rm SED}$ (L$_\odot$) & $193\pm30$ & $657\pm103$ \\
         L$_{\rm PLC}$ (L$_\odot$) & $382\pm83$ & -- \\\hline
    \end{tabular}\\
    \textbf{Notes:} $P_\textrm{puls}$ is the fundamental pulsation period, $P_\textrm{orb}$ is the orbital period, L$_{\rm SED}$ is the luminosity derived from the SED fitting, L$_{\rm PLC}$ is the luminosity calculated from the PLC relation. The periods and the detection of RVb phenomenon are acquired from the following catalogues: $^a$\cite{kluska2022GalacticBinaries}, $^b$\cite{manick2019RVTauDFCyg}, $^c$\cite{pawlak2019ASAS}, $^d$\cite{kiss2017RVTau}.\\
\end{table}

\subsection{SZ Mon}
SZ Mon is a spectral type F8 star showing broad infrared excess in the SED. The very first studies of SZ~Mon \citep{lloyd1968szmon, stobie1970PeriodInconsistency} classified this object as a Type II Cepheid with a double period\footnote{Double period is the empirically best-fit value for phase curves of Type II Cepheids, yet the fundamental pulsation period is twice shorter \citep{stobie1970PeriodInconsistency}.} of 32.686 days. \cite{maas2007t2cep} classified SZ~Mon as RVa photometric type, i.e., an RV Tau variable without a detected secondary variation of its phase curve with significant amplitude and long period $\sim1000$d (referred to as the RVb phenomenon).

In the recent studies \citep{maas2007t2cep, oomen2019depletion}, SZ Mon was classified as a post-AGB star with a disc-type SED. \cite{oomen2019depletion} presented the SED modelling for SZ~Mon and derived the SED luminosity $L_\textrm{SZ Mon, SED} = 2400^{+2000}_{-1000}L_\odot$.

Furthermore, chemical abundance studies by \cite{maas2007t2cep} showed that SZ~Mon is an evolved post-AGB binary star showing photospheric chemical depletion -- characteristic of post-AGB and post-RGB binary stars (see Section~\ref{sec:int_paper1}).

We note that, while SZ~Mon was classified as a post-AGB RV Tauri (RV Tau) variable in the literature, in this study we re-classified the evolutionary nature of this target. Based on the newly derived luminosity of SZ~Mon (see Section~\ref{ssec:obslum_paper1}), its disc-type SED, and the confirmation of photospheric chemical depletion (see Section~\ref{ssec:sanabs_paper1}), we conclude that SZ~Mon is a post-RGB binary star. Furthermore, given that the observed fundamental pulsation period is less than 20 days \citep[i.e., 16.336 days;][]{soszynski2008OGLE}, SZ~Mon rather belongs to the W Virginis (W Vir) subclass of Type II Cepheid family (instead of RV Tau subclass). We adopted this new classification of SZ~Mon (i.e., a post-RGB binary star and W Vir variable) for the rest of this study.

\subsection{DF Cyg}
DF Cyg is a star of G6/7Ib/IIa spectral type. DF~Cyg exhibits a broad infrared excess in the SED \citep[i.e., disc-type][]{deruyter2006discs}. \cite{manick2019RVTauDFCyg} derived the luminosity of DF~Cyg from SED fitting, wherein $L_\textrm{DF Cyg, SED} = 1010^{+150}_{-140}L_\odot$).

\cite{giridhar2005rvtau} conducted an extensive abundance analysis of DF~Cyg and identified a rather steep depletion pattern with an unusually negative volatile-refractory abundance ratio ([Zn/Ti]=$-0.7$ dex). In our current study (see Section~\ref{ssec:sanabs_paper1}), we performed an independent abundance analysis, which resulted in a standard depletion pattern with ([Zn/Ti]=$+0.35$ dex).

DF~Cyg was classified as an RV Tau variable with a fundamental pulsation period of 24.91 days \citep{kiss2017RVTau}. Using the pulsation period, \cite{manick2019RVTauDFCyg} derived a more reliable estimate of DF~Cyg luminosity from period-luminosity-colour (PLC) relation for Type II Cepheids $L_\textrm{DF Cyg, PLC} = 990\pm190L_\odot$), and classified DF~Cyg as a post-RGB star.

Previous studies have suggested that DF~Cyg could be a binary system due to the very strong RVb phenomenon in the light curve, as discussed by \cite{bodi2016dfcyg} and \cite{vega2017dfcyg}, or due to the presence of a disc, as mentioned by \cite{deruyter2005discs} and explained in \cite{vanwinckel2018Binaries}. Orbital parameter analysis \citep{oomen2018OrbitalParameters} confirmed the binary nature of DF~Cyg with an orbital period of 784 days (see Table~\ref{tab:varpro_paper1}).

In this paper, we validated the binary post-RGB nature of DF~Cyg (see Section~\ref{ssec:obslum_paper1}).

\section{Data and observations}\label{sec:obs_paper1}
In this section, we present the photometric data used for SED fitting (see Section~\ref{ssec:obspht_paper1}) and luminosity derivation (see Section~\ref{ssec:obslum_paper1}), as well as the spectroscopic data used for deriving elemental abundances and isotopic ratios (see Section~\ref{ssec:obsspc_paper1}).

\subsection{Photometric data}\label{ssec:obspht_paper1}
In Table~\ref{tab:phomag_paper1}, we present the photometric magnitudes of SZ~Mon and DF~Cyg in various wavelength bands, including optical, NIR, and mid-infrared (MIR). The UBVRI photometry is taken from the Johnson-Cousins system \citep{johnson1953Filters, cousins1976Filters}, and the I-band filters from the Sloan Digital Sky Survey \citep[SDSS, ][]{york2000SDSSphotometry}. The Two Micron All Sky Survey (2MASS) provides magnitudes in the J, H, and K bands \citep[1.24, 1.66, and 2.16 $\mu$m, respectively; ][]{skrutskie20062MASS}. WISE \citep{wright2010WISE} contains the magnitudes in the W1, W2, W3, and W4 bands (3.4, 4.6, 12, and 22 $\mu$m, respectively). The mid-infrared (MIR) was complemented by fluxes from the MSX, AKARI, and IRAS catalogues \citep{egan2003MSX, ishihara2010AKARI, neugebauer1984IRAS}.

\begin{figure}[!ht]
    \centering
    \includegraphics[width=.49\linewidth]{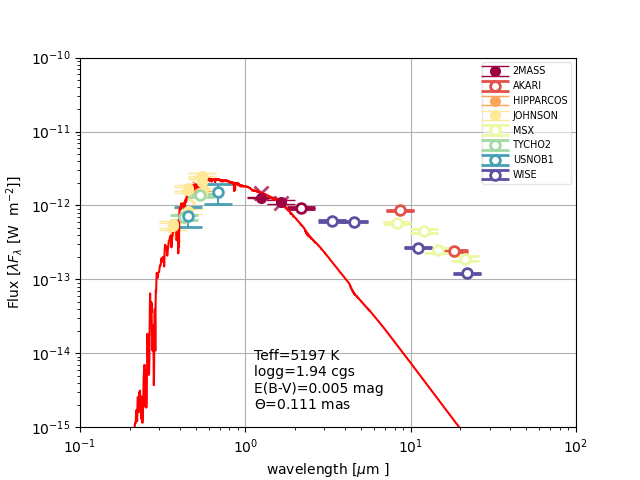}
    \includegraphics[width=.49\linewidth]{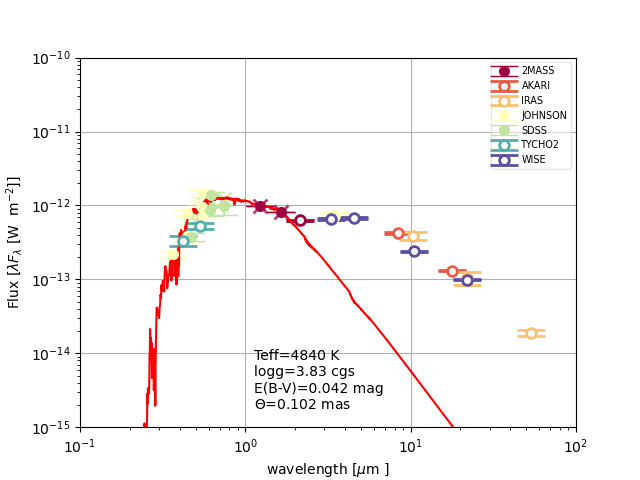}
    \caption[Spectral energy distribution of SZ Mon (top) and DF Cyg (bottom) suggests the possible presence of circumbinary discs in these targets (see Section~\ref{ssec:obspht_paper1})]{Spectral energy distribution of SZ Mon (top) and DF Cyg (bottom) suggests the possible presence of circumbinary discs in these targets (see Section~\ref{ssec:obspht_paper1}). Red fit is the appropriate Kurucz model atmosphere. We note that the photometric observations of both targets were obtained at slightly different pulsation phases. The legend for the symbols and colours used are included within the plot.}\label{fig:SEDs_paper1}
\end{figure}

The SEDs of SZ~Mon and DF~Cyg are shown in Fig.~\ref{fig:SEDs_paper1} (top and bottom, respectively). We note that in the case of DF~Cyg, we specifically excluded the $V$ and $I$ band magnitudes (which come from the ASAS Catalogue of Variable Stars in the Kepler Field), since they were obtained during the minima in the RVb period, which is out of phase with all the other photometric data points \citep[see Fig. 8 in][]{manick2018PLC}. This significantly improves the $\chi^2$ value of the SED fitting and provides a more accurate luminosity estimate (see Section~\ref{ssec:obslum_paper1}).

The SEDs of SZ~Mon and DF~Cyg exhibit characteristics of a disc type \citep{oomen2018OrbitalParameters}, suggesting the possible presence of a circumbinary disc, which is indicative of binarity in post-RGB and post-AGB stars \citep{kamath2016PostRGBDiscovery, kluska2022GalacticBinaries}.

\begin{table}[!ht]
    \centering
    \scriptsize
    \caption[Photometric data for SZ Mon and DF Cyg (see Section \ref{ssec:obspht_paper1})]{Photometric data for SZ Mon and DF Cyg (see Section \ref{ssec:obspht_paper1}). For each filter we provide the units and the central wavelengths in $\mu$m. This table is published in its entirety in the electronic edition of the paper. A portion is shown here for guidance regarding its form and content.}\label{tab:phomag_paper1}
    \begin{tabular}{|l|c|c|c|c|c|c|c|}\hline
        \textbf{Filter} & \textbf{JOHNSON.U} & \textbf{JOHNSON.B} & \textbf{...} & \textbf{AKARI.L18W} & \textbf{MSX.E} & \textbf{WISE.W4} & \textbf{IRAS.25um} \\
        & \textbf{(mag)} & \textbf{(mag)} & \textbf{...} & \textbf{(Jy)} & \textbf{(Jy)} & \textbf{(mag)} & \textbf{(Jy)} \\
        & \textbf{0.364} & \textbf{0.443} & \textbf{...} & \textbf{18.9} & \textbf{21.5} & \textbf{22.1} & \textbf{25.0} \\ \hline 
        SZ Mon & 11.1$\pm$0.1 & 11.33$\pm$0.08 & ... & 1.59$\pm$0.04 & 1.4$\pm$0.1 & 2.41$\pm$0.02 & -- \\
        DF Cyg & -- & 12.8$\pm$0.2 & ... & 0.88$\pm$0.02 & -- & 2.63$\pm$0.02 & 0.8$\pm$0.2 \\ \hline
    \end{tabular}
\end{table}

\subsection{Determination of luminosities from SED fitting and PLC relation}\label{ssec:obslum_paper1}
In this study, we calculated luminosities of SZ~Mon and DF~Cyg using two methods: i) by fitting the SED (SED luminosity, $L_{\rm SED}$); and ii) by using the PLC relation (PLC luminosity, $L_{\rm PLC}$). We briefly describe these two methods below. 

To compute the SED luminosities, we followed the procedure described in \cite{oomen2018OrbitalParameters, kluska2022GalacticBinaries}. In short, we started by assuming the appropriate Kurucz model atmospheres \citep{castelli2003ATLAS9} to fit the initial photometric data points. Following this, we accurately computed a dereddened SED model for SZ~Mon and DF~Cyg. This model took into account the overall reddening or extinction parameter $E(B-V)$ that yielded the lowest $\chi^2$ value from our extensive parameter grid search. The total reddening considered here encompassed both interstellar and circumstellar reddening. We made the assumption that the total reddening along the line of sight follows the wavelength dependence described by the interstellar-medium extinction law \citep{cardelli1989SEDextinction} with an $R_V$=3.1. We incorporated the more accurate Bailer-Jones geometric distances, denoted as $z_{\rm BJ}$, along with their respective upper and lower limits, represented as $z_{\rm BJU}$ and $z_{\rm BJL}$, sourced from the research conducted by \cite{bailerjones2021distances}. These geometric distances were computed based on Gaia EDR3 parallaxes, taking into consideration a direction-dependent prior distance. It is important to note that in all our calculations, we assumed that the emission of flux from the stars followed an isotropic radiation pattern. We also note that stellar variability was not considered, and this omission resulted in an increased $\chi^2$ value for high-amplitude variables.

To calculate the PLC luminosity, we used the PLC relation, which was obtained following the similar procedure as in \cite{manick2018PLC}, but updated with OGLE-IV data for RV Tau pulsating variables in the LMC (Menon et al., submitted). In short, this relation uses the colour-corrected V-band magnitude known as the Wesenheit index \citep[${\rm WI}=V-2.55(V-I)_0$; see][]{ngeow2005wesenheit} and is given as
\begin{equation}
    M_{bol, {\rm WI}} = m\times\log P_0 + c - \mu + BC + 2.55\times(V-I)_0,
\end{equation}
where $M_{bol, {\rm WI}}$ is the absolute Wesenheit magnitude, $P_0$ is the observed fundamental pulsation period in days, $\mu=18.49$ is the distance modulus for the LMC, $(V-I)_0$ is the intrinsic colour of each star. $m=-3.59$ is the slope, and $c=18.79$ is the intercept of the linear fit of the relation \citep[see Fig. 5 in][]{manick2018PLC}. Bolometric corrections $BC$ for SZ\,Mon and DF\,Cyg were calculated based on their observed effective temperatures, $T_{\rm eff}$ = 5\,460 K for SZ\,Mon (see Table~\ref{tab:fnlabsSZM_paper1}) and $T_{\rm eff}$ = 5\,770 K for DF\,Cyg (see Table~\ref{tab:fnlabsDFC_paper1}), following the tabulated relation from \cite{flower1996BoloCorr}, corrected by \cite{torres2010BoloCorrErrata}.

We would like to highlight that due to the exclusion of the only available I-band magnitude (as discussed in Section~\ref{ssec:obspht_paper1}), the calculation of the PLC luminosity for DF~Cyg is not feasible. Nevertheless, our refined photometric selection for the SED fit provides a more reliable luminosity in comparison to the SED and PLC luminosity values reported by \cite{manick2018PLC}.

In Table~\ref{tab:varpro_paper1}, we provide the final values and corresponding uncertainties in the SED and the PLC luminosities for SZ~Mon and the SED luminosity for DF~Cyg. The uncertainties in the SED luminosities were determined by calculating the standard deviation of the upper and lower luminosity limits, taking into consideration the uncertainties associated with distances ($z_{\rm BJU}$ and $z_{\rm BJL}$) and reddening ($E(B-V)$) for each respective target. On the contrary, the uncertainties in the PLC luminosities were dominated by the uncertainties of the reddening.

We note that we considered the PLC luminosity to be more precise and more reliable than the SED luminosity, mainly due to the significant $T_{\rm eff}$ fluctuations in our targets throughout the pulsation cycle (leading to significant changes in model fitting of stellar atmospheres shown as red lines in Fig. \ref{fig:SEDs_paper1}) and the uncertainty of distances obtained from parallax measurements, which were heavily affected by the binary orbital motion \citep{kamath2022GalacticSingles}.

\subsection{Spectroscopic data}\label{ssec:obsspc_paper1}
In this subsection, we provide details of the optical and NIR spectral observations used in our study.

\subsubsection{HERMES spectra}\label{sssec:obsspcopt_paper1}
The HERMES spectra were obtained as part of a long-term monitoring project (initiated in June 2009 and still ongoing) that made use of the HERMES spectrograph \citep[High Efficiency and Resolution Mercator Echelle Spectrograph;][]{raskin2011hermes} mounted on the 1.2 m Mercator telescope at the Roque de los Muchachos Observatory, La Palma. The project resulted in a substantial collection of high-resolution optical spectra of post-AGB systems, as documented by \cite{vanwinckel2018Binaries}.

High-resolution optical spectra from HERMES (R = $\lambda/\Delta\lambda\sim$ 85 000) covered a wavelength range from about 377 nm to 900 nm. The data were reduced through a dedicated pipeline with standard settings as described in \cite{raskin2011hermes}. To observe the whole pulsation cycle of SZ~Mon, a total of 88 spectra were obtained (see Table~\ref{tabA:szmvis_paper1}), with a total time span of 3700 days. Similarly, a total of 83 spectra over 4000 days were obtained for DF~Cyg (see Table~\ref{tabA:dfcvis_paper1}). 

\subsubsection{APOGEE spectra}\label{sssec:obsspcnir_paper1} 
In this study, we also used NIR spectra from the APOGEE survey, which was a large-scale stellar spectroscopic survey conducted in the NIR regime \cite[H-band;][]{majewski2017APOGEE}. APOGEE made use of two 300-fiber cryogenic spectrographs mounted in different hemispheres. The Northern Hemisphere was studied by the 2.5-metre Sloan Foundation Telescope and the 1-metre NMSU Telescope (for several bright sources) at Apache Point Observatory (APO) in New Mexico, United States. The Southern Hemisphere was observed by the 2.5-metre Ir\'{e}n\'{e}e du Pont Telescope of Las Campanas Observatory (LCO) in Atacama, Chile. APOGEE obtained the medium-resolution (R = 22\,500) spectra across the entire Milky Way as part of the Sloan Digital Sky Survey (SDSS).

SDSS-IV DR17 is the final release of APOGEE data. This release provided spectra, radial velocities, and stellar atmospheric parameters along with individual elemental abundances for more than $657\,000$ stars \citep{nidever2015ASPCAP, abdurrouf2022APOGEELastRelease}. In DR17, three types of APOGEE output files could be accessed: \texttt{apVisit/asVisit} (individual visit raw spectra), \texttt{apStar/asStar} (radial velocity-corrected individual visit and combined spectra), and \texttt{aspcapStar} (fully prepared spectra analysed by ASPCAP, see Section~\ref{sec:int_paper1}).

As mentioned by \cite{schultheis2020APOGEE-AGB}, ASPCAP was not suited to analyse the spectra of dusty giant stars, so the ASPCAP results (i.e., atmospheric parameters and elemental abundances) for post-AGB and post-RGB stars were out of corresponding grid bounds. Moreover, continuum normalisation routine used by ASPCAP performed poorly for our targets. As a result, we made the decision to omit the \texttt{aspcapStar} files from our study.

Instead, we performed our own spectral analyses (see Section~\ref{sec:san_paper1}) where we adopted the \texttt{apStar/asStar} files, which were found to be the most suitable choice for our study. The \texttt{apStar/asStar} files provided APOGEE spectra with a logarithmically-spaced wavelength scale: a common spacing of $\log\lambda_{i+1}-\log\lambda_{i} = 6\times10^{-6}$, starting from 1510.0802 nm.

We note that APOGEE wavelength scale was calibrated using vacuum wavelengths, while we used the NIR line list with transition data provided in standard temperature and pressure (S.T.P.). This lead to a wavelength difference between the spectra and NIR line lists on the order of $\approx0.45$ nm. As stated in APOGEE manual\footnote{Accessible via \url{https://www.sdss4.org/dr17/irspec/}}, the vacuum-to-air wavelength conversion was done according to \cite{1996ApOpt..35.1566C} given by
\begin{equation}
    \lambda_{\rm AIR} = \dfrac{\lambda_{\rm VAC}}{1 + \dfrac{5.792105\times10^{-2}}{238.0185 - [\dfrac{1000}{\lambda_{\rm VAC}}]^2} + \dfrac{1.67917\times10^{-3}}{57.362 - [\dfrac{1000}{\lambda_{\rm VAC}}]^2}},
\end{equation}
where $\lambda_{\rm AIR}$ and $\lambda_{\rm VAC}$ represent air and vacuum wavelengths in nm, respectively.

In addition to arrays of fluxes and flux errors, the \texttt{apStar/asStar} files also included calibration arrays (i.e., sky emission, telluric absorption lines, cosmic rays), which we used to remove the bad pixels from spectra before the analysis (see Section~\ref{ssec:saneis_paper1}).

\subsubsection{Epoch selection}\label{sssec:epochs_paper1}
Since our study focused on Type II Cepheid variables (SZ~Mon is a W Vir variable and DF~Cyg is an RV Tau variable, see Section~\ref{sec:tar_paper1}), we encountered an extra challenge in analysing their spectra because the atmospheric parameters of pulsating variables (i.e., $T_{\rm eff}$, $\log g$, and $\xi_{\rm t}$) experience notable variations throughout the pulsation cycle \citep{manick2019RVTauDFCyg}.

\begin{figure}[!ht]
    \centering
    \includegraphics[width=.49\linewidth]{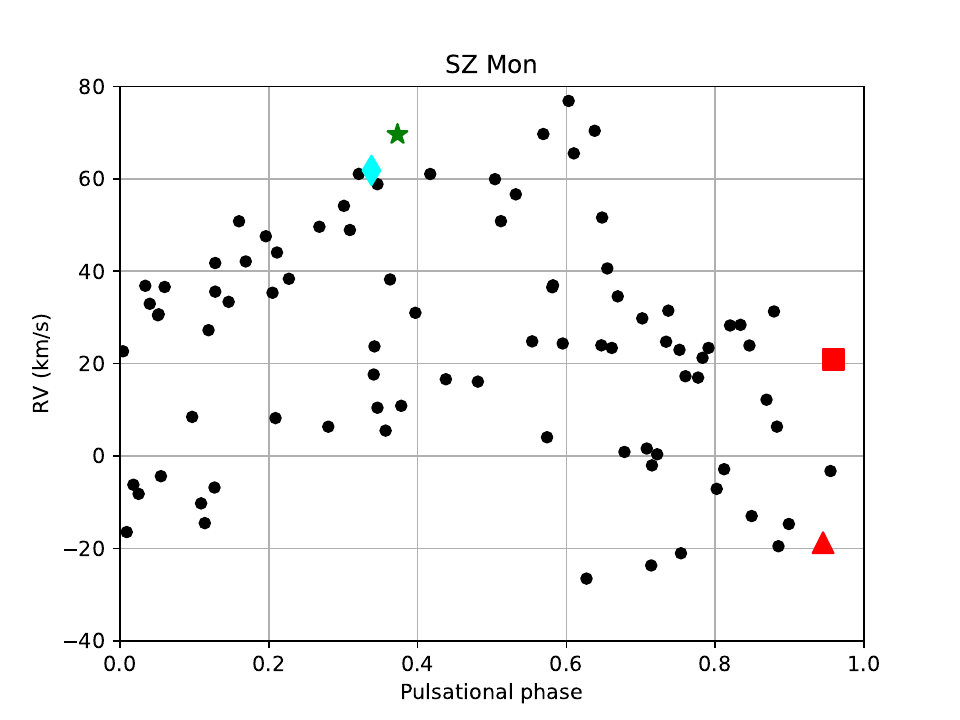}
    \includegraphics[width=.49\linewidth]{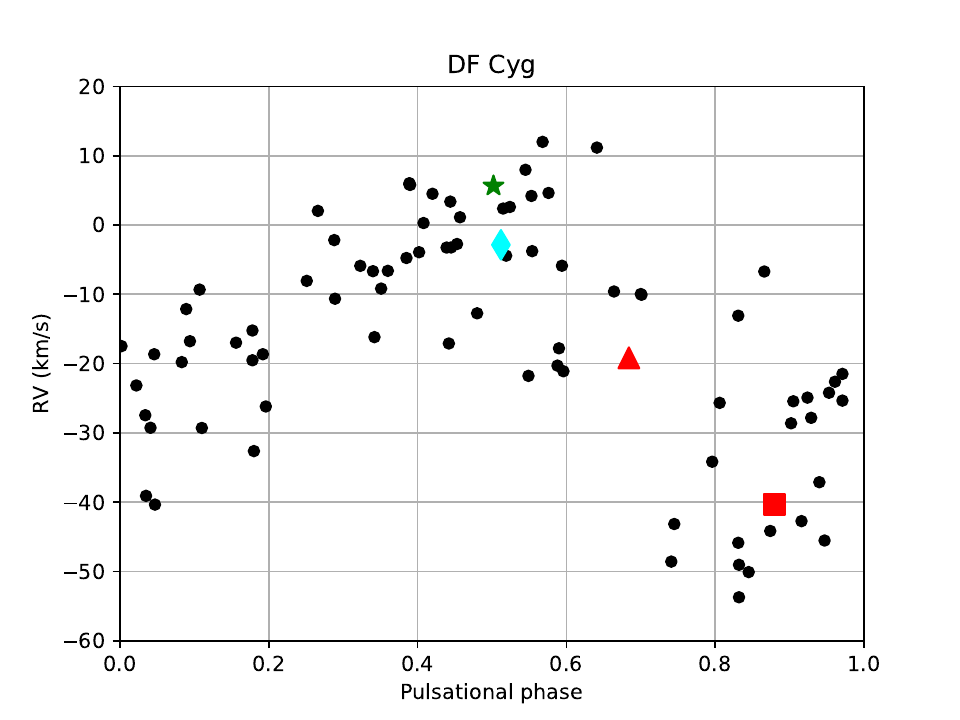}
    \caption[Radial velocity vs pulsation phase plots for SZ Mon (top; $P_\text{puls}$=16.336d) and DF Cyg (bottom; $P_\text{puls}$=25.04d)]{Radial velocity vs pulsation phase plots for SZ Mon (top; $P_\text{puls}$=16.336d) and DF Cyg (bottom; $P_\text{puls}$=25.04d). Green and blue triangles mark the spectra taken at close cooler phases of pulsations: near-infrared APOGEE visit and the optical HERMES visit, respectively. Red markers show the optical HERMES visits obtained at hotter phases of pulsations: triangle for the visit we use in our chemical analysis and square for the visit we use to test the obtained results. For more details see Section~\ref{sssec:epochs_paper1}).}\label{fig:RVs_paper1}
\end{figure}

Therefore, for each target we needed to ensure that the analysed optical and NIR spectra were observed at close pulsation phases. To do this, we acquired the radial velocities for each spectral visit of each target (see Tables~\ref{tabA:szmvis_paper1} and \ref{tabA:dfcvis_paper1}): for optical spectra, the radial velocities were derived using E-iSpec (see Section~\ref{ssec:saneis_paper1}), while for NIR spectra, we adopted the radial velocities derived by APOGEE's pipeline (see Section~\ref{sssec:obsspcnir_paper1}). The acquired radial velocities were then plotted against the corresponding pulsation phases, and the visits for the joint analysis were selected (depicted with blue and green triangles in Fig.~\ref{fig:RVs_paper1} for optical and NIR visits, respectively). To validate and extend the number of studied chemical species, we additionally selected two more optical visits for each target at hotter pulsation phases (depicted with red triangle and square in Fig.~\ref{fig:RVs_paper1} for main and secondary optical visits, respectively). The comprehensive explanation of epoch selection is provided in Appendix~\ref{app:epo_paper1}.

To summarise, based on the pulsation phases and S/N of the visits, we selected the following visits (see Table~\ref{tab:obslog_paper1}):
\begin{enumerate}
    \item the optical HERMES spectra to determine atmospheric parameters and elemental abundances from atomic lines: SH\#73, SH\#29, and SH\#47 for SZ~Mon; DH\#83, DH\#26, and DH\#54 for DF~Cyg\footnote{We note that we adopted the following naming convention: the first letter stands for the target (``S'' for SZ~Mon and ``D'' for DF~Cyg), the second letter stands for the instrument (``H'' for HERMES and ``A'' for APOGEE), and the number is the chronological order of the visit.};
    \item the NIR APOGEE spectra to determine elemental abundances and carbon isotopic ratios from molecular bands: SA\#2 for SZ~Mon; DA\#1 for DF~Cyg.
\end{enumerate}

\begin{sidewaystable}[ph!]
    \centering
    \small
    \caption[The details of spectroscopic observations of SZ Mon and DF Cyg]{The details of spectroscopic observations of SZ Mon and DF Cyg. The choice of shown visits and visit naming convention are explained in Section \ref{sssec:epochs_paper1}. For the full observing log, see Tables~\ref{tabA:szmvis_paper1} and \ref{tabA:dfcvis_paper1}).}\label{tab:obslog_paper1}
    \begin{tabular}{|ccccccc|}\hline
         \textbf{Visit} & \textbf{Obs. date} & \textbf{Obs. start} & \textbf{Obs. ID} & \textbf{Instrument +} & \textbf{Radial velocity} & \textbf{Pulsational}\\
         \textbf{number} & ~ & \textbf{(UT)} & ~ & \textbf{telescope} & \textbf{(km/s)} & \textbf{phase} \\ \hline
         \multicolumn{7}{|c|}{\textit{SZ Mon (2MASS J06512784-0122158)}} \\ \hline
         SH\#73$^{1}$ & 2018-02-18 & 01:43:40.80 & 00866282 & HERMES + Mercator$^{a}$ & --18.78$\pm$0.05 & 0.95 \\
         SH\#29$^{2}$ & 2012-03-24 & 22:40:48.00 & 00397261 & HERMES + Mercator$^{a}$ & 20.84$\pm$0.04 & 0.96 \\
         SH\#47$^{3}$ & 2013-04-10 & 21:04:19.20 & 00458609 & HERMES + Mercator$^{a}$ & 61.82$\pm$0.09 & 0.34 \\
         SA\#2$^{4}$ & 2016-01-19 & 06:24:28.80 & 8803.57406.58 & APOGEE-N + SFT$^{b}$ & 69.60$\pm$0.20 & 0.37 \\ \hline
         \multicolumn{7}{|c|}{\textit{DF Cyg (2MASS J19485394+4302145)}} \\ \hline
         DH\#83$^{1}$ & 2020-08-16 & 00:38:52.80 & 00972481 & HERMES + Mercator$^{a}$ & --19.17$\pm$0.04 & 0.68 \\
         DH\#26$^{2}$ & 2012-06-24 & 04:22:04.80 & 00412205 & HERMES + Mercator$^{a}$ & --40.4$\pm$0.2 & 0.88 \\
         DH\#54$^{3}$ & 2014-07-06 & 03:40:19.20 & 00574546 & HERMES + Mercator$^{a}$ & --2.86$\pm$0.06 & 0.51 \\
         DA\#1$^{4}$ & 2016-09-14 & 04:20:38.40 & 9129.57645.89 & APOGEE-N + SFT$^{b}$ & 5.64$\pm$0.05 & 0.50 \\ \hline
    \end{tabular}\\
    \textbf{Notes:} $^a$Mercator is the 1.2-metre telescope at the Observatorio del Roque de Los Muchachos, La Palma. $^b$SFT is the 2.5-metre Sloan Foundation Telescope located at Apache Point Observatory, New Mexico. $^{1,2}$Observations were taken at hotter pulsation phases which are ideal for studying atomic abundances (1) and confirming derived atmospheric parameters (2). $^{3,4}$Observations were taken at cooler pulsation phases which are ideal for deriving atmospheric parameters from HERMES spectra (3) and passing them to the analysis of molecular abundances from APOGEE spectra (4).
\end{sidewaystable}

\section{Spectral analysis}\label{sec:san_paper1}
To determine the atmospheric parameters, elemental abundances, and the carbon isotopic ratios for evolved stars, we developed E-iSpec -- a modified version of iSpec \citep{blancocuaresma2014, blancocuaresma2019}. Specifically, our modifications focused on improving the accuracy and reliability of elemental abundance measurements for objects with complex atmospheric conditions, and on adding the isotopic ratios derivation to the spectral analysis.

As a validation of our approach for epoch selection, we performed the chemical analysis for all selected visits of SZ~Mon and DF~Cyg (three optical and one NIR visits for each target, see Section~\ref{sssec:epochs_paper1}). We found that the differences in atmospheric parameters and elemental abundances were generally smaller than the corresponding uncertainties (except an expected change in the effective temperatures; see Table~\ref{tabA:tststp_paper1}).

In the following subsections, we briefly present iSpec and its limitations. Subsequently, we introduce E-iSpec and present the analyses and results for our target stars: SZ~Mon and DF~Cyg.

\subsection{iSpec and its limitations}\label{ssec:sanisp_paper1}
iSpec is an integrated spectroscopic software framework, which is capable of determining atmospheric parameters such as effective temperature, surface gravity, metallicity, micro/macroturbulence, rotation and individual elemental abundances for AFGKM stars. The full details of iSpec are presented in \cite{blancocuaresma2014, blancocuaresma2019}. In brief, iSpec integrates MARCS \citep{gustafsson2008MARCS} and different ATLAS9 \citep[i.e., Kurucz, Castelli \& Kurucz, KuruczODFNEW;][]{castelli2003ATLAS9} one-dimensional (1D) model atmospheres with assumed local thermodynamic equilibrium (LTE) together with the following radiative transfer codes: SPECTRUM \citep{gray1994spectrum}, Turbospectrum \citep{alvarez1998turbospectrum, turbospec2012ascl}, SME \citep{valenti1996SME}, MOOG \citep{sneden1973moog}, and Synthe/WIDTH9 \citep{kurucz1993SYNTHE, sbordone2004SYNTHE}. These radiative transfer codes are implemented in iSpec to work in two different approaches: equivalent widths (EW) method and synthetic spectral fitting (SSF) technique. Furthermore, to carry out the spectral analyses, iSpec provides several line lists with a wide wavelength coverage (overall, from 300\,nm to 4\,$\mu$m). The line lists included are: the VALD3 line list \citep{kupka2011vald}, the APOGEE line list \citep{smith2021APOGEELineList}, the Gaia-ESO line list \citep{heiter2021GaiaESO}, the ATLAS9 line lists \citep{kurucz1995linelist}, and the original SPECTRUM line list, which contains atomic and molecular lines obtained mainly from the NIST Atomic Spectra Database \citep{ralchenko2005nist}. Additionally, iSpec provides a collection of ready-to-use solar abundances from \cite{anders1989solar, grevesse1998solar, asplund2005solar, grevesse2007solar, asplund2009solar}.

We note that iSpec offers two usability modes: a GUI version and a Python version. E-iSpec (see Section~\ref{ssec:saneis_paper1}) exclusively utilises the Python version, which offers enhanced functionality.

While iSpec is capable of the spectral analysis of a wide range of targets, we note that it is not specifically designed for our goal. Firstly, iSpec routine for automatic continuum normalisation performs poorly for evolved stars, the spectra of which contain prominent molecular bands. Secondly, the molecular line lists presented in iSpec (especially for NIR regime) were not up-to-date. Thirdly, the error estimation of elemental abundances in iSpec only takes into account line-to-line (L2L) scatter completely ignoring the systematic errors caused by uncertainties of atmospheric parameters. Finally and most importantly, the isotopic ratios were not addressed in chemical analysis routines of iSpec. To overcome these limitations, we developed a Python wrapper for iSpec: E-iSpec.

\subsection{E-iSpec: Evolved stars solution of iSpec}\label{ssec:saneis_paper1}
E-iSpec\footnote{Accessible via \url{https://github.com/MaksTheUAstronomer/E-iSpec.git}} is a semi-automated spectral (optical and NIR) analysis tool implemented in Python, primarily derived from iSpec but enhanced with extended capabilities described below. E-iSpec allows for the determination of atmospheric parameters, elemental abundances (from both atomic lines and molecular lines), and isotopic ratios, particularly for evolved stars with complex atmospheres.

E-iSpec uses 1D LTE model atmospheres (from MARCS and ATLAS9 grids)\footnote{We note that spherically-symmetric MARCS model atmospheres generally perform better for cool giants \citep{meszaros2012ModelsAPOGEE}.} and relies on the following radiative transfer codes, which were chosen due to their performance and reliability: Turbospectrum \citep{turbospec2012ascl} for synthetic spectral fitting technique and Moog \citep{sneden1973moog} for equivalent width method. Line lists for atomic and molecular transitions, as detailed in Appendix~\ref{app:add_paper1}, are drawn from VALD3 for optical spectra ($\lambda\sim300-1100$ nm) and DR17 APOGEE for NIR spectra ($\lambda\sim1500-1700$ nm) \citep{kupka2011vald, hayes2022bacchusapogeelinelist}. Solar abundances are derived from LTE values provided by \cite{asplund2009solar}.

In summary, E-iSpec allows to prepare the stellar spectra for chemical analysis and to determine atmospheric parameters along with elemental abundances and isotopic ratios. In the following subsections, we describe the capabilities of E-iSpec. The final atmospheric parameters, elemental abundances, and carbon isotopic ratio for SZ~Mon and DF~Cyg derived with E-iSpec are provided in Tables~\ref{tab:fnlabsSZM_paper1} and \ref{tab:fnlabsDFC_paper1}, respectively. Additionally, we present an extensive benchmarking of E-iSpec in Appendix~\ref{app:tst_paper1}.

\subsubsection{Preparing data for spectral analyses}\label{sssec:saneisprep_paper1}
To prepare the spectra for the atmospheric parameter and elemental abundance analyses, E-iSpec offers a series of rigorous data pre-processing steps. These includes continuum normalisation, cosmic rays removal, radial velocity correction, and construction of the initial line list that is used in the stellar spectral analyses.

In E-iSpec, we provided two methods of continuum normalisation: manual and automatic. The manual normalisation involved fitting $10^{\rm th}$-order polynomial functions through interactively determined continuum points, typically spaced at intervals of $\approx$10\,nm. In contrast, the automatic normalisation involved fitting $3^{\rm rd}$-order splines through automatically determined continuum points. The continuum points in the automatic normalisation routine were determined using the iSpec functions \texttt{fit\_continuum} and \texttt{normalize\_spectrum}. In brief, \texttt{fit\_continuum} function performed a median filtering for every 3 data points and a maximum filtering for every 30 data points to the spectral data. The remaining data points were then fitted with the number of spline knots automatically determined by the normalisation routine. The selection of a normalisation procedure depended on the characteristics of the spectral features. In general, spectra containing atomic lines tended to yield the best results when a two-step approach is applied, involving manual normalisation followed by automatic normalisation. On the other hand, for spectra featuring molecular bands, manual continuum normalisation alone was recommended.  

Once the spectra were normalised, the cosmic rays could be removed using the iSpec function \texttt{create\_filter\_cosmic\_rays} with default settings (\texttt{resampling\_wave\_step} was the wavelength step, \texttt{window\_size} was 15, and \texttt{variation\_limit} was 0.5).

Then, the normalised and cleaned spectra were corrected for the radial velocity using the iSpec function \texttt{cross\_correlate\_with\_template}. In case of the HERMES spectra, the correction started with the spectrum smoothing by degrading its resolution from 85~000 to 50~000, and then cross-correlating the smoothed spectrum with the Arcturus template spectrum to determine the radial velocity shift. This step ensured precise radial velocity determination even for low-S/N ($<70$) spectra. For the APOGEE spectra, the radial velocity correction has already been performed by the APOGEE pipeline (see Section~\ref{ssec:obsspc_paper1}). The absolute difference of radial velocities between our method and the APOGEE pipeline was found to be below 1 km/s.

\begin{figure*}[!ht]
    \centering
    \includegraphics[height=.49\linewidth, angle=-90]{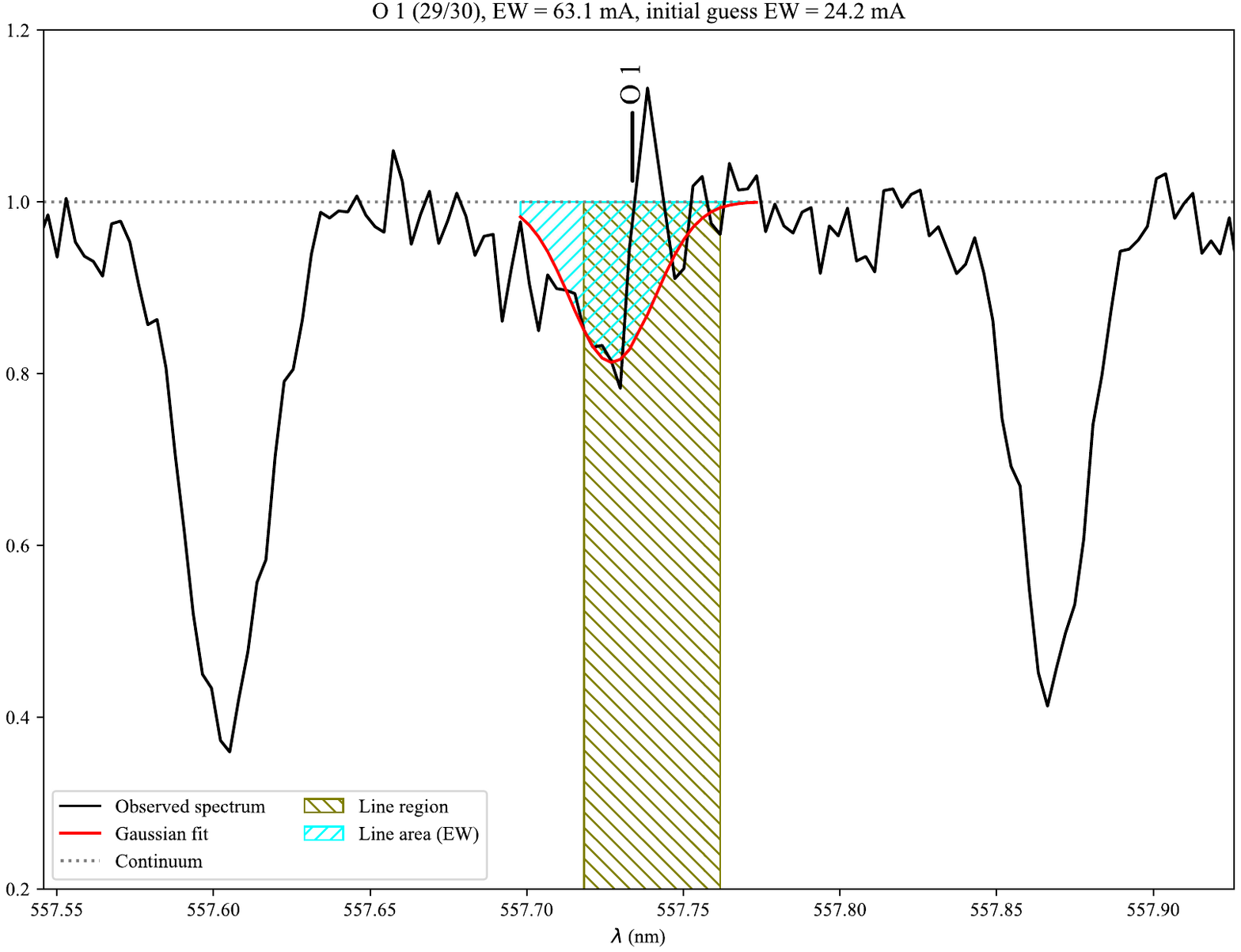}
    \includegraphics[height=.49\linewidth, angle=-90]{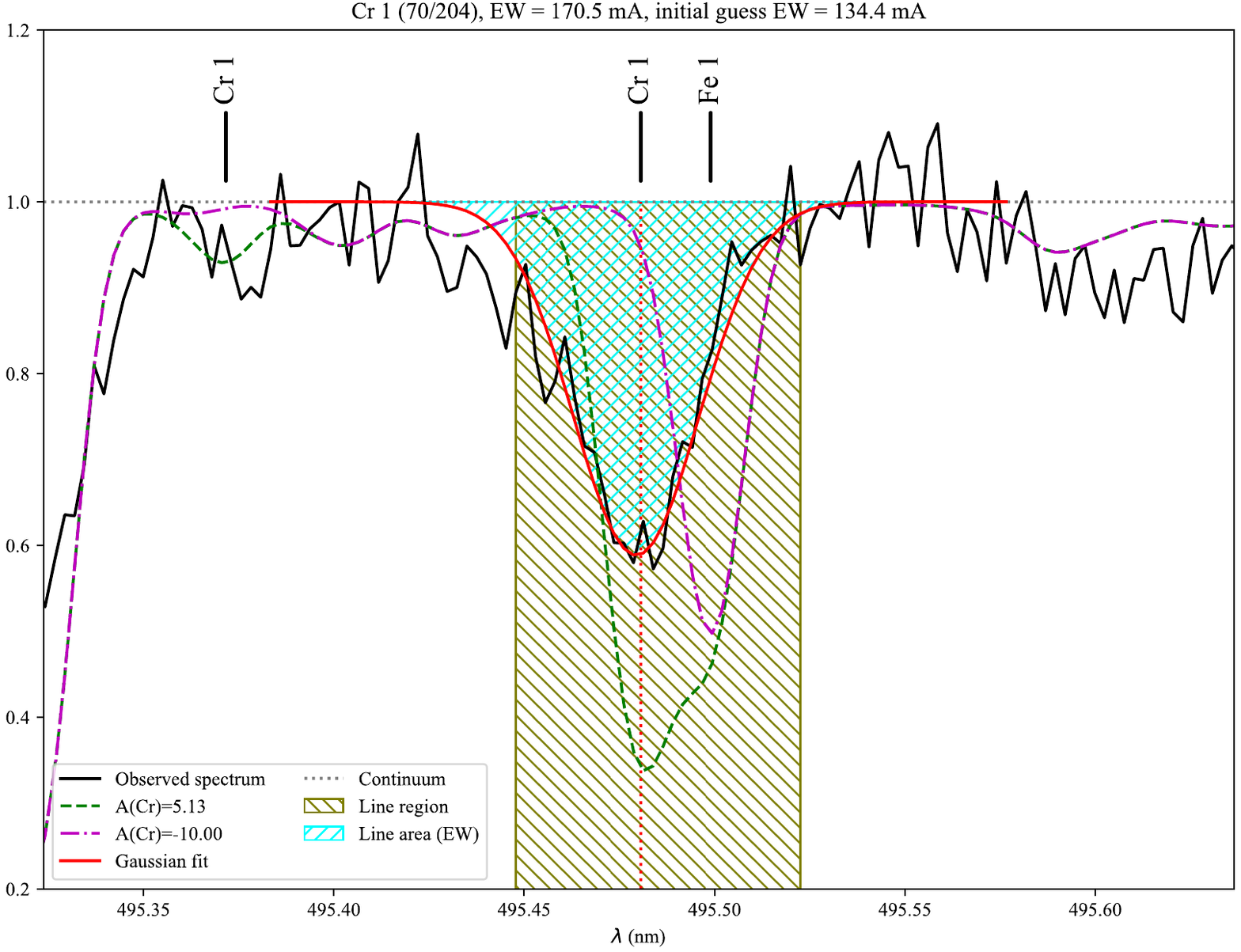}
    \includegraphics[height=.49\linewidth, angle=-90]{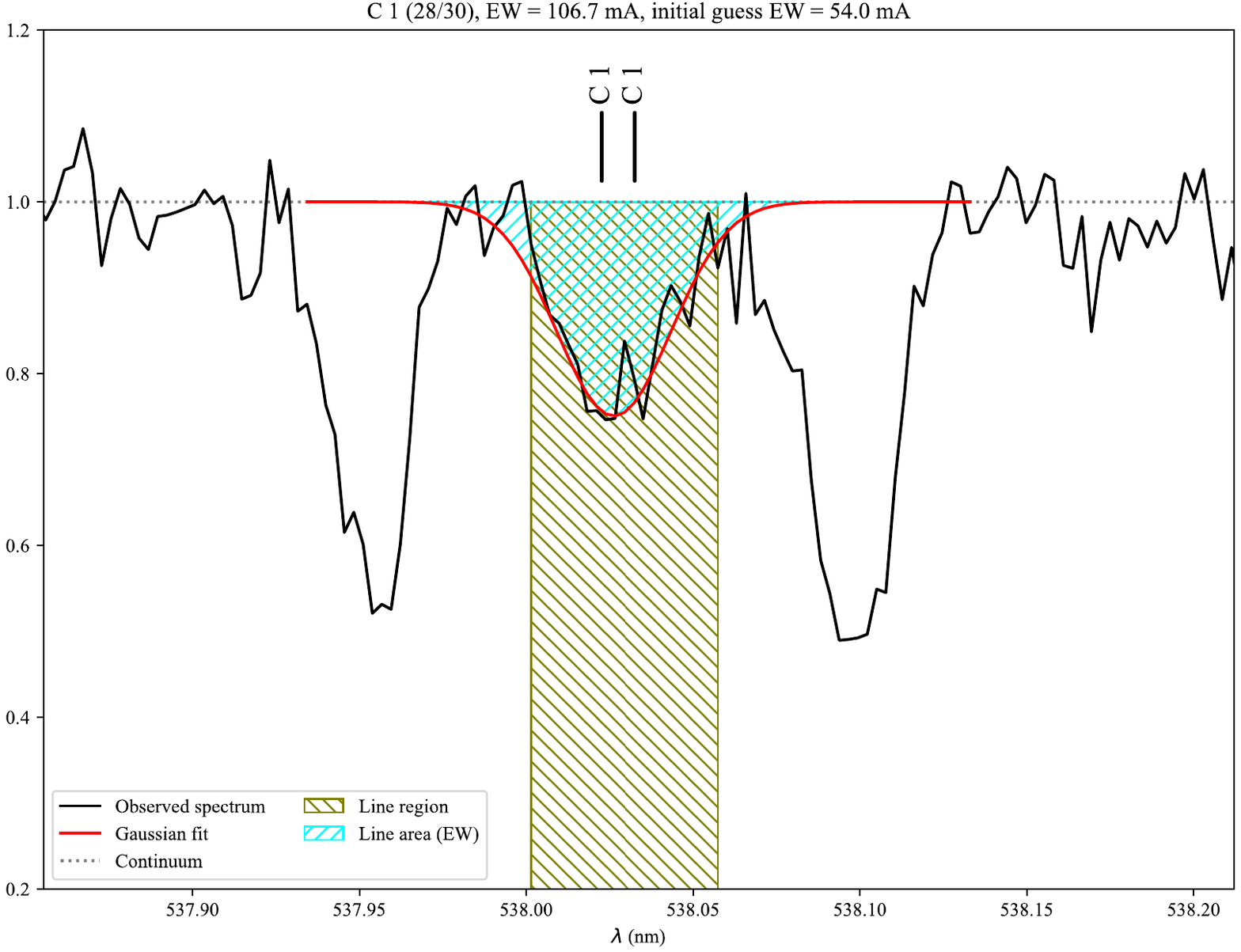}
    \includegraphics[height=.49\linewidth, angle=-90]{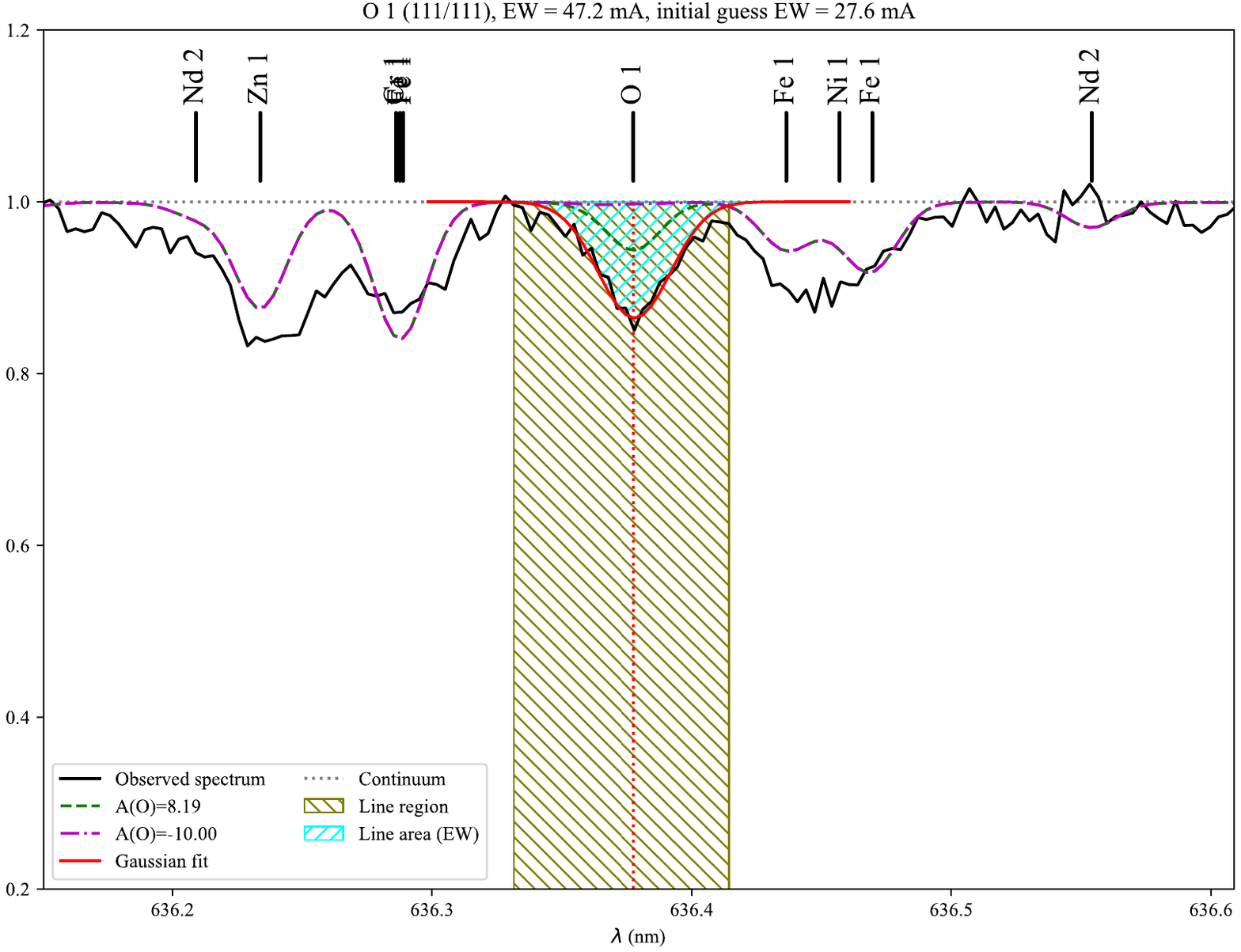}
    \caption[The examples of the spectral lines in different spectra of SZ~Mon showing the process of manual line selection]{The examples of the spectral lines in different spectra of SZ~Mon showing the process of manual line selection: from the automatically identified spectral lines we removed spectral lines affected by the cosmic hit (top left panel), blended spectral lines (top right panel), and ``self-blended'' spectral lines (bottom left panel). A typical spectral line used for the chemical analysis is shown in the bottom right panel. The legend for the symbols and colours used are included within the plot.}\label{fig:linsel_paper1}
\end{figure*}

Next, we employed the automatic line identification in iSpec (\texttt{fit\_lines}~function with \texttt{max\_atomic\_wave\_diff}~parameter being 0.005). This identification heavily depended on expected line strengths (defined as `theoretical depth' and `theoretical EW'), which were calculated for the initially specified set of the atmospheric parameters. Consequently, the closer the input atmospheric parameters were to the final results, the better spectral lines were identified in spectra. For our target sample, this effect was prominent only between visits obtained at slightly hotter and cooler phases (see Table~\ref{tab:obslog_paper1}).

Finally, we manually selected lines from the automatic line list (see Fig.~\ref{fig:linsel_paper1}). The process of line selection involved three steps: i) coarse filtering of poorly identified spectral lines (top left panel), ii) fine filtering of blended spectral lines (top right panel), and iii) final filtering of outliers (bottom left panel). To perform the fine filtering, we overplotted the observed spectrum with two synthetic spectra, which have atmospheric parameters fixed at literature values and most abundances fixed at metallicity-scaled solar values. The difference between synthetic spectra lied in the abundance of the element identified for the selected spectral feature: for one spectrum, $[$X/H$]$ was set at -10 dex to simulate the absence of the element; for the other spectrum, $[$X/H$]$ was set at metallicity-scaled solar value. We note that, while applying the first two filtering steps aided in excluding most of the misidentified lines and blends, the third step was needed to remove the blends of the element with itself (both same and different ionisation). Once the automatically selected lines were manually filtered (bottom right panel), the final line list was used to derive atmospheric parameters and elemental abundances.

\subsubsection{Determining atmospheric parameters and elemental abundances from atomic lines}\label{sssec:stepar_paper1}
To determine the atmospheric parameters ($T_{\rm eff}$, $\log g$, $[$Fe/H$]$, and $\xi_{\rm t}$) we used iSpec function \texttt{model\_spectrum\_from\_ew} with the following settings: radiative transfer code was chosen to be Moog, iterations were run until convergence with a maximum number of 10 (as proposed by \cite{blancocuaresma2014, blancocuaresma2019}), and automatic detection of outliers was almost switched off by setting the sigma clipping level at 10-$\sigma$. This function used EW method by comparing the observed spectral lines with the Gaussian fits. The errors of atmospheric parameters provided by this function were calculated from the covariance matrix constructed by the non-linear least-squares fitting algorithm.

The elemental abundances ($[$X/H$]$) were determined from atomic lines using function \texttt{model\_spectrum} with similar settings: Moog was the preferred radiative transfer code, and the iterations were run until convergence with a maximum number of 10. This function provided the errors of elemental abundances as standard deviations of the measured values (line-to-line scatter), so we needed to also account for the errors caused by the atmospheric parameters (systematic error; see Section~\ref{sssec:sanerr_paper1}). In Table~\ref{tabA:tstsmp_paper1} we provided the comparison of atmospheric parameters of chemically diverse sample of post-AGB stars: literature values were obtained from the manually selected spectral lines, while our values are obtained using automatic EW method.

\subsubsection{Deriving elemental abundances and isotopic ratios from molecular bands}\label{sssec:abuiso_paper1}
To derive the carbon isotopic ratios, we developed our own algorithm, which we discuss below. To test its performance, we expanded it to also derive elemental (atomic and molecular) abundances, which we could detect in APOGEE spectra (see Section~\ref{sssec:obsspcnir_paper1}). We note that our SSF algorithm for molecular bands was generally slower than iSpec's SSF routine for atomic lines. However, we benchmarked our algorithm for red giants (\cite{masseron2019APOGEE+BACCHUS}; see Table~\ref{tabA:tstabs_paper1}) and for different visits of SZ~Mon and DF~Cyg (optical counterparts were obtained using EW method; see Table~\ref{tabA:tststp_paper1}).

We started the abundance analysis with identification of spectral regions that were sensitive to specific elements and isotopes. For this, we synthesised a reference spectrum with all elemental abundances set to metallicity-scaled solar (MSS) values. We then created individual synthetic spectra for each element, with that element's abundance enhanced by 1 dex above MSS value while keeping all other elements at MSS values. Subtracting the reference spectrum from each enhanced spectrum resulted in a plot of spectral sensitivity for each selected chemical element. We note that for the degenerate broadening effects (spectral resolution, rotational velocity, and macroturbulent velocity), we used fixed values for resolution (R = 22500) and rotational velocity ($v\sin i$ = 0 km/s), and only the macroturbulence $v_{\rm mac}$ was a free parameter, accounting for all broadening effects.

We then manually filtered the regions in these spectral sensitivity plots where flux differences exceeded the overall standard deviation of the normalised flux in the corresponding enhanced spectrum. Upon this filtering, we obtained the final spectral regions used for abundance analysis.

Once we identified the final spectral windows, we synthesised spectra for different element abundances and used a $\chi^2$ minimisation procedure to determine the best-fit elemental abundances and isotopic ratios. In the current version of the script, we found the minimal $\chi^2$ values using the Fibonacci search method, which has a time complexity of $O(\log N)$. The $\chi^2$ for the selected spectral window is given by
\begin{equation}
    \chi^2 = \sum_\lambda\frac{(O_\lambda-S_\lambda)^2}{\sigma_S^2},
\end{equation}
where $O_\lambda$ and $S_\lambda$ represent the normalised fluxes of the observed and synthetic spectra, respectively. $\sigma_S$ represents the standard deviation of the synthetic flux within a specific spectral feature. This calculation is conducted across the wavelength range outlined in the NIR line lists, which are detailed in Tables~\ref{tabA:nirlstszm_paper1} and \ref{tabA:nirlstdfc_paper1} within Appendix~\ref{app:add_paper1}. These tables provide multicolumn wavelength ranges corresponding to different spectral features we used. The squared fractions ($\frac{(O_\lambda-S_\lambda)^2}{\sigma_S^2}$) are then summed over the entire wavelength range of the studied feature.

We note that in iSpec, the reference isotopic ratios are input in the SPECTRUM format $p(\textrm{X}_i) = N(\textrm{X}_i)/N(\textrm{X})$, where $N(\textrm{X}_i)$ and $N(\textrm{X})$ are number densities of element X (in specific isotopic form $\textrm{X}_i$ and in total, respectively). Similarly, the common convention for CNO isotopic ratios can be described as
\begin{equation}
    \frac{\textrm{X}_{\rm main}}{\textrm{X}_{\rm var}} = \frac{N(\textrm{X}_{\rm main})}{N(\textrm{X}_{\rm var})} = \frac{N(\textrm{X}_{\rm main})/N(\textrm{X})}{N(\textrm{X}_{\rm var})/N(\textrm{X})} = \frac{p(\textrm{X}_{\rm main})}{p(\textrm{X}_{\rm var})},
\end{equation}
where $\textrm{X}_{\rm main}$ and $\textrm{X}_{\rm var}$ are main and variant isotopic forms of element X, respectively.

We also note that E-iSpec inherits the Turbospectrum method of operating with the total [C/H] abundance for a selected $^{12}$C/$^{13}$C ratio. These two parameters should simultaneously fit the molecular bands of CO and CN, as well as their isotopologues $^{13}$CO and $^{13}$CN, respectively. For atomic lines, $^{12}$C and $^{13}$C contribute to the same spectral features, hence the [C/H] abundances derived from the optical spectra were crucial to validate the [C/H] abundances derived from the molecular spectra.

\begin{figure}[!ht]
    \centering
    \includegraphics[width=.49\linewidth]{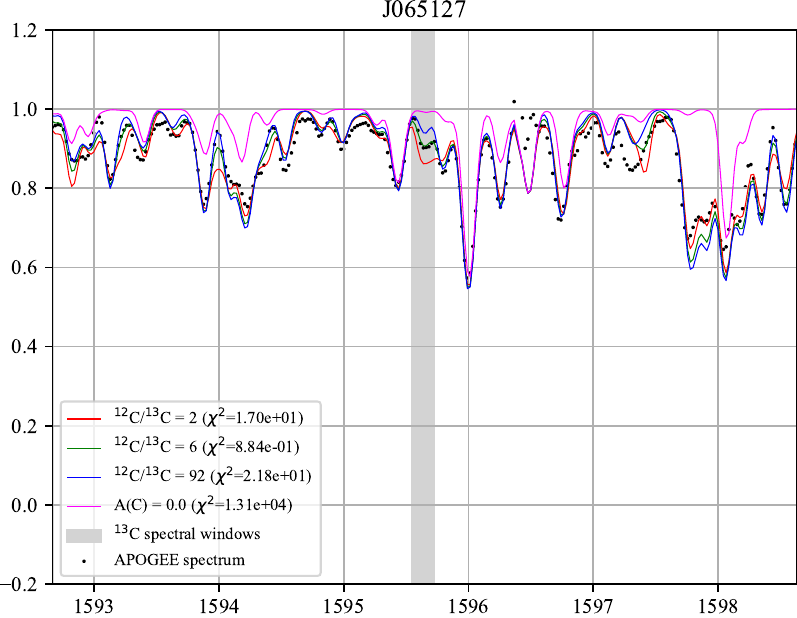}
    \includegraphics[width=.49\linewidth]{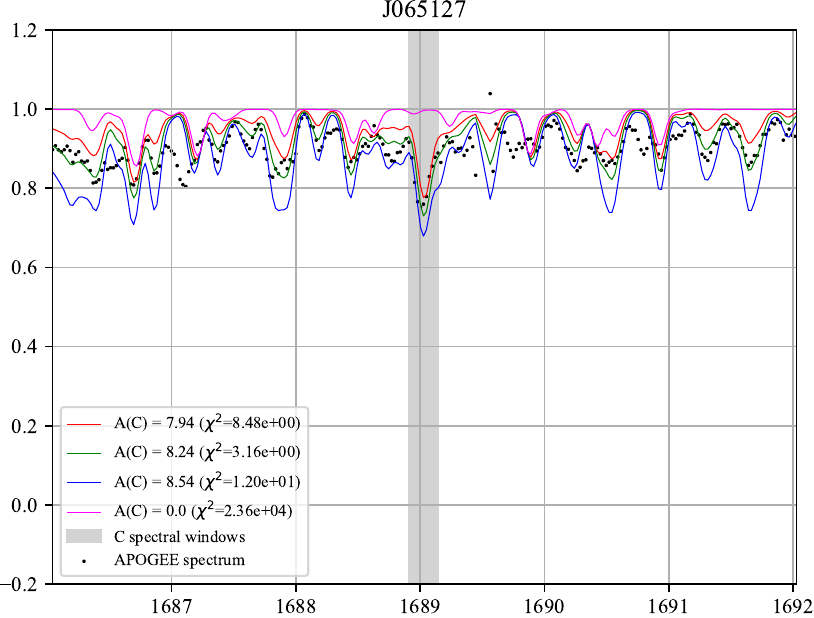}
    \includegraphics[width=.49\linewidth]{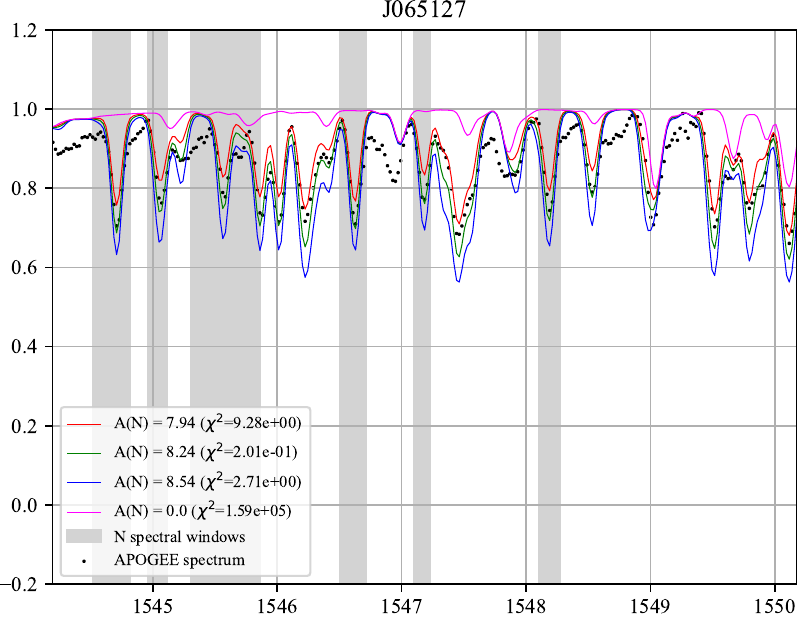}
    \includegraphics[width=.49\linewidth]{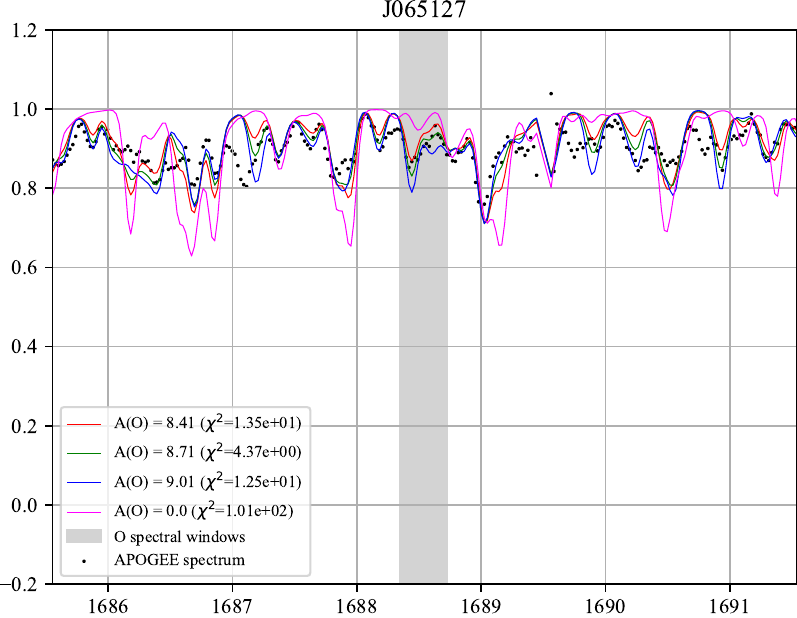}
    \caption[Example fits of spectral features sensitive to $^{12}$C/$^{13}$C (top left panel), {[C/H]} (top right panel), {[N/H]} (bottom left panel), and {[O/H]} (bottom right panel) in the spectrum of SZ~Mon (shown as shaded areas)]{Example fits of spectral features sensitive to $^{12}$C/$^{13}$C (top left panel), [C/H] (top right panel), [N/H] (bottom left panel), and [O/H] (bottom right panel) in the spectrum of SZ~Mon (shown as shaded areas). Observed spectrum is marked with black dots. Synthetic spectra with best-fit--0.3 dex, best-fit, best-fit+0.3 dex, and ``zero'' elemental abundances are graphed with magenta, green, red, and blue lines, respectively (``zero'' abundance is achieved by excluding the spectral features related to the element in question from the line list). Abundances of all other chemical elements in the synthetic spectra are fixed at metallicity scaled solar values. We highlight that the lower right panel demonstrates the well-known inverse relationship between oxygen abundance and the strength of spectral features associated with carbon-bearing molecules such as CN and C$_2$. The legend for the symbols and colours used are included within the plot. For more details see Section~\ref{sssec:abuiso_paper1}.}\label{fig:SpWnSZM_paper1}
\end{figure}

\begin{figure}[!ht]
    \centering
    \includegraphics[width=.49\linewidth]{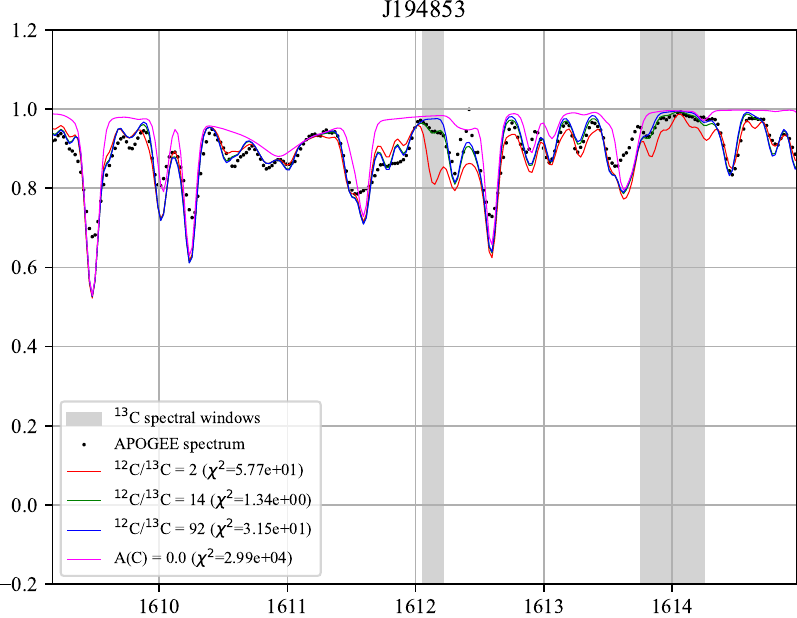}
    \includegraphics[width=.49\linewidth]{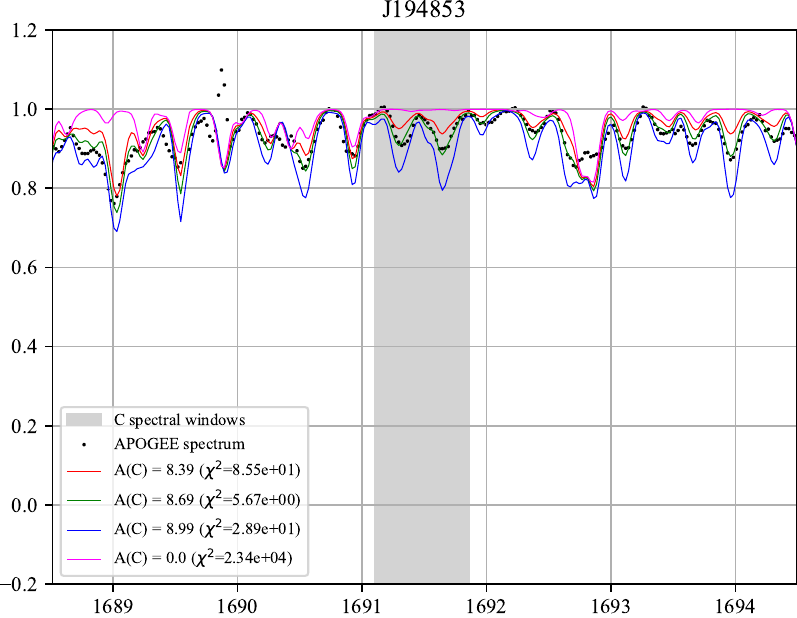}
    \includegraphics[width=.49\linewidth]{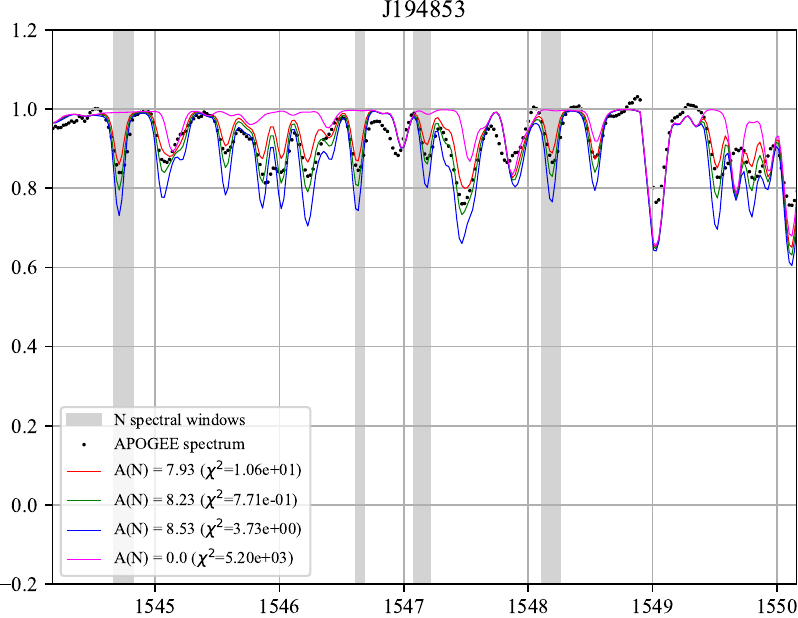}
    \includegraphics[width=.49\linewidth]{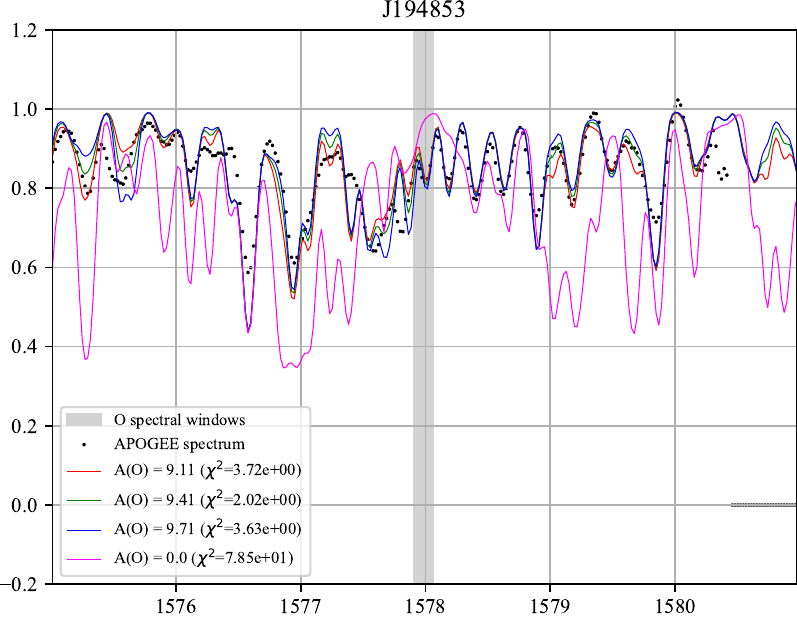}
    \caption[Example fits of spectral features sensitive to $^{12}$C/$^{13}$C (top left panel), [{C/H]} (top right panel), {[N/H]} (bottom left panel), and {[O/H]} (bottom right panel) in the spectra of DF~Cyg (shown as shaded areas)]{Example fits of spectral features sensitive to $^{12}$C/$^{13}$C (top left panel), [C/H] (top right panel), [N/H] (bottom left panel), and [O/H] (bottom right panel) in the spectra of DF~Cyg (shown as shaded areas). Observed spectrum is marked with black dots. Synthetic spectra with best-fit--0.3 dex, best-fit, best-fit+0.3 dex, and ``zero'' elemental abundances are graphed with magenta, green, red, and blue lines, respectively (``zero'' abundance is achieved by excluding the spectral features related to the element in question from the line list). Abundances of all other chemical elements in the synthetic spectra are fixed at metallicity scaled solar values. We highlight that the lower right panel demonstrates the well-known inverse relationship between oxygen abundance and the strength of spectral features associated with carbon-bearing molecules such as CN and C$_2$. The legend for the symbols and colours used are included within the plot. For more details see Section~\ref{sssec:abuiso_paper1}.}\label{fig:SpWnDFC_paper1}
\end{figure}

In Fig.~\ref{fig:SpWnSZM_paper1} for SZ~Mon and Fig.~\ref{fig:SpWnDFC_paper1} for DF~Cyg, we show examples of spectral regions (windows indicated in grey) sensitive to the $^{12}$C/$^{13}$C ratio (top left panel), [C/H] (top right panel), [N/H] (bottom left panel), and [O/H] (bottom right panel). The observed spectra are shown with black dotted lines, while the synthetic spectra are shown with pink, blue, green, and red solid lines. The synthetic spectra are calculated in the assumption of metallicity scaled solar abundances of all elements except the one in question. For $^{13}$C, pink spectra were synthesised for the assumed absence of the carbon (by removing all carbon-containing features from the line list), blue and red spectra set the boundaries of the solution search for $^{12}$C/$^{13}$C ratio (solar value of 92 and exotic value of 2, respectively), and green spectra were synthesised for the best-fit value of $^{12}$C/$^{13}$C ratio. For each of CNO elements, the spectra were synthesised for different elemental abundances: conditional absence of the element (pink; the abundance is set at the best-fit value, but the molecular bands containing this element and its atomic lines are excluded from the line list), best-fit decreased by 0.3 dex (blue), best-fit (green), and best-fit increased by 0.3 dex (red) abundances, respectively.

In case of DF~Cyg, there was one additional step in our analysis of CNO molecular bands. Given the near-solar metallicity of this target, spectral features in spectra of DF~Cyg (i.e., atomic lines and molecular bands) were significantly blended (compared to the spectra of more metal-poor SZ~Mon). This high level of blending lead to a degeneracy between T$_{\rm eff}$, $[$C/H$]$, and $[$O/H$]$, which was disentangled by running the iterative minimisation process: for different temperatures we fixed $[$C/H$]$ and $[$O/H$]$ in turns until all three parameters converged. We tested the range of temperatures from 4\,400\,K to 5\,400\,K (where we expected the solution to be, based on visual inspection of the aforementioned molecular bands) with increments of 100\,K, which were indicative of our uncertainties in T$_{\rm eff}$. As a result, this iteration process converged at T$_{\rm eff}$ = 4\,500\,K for DA\#1 and at T$_{\rm eff}$ = 5\,000\,K for DA\#2, highlighting our choice of APOGEE visit.

\subsubsection{Error estimation for atmospheric parameters and elemental abundances}\label{sssec:sanerr_paper1}

For the atmospheric parameters, we relied on iSpec's built-in error estimation (see Section~\ref{ssec:sanisp_paper1}).

For the elemental abundances, the error estimation procedure involved analysing the errors, which arose from the scattering of data points between spectral lines (random errors) and the uncertainties caused by the uncertainties of atmospheric parameters, such as temperature $T_{\rm eff}$, surface gravity $\log g$, metallicity $[$Fe/H$]$, and microturbulent velocity $\xi_{\rm t}$ (systematic errors).

We followed the procedure explained in the appendix of \cite{mcwilliam1995errors} with the assumption that atmospheric parameters are independent \citep{desmedt2014LeadMCs}. In short, the overall abundance uncertainty was determined by taking the quadratic sum of errors arising from line-to-line scatter ($\sigma_{\rm l2l}/\sqrt{N_l}$ where $N_l$ was the number of lines of an ionisation level of an element in question) and from atmospheric parameter uncertainties ($\sigma_{T_{\rm eff}}$, $\sigma_{\log g}$, $\sigma_{\xi_{\rm t}}$), given as
\begin{equation}
    \sigma_{\rm tot, [X/H]} = \sqrt{\left(\frac{\sigma_{\rm l2l}}{\sqrt{N_{l}}}\right)^2 + \left(\sigma_{T_{\rm eff}}\right)^2 + \left(\sigma_{\log g}\right)^2 + \left(\sigma_{\xi_{\rm t}}\right)^2}.
\end{equation}

We note that in cases when only one spectral line of an element was used for abundance derivation, we conventionally used 0.2 dex as an upper limit of the uncertainty.

\subsection{Spectral analysis results for SZ Mon and DF Cyg}\label{ssec:sanabs_paper1}
In this study, we used E-iSpec to determine atmospheric parameters, elemental abundances, and carbon isotopic ratios for two binary post-RGB stars: SZ~Mon and DF~Cyg. We found that the derived atmospheric parameters and elemental abundances are in good agreement with the previous studies of SZ~Mon \citep{maas2007t2cep} and DF~Cyg \citep{giridhar2005rvtau}, which is expected given the proximity of pulsation phases in our study and the corresponding literature (see Tables~\ref{tab:fnlabsSZM_paper1} and \ref{tab:fnlabsDFC_paper1}, as well as Fig.~\ref{fig:fnlabs_paper1}). We calculated more accurate elemental abundances and increased the amount of analysed species for SZ~Mon (9 newly studied species) and DF~Cyg (8 newly studied species), including $^{13}$C.

Similarly to \cite{maas2007t2cep} and \cite{giridhar2005rvtau}, we detected the chemical depletion profiles for both targets. To quantitatively describe the chemical depletion, we used [Zn/Ti] abundance ratio and the turn-off temperature $T_{\rm turn-off}$ \citep{oomen2019depletion, kluska2022GalacticBinaries}. Due to its volatile nature, Zn \citep[$T_{\rm cond,~Zn}=726$ K; ][]{lodders2003CondensationTemperatures} is anticipated to exhibit lesser depletion or non-depletion. In contrast, Ti \citep[$T_{\rm cond,~Ti}=1582$ K; ][]{lodders2003CondensationTemperatures} is a refractory element expected to be significantly more depleted for post-RGB/post-AGB binary stars, causing [Zn/Ti] ratio to grow up to 3.4 dex \citep[as in case of AF Crt;][]{kluska2022GalacticBinaries}. We found that our targets exhibit a rather mild depletion with [Zn/Ti] values of 0.4 dex for SZ~Mon and 0.3 dex for DF~Cyg, as opposed to more extreme literature values of 0.8 dex and --0.7 dex, respectively.

The turn-off temperature $T_{\rm turn-off}$ \citep{oomen2019depletion} marks the onset of chemical depletion pattern, separating the relatively non-depleted elements (i.e., volatile elements like Zn and S) and the elements affected by the depletion process (i.e., refractory elements like Fe and Ti). It is worth noting that when comparing the elemental abundances of SZ~Mon and DF~Cyg to those of the post-AGB sample (roughly selected from \cite{oomen2019depletion} by their SED luminosity $L_{\rm SED}>2\,500~L_\odot$; see Section~\ref{ssec:obslum_paper1}), we notice that $T_{\rm turn-off}$ in SZ~Mon and DF~Cyg ($\approx1\,400$~K; see Fig.~\ref{fig:fnlabs_paper1}) is higher than the median $T_{\rm turn-off}$ in binary post-AGB sample ($\approx1\,100$~K; see \cite{oomen2019depletion} and references therein).

Regarding carbon isotopic ratios, our study marks the first attempt at estimating them in post-RGB stars. Using E-iSpec, as explained in Section~\ref{sssec:abuiso_paper1}, we derived $^{12}$C/$^{13}$C=8$\pm$4 for SZ~Mon and $^{12}$C/$^{13}$C=12$\pm$3 for DF~Cyg. To better understand the implications, we compared these values with theoretical predictions from the ATON evolutionary models, outlined in Section~\ref{sec:mod_paper1}.

\section{Integration of E-iSpec with ATON stellar evolutionary models}\label{sec:mod_paper1}
Despite the expected chemical depletion pattern observed in our targets (see Section~\ref{ssec:sanabs_paper1}), the derived CNO abundances deviate from both volatile elements (e.g., S and Zn) and refractory elements (e.g., Fe, Ti, etc.). In this section, we aim to ascertain whether combined knowledge of luminosity, surface CNO abundances, and carbon isotopic ratios could serve as valuable indicators of the nucleosynthetic history during the RGB phase before transitioning to the post-RGB phase.

\begin{figure*}[!ht]
    \centering
    \includegraphics[width=.49\linewidth]{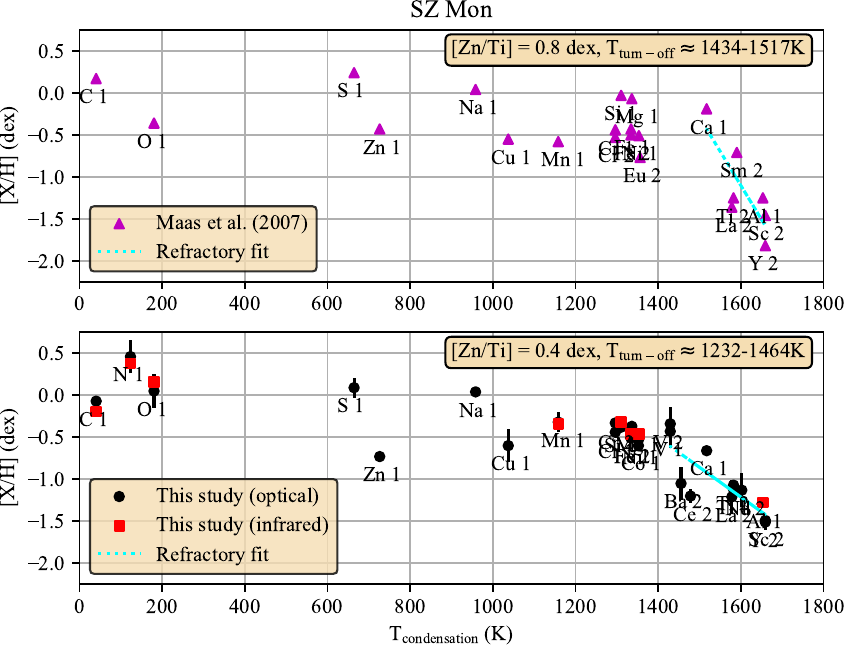}
    \includegraphics[width=.49\linewidth]{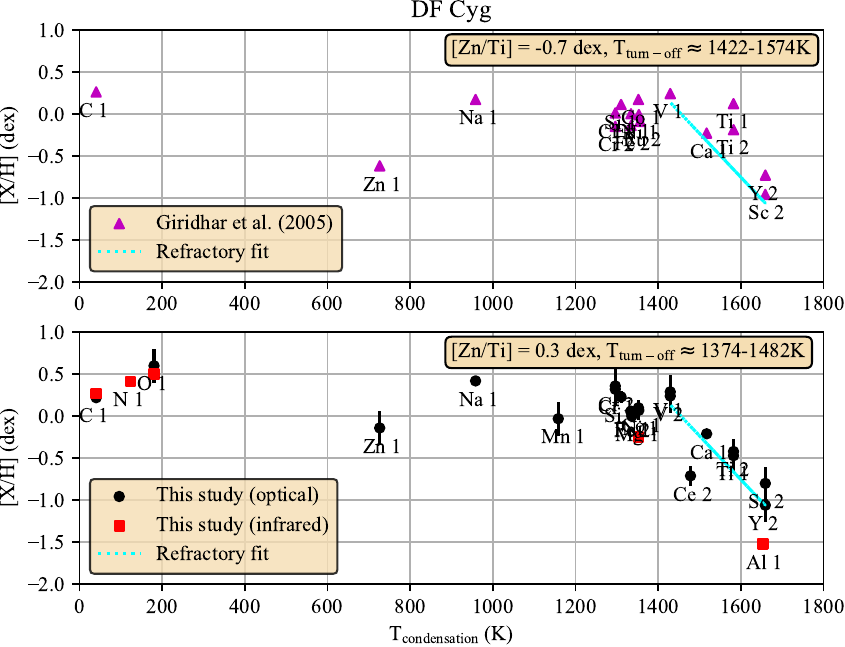}
    \caption[Abundances of SZ Mon (left) and DF Cyg (right) as functions of condensation temperature]{Abundances of SZ Mon (left) and DF Cyg (right) as functions of condensation temperature \citep{lodders2003CondensationTemperatures}. Top panels show the chemical analysis results from the literature \citep[][for SZ~Mon and DF~Cyg, respectively]{maas2007t2cep, giridhar2005rvtau}, bottom panels show the results from this study. Linear fits of the decrease in the abundances of the refractory elements are shown in cyan. The legend for the symbols and colours used are included within the plot.}\label{fig:fnlabs_paper1}
\end{figure*}

To do this, we compared the $^{12}$C/$^{13}$C ratio, [N/H], and C/O ratio of our two post-RGB targets (SZ~Mon and DF~Cyg, see Tables~\ref{tab:fnlabsSZM_paper1} and \ref{tab:fnlabsDFC_paper1}, and Fig.~\ref{fig:fnlabs_paper1}) presented above with results from stellar evolution modelling, obtained with the ATON code for stellar evolution \citep{ventura1998}. We considered evolutionary sequences of low-mass stars, running from the pre-Main Sequence stage through the core H-burning and the ascending of the RGB, until the ignition of the helium flash. We note that for both targets we assumed the initial mass of $1~M_{\odot}$ based on the evolutionary sequences for post-AGB/post-RGB single stars presented in \cite{kamath2023models}.

While the ATON evolutionary models and the models from \cite{kamath2023models} are primarily designed for single stars, we used these models to reproduce the chemical composition of our post-RGB binary targets and therefore investigate the extent, to which binary interactions affect the stellar chemical composition. Specifically, we examined whether the chemical composition, especially the CNO abundances, observed in the post-RGB phase, could reflect the nucleosynthesis occurring during the RGB phase before this phase is terminated by the binary interaction.

We chose the luminosity as an indicator of the evolutionary stage, as use of time would be of little help in this case, given the significant difference among the duration of the various evolutionary phases. Given that the luminosities of SZ~Mon and DF~Cyg are significantly below $\rm 1.5\times10^3~L_{\odot}$ (see Section~\ref{ssec:obslum_paper1}), we safely assumed that these objects never entered the thermal-pulse phase \citep{ventura2022InternalProcesses}. However, the possibility that SZ~Mon and DF~Cyg are currently evolving through the early AGB phase following the exhaustion of central helium can not be ruled out completely. If we consider the evolutionary phases running from the tip of the RGB $L_{\rm RGB\ tip}\approx2\,500 L_\odot$ \citep{kamath2016PostRGBDiscovery} until the first few thermal pulses, possible changes in the surface chemistry might be caused by the action of the second dredge-up in stars of initial mass above $\rm 3~M_{\odot}$, or by non-canonical deep mixing during the helium flash. The first hypothesis can be ruled out in the present context, as the luminosities at which the second dredge-up occurs are $\rm \sim 10^4~L_{\odot}$ or more, thus inconsistent with the values reported in Table~\ref{tab:varpro_paper1}. The possibility that SZ~Mon and DF~Cyg  are currently in a post-HB phase, and that their surface chemistry was altered by some deep mixing episode, such as those invoked by \citet{schwab2020HeFlashMixing} to explain the presence of lithium-rich giants, can not be excluded. However, we consider this option unlikely, as the deep mixing would favour lithium and $^{12}$C enrichment, whereas the valued reported in Tables~\ref{tab:fnlabsSZM_paper1} and \ref{tab:fnlabsDFC_paper1} indicate the typical effects of deep mixing of the surface convection with material processed by CNO cycling, typical of the RGB evolution. Therefore, we believe that the modelling of the RGB phase sufficiently characterises the two post-RGB targets.

The standard evolutionary sequences used in this study were calculated by using a diffusive approach to couple mixing of chemicals in regions unstable to convective motions and nuclear burning, following the classic scheme proposed by \citet{cloutman1976diffus}, where the diffusion coefficient $\rm D_{conv}$ is assumed to be proportional to convective velocities. To model the effects of deep extra mixing connected to the thermohaline instability, we also calculated evolutionary sequences where after the RGB bump the diffusion coefficient is estimated by adding to $\rm D_{conv}$ a term proportional to the gradient of the molecular weight \citep[$\rm D_t$; see Eq.~3 in][]{charbonnel2010thermohaline}. As a first try, we used the same value of the proportional coefficient for the molecular weight \citep[$\rm C_t$; see Eq.~4 in][]{charbonnel2010thermohaline}, namely $\rm C_t=1000$. To explore the sensitivity of the results obtained from this assumption, we also considered higher values of $\rm C_t=3000$, thus simulating the effects of very deep mixing experienced by the stars after the RGB bump.

\begin{table}[!ht]
    \centering
    \scriptsize
    \caption[Final chemical analysis results for SZ Mon]{Final chemical analysis results for SZ Mon. Atmospheric parameters and elemental abundances were derived from SH\#73, unless stated otherwise. Chemical analysis results derived from other visits are specified in Table \ref{tabA:tststp_paper1}.}\label{tab:fnlabsSZM_paper1}
    \begin{tabular}{|cccc|}\hline
        \multicolumn{4}{|c|}{\textbf{SZ Mon}} \\\hline
        \multicolumn{2}{|c}{$T_{\rm eff}$ = 5460$\pm$60 K} & \multicolumn{2}{c|}{$\log g$ = 0.93$\pm$0.10 dex} \\
        \multicolumn{2}{|c}{[Fe/H] = --0.50$\pm$0.05 dex} & \multicolumn{2}{c|}{$\xi_t$ = 4.37$\pm$0.08 km/s} \\\hline
        \textbf{Element} & \begin{tabular}{c} \textbf{A09}$^a$\\ \boldmath$\log\varepsilon_\odot$ \textbf{(dex)} \end{tabular} & \begin{tabular}{c} \textbf{This study} \\ \boldmath$[{\rm X/H}]$ \textbf{(dex)} \end{tabular} & \begin{tabular}{c} \textbf{M07}$^b$ \\ $[{\rm X/H}]$ \textbf{(dex)} \end{tabular} \\ \hline
        \ion{C}{i} & 8.43 & --0.07$\pm$0.05 & +0.17\\
        \ion{N}{i} & 7.83 & +0.38$\pm$0.06$^c$ & --\\
        \ion{O}{i} & 8.69 & +0.16$\pm$0.14$^c$ & --0.36\\
        \ion{Na}{i} & 6.24 & +0.04$\pm$0.05 & +0.04\\
        \ion{Mg}{i} & 7.60 & --0.37$\pm$0.04 & --0.07\\
        \ion{Al}{i} & 6.45 & --1.29$\pm$0.01$^c$ & --1.25\\
        \ion{Si}{i} & 7.51 & --0.38$\pm$0.05 & --0.03\\
        \ion{S}{i} & 7.12 & +0.09$\pm$0.12 & +0.24\\
        \ion{Ca}{i} & 6.34 & --0.66$\pm$0.03 & --0.19\\
        \ion{Sc}{ii} & 3.15 & --1.50$\pm$0.07 & --1.46\\
        \ion{Ti}{i} & 4.95 & --1.12$\pm$0.01 & --\\
        \ion{Ti}{ii} & 4.95 & --1.07$\pm$0.06 & --1.25\\
        \ion{V}{i} & 3.93 & --0.43$\pm$0.20 & --\\
        \ion{V}{ii} & 3.93 & --0.34$\pm$0.20 & --\\
        \ion{Cr}{i} & 5.64 & --0.45$\pm$0.02 & --0.44\\
        \ion{Cr}{ii} & 5.64 & --0.34$\pm$0.01 & --0.53\\
        \ion{Mn}{i} & 5.43 & --0.32$\pm$0.12 & --0.58\\
        \ion{Fe}{i} & 7.50 & --0.51$\pm$0.05 & --0.43\\
        \ion{Fe}{ii} & 7.50 & --0.51$\pm$0.02 & --0.50\\
        \ion{Co}{i} & 4.99 & --0.60$\pm$0.06 & --\\
        \ion{Ni}{i} & 6.22 & --0.52$\pm$0.08 & --0.51\\
        \ion{Cu}{i} & 4.19 & --0.60$\pm$0.20 & --0.55\\
        \ion{Zn}{i} & 4.56 & --0.73$\pm$0.04 & --0.43\\
        \ion{Y}{ii} & 2.21 & --1.51$\pm$0.09 & --1.82\\
        \ion{Ba}{ii} & 2.18 & --1.05$\pm$0.20 & --\\
        \ion{La}{ii} & 1.10 & --1.21$\pm$0.20 & --1.36\\
        \ion{Ce}{ii} & 1.58 & --1.20$\pm$0.08 & --\\
        \ion{Nd}{ii} & 1.42 & --1.13$\pm$0.20 & --\\
        \ion{Sm}{ii} & 1.01 & -- & --0.71\\
        \ion{Eu}{ii} & 0.52 & -- & --0.77\\\hline
        $^{12}$C/$^{13}$C & $\sim$89 & 8$\pm$4 & -- \\\hline
    \end{tabular}\\
    \textbf{Note:} The given value comes from: $^a$\cite{asplund2009solar}, $^b$\cite{maas2007t2cep}, $^c$SA\#2.
\end{table}

\begin{table}[!ht]
    \centering
    \scriptsize
    \caption[Final chemical analysis results for DF Cyg]{Final chemical analysis results for DF Cyg. Atmospheric parameters and elemental abundances were derived from DH\#83, unless stated otherwise. Chemical analysis results derived from other visits are specified in Table \ref{tabA:tststp_paper1}.}\label{tab:fnlabsDFC_paper1}
    \begin{tabular}{|p{1cm}p{0.5cm}p{1.75cm}p{1.75cm}|}\hline
        \multicolumn{4}{|c|}{\textbf{DF Cyg}} \\\hline
        \multicolumn{2}{|c}{$T_{\rm eff}$ = 5770$\pm$70 K} & \multicolumn{2}{c|}{$\log g$ = 1.92$\pm$0.09 dex} \\
        \multicolumn{2}{|c}{[Fe/H] = 0.05$\pm$0.05 dex} & \multicolumn{2}{c|}{$\xi_t$ = 3.97$\pm$0.03 km/s} \\\hline
        \textbf{Element} & \begin{tabular}{c} \textbf{A09}$^a$\\ \boldmath$\log\varepsilon_\odot$ \textbf{(dex)} \end{tabular} & \begin{tabular}{c} \textbf{This study} \\ \boldmath$[{\rm X/H}]$ \textbf{(dex)} \end{tabular} & \begin{tabular}{c} \textbf{G05}$^c$ \\ \boldmath$[{\rm X/H}]$ \textbf{(dex)} \end{tabular} \\ \hline
        \ion{C}{i} & 8.43 & +0.22$\pm$0.04 & +0.26\\
        \ion{N}{i} & 7.83 & +0.41$\pm$0.08$^c$ & --\\
        \ion{O}{i} & 8.69 & +0.60$\pm$0.20 & --\\
        \ion{Na}{i} & 6.24 & +0.42$\pm$0.02 & +0.17$\pm$0.18\\
        \ion{Mg}{i} & 7.60 & +0.00$\pm$0.06 & --\\
        \ion{Al}{i} & 6.45 & --1.52$\pm$0.10$^c$ & --\\
        \ion{Si}{i} & 7.51 & +0.23$\pm$0.07 & +0.11$\pm$0.06\\
        \ion{Ca}{i} & 6.34 & --0.21$\pm$0.02 & --0.23$\pm$0.12\\
        \ion{Sc}{ii} & 3.15 & --0.80$\pm$0.20 & --0.96$\pm$0.32\\
        \ion{Ti}{i} & 4.95 & --0.47$\pm$0.20 & +0.12$\pm$0.16\\
        \ion{Ti}{ii} & 4.95 & --0.42$\pm$0.01 & --0.19\\
        \ion{V}{i} & 3.93 & +0.29$\pm$0.20 & +0.24$\pm$0.19\\
        \ion{V}{ii} & 3.93 & +0.24$\pm$0.20 & --\\
        \ion{Cr}{i} & 5.64 & +0.32$\pm$0.20 & +0.01$\pm$0.13\\
        \ion{Cr}{ii} & 5.64 & +0.36$\pm$0.20$^d$ & --0.15$\pm$0.07\\
        \ion{Mn}{i} & 5.43 & --0.03$\pm$0.20$^d$ & --\\
        \ion{Fe}{i} & 7.50 & +0.06$\pm$0.05 & +0.03$\pm$0.17\\
        \ion{Fe}{ii} & 7.50 & +0.05$\pm$0.02 & --0.11$\pm$0.18\\
        \ion{Co}{i} & 4.99 & +0.10$\pm$0.06 & +0.17$\pm$0.06\\
        \ion{Ni}{i} & 6.22 & +0.07$\pm$0.12 & --0.01$\pm$0.08\\
        \ion{Zn}{i} & 4.56 & --0.14$\pm$0.20 & --0.62\\
        \ion{Y}{ii} & 2.21 & --1.06$\pm$0.20 & --0.73\\
        \ion{Ce}{ii} & 1.58 & --0.71$\pm$0.12 & --\\
        \ion{Eu}{ii} & 0.52 & -- & --0.09\\\hline
        $^{12}$C/$^{13}$C & $\sim$89 & 12$\pm$3 & -- \\\hline
    \end{tabular}\\
    \textbf{Note:} The given value comes from: $^a$\cite{asplund2009solar}, $^b$\cite{giridhar2005rvtau}, $^c$DA\#1, $^d$DH\#26.\\
\end{table}

\subsection{SZ Mon}\label{ssec:conszm_paper1}
\begin{figure}[!ht]
    \centering
    \includegraphics[width=.70\linewidth]{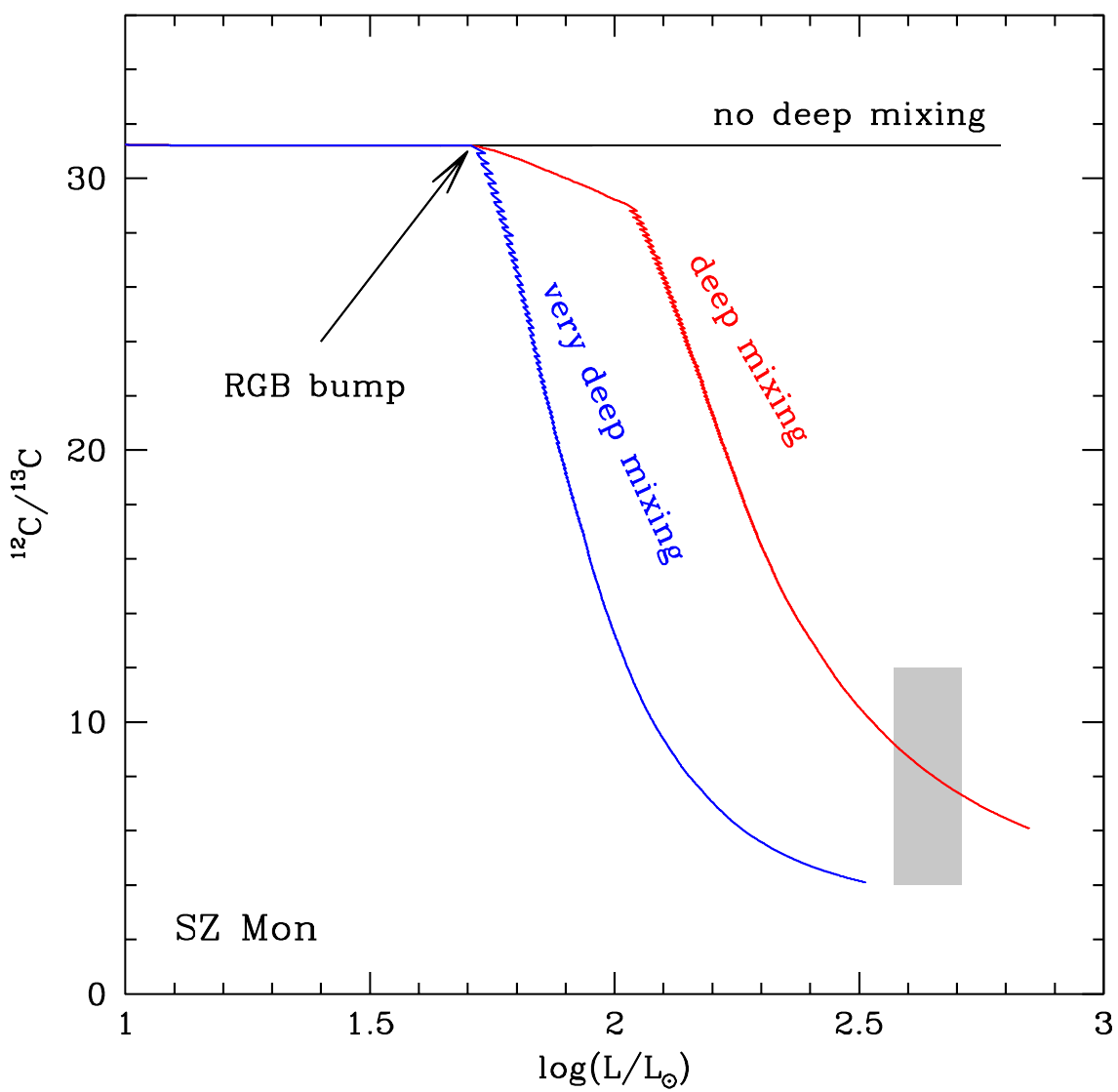}
    \includegraphics[width=.49\linewidth]{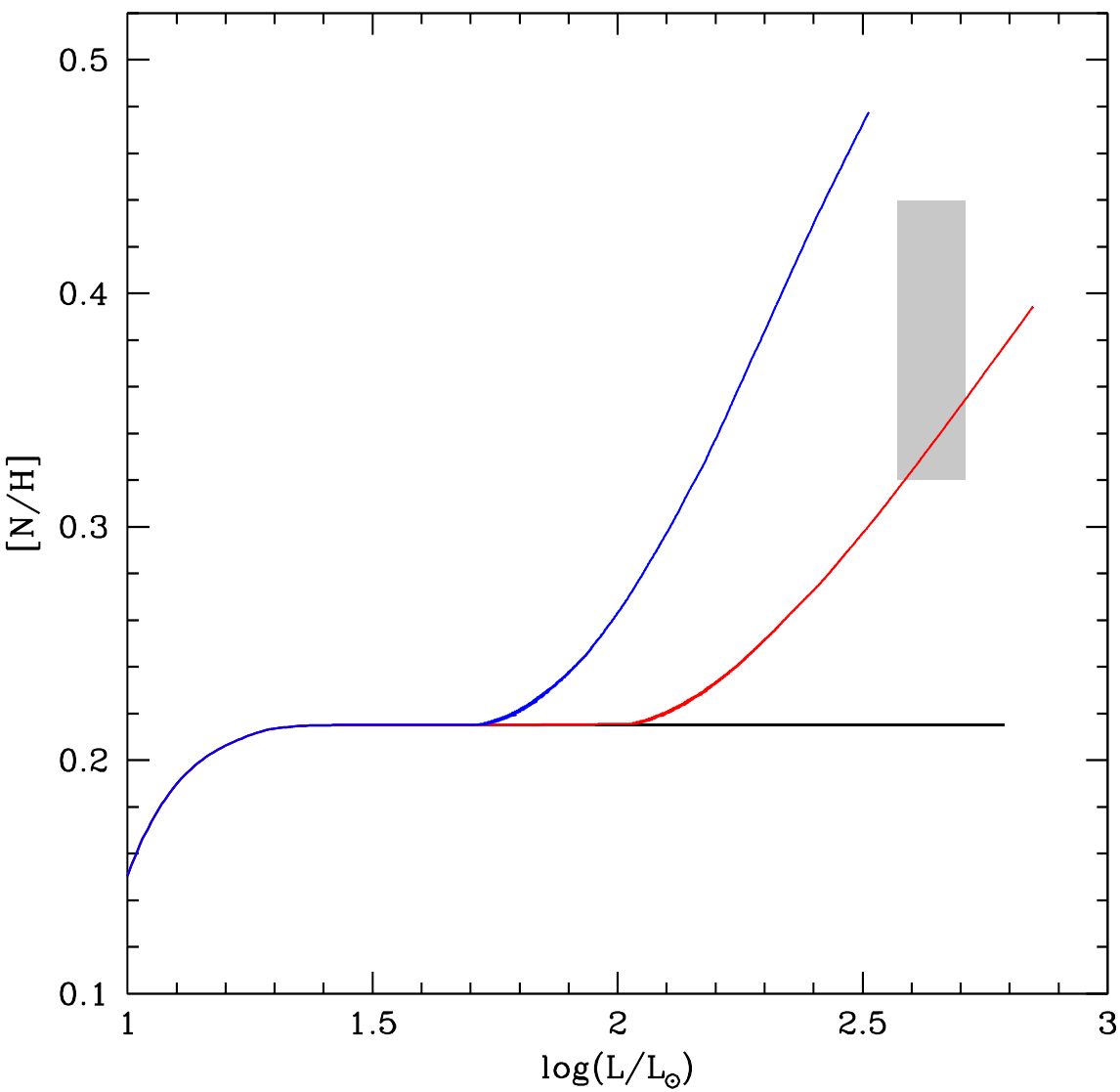}
    \includegraphics[width=.49\linewidth]{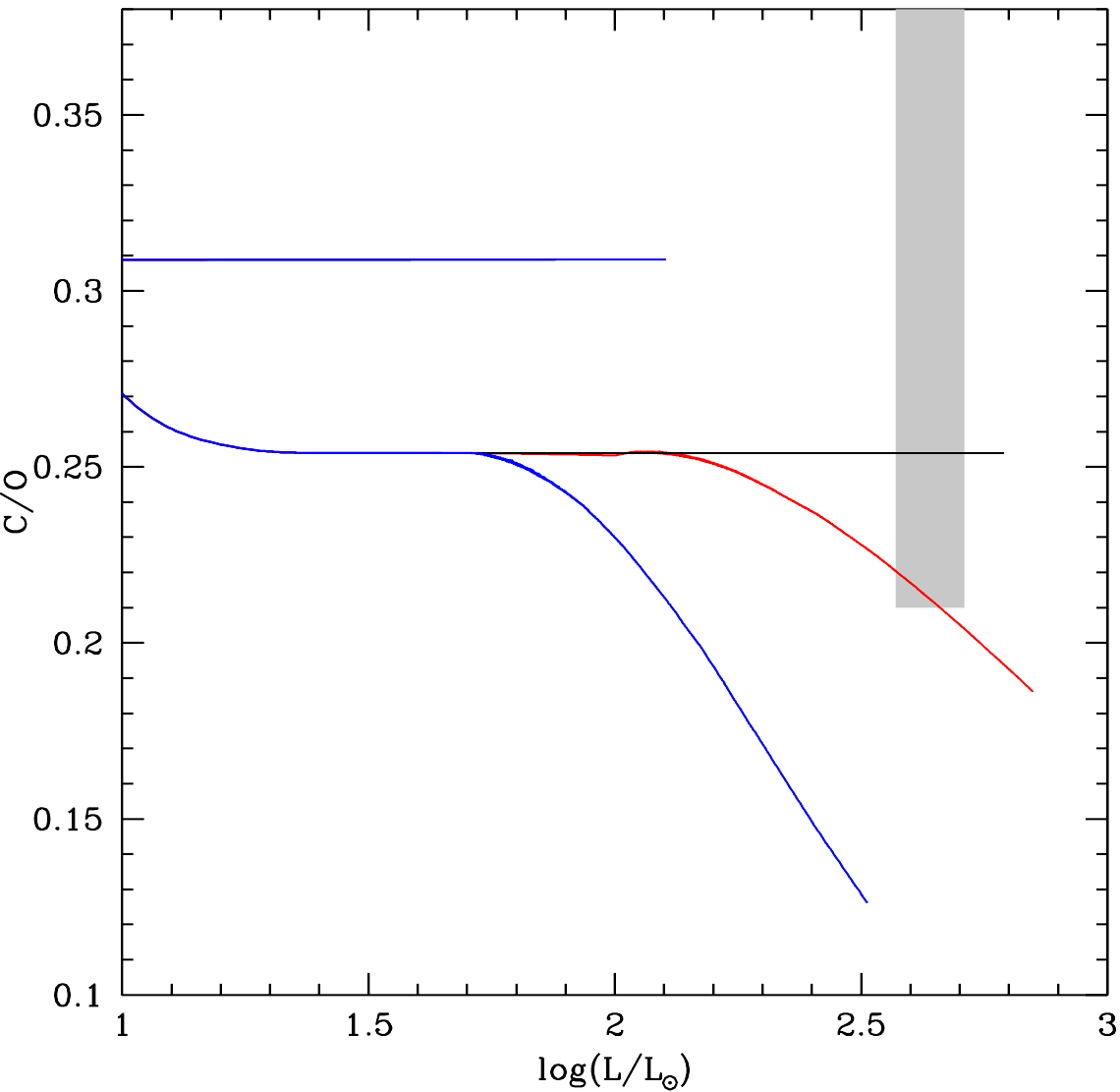}
    \caption[ATON evolutionary tracks of luminosity vs $^{12}$C/$^{13}$C (upper panel), luminosity vs {[N/H]} (lower left panel), and luminosity vs C/O (lower right panel) for a 1 M$_\odot$, Z=0.004 star]{ATON evolutionary tracks of luminosity vs $^{12}$C/$^{13}$C (upper panel), luminosity vs [N/H] (lower left panel), and luminosity vs C/O (lower right panel) for a 1 M$_\odot$, Z=0.004 star. The solid lines are the evolutionary tracks of CNO abundances and carbon isotopic ratios for models assuming FDU and different levels of extra mixing: no deep mixing (black), moderately deep mixing (red), and very deep mixing (blue). We note that the flat line at C/O$\sim$0.31 (see lower right panel) represents the starting part (pre-MS) of the C/O evolutionary track. The grey boxes indicate the observed values for SZ~Mon within corresponding uncertainties (see Table~\ref{tab:fnlabsSZM_paper1}).}\label{fig:modszm_paper1}
\end{figure}

To interpret the CNO abundances of SZ Mon and investigate its evolution, we modelled the evolution of $\rm 1~M_{\odot}$ star of metallicity $\rm Z=0.004$, which corresponded to initial metallicity of $\approx-0.7$ dex as traced by observed [S/H] and [Zn/H] (see Fig.~\ref{fig:fnlabs_paper1}). We note that the horizontal part of the evolutionary tracks shown in Fig.~\ref{fig:modszm_paper1} represents the initial pre-Main Sequence phase, during which the luminosity decreases and the surface chemistry is unchanged. During the RGB phase the luminosity increases, and the different sequences bifurcate after the RGB bump, when changes (due to extra mixing) in the surface chemical composition take place. In Fig.~\ref{fig:modszm_paper1}, we show the calculated models with no extra mixing (black lines), in which the only modification of the surface chemistry took place after the first dredge-up (FDU). We also assumed some further mixing after the bump, either moderate (red lines) or deep (blue lines). The grey boxes represent the corresponding observed values for SZ~Mon within their uncertainties.

The upper subplot of Fig.~\ref{fig:modszm_paper1} shows the evolution of the surface carbon isotopic ratio. It is clear that the assumption that no extra-mixing takes place after the RGB bump, thus the surface chemistry of the star is unchanged after the FDU (black line), is not consistent with the measured value (indicated with a grey box). The agreement of theoretical and experimental values of $^{12}$C/$^{13}$C becomes satisfactory, when moderately deep extra mixing is considered.

The lower left subplot of Fig.~\ref{fig:modszm_paper1} shows the evolution of the nitrogen abundance. We find that some extra mixing is required to fit the observations, as the FDU can account for a N enhancement of only $\sim0.2$ dex, against the $\sim0.4$ dex range derived from the spectroscopic analysis. Given that the efficiency of deep mixing affects the luminosity, at which a given surface nitrogen is reached, we deduce that use of a moderate deep mixing leads to a better agreement with the observations for [N/H].

Finally, in the lower right subplot of Fig.~\ref{fig:modszm_paper1}, we show the evolution of C/O ratio. The central value is more consistent with models without extra mixing, but the significant uncertainty in C/O of 0.11 (which is mostly caused by the uncertainty in oxygen abundance of 0.14 dex) does not allow us to draw any certain conclusions from this tracer.

Overall, the results presented so far indicate that SZ Mon evolved along the RGB until the core mass grew to $\rm \sim 0.36-0.37~M_{\odot}$, and the luminosity reached the nowadays value, in the $\rm 400-500~L_{\odot}$ range. At this evolutionary stage, mass transfer to the companion favoured the loss of the external mantle, an early departure from the RGB, and the beginning of the general contraction of the structure, until the current status was reached. The derived surface chemical composition, particularly the carbon isotopic ratio and nitrogen, suggests the combined effects of the FDU and deep mixing occurred after the RGB bump.

An alternative possibility is that the contraction to the blue started during the early AGB phase, after the exhaustion of central helium, before the occurrence of the first thermal pulse. Nevertheless, we contend that the evolution of both our targets was terminated during the RGB phase rather than the early-AGB phase. This is plausible, given the larger radius  of a star at the tip of the RGB compared to an early AGB star. Additionally, RGB timescales are about three times longer than early AGB timescales. Furthermore, if the star evolved until the tip of the RGB, the mixing mechanism would act for a longer duration, favouring a more pronounced modification of surface chemistry, such as a smaller carbon isotopic ratio and a higher abundance of nitrogen.

\subsection{DF Cyg}\label{ssec:condfc_paper1}
\begin{figure}[!ht]
    \centering
    \includegraphics[width=.70\linewidth]{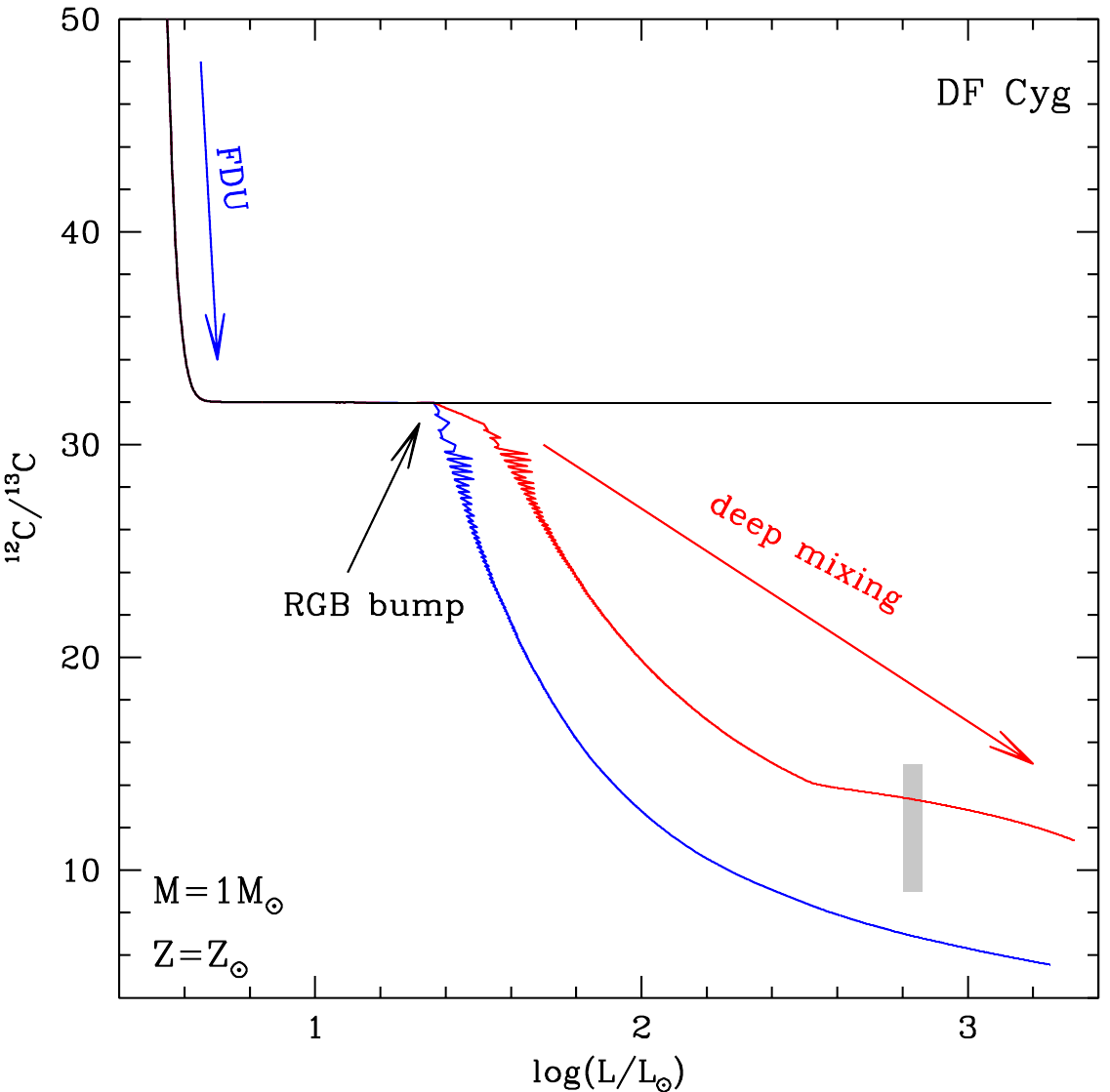}
    \includegraphics[width=.49\linewidth]{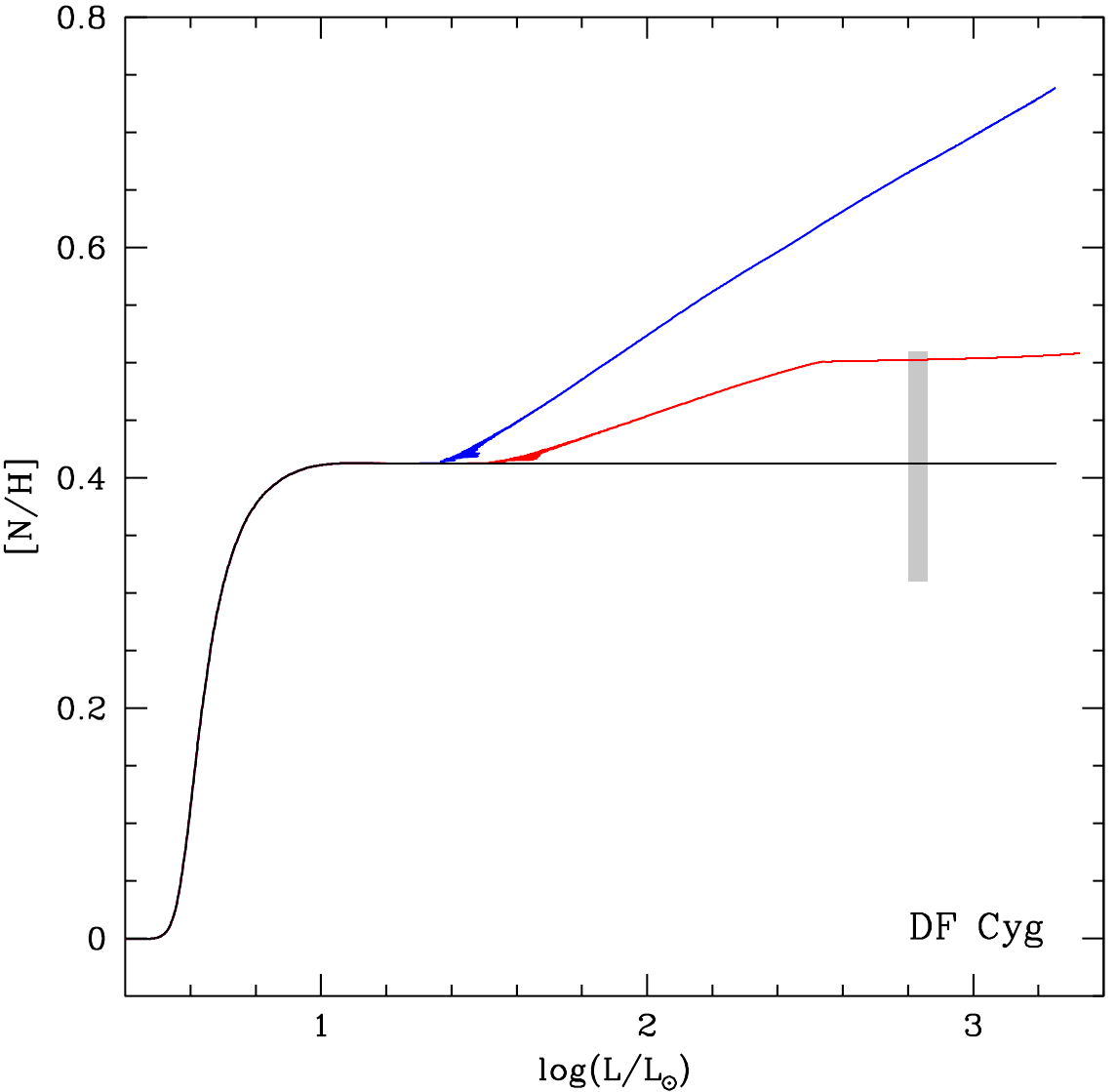}
    \includegraphics[width=.49\linewidth]{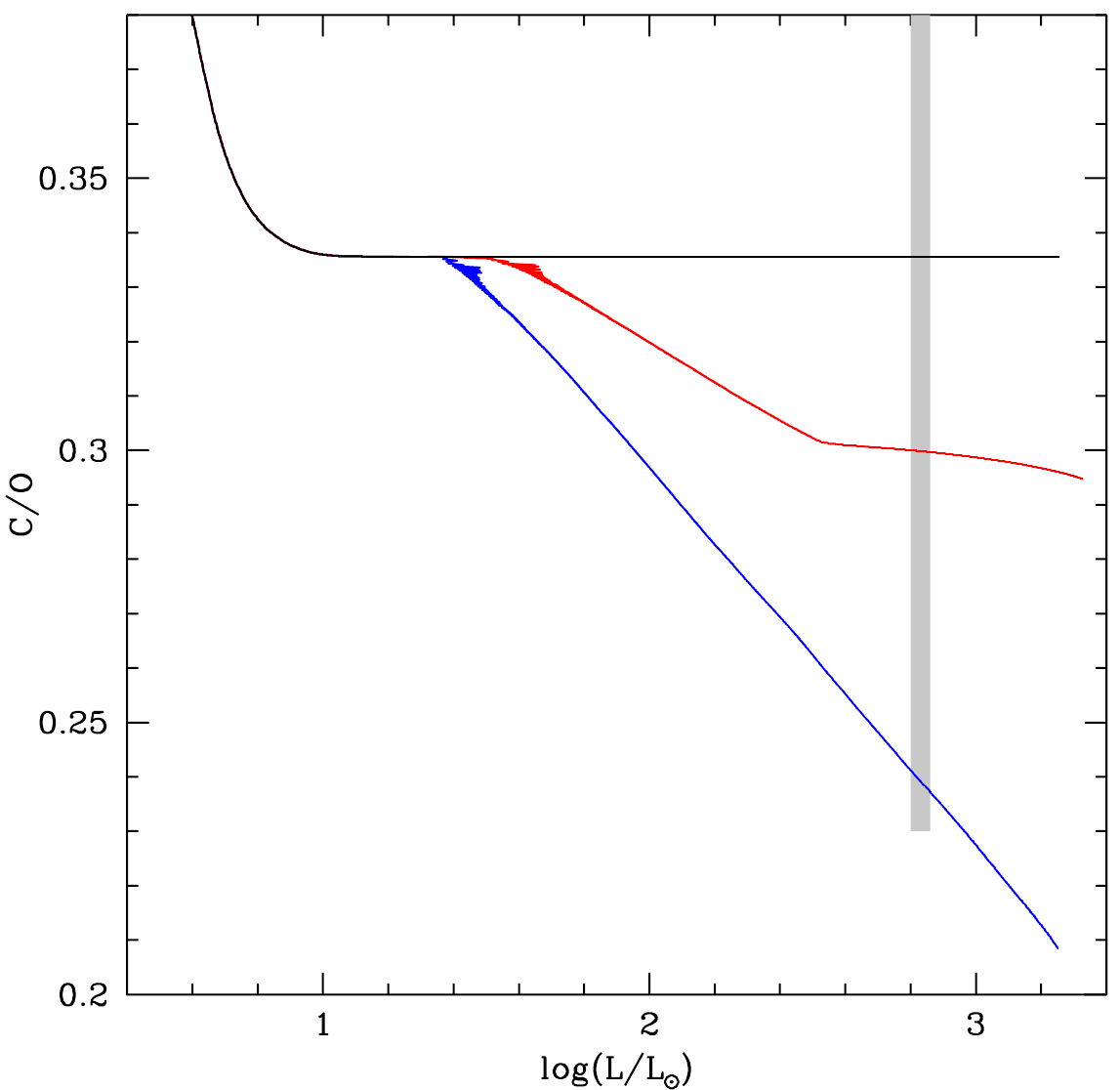}
    \caption[ATON evolutionary tracks of luminosity vs $^{12}$C/$^{13}$C (upper panel), luminosity vs {[N/H]} (lower left panel), luminosity vs C/O (lower right panel) for a 1 M$_\odot$, Z=Z$_\odot$ star]{ATON evolutionary tracks of luminosity vs $^{12}$C/$^{13}$C (upper panel), luminosity vs [N/H] (lower left panel), luminosity vs C/O (lower right panel) for a 1 M$_\odot$, Z=Z$_\odot$ star. The solid lines are the evolutionary tracks of CNO abundances and carbon isotopic ratios for models assuming FDU and different levels of extra mixing: no deep mixing (black), moderately deep mixing (red), and very deep mixing (blue). The grey boxes indicate the observed values for DF~Cyg within corresponding uncertainties (see Table~\ref{tab:fnlabsDFC_paper1}).}\label{fig:moddfc_paper1}
\end{figure}

To model the evolution of DF~Cyg, we constructed evolutionary sequences of a 1 $M_\odot$ star of solar metallicity. Similarly to SZ~Mon, the initial metallicity of DF~Cyg ($\approx0$ dex) was chosen according to [Zn/H] serving as a proxy (see Fig.~\ref{fig:fnlabs_paper1}). We note that the horizontal part of the evolutionary tracks shown in Fig.~\ref{fig:moddfc_paper1} represents the initial pre-Main Sequence phase, during which the luminosity decreases and the surface chemistry is unchanged. In Fig.~\ref{fig:moddfc_paper1}, the black lines indicate the results obtained by assuming only FDU, without any further extra mixing. The red and blue solid lines indicate the models involving extra mixing (moderately deep and very deep, respectively). The grey boxes in the different panels represent the observed range of the corresponding values within the errors.

The upper subplot of Fig.~\ref{fig:moddfc_paper1} shows the evolution of the carbon isotopic ratio. The observed value of DF~Cyg agrees well with the model predictions when some moderately deep extra mixing is considered (red line).

The lower left subplot of Fig.~\ref{fig:moddfc_paper1} shows the evolution of the nitrogen abundance. The substantial uncertainty of observed [N/H] for DF~Cyg (0.11 dex) prevents us from making definitive conclusions. However, we can deduce that experimental values of [N/H] agree with theoretical predictions from models involving either no extra mixing or moderately deep extra mixing.

In the lower right subplot of Fig.~\ref{fig:moddfc_paper1}, we show the evolution of the C/O with luminosity. Significant uncertainty in C/O (caused by uncertainty in [O/H] of 0.2 dex) does not allow us to use C/O as an indicator of extra mixing efficiency. However, for DF Cyg, the surface carbon was found to be super-solar (see Table~\ref{tab:fnlabsDFC_paper1} and Fig.~\ref{fig:fnlabs_paper1}). Since the surface carbon content is known to decrease during the RGB phase, we conclude that the matter, from which the star formed was slightly enhanced in carbon, with $\rm [C/Zn]_0\approx+0.3$. This value of initial carbon enhancement is within the scatter for Galactic solar type stars, as observed in the Galactic chemical history of carbon by \cite{nissen2020SolarTypeAbunds}.

We find an overall consistency between the results from stellar evolution modelling and the observations. The analysis of the behaviour of the surface carbon suggests that even a higher initial carbon would be consistent with the derived carbon abundance. However this would create some tension with the behaviour of nitrogen, which in that case would grow to values in excess of those consistent with the spectroscopic analysis.

Similarly to SZ~Mon, we argue that DF~Cyg underwent an early departure of the RGB, after the luminosity reached $\rm \sim 600-700~L_{\odot}$. In comparison with the evolution of SZ~Mon, we note that the effects of deep mixing are softer here, as clear in the slope of the $^{12}$C/$^{13}$C vs luminosity trends of the two stars. This is explained by the difference in the metallicity, as the temperature gradients in lower metallicity stars are higher, resulting in a more efficient transport process \citep{lagarde2012ExMixMod}.


\section{Summary and Conclusions}\label{sec:con_paper1}
In this study, we used dusty post-RGB binary stars as tracers of the evolutionary processes and nucleosynthesis that occur during the RGB phase. To do this, we acquired multi-wavelength spectroscopic observations of SZ~Mon and DF~Cyg: high-resolution optical spectra obtained with the HERMES at Mercator telescope and mid-resolution NIR spectra from the APOGEE survey.

For our study, we developed E-iSpec, a semi-automatic Python-based tool, primarily to address the need of combining and analysing the HERMES (optical) and APOGEE (NIR) spectra. Existing optical spectral analysis tools (as well as APOGEE's internal pipeline ASPCAP) had significant limitations when dealing with spectra of evolved stars with complex atmospheres. E-iSpec, built upon iSpec (v2023.08.04), was specifically developed for spectral analyses of evolved stars. Updates included semi-automatic continuum normalisation, an updated molecular line list, error estimation for elemental abundances, and isotopic ratio calculations, substantially widened the usability of the tool. In this study, the atmospheric parameters and elemental abundances were derived from the atomic lines using EW method, while the carbon isotopic ratios were calculated from the molecular bands using SSF technique.

Using E-iSpec, we performed the chemical analysis of SZ~Mon and DF~Cyg (see Tables~\ref{tab:fnlabsSZM_paper1} and \ref{tab:fnlabsDFC_paper1}, respectively). We found that the depletion patterns of our post-RGB targets were similar to those of post-AGB binary stars from the sample of \cite{kluska2022GalacticBinaries}. However, for SZ~Mon and DF~Cyg, the chemical depletion pattern starts at higher condensation temperatures $T_{\rm turn-off,~post-RGB}\approx1\,400$~K than for the binary post-AGB sample ($T_{\rm turn-off,~post-AGB}\approx1\,100$~K). Even though our study is the first one to detect this prominent difference in $T_{\rm turn-off}$, a statistically comprehensive and methodologically homogeneous study with a larger sample of targets is required to validate whether the onset of the chemical depletion pattern occurs at higher condensation temperatures in post-RGB stars compared to their high-luminosity analogues, i.e, the post-AGB binary stars. This difference, if detected for the large samples, may hint at the different binary-disc interaction processes for post-AGB and post-RGB binary stars. However, this question is out of the scope of the current study and will be addressed in the future work.

Additionally, we derived the first carbon isotopic ratios in the atmospheres of the post-RGB stars ($^{12}$C/$^{13}$C$_{\rm SZ~Mon}=8\pm4$, $^{12}$C/$^{13}$C$_{\rm DF~Cyg}=12\pm3$). We investigated whether the binary interaction affected the CNO nucleosynthesis that occurred on the RGB and until the binary interaction prematurely terminated the RGB evolution of our targets. We investigated this by comparing the derived CNO abundances and the carbon isotopic ratios of our post-RGB binary targets with relevant theoretical yields from dedicated single star models calculated using the ATON stellar evolutionary code.

We found that CNO elemental abundances (i.e., [N/H] and C/O) and carbon isotopic ratios agreed well with the predictions from models, which experienced FDU with deep extra mixing. This agreement emphasised that in post-RGB binary targets, the observed CNO abundances reflect the chemical composition expected from single star nucleosynthesis (i.e., convective and non-convective mixing processes) occurring during the RGB phase before it is terminated by binary interaction. However, we stress that our findings are based on a limited sample size. In our ongoing study, our aim is to broaden our sample of targets, thereby facilitating a statistically comprehensive investigation, enabling us to validate and confirm the aforementioned results.

\section*{Acknowledgements}\label{sec:ack_paper1}
The spectroscopic results presented in this paper are based on observations made with the Mercator Telescope, operated on the island of La Palma by the Flemish Community, at the Spanish Observatorio del Roque de los Muchachos of the Instituto de Astrofisica de Canarias. This paper also includes data collected by the APOGEE survey.

MM1 acknowledges the International Macquarie Research Excellence Scholarship (iMQRES) program for the financial support during the research. MM1, DK, and MM2 acknowledge the ARC Centre of Excellence for All Sky Astrophysics in 3 Dimensions (ASTRO 3D), through project CE170100013. PV acknowledges the support received from the PRIN INAF 2019 grant ObFu 1.05.01.85.14 (“Building up the halo: chemo-dynamical tagging in the age of large surveys”, PI. S. Lucatello). HVW acknowledges support from the Research Council, KU Leuven under grant number C14/17/082. DAGH acknowledges the support from the State Research Agency (AEI) of the Spanish Ministry of Science and Innovation (MCIN) under grant PID2020-115758GB-I00/AEI/10.13039/501100011033. This article is based upon work from European Cooperation in Science and Technology (COST) Action NanoSpace, CA21126, supported by COST. TM acknowledges financial support from the Spanish Ministry of Science and Innovation (MICINN) through the Spanish State Research Agency, under the Severo Ochoa Program 2020-2023 (CEX2019-000920-S)
\begin{savequote}[75mm]
\foreignlanguage{ukrainian}{
``Єдиний, хто не втомлюється, -- час. А ми -- живі, нам треба поспішати...''}
\qauthor{\foreignlanguage{ukrainian}{---Ліна Костенко (нар. 1930), \\українська поетеса, журналістка, письменниця, публіцистка та дисидентка}}
``The only one who never tires is time. But we're alive, hence we must hurry...''
\qauthor{---Lina Kostenko (born 1930), \\Ukrainian poet, journalist, writer, publisher, and former Soviet dissident}
\end{savequote}

\chapter{Photospheric depletion in post-AGB/post-RGB binaries \texorpdfstring{\\with transition discs}{}}\label{chp:pap2}
\graphicspath{{ch_paper2/figures/}} 

\clearpage

\textit{This chapter was originally published as:}\\
\textbf{Chemical Depletion in Evolved Binaries with Second-Generation Protoplanetary Discs}\\
Mohorian M., Kamath D., Menon M., Amarsi A.~M., Van Winckel H., Fava C., Andrych K.\\
\textit{Accepted to Monthly Notices of the Royal Astronomical Society}, 2025\\
\\
\textit{In this chapter, we examine the depletion of refractory elements in the photospheres of 12 post-AGB and post-RGB binary systems surrounded by transition discs. High-resolution optical spectra from HERMES/Mercator and UVES/VLT are analysed using 1D LTE techniques, with additional 1D NLTE corrections applied to key chemical elements spanning from C to Fe. We confirm a significantly strong depletion exhibited by transition disc systems (indicated by [S/Ti]\,>\,0.7\,dex), compared to the broader sample of post-AGB/post-RGB binaries. Correlations between depletion signatures and binary orbital properties are also explored, alongside comparisons to abundance trends observed in the interstellar medium and young stars with transition discs. These results shed new light on the mechanisms influencing chemical processing in circumbinary environments around evolved stars.}\\
\\
\textbf{\textit{Author Contributions}}\\
M.\,Mohorian performed the reduction and analysis of high-resolution optical spectra from HERMES/Mercator and UVES/VLT, and was primarily responsible for interpretation of the results presented in this chapter. D.\,Kamath provided feedback and engaged in regular discussions throughout each stage of the research process. H.\,Van\,Winckel and D.\,Kamath are the principal investigators of the proposals that resulted in the data used in this study. M.\,Menon and A.\,Amarsi contributed to the development of the E-iSpec and NLTE analysis methodology, respectively. C.\,Fava and K.\,Andrych contributed ideas and feedback for paper discussion. The chapter was written by M.\,Mohorian, with all co-authors providing feedback and comments.

\begin{tcolorbox}[colback=blue!10!white, boxrule=0mm]
    \section*{Original paper abstract}
    
    The mechanisms responsible for chemical depletion across diverse astrophysical environments are not yet fully understood. In this paper, we investigate chemical depletion in post-AGB/post-RGB binary stars hosting second-generation transition discs using high-resolution optical spectra from HERMES/Mercator and UVES/VLT. We performed a detailed chemical abundance analysis of 6 post-AGB/post-RGB stars and 6 post-AGB/post-RGB candidates with transition discs in the Galaxy and in the Large Magellanic Cloud. The atmospheric parameters and elemental abundances were obtained through 1D LTE analysis of chemical elements from C to Eu, and 1D NLTE corrections were incorporated for elements from C to Fe. Our results confirmed that depletion efficiency, traced by the [S/Ti] abundance ratio, is higher in post-AGB/post-RGB binaries with transition discs compared to the overall sample of post-AGB/post-RGB binaries. We also examined correlations between derived abundances and binary system parameters (astrometric, photometric, orbital, pulsational). Additionally, we compared the depletion patterns in our sample to those observed in young stars with transition discs and in the interstellar medium. We confirmed that the depletion is significantly stronger in post-AGB/post-RGB binaries with transition discs than in young stars with transition discs. Furthermore, we found that [X/Zn] abundance ratio trends of volatile and refractory elements in post-AGB/post-RGB binaries with transition discs generally resemble similar trends in the interstellar medium (except for trends of [Si/Zn] and [Mg/Zn] ratios). These findings, although based on a limited sample, provide indirect constraints for depletion mechanism in circumbinary discs around post-AGB/post-RGB stars.
\end{tcolorbox}    

\section{Introduction}\label{sec:int_paper2}
One the most complex aspects of binary star evolution is the interaction between binaries and circumbinary discs \citep{heath2020DiscBinaryInteraction, penzlin2022DiscBinaryInteraction, coleman2022DiscBinaryInteractionPlanets, zagaria2023DiscBinaryInteractionPlanets}. These discs, characterised by intricate dynamics and varying dust content, significantly influence observable properties of the host binary, such as infrared excess \citep{itoh2015DiscBinaryInteractionIRexcess}, accretion rates \citep{izzard2023DiscBinaryInteractionAccretion}, orbital eccentricity \citep{heath2020DiscBinaryInteraction}, and jet activity \citep{bollen2022Jets}. The diversity of dust content in circumbinary discs results from dust condensation being dependent on the chemical composition of the condensing mixture and the physical conditions of the environment, such as pressure and temperature \citep{lodders2003CondensationTemperatures, wood2019CondensationTemperatures}. While the process of dust condensation is well-studied for protoplanetary discs around young single stars \citep{lagage1994DustDepletionPlanetIndicator, birnstiel2016DustEvolutionAndPlanetesimals}, the effect of binarity on dust condensation in circumbinary discs around evolved binary stars remains poorly explored \citep{oomen2020MESAdepletion, miguel2024DiscBinaryInteractionDustFormation}.

During the evolution along the asymptotic giant branch (AGB) or red giant branch (RGB), low- and intermediate-mass stars  ($M\sim0.8-8M_\odot$) in a binary system can fill their Roche lobe, leading to mass loss that terminates AGB/RGB evolution. The primary star then transitions to the post-asymptotic giant branch (post-AGB) or post-red giant branch (post-RGB) stage \citep{vanwinckel2003Review, kamath2016PostRGBDiscovery, vanwinckel2018Binaries, kluska2022GalacticBinaries}. While the binary interactions between the post-AGB/post-RGB primary star and the secondary component \citep[often a main sequence star;][]{oomen2018OrbitalParameters} remain poorly understood, observational studies reveal that these interactions often result in the formation of a stable disc of circumstellar gas and dust \citep[with radii <\,1\,000\,AU;][]{deruyter2006discs, deroo2006CompactDisc, bujarrabal2015KeplerianRotation, hillen2016IRAS08, bujarrabal2018IRAS08, kluska2022GalacticBinaries, corporaal2023DiscParameters, gallardocava2023thesis}.

The presence of the disc around post-AGB binaries was observationally established by the distinct pattern in spectral energy distribution \citep[SED;][]{deruyter2005discs, deruyter2006discs, kamath2014SMC, gezer2015WISERVTau, kamath2015LMC, kluska2022GalacticBinaries}. This pattern includes a near-infrared (near-IR) dust excess, indicative of hot dust in the system \citep{vanwinckel2003Review, oomen2018OrbitalParameters}. Interferometric imaging studies resolved the inner rim of circumbinary discs in several post-AGB/post-RGB binary systems, which in many cases is close to the dust sublimation radius \citep[typically, $\sim$5--30 AU;][]{kluska2019DiscSurvey, corporaal2023DiscParameters}. In addition, high-resolution polarimetric imaging studies with SPHERE revealed complex substructures in circumbinary discs around post-AGB/post-RGB stars, including rings, spirals, and arc-like features \citep{ertel2019Imaging, andrych2023Polarimetry}. Moreover, the outer regions of these discs are known to display crystallisation \citep{gielen2011silicates, hillen2015ACHerMinerals} and grain growth \citep{scicluna2020GrainGrowth}. Additionally, circumbinary discs around post-AGB/post-RGB binaries display Keplerian rotation based on position–velocity maps of $^{12}$CO \citep{bujarrabal2015KeplerianRotation, gallardocava202389HerNebula}.

Recently, \citet{kluska2022GalacticBinaries} compiled a comprehensive sample of 85 Galactic post-AGB binaries and their low-luminosity analogues, dusty post-RGB stars. Using the near-IR and mid-IR colours (2MASS $H-K$ and WISE $W_1-W_3$, respectively), \citet{kluska2022GalacticBinaries} categorised the discs around Galactic post-AGB/post-RGB binaries in three groups: i) full discs, where the dust in the disc extends from the dust sublimation radius outward, ii) transition discs, where IR colours suggest the presence of large dust cavities in the inner discs, and iii) discs with significant lack of IR excess, which points to virtual absence of circumbinary dust (we refer to these discs as dust-poor, though this group includes gas-poor debris discs). A subsequent mid-IR interferometric study of full and transition disc candidates in the Galaxy by \citet{corporaal2023DiscParameters} confirmed the presence of this gap (the dust inner rims in transition disc systems are 2.5--7.5 times larger than the corresponding dust sublimation radii). This subset of transition disc targets is the centrepiece of this study (see Section~\ref{sec:tar_paper2}).

The interaction between the binary star and the circumbinary disc significantly affects the surface composition of the post-AGB/post-RGB star. In particular, the post-AGB/post-RGB binaries with circumbinary discs exhibit photospheric chemical depletion (hereafter referred to as depletion) with a notable underabundance of refractory elements (i.e. those with condensation temperatures $T_{\rm cond}>1250$ K, such as Al, Fe, Ti, and the majority of the slow neutron capture process elements) relative to volatile elements \citep[i.e. those with condensation temperatures $T_{\rm cond}<1250$ K, like S, Zn, Na, and K;][]{gielen2009Depletion, gezer2015WISERVTau, kamath2019depletionLMC}. The exact mechanism behind chemical depletion in post-AGB/post-RGB stars is not yet fully understood. However, it is believed to result from the chemical fractionation of gas and dust in the circumbinary disc and subsequent accretion of a small portion of this pure gas onto the photosphere of the primary component \citep{waters1992GasDustFractionation, oomen2020MESAdepletion}. The efficiency of the gas-dust fractionation is specific to each chemical element and depends on its condensation temperature,  $T_{\rm cond}$ \citep{lodders2003CondensationTemperatures, wood2019CondensationTemperatures}. Consequently, mostly refractory dust particles are settling in the mid-plane of the disc, while mostly volatile gas is partially re-accreted onto the binary \citep{mosta2019ReaccretionInnerRim, munoz2019ReaccretionInnerRim}.

Observational studies of depletion in post-AGB/post-RGB binaries in the Galaxy and the Magellanic Clouds \citep{giridhar1998RVTauVars, maas2002RUCenSXCen, deruyter2005discs, giridhar2005rvtau, deruyter2006discs, maas2007t2cep, vanwinckel2012AFCrt, desmedt2012j004441, rao2014RVTauAbundances, desmedt2014LeadMCs, desmedt2015LMC2sEnrichedPAGBs, desmedt2016LeadMW} showed that the relative abundances [X/H]\footnote{$[$X/H$]\,=\,\log\frac{N(X)}{N(H)}-\log\frac{N(X)_\odot}{N(H)_\odot}+12$, where N(X) and N(H) are the number abundances of element X and hydrogen, respectively. Symbol $\odot$ denotes the corresponding solar abundances.} are prominently decreased for those elements, which have high condensation temperatures $T_{\rm cond}$. This leads to a prominent break in the plots of condensation temperature $T_{\rm cond}$ vs. relative abundance [X/H] for post-AGB/post-RGB binaries \citep{oomen2019depletion}. The condensation temperature at the break -- the turn-off temperature $T_{\rm turn-off}$ -- has a wide range of values within the Galactic subsample \citep[from 800 to 1500 K;][]{kluska2022GalacticBinaries}.

The rate (or efficiency) of the chemical depletion in post-AGB/post-RGB binary stars is traced by a volatile-to-refractory abundance ratio, usually [Zn/Ti] ratio \citep{gezer2015WISERVTau, oomen2019depletion}. Depletion efficiency may be categorised into the following groups: mild ([Zn/Ti] < 0.5 dex), moderate (0.5 dex < [Zn/Ti] < 1.5 dex), or strong ([Zn/Ti] > 1.5 dex). \citet{kluska2022GalacticBinaries} showed that the observed depletion efficiency ([Zn/Ti] ratio) is generally the highest in the subsample of transition disc candidates. Additionally, the high-temperature end of depletion profile in the $T_{\rm cond}$-[X/H] plots (i.e. for elements with $T_{\rm cond}>T_{\rm turn-off}$) may follow a linear trend (``saturated'' profile) or a two-piece linear fit with a horizontal plateau at higher condensation temperatures \citep[``plateau'' profile;][]{waelkens1991depletion, oomen2019depletion, oomen2020MESAdepletion}. Theoretical studies using the detailed MESA stellar evolution models confirmed that the observed depletion profiles in post-AGB binaries (including the breaks at $T_{\rm turn-off}$ and ``plateau'' start) may be qualitatively reproduced by dilution of re-accreted metal-poor gas from the disc with the pristine composition of the stellar surface \citep{oomen2019depletion, oomen2020MESAdepletion}.

C, N, and O (CNO elements) are generally excluded from the depletion profiles, because it is difficult to separate their 
depletion from the effects of nucleosynthetic and mixing processes that occur during AGB/RGB phase \citep{mohorian2024EiSpec, menon2024EvolvedBinaries}. Current stellar evolution models predict that surface abundances of C and N are significantly modified by mixing processes on AGB/RGB, while surface abundance of O is unaffected by mixing processes in low-mass ($M<2M_\odot$) AGB stars and in RGB stars \citep{ventura2008aton3, karakas2014dawes, ventura2020CNOinAGB, kobayashi2020OriginOfElements, kamath2023models}. This highlights the complexity of disentangling chemical impacts on CNO abundances from evolution and re-accretion.

In this study, we further explore the disc-binary interaction by systematically investigating chemical depletion in post-AGB/post-RGB binary stars. Our focus is on the subset of post-AGB/post-RGB binaries hosting second-generation transition discs, as these systems share key similarities with young stellar objects (YSOs) that host planet-forming discs. The goals of this study are: i) to homogeneously derive the depletion profiles of transition disc targets, ii) to examine the connection between depletion and other observational parameters, and iii) to establish a comparative study between depletion patterns observed in post-AGB/post-RGB binary stars, YSOs, and the interstellar medium (ISM). To achieve our goals, we derived precise atmospheric parameters and elemental abundances using 1D local thermodynamic equilibrium (LTE) models for chemical elements from C to Eu. In addition, we accounted for 1D non-LTE (NLTE) effects for a representative set of chemical elements (C, N, $\alpha$-elements, and Fe). In Section~\ref{sec:tar_paper2}, we provide an overview of our target sample. In Section~\ref{sec:dob_paper2}, we introduce the photometric and spectroscopic data used in our research. In Section~\ref{sec:san_paper2}, we present the methodology of deriving atmospheric parameters and elemental abundances. In Section~\ref{sec:res_paper2}, we present results of our detailed abundance analysis of transition disc targets. In Section~\ref{sec:dpl_paper2}, we correlate the obtained chemical depletion profiles with other parameters of the studied binaries (astrometric, photometric, spectroscopic, pulsational, and orbital parameters) and compare with chemical depletion in young stars and ISM. Finally, in Section~\ref{sec:con_paper2}, we present our conclusions.

\section{Target sample}\label{sec:tar_paper2}
\begin{figure}[!ht]
    \centering
    \includegraphics[width=.99\linewidth]{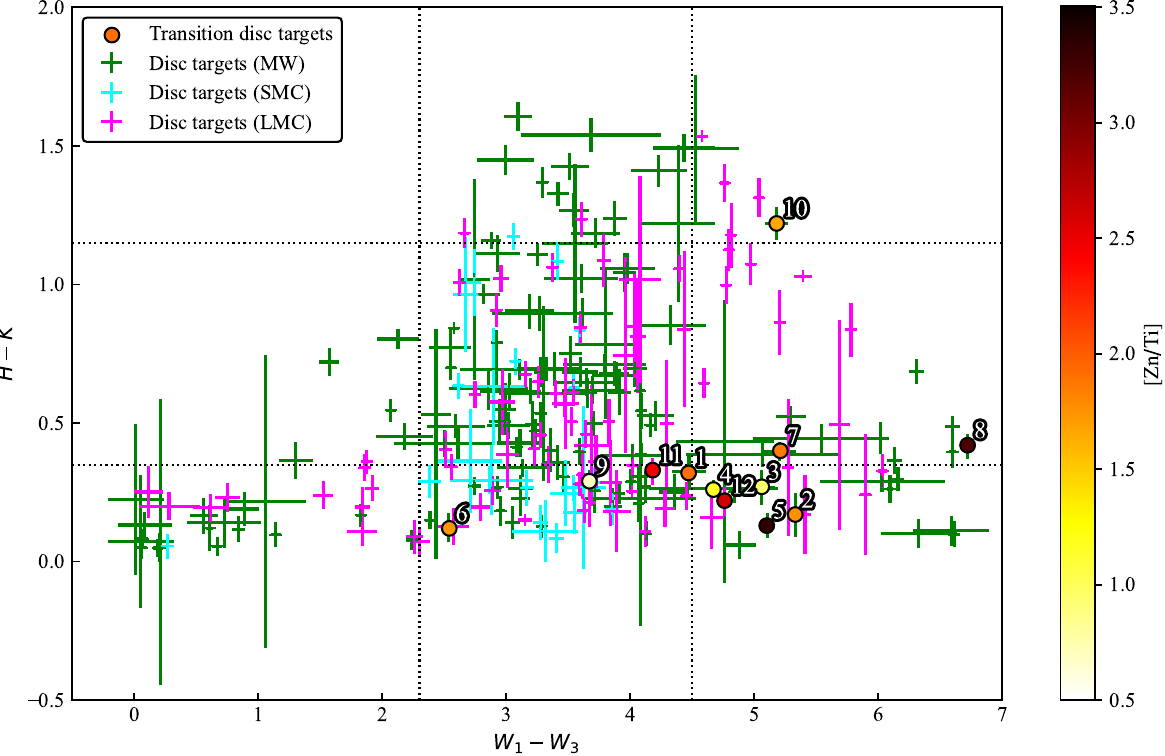}
    \caption[Updated IR colour-colour plot of post-AGB/post-RGB binary stars in the Galaxy and in the Magellanic Clouds]{Updated IR colour-colour plot of post-AGB/post-RGB binary stars in the Galaxy and in the Magellanic Clouds. NIR magnitudes ($H$ and $K$) are adopted from 2MASS 6X, while MIR magnitudes ($W_1$ and $W_3$) are adopted from AllWISE (for more details, see Section~\ref{sec:tar_paper2}). Green error bars represent disc targets in the Galaxy \citep{kluska2022GalacticBinaries}, cyan error bars represent disc targets in the SMC \citep{kamath2014SMC}, magenta error bars represent disc targets in the LMC \citep{kamath2015LMC}. Our sample of transition disc targets is marked with circles coloured by [Zn/Ti] abundance ratios from this study (see Table~\ref{tab:fnlabu_paper2}). Adopted contours represent the rough demarcation between different disc categories \citep{kluska2022GalacticBinaries}.}\label{fig:colplt_paper2}
\end{figure}

In this study, we focus on post-AGB/post-RGB binary stars with disc-type SED\footnote{The disc-type SED is characteristic of the binarity of the host star \citep{gezer2015WISERVTau, kamath2016PostRGBDiscovery, oomen2018OrbitalParameters, kluska2022GalacticBinaries}.} of the transition type (see Fig.~\ref{fig:colplt_paper2} and Table~\ref{tab:tarlst_paper2}). \citet{kluska2022GalacticBinaries} presented seven high-probability ($W_1-W_3>4.5$; category 2) and three moderate-probability ($2.3<W_1-W_3<4.5$, $H-K<0.3$, [Zn/Ti]>0.7 dex; category 3) transition disc candidates in the Galaxy. A mid-IR interferometric study by \citet{corporaal2023DiscParameters} confirmed the inner gaps in six of these Galactic candidates -- hence, they are referred to as transition disc stars (see targets \#1--\#6 in Table~\ref{tab:tarlst_paper2}). The remaining subsample of transition disc targets (targets \#7--\#10) are referred to as transition disc candidates for the rest of the paper.

We complemented the Galactic subsample of transition disc targets with two moderate-probability transition disc candidates (targets \#11 and \#12) in the Large Magellanic Cloud (LMC) following the procedure outlined in \citet{kluska2022GalacticBinaries}. In brief, we used the near-IR colours \citep[$H-K$ from 2MASS;][]{skrutskie20062MASS} and mid-IR colours \citep[$W_1-W_3$ from WISE;][]{wright2010WISE} to select new targets from the overall sample of post-AGB/post-RGB binary stars with disc-type SEDs in the Small Magellanic Cloud (SMC) \citep{kamath2014SMC} and in the LMC \citep{kamath2015LMC}. From all SMC and LMC targets in categories 2 and 3, we selected those for which high-resolution optical spectra were available (see Fig.~\ref{fig:colplt_paper2}).

We note that we excluded SS Lep and V777 Mon (Red Rectangle) from our preliminary sample despite their high mid-IR excesses because of the large corresponding uncertainties ($(W_1-W_3)_{\rm SS~Lep}\,=\,4.76\pm0.39$ mag, $(W_1-W_3)_{\rm V777~Mon}\,=\,4.53\pm0.35$ mag) as compared to the other 12 targets, which had the uncertainty in $W_1-W_3$ below 0.08 mag. We also note that transition disc candidates AF Crt (\#8) and V1504 Sco (\#10) are observed edge-on \citep{vanwinckel2012AFCrt, kluska2022GalacticBinaries}, which makes their luminosity estimates (see Section~\ref{ssec:doblum_paper2}) less reliable. Nevertheless, despite high inclinations, the depletion profiles of AF Crt (\#8) and V1504 Sco (\#10) are similar to those of other transition disc targets.

Our final sample consists of 12 transition disc targets: six confirmed transition disc stars (all in the Galaxy) and six transition disc candidates (four in the Galaxy and two in the LMC). We note that the surface composition of all 12 targets in our sample was previously studied by different groups using different methods (see Table~\ref{tab:litpar_paper2}). In this study, we perform a homogeneous analysis of our targets within the context of disc-binary interaction, accounting for NLTE effects, and re-define depletion efficiency in post-AGB/post-RGB binaries with transition discs. We also note that all targets are Type II Cepheid variables, which allowed us to calculate their luminosities using period-luminosity-colour (PLC) relation (see Section~\ref{ssec:doblum_paper2}). We present the target details individually in Appendix~\ref{app:lit_paper2}.

\begin{sidewaystable}[ph!]
    \centering
    \footnotesize
    \caption[Names, coordinates, and selection criteria of the target sample (see Section~\ref{sec:tar_paper2})]{Names, coordinates, and selection criteria of the target sample (see Section~\ref{sec:tar_paper2}). $H$ and $K$ magnitudes were adopted from the 2MASS 6X catalogue; $W_1$ and $W_3$ magnitudes were adopted from the AllWISE catalogue. [Zn/Ti] abundance ratio is a proxy for the efficiency of depletion process (see Section~\ref{sec:int_paper2}).} \label{tab:tarlst_paper2}
    \begin{tabular}{|c|ccc|cc|ccc|}
    \hline
        ~ & \multicolumn{3}{|c|}{\textbf{Names}} & \multicolumn{2}{|c|}{\textbf{Coordinates}} & \multicolumn{3}{|c|}{\textbf{Selection criteria}} \\
        \textbf{ID} & \textbf{IRAS/OGLE} & \textbf{2MASS} & \textbf{Variable} & \textbf{R.A.} & \textbf{Dec.} & \boldmath$H-K$ & \boldmath$W_1-W_3$ & \textbf{[Zn/Ti]} \\
        ~ & ~ & ~ & ~ & \textbf{(deg)} & \textbf{(deg)} & \textbf{(mag)} & \textbf{(mag)} & \textbf{(dex)} \\\hline
        \multicolumn{9}{|c|}{\textit{Transition disc stars \citep{corporaal2023DiscParameters}}} \\\hline
        1 & 06072+0953 & J06095798+0952318 & CT Ori & 092.491630 & +09.875519 & 0.324 & 4.473 & 1.9$^a$ \\
        2 & 06472--3713 & J06485640--3716332 & ST Pup & 102.235100 & --37.275900 & 0.168 & 5.329 & 2.1$^b$ \\
        3 & 12067--4508 & J12092381--4525346 & RU Cen & 182.348800 & --45.426400 & 0.268 & 5.059 & 1.0$^c$ \\
        4 & 18281+2149 & J18301623+2152007 & AC Her & 277.567500 & +21.866670 & 0.263 & 4.672 & 0.7$^d$ \\
        5 & 18564--0814 & J18590869--0810140 & AD Aql & 284.786283 & --08.170671 & 0.132 & 5.103 & 2.5$^d$ \\
        6 & 19163+2745 & J19181955+2751031 & EP Lyr & 289.581300 & +27.850890 & 0.123 & 2.537 & 1.3$^e$ \\ \hline
        \multicolumn{9}{|c|}{\textit{Transition disc candidates \citep[categories 2 and 3 in][]{kluska2022GalacticBinaries}}} \\\hline
        7 & 06034+1354 & J06061490+1354191 & DY Ori & 091.562130 & +13.905310 & 0.398 & 5.206 & 2.1$^e$ \\
        8 & 11472--0800 & J11494804--0817204 & AF Crt & 177.450151 & --08.289059 & 0.417 & 6.721 & 3.4$^f$ \\
        9 & 16278--5526 & J16315414--5533074 & GZ Nor & 247.975624 & --55.552108 & 0.288 & 3.669 & 0.8$^g$ \\
        10 & 17233--4330 & J17265864--4333135 & V1504 Sco & 261.744331 & --43.553756 & 1.221 & 5.179 & 1.4$^h$ \\
        11 & LMC--029$^1$ & J05030498--6840247 & LMC V0770 & 075.770630 & --68.673500 & 0.326 & 4.181 & 2.3$^i$ \\
        12 & LMC--147$^1$ & J05315099--6911463 & LMC V3156 & 082.962495 & --69.196213 & 0.218 & 4.758 & 2.5$^j$ \\ \hline
    \end{tabular}\\
    \textbf{Notes:} $^1$OGLE names of two LMC targets were shortened from OGLE LMC-T2CEP-029 and OGLE LMC-T2CEP-147 to LMC-029 and LMC-147, respectively. The superscripts of the [Zn/Ti] values indicate the individual chemical abundance studies: $^a$\cite{gonzalez1997CTOri}; $^b$\cite{gonzalez1996STPup}; $^c$\cite{maas2002RUCenSXCen}; $^d$\cite{giridhar1998RVTauVars}; $^e$\cite{gonzalez1997EPLyrDYOriARPupRSgt}; $^f$\cite{vanwinckel2012AFCrt}; $^g$\cite{gezer2019GKCarGZNor}; $^h$\cite{maas2005DiscPAGBs}; $^i$\cite{kamath2019depletionLMC}; $^j$\cite{reyniers2007LMC147}.
\end{sidewaystable}
\begin{sidewaystable}[ph!]
    \centering
    \footnotesize
    \caption{Literature data on photometric parameters, orbital parameters, luminosity estimates, and depletion parameters of transition disc targets (see Section~\ref{sec:tar_paper2}).} \label{tab:litpar_paper2}
    \begin{tabular}{|cc|cc|cc|ccc|cc|}
    \hline
        && \multicolumn{2}{|c|}{\textbf{Photometric parameters}} & \multicolumn{2}{|c|}{\textbf{Orbital parameters}} & \multicolumn{3}{|c|}{\textbf{Luminosity estimates}} & \multicolumn{2}{|c|}{\textbf{Depletion parameters}} \\
        \textbf{ID} & \textbf{Name} & \textbf{RVb} & \boldmath$P_{\rm puls}$ & \boldmath$P_{\rm orb}$ & \boldmath$e$ & \boldmath$L_{\rm SED}$ & \boldmath$L_{\rm PLC}$ & \boldmath$L_{\rm IR}/L_\ast$ & \boldmath$T_{\rm turn-off}$ & \textbf{Profile} \\
        ~ & ~ & ~ & \textbf{(d)} & \textbf{(d)} & ~ & \textbf{(}\boldmath$L_\odot$\textbf{)} & \textbf{(}\boldmath$L_\odot$\textbf{)} & ~ & \textbf{(K)} & ~ \\\hline
        1 & CT Ori & no & 33.65$^f$ & - & - & 15100$^c$ & - & 0.55$^d$ & 1200 & S \\
        2 & ST Pup & no & 18.73$^e$ & 406$^b$ & 0.00$^b$ & 690$^c$ & - & 0.72$^d$ & 800 & S \\
        3 & RU Cen & no & 32.37$^g$ & 1489$^b$ & 0.62$^b$ & 1100$^c$& - & 0.40$^d$ & 800 & P \\
        4 & AC Her & no & 37.73$^h$ & 1189$^b$ & 0.00$^b$ & 2400$^c$ & 3600$^i$ & 0.21$^d$ & 1200 & U \\
        5 & AD Aql & no & 32.7$^h$ & - & - & 11500$^c$ & - & 0.51$^d$ & 1000 & S \\
        6 & EP Lyr & no & 41.59$^g$ & 1151$^b$ & 0.39$^b$ & 5500$^c$ & 7100$^i$ & 0.02$^d$ & 800 & P \\ \hline
        7 & DY Ori & no & 30.155$^a$ & 1248$^b$ & 0.22$^b$ & 21500$^c$ & - & 0.55$^d$ & 1000 & U \\
        8 & AF Crt & no & 31.5$^f$ & - & - & 280$^c$ & - & 1.83$^d$ & 1000 & S \\
        9 & GZ Nor & no & 36.2$^l$ & - & - & 1400$^c$ & - & 0.22$^d$ & 800 & P \\
        10 & V1504 Sco & yes & 22.0$^f$ & 735$^f$ & - & 1100$^c$ & - & 4.69$^d$ & 1000 & S \\
        11 & LMC V0770 & no & 31.245$^j$ & - & - & 3300$^m$ & 2629$^j$ & 0.63$^k$ & -- & -- \\
        12 & LMC V3156 & no & 46.795$^j$ & - & - & 5900$^j$ & 6989$^j$ & 0.84$^k$ & -- & -- \\ \hline
    \end{tabular}\\
    \textbf{Notes:} Literature data on depletion profiles is adopted from \cite{oomen2019depletion}: `S' means `saturated', `P' means `plateau', `U' means `uncertain'. The source list: $^a$\cite{pawlak2019ASAS}, $^b$\cite{oomen2018OrbitalParameters}, $^c$\cite{oomen2019depletion}, $^d$\cite{kluska2022GalacticBinaries}, $^e$\cite{walker2015STPup}, $^f$\cite{kiss2007T2Cepheids}, $^g$\cite{bodi2019RVTauVars}, $^h$\cite{giridhar1998RVTauVars}, $^i$\cite{bollen2022Jets}, $^j$\cite{manick2018PLC}, $^k$\cite{vanaarle2011PAGBsInLMC}, $^l$\cite{gezer2019GKCarGZNor}, $^m$\cite{kamath2015LMC}.\\
\end{sidewaystable}

\section{Data and observations}\label{sec:dob_paper2}
In this section, we present the photometric data (see Section~\ref{ssec:dobpht_paper2}) used to derive luminosities of transition disc targets. The luminosities were derived using SED fitting and PLC relation for Type II Cepheids (see Section~\ref{ssec:doblum_paper2}). We also present the spectroscopic data used to calculate atmospheric parameters and elemental abundances (see Section~\ref{ssec:dobspc_paper2}).

\subsection{Photometric data}\label{ssec:dobpht_paper2}
To obtain the SEDs of our target sample (see Appendix~\ref{app:sed_paper2}), we followed the procedure originally developed by \citet{degroote2013SEDfitting} and recently presented in \citet{mohorian2024EiSpec}. In brief, we collected the photometric magnitudes across various wavelength bands, which span from optical to far-infrared (far-IR; see Table~\ref{tabA:phomag_paper2}), including data from Johnson-Cousins system \citep{johnson1953Filters,cousins1976Filters}, Tycho-2 catalogue \citep{hog2000Tycho2}, Sloan Digital Sky Survey \citep[SDSS, ][]{york2000SDSSphotometry}, Two Micron All Sky Survey \citep[2MASS;][]{skrutskie20062MASS}, WISE \citep{wright2010WISE}, AKARI \citep{ishihara2010AKARI}, Infrared Astronomical Satellite \citep[IRAS;][]{neugebauer1984IRAS}, Photodetector Array Camera and Spectrometer \citep[PACS;][]{poglitsch2010PACS}, and Spectral and Photometric Imaging REceiver \citep[SPIRE;][]{griffin2010SPIRE}. In Appendix~\ref{app:sed_paper2}, we present the SEDs of transition disc targets, fitted with updated \textit{Gaia} DR3 distances (see Section~\ref{ssec:doblum_paper2}).

\subsection{Determination of luminosities from SED fitting and PLC relation}\label{ssec:doblum_paper2}
In this study, we determined the luminosities of the target sample using two methods: i) through SED fitting (referred to as SED luminosity, $L_{\rm SED}$) following the methodology outlined in \citet{mohorian2024EiSpec}; and ii) employing the PLC relation (referred to as PLC luminosity, $L_{\rm PLC}$) following the procedure outlined in \citet{menon2024EvolvedBinaries}. In this subsection, we provide a brief overview of these methods. In Table~\ref{tab:fnlvls_paper2}, we present the estimated SED and PLC luminosities.

To determine the SED luminosities, we selected appropriate Kurucz model atmospheres \citep{castelli2003ATLAS9} to fit the initial photometric data points (the bolometric IR luminosity $L_{\rm IR}$ was obtained through integration of star-subtracted IR excess) and computed the de-reddened model atmospheres for each target through an extensive parameter grid search \citep{mohorian2024EiSpec}. The search was performed by the minimisation of the $\chi^2$ value in the parameter space of four variables: effective temperature $T_{\rm eff}$, surface gravity $\log g$, total reddening (extinction parameter) $E(B-V)$, and angular diameter of the star $\theta$. The total reddening comprised both interstellar and circumstellar contributions. For interstellar reddening, we adopted the wavelength-dependent extinction law \citep{cardelli1989SEDextinction}, assuming an $R_V$\,=\,3.1. Additionally, we used the Bailer-Jones geometric distances, denoted as $z_{\rm BJ}$, along with their corresponding lower and upper limits, $z_{\rm BJL}$ and $z_{\rm BJU}$ \citep{bailerjones2021distances}. These more precise geometric distances were computed based on \textit{Gaia} DR3 parallaxes, incorporating a direction-dependent prior distance. We note that throughout our computations, we assumed isotropic radiation emission from the stars. We also note that stellar variability was not accounted for, resulting in an increased $\chi^2$ value for high-amplitude variables. Once the solution for the de-reddened model atmosphere was found, we integrated it to obtain the SED luminosity $L_{\rm SED}$ and the relative bolometric IR luminosity $L_{\rm IR}/L_\ast\,=\,L_{\rm IR}/L_{\rm SED}$. The luminosity uncertainties were derived for each respective target by computing the standard deviation of the lower and upper luminosity bounds, caused by uncertainties of the geometric distances and the photometric data points.

To derive the PLC luminosity, we used the calibrated relation \citep{menon2024EvolvedBinaries} given by
\begin{equation}
    M_{bol} = m\times\log P_0 + c - \mu + BC + 2.55\times(V-I)_0,
\end{equation}
where $M_{bol}$ represents the absolute bolometric magnitude obtained using Wesenheit (colour-corrected) V-band magnitude \citep[$WI_V\,=\,V-2.55(V-I)_0$; see][]{ngeow2005wesenheit}, while the parameters $m\,=\,-3.59$ and $c\,=\,18.79$ correspond to the calibrated slope and intercept of the linear fit, respectively. $P_0$ represents the observed fundamental pulsation period in days, $\mu\,=\,18.49$ denotes the distance modulus to the LMC, $BC$ signifies the bolometric correction derived from the effective temperature \citep{flower1996BoloCorr, torres2010BoloCorrErrata}, and $(V-I)_0$ denotes the intrinsic (de-reddened) colour of each star (reddening value is adopted from SED fits). The uncertainties of PLC luminosity are primarily influenced by the uncertainties of reddening.

We note that SED luminosities require \textit{Gaia} parallax measurements (underlying the derivation of geometric distances), which are plagued by orbital motion in case of binary systems \citep{kluska2022GalacticBinaries}. Additionally, since the targets in this study are Type II Cepheid pulsating variables with periods ranging from 18.73 to 46.7 days (see Table~\ref{tab:litpar_paper2}), their atmospheric parameters, particularly effective temperature $T_{\rm eff}$, surface gravity $\log g$, and microturbulent velocity $\xi_t$, undergo significant variations throughout the pulsation cycle \citep{mohorian2024EiSpec}. This causes a considerable scatter in the photometric data points leading to increased uncertainties of SED luminosity (see Appendix~\ref{app:sed_paper2}). Hence, we regard the PLC luminosity to be more precise and reliable compared to the SED luminosity in our targets.

\begin{sidewaystable}[ph!]
    \centering
    \footnotesize
    \caption[Derived luminosities and atmospheric parameters of transition disc targets]{Derived luminosities and atmospheric parameters of transition disc targets. The columns are as follows: col. 1: Target ID; col. 2: Target name; col. 3: SED luminosity (see Section~\ref{ssec:doblum_paper2}); col. 4: infrared luminosity; col. 5: PLC luminosity (adopted; see Section~\ref{ssec:doblum_paper2}); col. 6, 7, 8, 9: derived atmospheric parameters (see Section~\ref{ssec:respro_paper2}).}\label{tab:fnlvls_paper2}
    \begin{tabular}{|cc|ccc|cccc|}\hline
        \multirow{2}{*}{\textbf{ID}} & \multirow{2}{*}{\textbf{Name}} & \multirow{2}{*}{\boldmath$\log\dfrac{L_{\rm SED}}{L_\odot}$} & \multirow{2}{*}{\boldmath$\log\dfrac{L_{\rm IR}}{L_{\rm SED}}$} & \multirow{2}{*}{\boldmath$\log\dfrac{L_{\rm PLC}}{L_\odot}$} & \boldmath$T_{\rm eff}$ & \boldmath$\log g$ & \textbf{[Fe/H]} & \boldmath$\xi_{\rm t}$ \\
        &&&&& \textbf{(K)} & \textbf{(dex)} & \textbf{(dex)} & \textbf{(km/s)} \\ \hline
        1 & CT Ori & 3.41$\pm$0.13 & --0.26$\pm$0.13 & 3.26$\pm$0.16 & 5940$\pm$120 & 1.01$\pm$0.18 & --1.89$\pm$0.11 & 3.37$\pm$0.10 \\
        2 & $^\ast$ST Pup & 2.87$\pm$0.08 & --0.14$\pm$0.08 & 2.96$\pm$0.17 & 5340$\pm$80 & 0.20$\pm$0.10 & --1.92$\pm$0.08 & 2.83$\pm$0.03 \\
        3 & RU Cen & 4.00$\pm$0.22 & --0.40$\pm$0.22 & 3.50$\pm$0.27 & 6120$\pm$80 & 1.46$\pm$0.15 & --1.93$\pm$0.08 & 3.26$\pm$0.10 \\
        4 & AC Her & 3.79$\pm$0.11 & --0.68$\pm$0.11 & 3.71$\pm$0.21 & 6140$\pm$100 & 1.27$\pm$0.16 & --1.47$\pm$0.08 & 3.92$\pm$0.12 \\
        5 & AD Aql & 3.41$\pm$0.19 & --0.29$\pm$0.19 & 2.84$\pm$0.29 & 6200$\pm$170 & 1.67$\pm$0.45 & --2.20$\pm$0.09 & 2.98$\pm$0.36 \\
        6 & EP Lyr & 3.74$\pm$0.16 & --1.70$\pm$0.16 & 3.96$\pm$0.23 & 6270$\pm$160 & 1.24$\pm$0.18 & --2.03$\pm$0.17 & 2.48$\pm$0.10 \\ \hline
        7 & $^\ast$DY Ori & 2.81$\pm$0.10 & --0.26$\pm$0.13 & 3.50$\pm$0.17 & 6160$\pm$70 & 0.88$\pm$0.14 & --2.03$\pm$0.04 & 2.48$\pm$0.09 \\
        8 & $^\ast$AF Crt & -- & 0.26$\pm$0.08 & 2.51$\pm$0.17 & 6110$\pm$110 & 0.96$\pm$0.21 & --2.47$\pm$0.05 & 4.87$\pm$0.16 \\ 
        9 & $^\ast$GZ Nor & 3.24$\pm$0.15 & --0.66$\pm$0.22 & 2.71$\pm$0.29 & 4830$\pm$20 & 0.00$\pm$0.18 & --1.89$\pm$0.11 & 5.95$\pm$0.18 \\
        10 & V1504 Sco & -- & 0.67$\pm$0.11 & 3.71$\pm$0.35 & 5980$\pm$90 & 0.98$\pm$0.17 & --1.05$\pm$0.07 & 4.29$\pm$0.05 \\ 
        11 & $^\ast$LMC V0770 & 3.42$\pm$0.11 & --0.20$\pm$0.19 & 3.46$\pm$0.20 & 5750$\pm$100 & 0.00$\pm$0.18 & --2.61$\pm$0.05 & 2.20$\pm$0.01 \\
        12 & $^\ast$LMC V3156 & 3.75$\pm$0.08 & --0.08$\pm$0.16 & 3.99$\pm$0.19 & 6160$\pm$130 & 1.38$\pm$0.20 & --2.48$\pm$0.04 & 4.28$\pm$0.12 \\ \hline
    \end{tabular}\\
    \textbf{Notes:} We highlight with asterisks ($^\ast$) those targets, for which we used the ATLAS9 model atmospheres (see Section~\ref{ssec:sanlte_paper2}). For the uncertainty of infrared luminosity, we assume that the error of SED luminosity, primarily due to reddening uncertainty, dominates over the error from fitting the infrared bump, which is influenced by the uncertainty in photometric data points. SED luminosities of edge-on targets AF Crt (\#8) and V1504 Sco (\#10) were removed from the table as SED fitting is not applicable when the optical light is dominated by scattering.
\end{sidewaystable}

\subsection{Spectroscopic data}\label{ssec:dobspc_paper2}
In this subsection, we present the optical high-resolution spectra used in this study, which were obtained from the High Efficiency and Resolution Mercator Echelle Spectrograph mounted on Mercator telescope (HERMES/Mercator; see Section~\ref{sssec:dobspcher_paper2}) and Ultraviolet and Visual Echelle Spectrograph mounted on Very Large Telescope (UVES/VLT; see Section~\ref{sssec:dobspcuvs_paper2}). In Table~\ref{tab:obslog_paper2}, we present the final selection of spectral visits (for all considered visits, see Appendix~\ref{app:vis_paper2}). In Fig.~\ref{fig:spcmon_paper2}, we show exemplary spectral regions for all targets, focusing on the S and Zn lines, which are crucial for deriving the initial metallicity [M/H]$_{\rm 0,min}$, because Fe is depleted in transition disc targets (see Section~\ref{sec:san_paper2}).

To ensure the accuracy of the abundance analysis of the pulsating transition disc targets, our selection of optical visits depended on the time span of observational dataset for each target. Our selection strategy is as follows (see Table~\ref{tab:obslog_paper2}):
\begin{enumerate}
    \item ST Pup (\#2), RU Cen (\#3), GZ Nor (\#9), V1504 Sco (\#10): We used the only available spectrum for each of these targets.
    \item LMC V3156 (\#11) and LMC V0770 (\#12): Multiple spectra of these targets were taken in less than two-day span (1.08 days and 0.07 days, respectively). Since the pulsation periods of these targets are greater than 30 days, we merged the RV-corrected spectra of each target to increase the S/N ratio.
    \item CT Ori (\#1), AC Her (\#4), AD Aql (\#5), EP Lyr (\#6), DY Ori (\#7), and AF Crt (\#8): These targets are part of long-term spectral monitoring program using HERMES (see Section~\ref{sssec:dobspcher_paper2}). Since the observations of these targets were generally taken with a gap of at least few days, we selected the individual visits of these targets with the highest S/N values.
\end{enumerate}

\begin{table}[!ht]
    \centering
    \footnotesize
    \caption[Observation log of the target sample]{Observation log of the target sample. For more information on the selection criteria for the spectral visits, see Section~\ref{sec:tar_paper2}. In Appendix~\ref{app:vis_paper2}, we provide the full observational log of transition disc targets.} \label{tab:obslog_paper2}
    \begin{tabular}{|cc|cc|cc|}
    \hline
        \textbf{ID} & \textbf{Name} & \textbf{Facility} & \textbf{MJD} & \textbf{RV (km/s)} & \textbf{S/N} \\ \hline
        1 & CT Ori & H/M & 56254.1414 & 47.12$\pm$0.96 & 45 \\
        2 & ST Pup & H/M & 51085.2822 & 23.03$\pm$0.38 & 50 \\
        3 & RU Cen & H/M & 51627.1188 & -0.40$\pm$0.28 & 55 \\
        4 & AC Her & H/M & 57876.1807 & -54.18$\pm$0.57 & 45 \\
        5 & AD Aql & H/M & 56870.9410 & 54.36$\pm$0.96 & 40 \\
        6 & EP Lyr & H/M & 55005.1064 & 29.37$\pm$1.29 & 45 \\ \hline
        7 & DY Ori & H/M & 58452.1440 & 11.67$\pm$0.90 & 40 \\
        8 & AF Crt & H/M & 56298.2332 & 27.72$\pm$1.08 & 40 \\
        9 & GZ Nor & U/V & 56819.4194 & -121.61$\pm$0.26 & 50 \\
        10 & V1504 Sco & H/M & 51626.3207 & 28.43$\pm$0.25 & 50 \\
        11 & LMC V0770 & U/V & 56589.1761 & 288.96$\pm$3.37 & 45 \\
        12 & LMC V3156 & U/V & 53409.9187 & 280.06$\pm$1.26 & 45 \\ \hline
    \end{tabular}\\
    \textbf{Notes:} H/M means HERMES/Mercator, U/V means UVES/VLT. RV is the radial velocity of the spectrum used in the analyses. S/N is the average signal-to-noise ratio of the spectrum.
\end{table}
\begin{figure}[ph!]
    \centering
    \includegraphics[width=.59\linewidth]{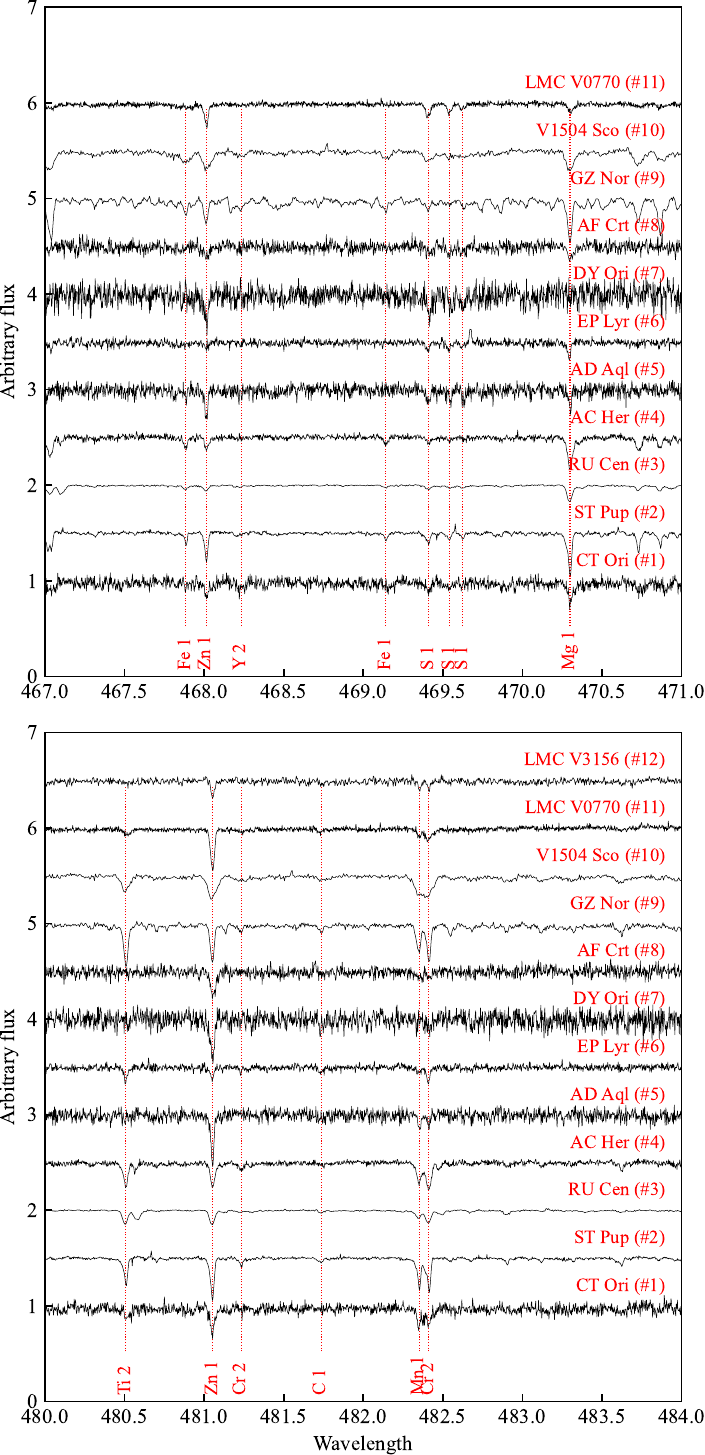}
    \caption[Comparison of the normalised spectra of all sample stars in two spectral regions containing lines of volatile elements S and Zn]{Comparison of the normalised spectra of all sample stars in two spectral regions containing lines of volatile elements S and Zn. All spectra are normalised, corrected for radial velocity, and shifted in flux for clarity. The dashed red vertical lines indicate positions of line peaks. The names of the stars are provided in the plot. The observed spectrum of LMC V3156 (\#12) does not contain the region between 467.5 and 470.5 nm (for more details, see Section~\ref{ssec:dobspc_paper2}).}\label{fig:spcmon_paper2}
\end{figure}

\subsubsection{HERMES spectra}\label{sssec:dobspcher_paper2}
For CT Ori (\#1), ST Pup (\#2), RU Cen (\#3), AC Her (\#4), AD Aql (\#5), EP Lyr (\#6), DY Ori (\#7), AF Crt (\#8), and V1504 Sco (\#10), we used the high-resolution (R\,=\,$\lambda/\Delta\lambda\sim$ 85 000) optical spectra obtained within an extensive monitoring initiative (June 2009--ongoing). This initiative involved HERMES \citep{raskin2011hermes} installed on the 1.2-m Mercator telescope at the Roque de los Muchachos Observatory, La Palma. 

This monitoring program resulted in the collection of a substantial dataset of high-resolution optical spectra for post-AGB systems, thoroughly detailed in \citet{vanwinckel2018Binaries}. The HERMES spectra of transition disc targets were reduced using the standard pipeline, as outlined in \citet{raskin2011hermes}. The complete log of HERMES observations can be found in Appendix~\ref{app:vis_paper2}, with the selected visits listed in Table~\ref{tab:obslog_paper2}.

\subsubsection{UVES spectra}\label{sssec:dobspcuvs_paper2}
For GZ Nor (\#9), LMC V0770 (\#11), and LMC V3156 (\#12) we employed high-resolution optical spectra (R\,=\,$\lambda/\Delta\lambda\sim$ 80 000 in the Blue arm; R\,=\,$\lambda/\Delta\lambda\sim$ 110 000 in the Red arm) obtained with the UVES \citep{dekker2000UVES}. UVES is mounted on the 8-m UT2 Kueyen Telescope at the VLT located at the Paranal Observatory of ESO in Chile.

To reduce the UVES spectra, we followed the standard steps for UVES reduction pipeline \citep[frame extraction, flat-field correction, wavelength calibration, cosmic clipping; see][]{dekker2000UVES}. The full observational log of UVES spectra can be found in Appendix~\ref{app:vis_paper2}, and the selected visits are listed in Table~\ref{tab:obslog_paper2}.

\section{Spectral analysis}\label{sec:san_paper2}
In this study, we used E-iSpec \citep[explained in detail in][]{mohorian2024EiSpec} -- a modified version of iSpec \citep{blancocuaresma2014, blancocuaresma2019} to investigate the chemical composition of the target sample. Our chemical analysis included precise derivation of atmospheric parameters (effective temperature $T_{\rm eff}$, surface gravity $\log g$, metallicity [Fe/H], and microturbulent velocity $\xi_{\rm t}$) and abundances [X/H] of all elements, which had detectable spectral features in the spectra of transition disc targets. Additionally, we used \texttt{Balder} \citep{amarsi2018Balder} to calculate NLTE corrections for the abundances of a representative set of 11 chemical elements: C, N, O, Na, Mg, Al, Si, S, K, Ca, and Fe. In the following subsections, we present the methodology of abundance analysis of transition disc stars and candidates (for individual depletion profiles, see Appendix~\ref{app:dpl_paper2}).

\subsection{LTE analysis using E-iSpec}\label{ssec:sanlte_paper2}
To calculate the atmospheric parameters and elemental abundances from atomic spectral lines, we followed the procedure presented in \citet{mohorian2024EiSpec}. In brief, we used the Moog radiative transfer code \citep[equivalent width method;][]{sneden1973moog} with the VALD3 line list \citep{kupka2011vald}, the recently updated solar abundances from \citet{asplund2021solar}, and LTE model atmospheres: spherically-symmetric MARCS models \citep{gustafsson2008MARCS} and plane-parallel ATLAS9 models \citep{castelli2003ATLAS9}\footnote{In general, spherically-symmetric models are known to perform better for giant stars due to their extended atmospheres \citep{meszaros2012ModelsAPOGEE}. However, half of our target sample had atmospheric parameters outside of MARCS parameter grid values; for such targets we used the ATLAS9 model atmospheres instead.}.

\begin{figure}[!ht]
    \centering
    \includegraphics[width=.38\linewidth]{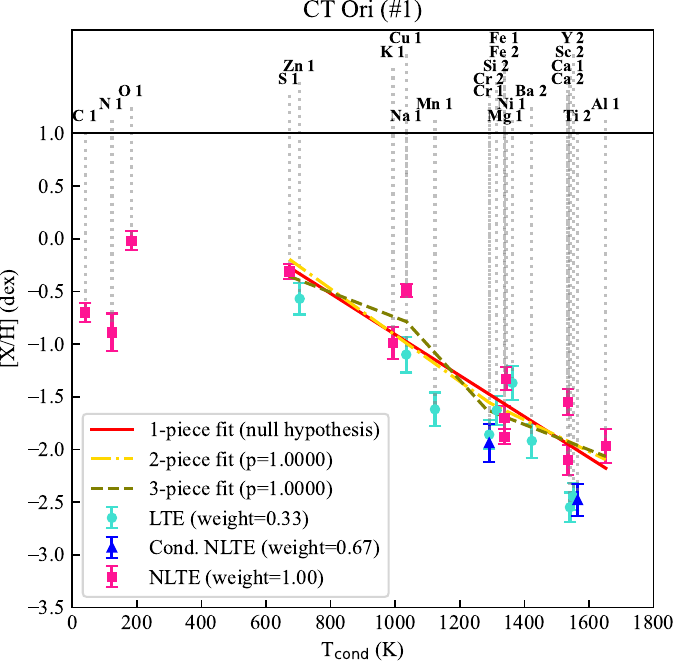}
    \includegraphics[width=.38\linewidth]{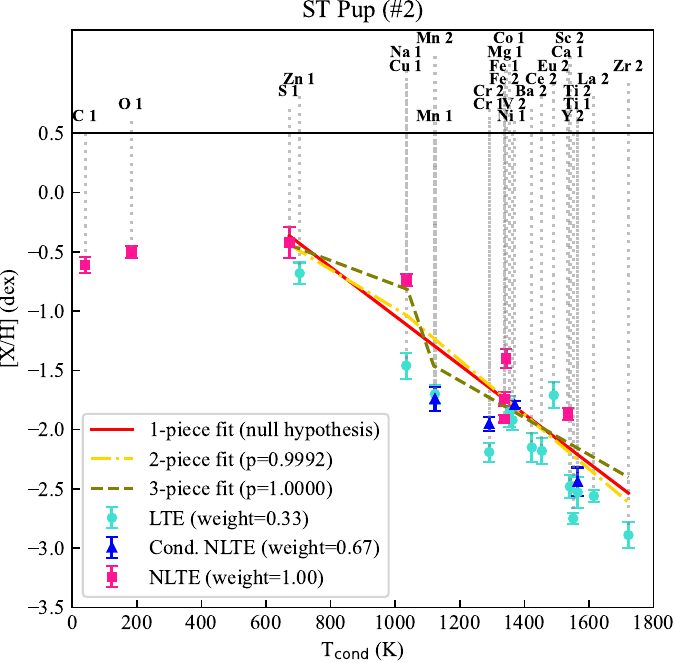}
    \includegraphics[width=.38\linewidth]{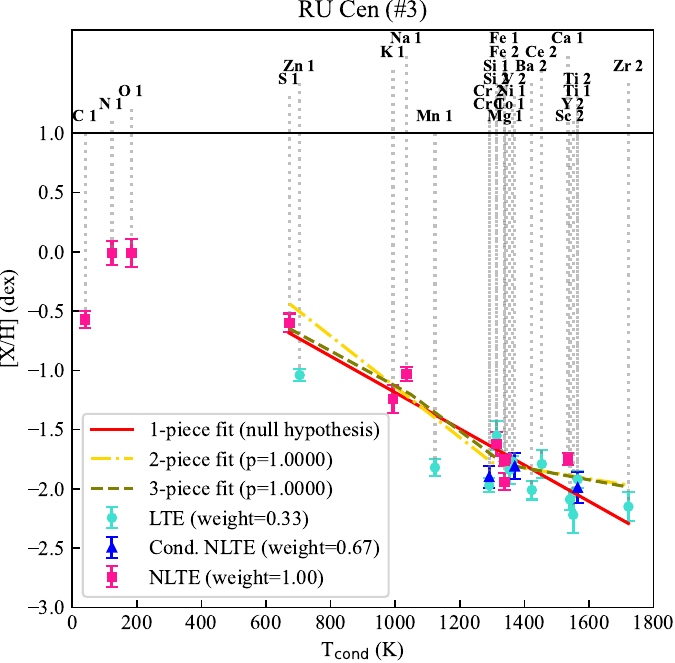}
    \includegraphics[width=.38\linewidth]{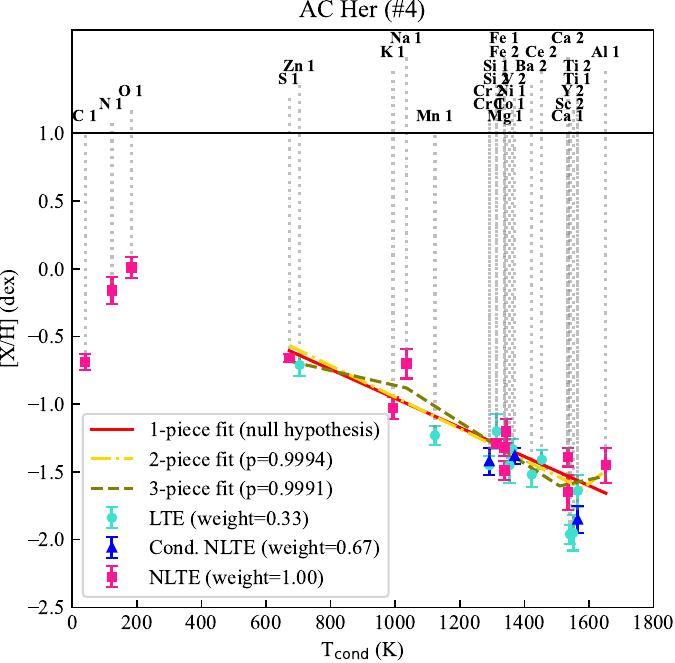}
    \includegraphics[width=.38\linewidth]{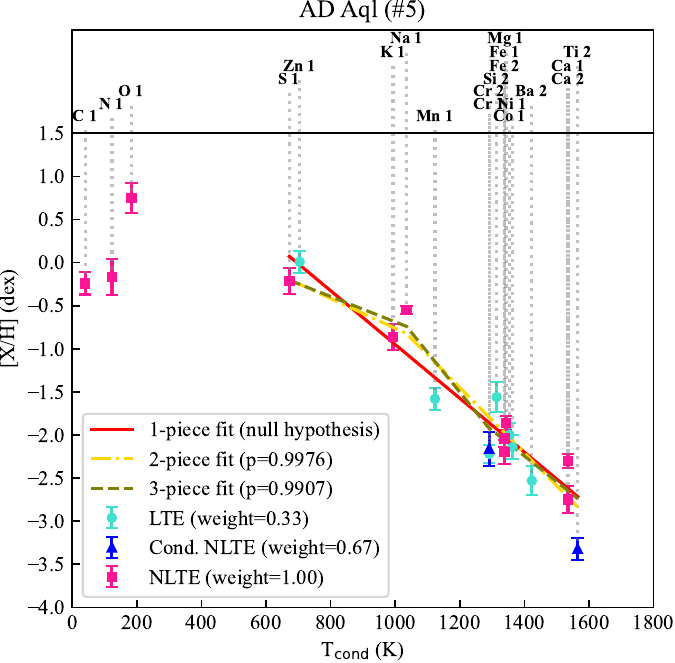}
    \includegraphics[width=.38\linewidth]{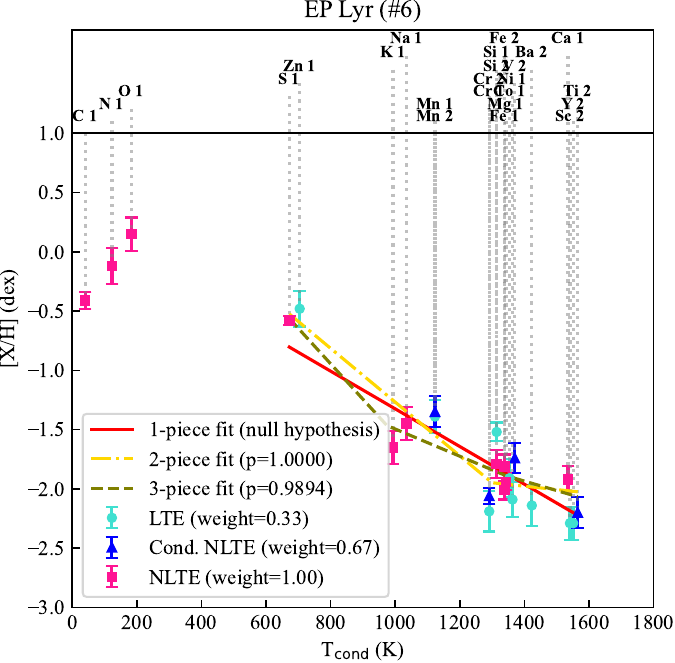}
    \caption[Elemental abundances of transition disc stars (CT Ori, ST Pup, RU Cen, AC Her, AD Aql, and EP Lyr) as functions of condensation temperature \citep{lodders2003CondensationTemperatures, wood2019CondensationTemperatures}]{Elemental abundances of transition disc stars (CT Ori, ST Pup, RU Cen, AC Her, AD Aql, and EP Lyr) as functions of condensation temperature \citep{lodders2003CondensationTemperatures, wood2019CondensationTemperatures}. The legend for the symbols and colours used is included within the plot. ``Cond. NLTE'' means conditionally NLTE abundance (derived from spectral lines of \ion{Ti}{ii}, \ion{V}{ii}, \ion{Cr}{ii}, or \ion{Mn}{ii}; for more details, see Section~\ref{ssec:respro_paper2}).}\label{fig:dplstr_paper2}
\end{figure}

\begin{figure}[!ht]
    \centering
    \includegraphics[width=.38\linewidth]{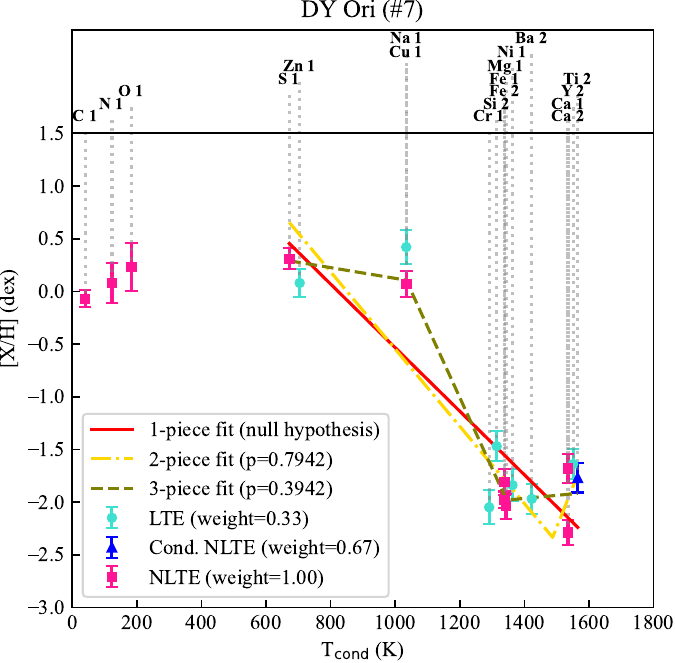}
    \includegraphics[width=.38\linewidth]{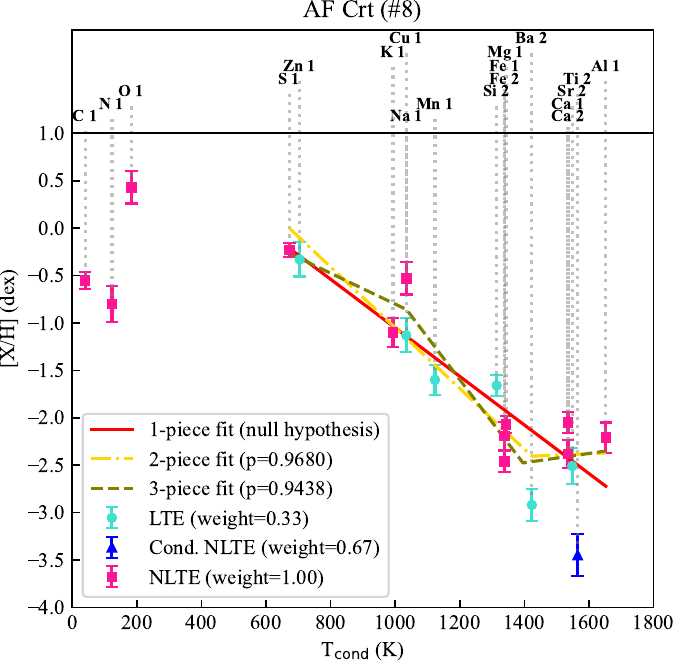}
    \includegraphics[width=.38\linewidth]{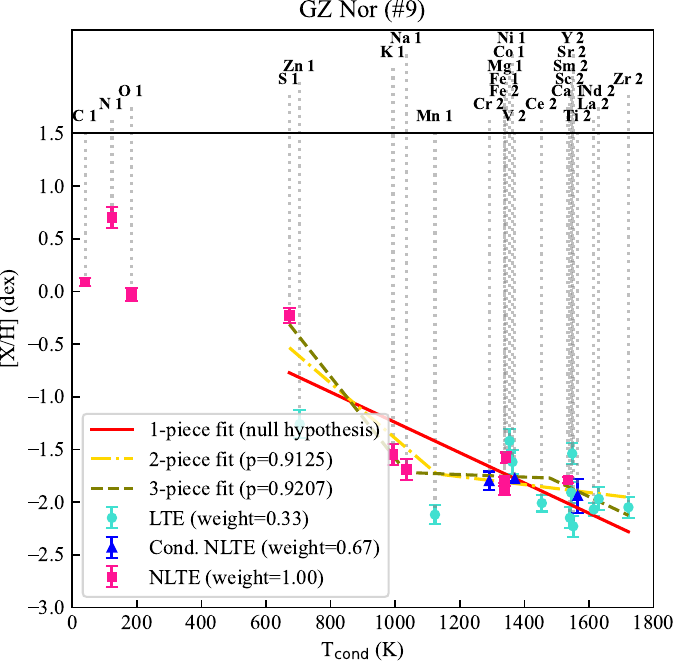}
    \includegraphics[width=.38\linewidth]{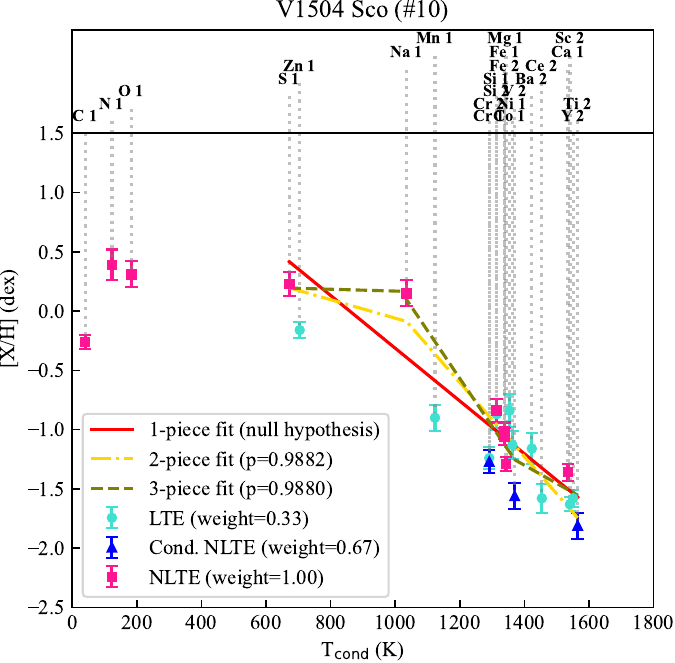}
    \includegraphics[width=.38\linewidth]{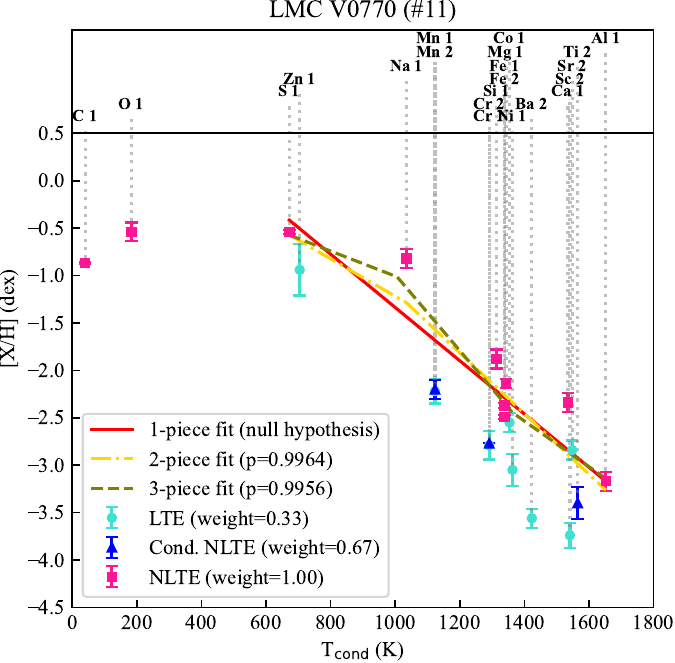}
    \includegraphics[width=.38\linewidth]{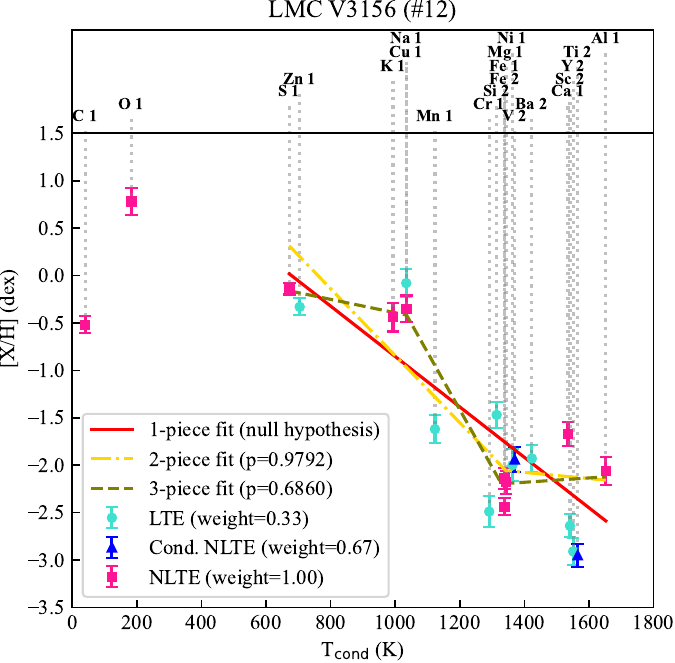}
    \caption[Elemental abundances of transition disc candidates (DY Ori, AF Crt, GZ Nor, V1504 Sco, LMC V0770, and LMC V3156) as functions of condensation temperature \citep{lodders2003CondensationTemperatures, wood2019CondensationTemperatures}]{Elemental abundances of transition disc candidates (DY Ori, AF Crt, GZ Nor, V1504 Sco, LMC V0770, and LMC V3156) as functions of condensation temperature \citep{lodders2003CondensationTemperatures, wood2019CondensationTemperatures}. The legend for the symbols and colours used is included within the plot. ``Cond. NLTE'' means conditionally NLTE abundance (derived from spectral lines of \ion{Ti}{ii}, \ion{V}{ii}, \ion{Cr}{ii}, or \ion{Mn}{ii}; for more details, see Section~\ref{ssec:respro_paper2}.}\label{fig:dplcnd_paper2}
\end{figure}

In the original version of E-iSpec \citep{mohorian2024EiSpec}, the uncertainties were calculated manually. In this study, we updated E-iSpec by automating the process of uncertainty calculation for elemental abundances, while following the same approach as in manual method. In brief, the total uncertainty is calculated as the sum in quadrature of random and systematic uncertainties. We note that we set the random uncertainty to be 0.1 dex for those ionisations, for which we used only one spectral line to derive the elemental abundance. We assume that the abundance change caused by metallicity affects [X/Fe], but does not affect [X/H]. In Fig.~\ref{fig:dplstr_paper2} and \ref{fig:dplcnd_paper2}, we show the derived elemental abundances with uncertainties for all transition disc stars and candidates, respectively. Furthermore, we updated E-iSpec with a functionality to recalculate atmospheric parameters and elemental abundances using NLTE abundance corrections from \texttt{Balder} (see Section~\ref{ssec:sannlte_paper2}).

In Table~\ref{tab:fnlvls_paper2}, we present the estimated SED and PLC luminosities (see Section~\ref{ssec:doblum_paper2}) and the derived atmospheric parameters of the target sample (the targets, for which we used ATLAS9 model atmospheres, are highlighted with asterisks). In Tables~\ref{tab:fnlabu_paper2} and \ref{tabA:fnlabu_paper2}, we provide the results of our abundance analysis using E-iSpec: the selected abundance ratios and the elemental abundances, respectively. The combined line list for all 12 targets is provided in Appendix~\ref{app:lst_paper2}.

\begin{table}[!ht]
    \centering
    \footnotesize
    \caption[Selected abundance ratios of transition disc targets (see Table~\ref{tabA:fnlabu_paper2} for a full list of {[X/H]} abundances)]{Selected abundance ratios of transition disc targets (see Table~\ref{tabA:fnlabu_paper2} for a full list of [X/H] abundances). The columns are as follows: col. 1: Target ID; col. 2: Target name; col. 3, 4, 5: proxies for depletion efficiency (we adopt the NLTE-corrected [S/Ti] ratio, see Section~\ref{ssec:reseff_paper2}); col. 6: NLTE-corrected C/O ratio; col. 7: minimal initial metallicity (see Section~\ref{ssec:reseff_paper2}); col. 8: depletion strength (see Section~\ref{ssec:dplyso_paper2}).}\label{tab:fnlabu_paper2}
    \begin{tabular}{|cc|cc|cccc|}\hline
        ~ & ~ & \textbf{LTE} & \textbf{LTE} & \textbf{NLTE} & \textbf{NLTE} & \textbf{NLTE} & \textbf{NLTE} \\
        \textbf{ID} & \textbf{Name} & \textbf{[Zn/Ti]} & \textbf{[Zn/Fe]} & \textbf{[S/Ti]} & \textbf{C/O} & \textbf{[M/H]}\boldmath$_{\rm0,min}$ & \boldmath$\Delta_{\rm g/d}$ \\
        ~ & ~ & \textbf{(dex)} & \textbf{(dex)} & \textbf{(dex)} & ~ & \textbf{(dex)} & ~ \\ \hline
        1 & CT Ori & 1.91$\pm$0.30 & 1.31$\pm$0.23 & 2.17$\pm$0.25 & 0.12$\pm$0.04 & --0.31$\pm$0.10 & 3720 \\
        2 & ST Pup & 1.76$\pm$0.21 & 1.23$\pm$0.15 & 2.02$\pm$0.35 & 0.46$\pm$0.11 & --0.42$\pm$0.23 & 3090 \\
        3 & RU Cen & 0.95$\pm$0.18 & 0.90$\pm$0.14 & 1.39$\pm$0.28 & 0.16$\pm$0.06 & --0.60$\pm$0.15 & 2190 \\
        4 & AC Her & 1.14$\pm$0.18 & 0.78$\pm$0.18 & 1.19$\pm$0.14 & 0.12$\pm$0.03 & --0.66$\pm$0.04 & 680 \\
        5 & AD Aql & 3.33$\pm$0.26 & 2.21$\pm$0.28 & 3.11$\pm$0.34 & 0.06$\pm$0.03 & --0.21$\pm$0.21 & 9770 \\
        6 & EP Lyr & 1.72$\pm$0.28 & 1.53$\pm$0.31 & 1.62$\pm$0.20 & 0.16$\pm$0.06 & --0.58$\pm$0.07 & 2690 \\ \hline
        7 & DY Ori & 1.85$\pm$0.27 & 2.06$\pm$0.22 & 2.08$\pm$0.35 & 0.30$\pm$0.15 & 0.31$\pm$0.21 & 19500 \\
        8 & AF Crt & 3.12$\pm$0.40 & 2.13$\pm$0.30 & 3.22$\pm$0.31 & 0.06$\pm$0.03 & --0.23$\pm$0.09 & 16980 \\
        9 & GZ Nor & 0.68$\pm$0.29 & 0.63$\pm$0.22 & 1.71$\pm$0.28 & 0.78$\pm$0.16 & --0.23$\pm$0.12 & 4570 \\
        10 & V1504 Sco & 1.65$\pm$0.18 & 0.90$\pm$0.20 & 2.04$\pm$0.11 & 0.16$\pm$0.05 & 0.23$\pm$0.00 & 1950 \\
        11 & LMC V0770 & 2.46$\pm$0.44 & 1.55$\pm$0.32 & 2.86$\pm$0.19 & 0.28$\pm$0.06 & --0.54$\pm$0.02 & 8910 \\
        12 & LMC V3156 & 2.62$\pm$0.21 & 2.11$\pm$0.19 & 2.81$\pm$0.19 & 0.03$\pm$0.01 & --0.14$\pm$0.07 & 19950 \\ \hline
    \end{tabular}
\end{table}

\subsection{NLTE abundance corrections using Balder}\label{ssec:sannlte_paper2}
The departures from LTE typically (but not always) grow with increasing temperature $T_{\rm eff}$, decreasing surface gravity $\log g$, or decreasing metallicity [Fe/H] \citep{lind2012NLTE}. These departures primarily stem from the intense radiation field at shorter wavelengths, which is not effectively compensated by thermal collisions in the stellar photosphere. Among the various species, neutral and relatively low-ionisation elements like \ion{Fe}{i} or \ion{Ti}{i} experience overionisation, resulting in weakened spectral lines. Conversely, the dominant ionisation stage like \ion{Fe}{ii} or \ion{Ti}{ii} closely follows the Saha distribution, and remains mostly unaffected by NLTE effects \citep{amarsi2016NLTEiron, amarsi2022NLTEiron}. Thus, we adopted [Fe/H] abundance derived from \ion{Fe}{ii} lines as the metallicity and applied NLTE corrections to elemental abundances \citep[see][and references therein]{amarsi2020NLTEgalah}. 

To study the chemical depletion in post-AGB/post-RGB binary stars with transition discs, we selected C, N, O, Na, Mg, Al, Si, S, K, Ca, and Fe as a representative set of chemical elements. We calculated the NLTE corrections for each individual studied spectral line of these elements using the code \texttt{Balder}, which is based on \texttt{Multi3D} \citep{leenaarts2009Multi3D}.

The method of NLTE correction calculation closely follows that described in \citet{amarsi2020NLTEgalah}. We list the model atoms used in this work in Table~\ref{tab:nonref_paper2}. The NLTE calculations were performed on a small number of LTE spherically-symmetric MARCS model atmospheres \citep{gustafsson2008MARCS} using two small grids: the hotter grid spanned 5\,500\,K\,$\leq\,T_{\rm eff}\,\leq$\,6\,250\,K, 1.0\,dex\,$\leq\,\log g\,\leq$\,2.0\,dex, and --2.5\,dex\,$\leq\,[{\rm Fe/H}]\,\leq$\,--1.0\,dex, while the cooler grid spanned 4\,750\,K\,$\leq\,T_{\rm eff}\,\leq$\,5\,250\,K, $\log g\,=\,0.5$\,dex, and --2.0\,dex\,$\leq\,[{\rm Fe/H}]\,\leq$\,--1.5\,dex. Calculations were performed for a range of abundances from $\mathrm{[X/H]\,=\,-4.5}$ to $+2.0$ in steps of $0.5\,\mathrm{dex}$, and microturbulences $\xi_{\rm t}$\,=\,2.0 km/s and 5.0 km/s. The theoretical equivalent widths were used to calculate NLTE abundance corrections for each individual spectral line $i$, denoted as $\Delta_{i}^{\rm diff.}$. These were interpolated onto the stellar parameters of interest, with edge values adopted for stars outside of this theoretical grid. The resulting line-by-line NLTE abundances were given by:
\begin{equation}
    [{\rm X/H}]_{i}^{\rm NLTE} = [{\rm X/H}]_{i}^{\rm LTE} + \Delta_{i}^{\rm diff.},
\end{equation}
where superscript ``diff.'' highlights that we use the relative abundance scale [X/H], which requires $\Delta_{i}^{\rm diff.}$ to also include the NLTE corrections to the solar absolute abundances of the corresponding elements. We note that we assume the differences between the ATLAS9 model atmospheres (used for a subsample of targets) and the MARCS models (used in NLTE calculations) to be of secondary importance compared to the total uncertainties of the LTE abundances as well as the uncertainties of the NLTE models. For the uncertainties of NLTE abundances, we assume the same systematic component as in the LTE abundances, and recalculate the random component in similar way (for one spectral line, $\sigma_{\rm random}\,=\,0.1$\,dex; for more spectral lines $\sigma_{\rm random}$ is the standard deviation).

\section{Elemental abundances of transition disc targets}\label{sec:res_paper2}
In this section, we present the results of our LTE and NLTE abundance analysis of optical spectra of transition disc targets (see Section~\ref{ssec:respro_paper2}), define depletion indicators used for our sample (see Section~\ref{ssec:reseff_paper2}), and compare derived depletion profiles with those from previous studies (see Section~\ref{ssec:rescmp_paper2}).

\subsection{Observed depletion profiles}\label{ssec:respro_paper2}
In Fig.~\ref{fig:dplstr_paper2} and \ref{fig:dplcnd_paper2}, we present the depletion patterns together with the corresponding fits for each target. We mark elemental abundances corrected for NLTE effects with pink squares. For Fe-peak elements beyond the $\alpha$-process (from Ti to Fe), we considered the abundances derived from the ionised lines to be likely less sensitive to NLTE effects (i.e., conditionally NLTE; \ion{Ti}{ii}, \ion{V}{ii}, \ion{Cr}{ii}, \ion{Mn}{ii}). We denote these abundances with blue triangles. We mark the LTE abundances of the all remaining ionisations with light blue circles. The depletion profiles were fitted from S to Zr based on the arbitrary reliability of the abundance measurement: NLTE abundances were weighted as 1, conditionally NLTE abundances were weighted as 0.67, and LTE abundances were weighted as 0.33. In this approach, the abundance of each ionisation was treated separately, which provided more weight to those elements, for which the abundances of two ionisations were measured. We note that the chosen weighting is an arbitrary measure used to emphasise the difference between fully-LTE depletion profile and NLTE-corrected depletion profile.

One-piece linear fits (red solid lines) represent the homogeneous depletion profile without any break temperatures caused by the onset of depletion ($T_{\rm turn-off}$) or plateau ($T_{\rm plateau}$). Two-piece and three-piece linear fits (yellow dashed-dotted and green dashed lines, respectively) represent the depletion with any or both of these break temperatures, respectively. Two-piece and three-piece fits were tested against the one-piece linear fit (which was considered null hypothesis) using the likelihood ratio test: if the p-value of the two- or three-piece linear fit is less than 0.05, this fit offers significantly better\footnote{Meaning that the decrease in the $\chi^2$ value for a more elaborate model is large enough to justify the reduction in degrees of freedom by invoking new constraints (parameters).} goodness-of-fit for the depletion profile than the one-piece linear fit.

The major differences between LTE and NLTE-corrected abundances in transition disc targets are summarised below:
\begin{enumerate}
    \item For our targets, NLTE corrections decrease C/O ratio by up to $\sim$30\% (mainly due to NLTE corrections for the high-excitation \ion{C}{i} lines used here\footnote{For all targets except GZ Nor (\#9), the O abundances are based on the low-excitation forbidden [\ion{O}{i}] lines that are insensitive to NLTE effects (e.g. \citep{amarsi2016oxygen}.}). However, for GZ Nor (\#9), NLTE corrections increase C/O ratio by two times.
    \item The average line-to-line scatter of NLTE abundances is generally lower than the average line-to-line scatter of LTE abundances, with the most prominent reduction of the scatter for Na (0.04 dex), Al (0.05 dex), Si (0.07 dex), K (0.04 dex), and Ca (0.04 dex).
    \item The final depletion profiles of transition disc targets are well-fitted by one-piece linear trends. This result highlights that derived depletion profiles of all transition disc targets are saturated.
    \item There are few prominent (but statistically insignificant) deviations from one-piece fits of depletion profiles: i) [Na/H] and [Cu/H] in DY Ori (\#7), ii) [S/H] in GZ Nor (\#9), iii) [Na/H] in V1504 Sco (\#10), and iv) [Mn/H] in RU Cen (\#3) and GZ Nor (\#9). The deviations in [Na/H], [Cu/H], and [S/H] abundances may be caused by the differences in chemical composition and conditions between the transition disc targets and the solar-mixture gas assumed in chemical equilibrium calculations \citet{wood2019CondensationTemperatures} (especially, for S in the least O-rich transition disc target from our sample, C/O$_{\rm GZ~Nor}$\,=\,0.78). We note that RU Cen (\#3) and GZ Nor (\#9) clearly show saturation if we temporarily set aside the [Mn/H] abundance. The explanation for this is that the NLTE correction for Mn lines in metal-poor giants generally are positive \citep[{[Mn/H]}$_{\rm NLTE}$ – {[Mn/H]}$_{\rm LTE}\,\sim\,$+0.6 dex for {[Fe/H]} = –-3 dex;][]{bergemann2008MnNLTE, amarsi2020NLTEgalah}, bringing [Mn/H] abundance closer to the linear decline of the depletion profile. We also note that the behaviour of the condensation temperatures for environments with different C/O ratios is out of the scope of this work, though it is a promising path for the future study.
\end{enumerate}

Finally, we also note that an alternative explanation for the enhancement of [Na/H]\,=\,0.15 dex in V1504 Sco involves the first and second dredge-ups, which may be responsible for [Na/H] enhancement in progenitors with intermediate masses $M_\ast>4.5M_\odot$ \citep{karakas2014dawes}.

\subsection{Definition and rationale for depletion indicators}\label{ssec:reseff_paper2}
To characterise the depletion profiles, we defined [S/H]$_{\rm NLTE}$ as the lower limit of the initial metallicity [M/H]$_{\rm 0,min}$, and [S/Ti]$_{\rm NLTE}$ as the NLTE depletion scale (we use [Ti/H]$_{\rm LTE}$ derived from \ion{Ti}{ii} lines as [Ti/H]$_{\rm NLTE}$). Our reasoning for selecting S as volatile indicator and Ti as refractory indicator is as follows:
\begin{enumerate}
    \item C, N, and O in post-AGB/post-RGB stars are modified to an unknown extent by the convective and non-convective mixing processes on AGB/RGB, such as dredge-ups \citep{kobayashi2011IsotopeEvolutionModels, ventura2020CNOinAGB, kamath2023models, mohorian2024EiSpec}. Moreover, CNO elements may be partially depleted, as seen in protoplanetary discs, through a poorly constrained process of CO and N$_2$ molecules (the major carriers of volatile CNO elements in the disc) converting into CO$_2$ and NH$_3$ ice and freezing-out onto dust grains \citep[possible explanations include dispersal of gas disc, interactions between the gas and the dust, and chemical reprocessing; see][and references therein]{reboussin2015Odepletion, krijt2016Odepletion, bai2016Odepletion, xu2017Odepletion, francis2022Odepletion, furuya2022CNdepletion}.
    \item In protoplanetary discs and in the ISM, S is depleted into sulphide minerals to an unknown extent (e.g., FeS or FeS$_2$) \citep{kama2019Sdepletion, konstantopoulou2022ISMdepletion}. However, given that S follows the abundance profiles of our target sample, this element generally is the least depleted after the exclusion of C, N, and O.
    \item S and Ti are $\alpha$-elements with multiple spectral features available in optical range and the most different condensation temperature (Sc is more refractory, yet has significantly less optical spectral lines), but these elements share similar nucleosynthetic history so that the intrinsic [S/Ti] ratio is supposed to be close to zero in the absence of depletion \citep{kobayashi2020OriginOfElements}. Additionally, our approach allows for the first time to use NLTE [S/Ti] ratio instead of LTE [Zn/Ti] ratio.
\end{enumerate}

Finally, the turn-off temperature $T_{\rm turn-off}<1000$ K for all transition disc targets (see Fig.~\ref{fig:dplstr_paper2} and \ref{fig:dplcnd_paper2}). Since the one-piece fits are statistically preferred for each target, and there are only few derived elemental abundances with condensation temperatures $T_{\rm cond}<1000$ K, setting $T_{\rm turn-off}$ becomes less trivial. Hence, we define the upper limit of turn-off temperatures in the transition disc sample to be located between S and Zn condensation temperatures ($T_{\rm turn-off}\,=\,700$ K).

\subsection{Comparison with literature depletion profiles}\label{ssec:rescmp_paper2}
In Appendix~\ref{app:dpl_paper2}, we provide a detailed comparison of depletion profiles of transition disc targets from the literature and from our homogeneous analysis. The key findings of this comparison are summarised below:
\begin{enumerate}
    \item Depletion profiles for RU Cen (\#3), EP Lyr (\#6), DY Ori (\#7), and GZ Nor (\#9), previously classified as ``plateau'' profiles based on LTE abundances \citep{oomen2019depletion}, become ``saturated'' profiles when NLTE corrections are applied (as indicated by p-values).
    \item In previous studies, the onset of depletion, characterised by $T_{\rm turn-off}$, was reported at varying temperatures for our sample, ranging from 800 K for four out of 10 Galactic targets to 1200 K for CT Ori (\#1) and AC Her (\#4) \citep{kluska2022GalacticBinaries}. In contrast, our analysis reveals a consistently minimal $T_{\rm turn-off}$ across all transition disc targets ($T_{\rm turn-off}$\,=\,700 K).
\end{enumerate}

Overall, the spectral data collected in this study enabled the determination of a more extensive set of elemental abundances compared to those reported in the literature for transition disc targets (see Table~\ref{tab:litpar_paper2}). The derived elemental abundances are discussed in the context of depletion in Section~\ref{sec:dpl_paper2}.

\begin{table}[!ht]
    \centering
    \footnotesize
    \caption{The references for model atoms used in this study (see Section~\ref{ssec:sannlte_paper2}).} \label{tab:nonref_paper2}
    \begin{tabular}{|c|c|}
    \hline
        \textbf{Element} & \textbf{Reference} \\ \hline
        C & \cite{amarsi2019NLTEcarbon} \\
        N & \cite{amarsi2020NLTEnitrogen} \\
        O & \cite{amarsi2018NLTEoxygen} \\
        Na & \cite{lind2011NLTEsodium} \\
        Mg & \cite{asplund2021solar} \\
        Al & \cite{nordlander2017NLTEaluminium} \\
        Si & \cite{amarsi2017NLTEsilicon} \\
        S & Amarsi et al. (in prep.) \\
        K & \cite{reggiani2019NLTEpotassium} \\
        Ca & \cite{asplund2021solar} \\
        Fe & \cite{amarsi2022NLTEiron} \\ \hline
    \end{tabular}
\end{table}

\section{Discussion}\label{sec:dpl_paper2}
In this section, we analyse the obtained results to gain a deep understanding of the conditions and factors driving the depletion process in post-AGB/post-RGB binaries with transition discs. We do this by comparing the chemical depletion parameters in our targets with: i) other observational parameters of our target sample, ii) chemical depletion parameters in transition disc YSOs, and iii) chemical depletion parameters in the interstellar medium (ISM).

\subsection{Correlation analysis of known parameters in post-AGB/post-RGB binaries}\label{ssec:dplcor_paper2}
To comprehensively investigate our diverse sample of the most chemically depleted subclass of post-AGB/post-RGB binaries (transition disc targets) we conducted a correlation analysis on a representative selection of observational parameters to address specific questions, including:
\begin{enumerate}
    \item Photometric parameters (IR colours $H-K$ and $W_1-W_3$, SED luminosity $L_{\rm SED}$, PLC luminosity $L_{\rm PLC}$, dust-to-star luminosity ratio $L_{\rm IR}/L_\ast$): to explore potential connections between the current IR excess and depletion efficiency.
    \item Orbital parameters (orbital period $P_{\rm orb}$, eccentricity $e$): to explore whether certain orbital parameter configurations are more prone to depletion.
    \item Pulsational parameter (fundamental pulsation period $P_{\rm puls}$): to investigate the impact of pulsations on the depletion profile.
    \item Astrometric parameters (coordinates R.A. and Dec.): to study the spatial distribution of transition disc targets.
    \item Spectroscopic parameters (effective temperature $T_{\rm eff}$, surface gravity $\log g$, metallicity [Fe/H], C/O ratio, initial metallicity [M/H]$_{\rm 0,min}$, [Zn/Ti] and [S/Ti] abundance ratios): to analyse the shape and the scale of depletion patterns in our targets.
\end{enumerate}

The resulting correlation matrix is depicted in Fig.~\ref{fig:cormat_paper2} (see Appendix~\ref{app:cor_paper2} for individual dependence plots). To examine the behaviour of the dependencies, a logarithmic scale was applied to all parameters except for coordinates and eccentricity. In the following discussion, we address families of correlations, ordered from the most expected to the least.

We found several obvious physical correlations and anti-correlations (Spearman's coefficient $|\rho|\geq0.6$), including: i) spectroscopic parameters related to depletion, such as [Zn/Ti], [S/Ti], [S/Ti]$_{\rm NLTE}$, [Zn/Fe], and [Fe/H]; ii) orbital and other parameters (P$_{\rm orb}$ and $e$ are known for six and five targets, respectively), and iii) surface gravity $\log g$, effective temperature $T_{\rm eff}$, luminosity $L_\ast/L_\odot$, and $\log$(C/O) ratio.

The correlation between $\log L_{\rm IR}/L_\ast$ and [S/Ti]$_{\rm NLTE}$ abundance ratio is rather unexpected. By excluding the edge-on targets AF Crt (\#8) and V1504 Sco (\#10), and relatively dust-poor target EP Lyr (\#6), the correlation between the IR luminosity and the depletion scale becomes even more prominent. However, this correlation is likely caused by the sample bias.

Another notable finding was a strong anti-correlation of the fundamental pulsation period with $W_1-W_3$ colour. However, the circumbinary dust around our targets is produced from a previous phase of mass loss rather than pulsations \citep{vanwinckel2003Review}, so this correlation may be caused by sample bias as well.

Finally, we note that the adopted luminosities (see PLC luminosities in Table~\ref{tab:fnlvls_paper2}) infer the post-RGB evolutionary status of ST Pup (\#2), AD Aql (\#5), AF Crt (\#8), and GZ Nor (\#9). We found that the surface [O/H] abundance in these four post-RGB binaries generally deviates the most from the corresponding value predicted by the linear fit of the depletion profile. However, the surface [O/H] abundance in post-AGB binaries generally aligns with the corresponding linear fit of the depletion profile (see Fig.~\ref{fig:dplstr_paper2} and \ref{fig:dplcnd_paper2}). To quantify this deviation, we used a ``de-scaled'' [O/S] abundance ratio using the following equation:
\begin{equation}
    {\rm [O/S]_{NLTE}^{descaled}} = {\rm \frac{[O/S]_{obs}}{[O/S]_{calc}}},
\end{equation}
where [O/S]$_{\rm obs}$ is the observed [O/S] abundance ratio, while [O/S]$_{\rm calc}\,=\,{\rm [S/Ti]\cdot\frac{\textit{T}_{cond,O}-\textit{T}_{cond,S}}{\textit{T}_{cond,S}-\textit{T}_{cond,Ti}}}$ is the [O/S] abundance ratio calculated by scaling the corresponding [S/Ti] abundance ratio. In our sample, the values of ${\rm [O/S]_{NLTE}^{descaled}}$ range from $\sim0$ (when [O/H] and [S/H] abundances are similar) to 1 (when [O/H] abundance follows the linear depletion trend).

The correlation between ``de-scaled'' abundance ratio ${\rm [O/S]_{NLTE}^{descaled}}$ and luminosity $L_{\ast}$ is not significant for the whole sample, but it is prominent for the subsample of confirmed transition disc targets (\#1--\#6; see Fig.~\ref{figA:allcor2_paper2}). This connection may hint at O depletion being higher in post-AGB binaries than in post-RGB binaries (in other words, $T_{\rm turn-off}$ is lower for depletion profiles of post-AGB binaries rather than for those of post-RGB binaries). This is consistent with the results from \citet{mohorian2024EiSpec}, where the average turn-off temperature for two post-RGB binaries SZ~Mon and DF~Cyg ($T_{\rm turn-off,~post-RGB}\approx1300$K) was found to be higher than the average value for the post-AGB binary sample \citep[$T_{\rm turn-off,~post-AGB}\approx1100$K;][]{oomen2019depletion}. Our investigation will be extended in future work, as it is beyond the scope of this study.

\begin{figure}[!ht]
    \centering
    \includegraphics[width=1.15\linewidth]{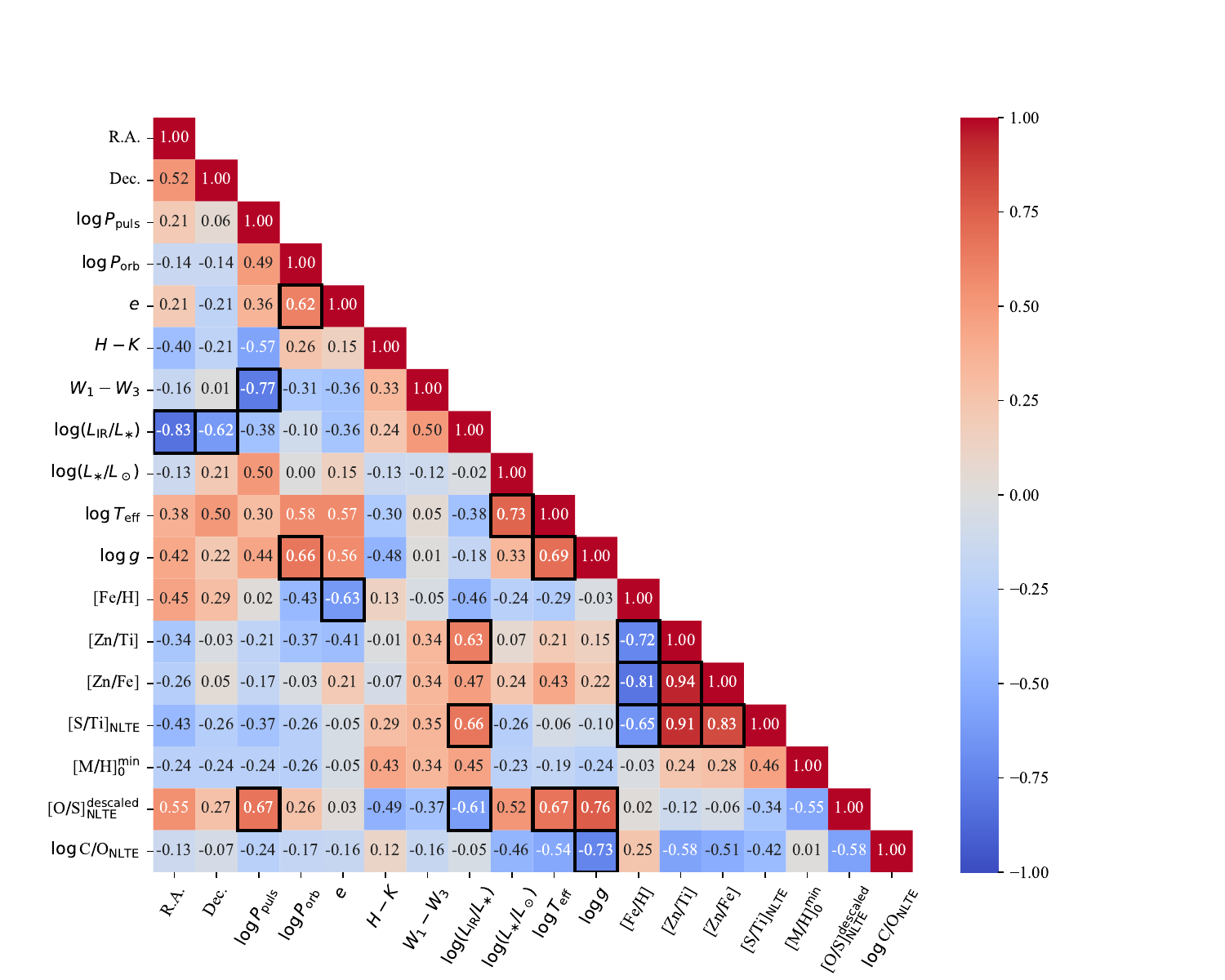}
    \caption[Correlation matrix of various stellar parameters of transition disc targets]{Correlation matrix of various stellar parameters of transition disc targets. The strong correlations and anti-correlations (i.e., with Spearman's correlation coefficients $|\rho|\geq0.6$) are highlighted with black boxes (for more details, see Section~\ref{ssec:dplcor_paper2}).}\label{fig:cormat_paper2}
\end{figure}

\subsection{Parallels with chemical depletion in young stars hosting transition discs}\label{ssec:dplyso_paper2}
Transition discs around young T~Tauri ($M_{\rm T~Tauri}\,<\,2\,M_\odot$) and Herbig Ae/Be ($2\,M_\odot\,<\,M_{\rm Herbig~Ae/Be}\,<\,10\,M_\odot$) stars are similar to post-AGB/post-RGB transition discs in structure and many physical properties, including a broad near-IR excess indicative of hot dust in the disc, Keplerian rotation, dust disc mass, and dust mineralogy \citep[see, e.g.,][and references therein]{follette2017PPDGrainEvolution, deruyter2005discs, corporaal2023DiscParameters, andrych2023Polarimetry, andrych2024IRAS08}. 

Interestingly, similar to post-AGB/post-RGB binaries, a subclass of young T~Tauri and Herbig Ae/Be stars hosting transition discs also exhibits photospheric depletion of refractory elements, a phenomenon known as the $\lambda$ Boo phenomenon \citep[see, e.g.,][]{andrievsky2002lambdaBooAbundances, jura2015lambdaBoo, kama2015DiscDepletionLinkinYSOs, jermyn2018Depletion, murphy2020lambdaBoo}. While both young T~Tauri and Herbig Ae/Be stars with $\lambda$ Boo-like depletion and post-AGB/post-RGB binaries show similar photospheric underabundances of refractory elements, the chemical depletion process is significantly more efficient in post-AGB/post-RGB systems (as indicated by volatile-to-refractory abundance ratios). For instance, the young stars with $\lambda$ Boo-like depletion show underabundances of Mg, Si, and Fe in the range from 0 to 1 dex \citep{kama2015DiscDepletionLinkinYSOs}, whereas post-AGB/post-RGB binaries display [Zn/Ti] abundance ratios in the range from 0 to 3.5 dex \citep{kluska2022GalacticBinaries}.

In post-AGB/post-RGB binaries, the separation of volatile-rich gas and refractory-rich dust remains poorly understood. In contrast, in young T~Tauri and Herbig Ae/Be stars with transition discs and displaying the $\lambda$ Boo phenomenon, the dust-gas separation is linked to several theoretical mechanisms, including grain growth \citep{dullemond2001GrainGrowth}, photoevaporation \citep{alexander2006Photoevaporation}, dead zones \citep{regaly2012DeadZones}, and embedded giant planets \citep{birnstiel2010EmbeddedPlanets}. Furthermore, \citet{folsom2012LambdaBooFraction} suggested that up to a third of Herbig Ae/Be stars hosting protoplanetary discs show signs of depletion and harbour giant planets. Given the structural and chemical parallels between transition discs in $\lambda$ Boo stars and those in post-AGB/post-RGB binaries, we explore the potential role of giant planets in carving the inner gaps in transition discs around post-AGB/post-RGB binaries.

To study the depletion efficiency in protoplanetary transition discs around young stars, \citet{kama2015DiscDepletionLinkinYSOs} used the photospheric composition of Herbig Ae/Be single stars as a proxy for the chemical composition of the accreted matter assuming that this matter (with accretion rates of $\sim10^{-9}-10^{-6}M_\odot$/yr) quickly dominates the original surface chemistry of a star. Previously, \citet{turcotte2002AccretionDomination} showed that for a young star at an age of 10$^6$ years, the domination of accreted matter may be achieved with the accretion rates of as low as $\sim10^{-11}M_\odot$/yr.

To investigate the depletion efficiency in circumbinary discs around post-AGB/post-RGB stars, \citet{oomen2019depletion} modelled the accretion rate onto the binary from a viscously evolving disc for a range of accretion rates and disc masses. They showed that to fit the observed parameters, re-accretion in post-AGB/post-RGB stars require significantly larger initial accretion rates than in young stars ($>3\times10^{-7}M_\odot$/yr). Following the approach from \citet{turcotte2002AccretionDomination}, we estimate that the re-accreted matter should dominate the original surface material in post-AGB/post-RGB binaries within $\sim100$ years. However, this approach is limited by the assumptions of the negligible impact of binary interaction and the simplified chemical composition of the accreted matter \citep[see Fig.~2 in][]{oomen2019depletion}.

To compare the depletion scales in our sample and in YSOs, we calculated the depletion strength $\Delta_{\rm g/d}$ \citep{kama2015DiscDepletionLinkinYSOs} given by
\begin{equation}
    \Delta_{\rm g/d} = 100\times10^{\rm [V/H]-[R/H]} = 100\times10^{\rm [V/R]},
\end{equation}
where [V/H] and [R/H] are the abundances of volatile and refractory tracing elements, respectively. By definition, the solar composition corresponds to depletion strength $\Delta_{\rm g/d,\ \odot}\,=\,100$.

For the [V/H] and [R/H] abundances, \citet{kama2015DiscDepletionLinkinYSOs} combined volatile [C/H] and [O/H], and refractory [Fe/H], [Mg/H], and [Si/H], respectively. As mentioned in Section~\ref{sec:res_paper2}, we considered volatile [S/H] and refractory [Ti/H] to be a more reliable scale of dust depletion. However, for comparison consistency, we used the NLTE-corrected abundances [S/H]$_{\rm NLTE}$ and [Fe/H]$_{\rm NLTE}$, as the depletion tracers in this subsection. Therefore, the expression for the depletion strength $\Delta_{\rm g/d}$ in transition disc targets is given by
\begin{equation}
    \Delta_{\rm g/d} = 100\times10^{\rm [S/H]_{NLTE}-[Fe/H]_{NLTE}} = 100\times10^{\rm [S/Fe]_{NLTE}}.
\end{equation}

For young stars hosting transition discs, the depletion strength $\Delta_{\rm g/d}$ was found to be below $\approx10^3$ \citep[see Fig.~2 in][]{kama2015DiscDepletionLinkinYSOs}. However, for post-AGB/post-RGB targets hosting transition discs, our calculated values of $\Delta_{\rm g/d}$ lie in the range of higher values (700--20\,000; see Table~\ref{tab:fnlabu_paper2}). This notable increase of the depletion strength $\Delta_{\rm g/d}$ in our transition disc targets points at an increased dust depletion efficiency, which may hint at a more effective dust fractionation in the inner circumbinary disc. An alternative explanation could be the dilution efficiency being higher in post-AGB/post-RGB stars due to their smaller atmospheres \citep{vanwinckel2003Review}. However, to solidify our qualitative comparison, there is a clear need to model discs around post-AGB/post-RGB binaries incorporating more sophisticated disc dynamics and more accurate stellar luminosities. Moreover, the assumed chemical composition of the re-accreted matter should be revised taking into account observed patterns, as demonstrated in the present study.

\subsection{Parallels with chemical depletion in ISM}\label{ssec:dplism_paper2}
The ISM is enriched by various sources, including stellar mass-loss, star formation, and supernovae  \citep{zhukovska2008ISMCompositionFromStars, bierbaum2011ISMcomposition, hofner2018MassLossAGB, saintonge2022ISMSourcesOfEnrichment}. The gas and dust in the ISM have different chemical compositions, and this difference, known as depletion, is studied by measuring ion column densities in the gas phase \citep{jenkins2009ISMdepletion}. The gas-phase ISM abundances can vary due to factors like differences in star formation, in nucleosynthetic history, or in the condensation of metals into dust grains \citep{decia2016ISMdepletion, konstantopoulou2022ISMdepletion}. Distinguishing between these factors, especially at low metallicities, is crucial for studying depletion in the ISM \citep{jenkins2014DustDepletionInISM}.

\begin{figure}[!ht]
    \centering
    \includegraphics[width=.45\linewidth]{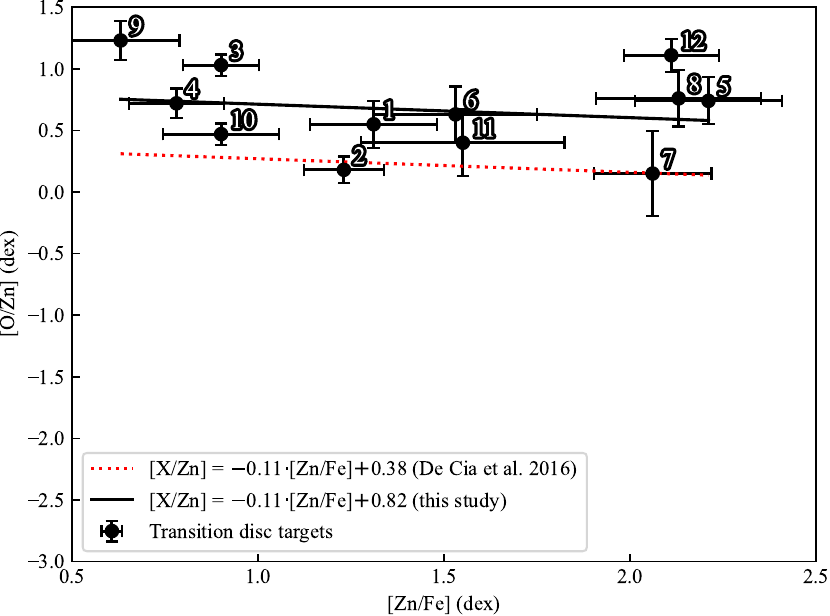}
    \includegraphics[width=.45\linewidth]{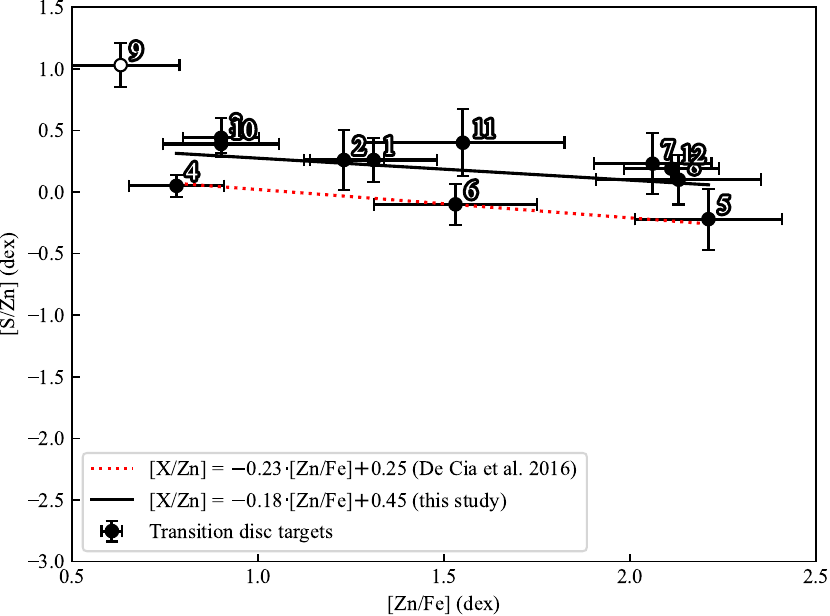}
    \includegraphics[width=.45\linewidth]{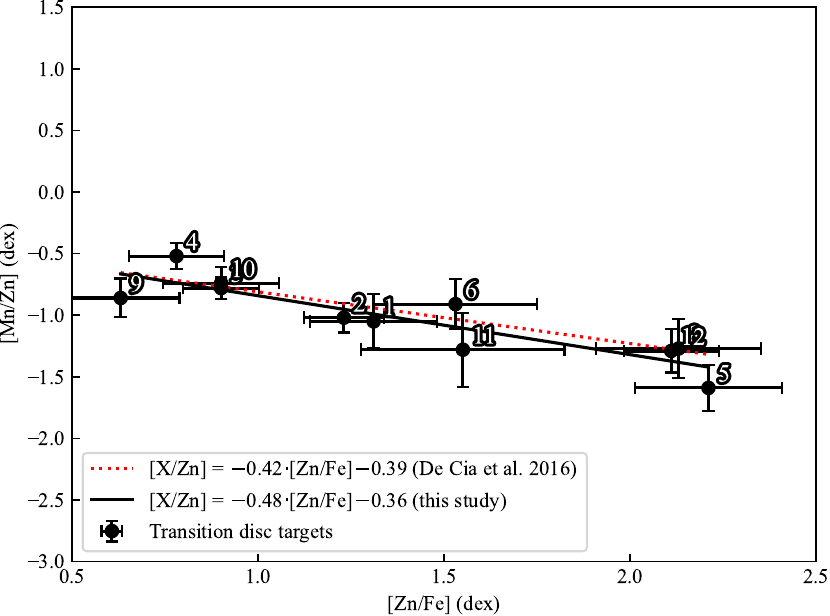}
    \includegraphics[width=.45\linewidth]{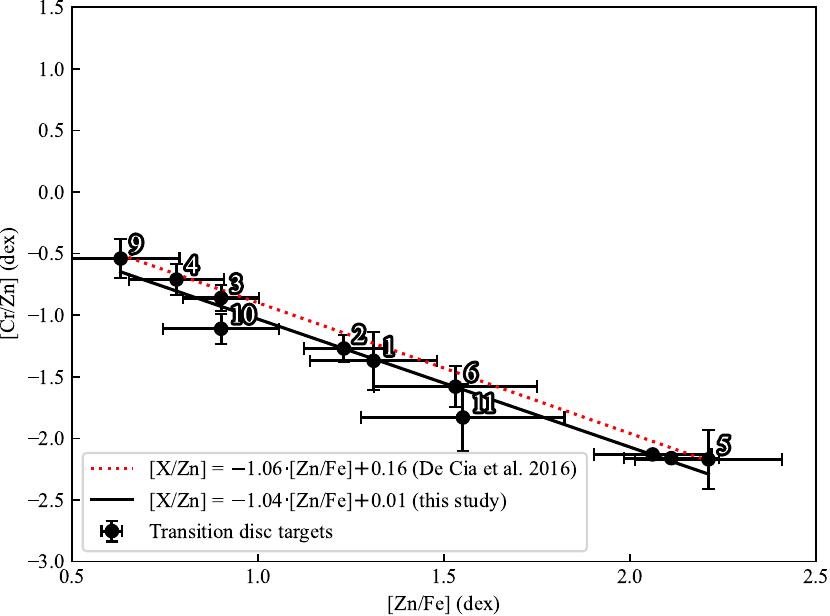}
    \includegraphics[width=.45\linewidth]{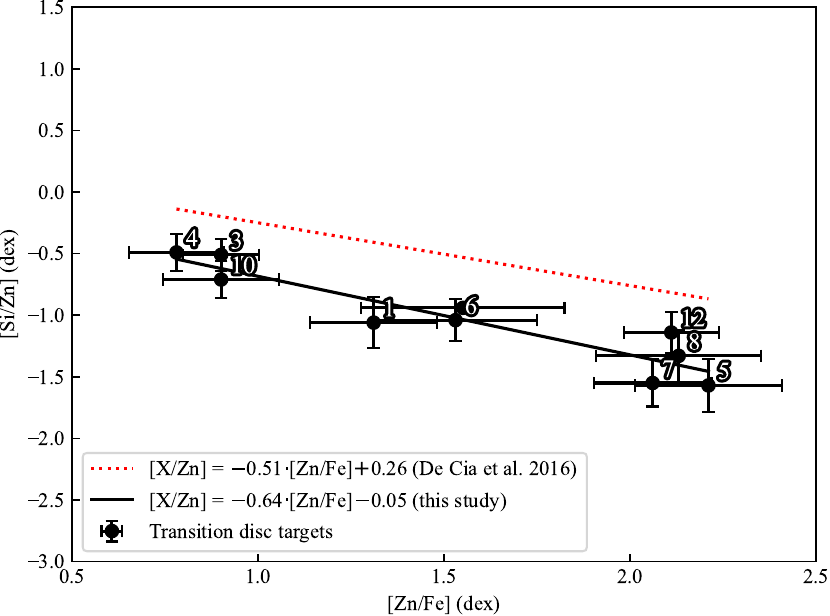}
    \includegraphics[width=.45\linewidth]{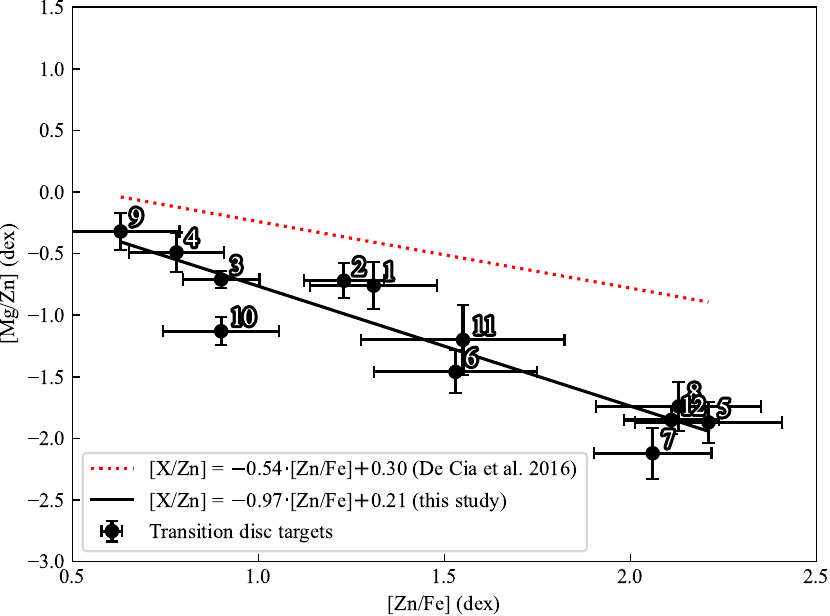}
    \caption[Comparison of {[X/Zn]} ratio trends for \ion{O}{i}, \ion{S}{i}, \ion{Mn}{i}, \ion{Cr}{i}, \ion{Si}{ii}, and \ion{Mg}{i} between transition disc targets and the ISM \citep{decia2016ISMdepletion}]{Comparison of [X/Zn] ratio trends for \ion{O}{i}, \ion{S}{i}, \ion{Mn}{i}, \ion{Cr}{i}, \ion{Si}{ii}, and \ion{Mg}{i} between transition disc targets and the ISM \citep{decia2016ISMdepletion}. [Zn/Fe] was calculated based on abundances from \ion{Zn}{i} and \ion{Fe}{ii} spectral lines. The legend for the symbols and colours used is included within the plot. We note that for [S/Zn] subplot we excluded GZ Nor (\#9) from fitting because of the unique depletion profile of this target (see Fig.~\ref{fig:dplcnd_paper2}). We also note that for [Cr/Zn] ratio of GZ Nor (\#9) we used the abundance derived from \ion{Cr}{ii} lines, while for [Si/Zn] ratio of LMC V0770 (\#11) we used the abundance derived from \ion{Si}{i} lines (for more details, see Section~\ref{ssec:dplism_paper2}).}\label{fig:cmpism_paper2}
\end{figure}

A homogeneous research on depletion across various ISM environments, from the Galaxy to damped Ly-$\alpha$ absorbers, was conducted by \citet{decia2016ISMdepletion}. In their study, the ISM sites were distinguished not by the location, but by [Zn/Fe] abundance ratio: their pointings in the Galaxy and in the Magellanic Clouds occupied the region of [Zn/Fe]>0.5 dex, while their pointings towards damped Ly-$\alpha$ absorbers covered the region of [Zn/Fe]<1 dex. In Fig.~\ref{fig:cmpism_paper2}, the ISM trends of [X/Zn] abundance ratios for O, S, Mn, Cr, Si, and Mg are denoted with dotted red lines and the corresponding abundance ratios in transition disc targets are shown in black circles and are fitted with black lines. We note that we used NLTE abundances of O, S, and Mg, conditionally NLTE abundances of Cr and Fe (derived from spectral lines of \ion{Cr}{ii} and \ion{Fe}{ii}, respectively), and LTE abundances of Mn, Si, and Zn.

We found that the variation in depletion efficiencies of different chemical elements between our transition disc targets and the ISM depends on the volatility of these elements (traced by the corresponding condensation temperatures):

\begin{enumerate}
    \item O ($T_{\rm cond}$\,=\,183 K; highly volatile): The [O/Zn] ratio in transition disc targets shows a similar slope as in the ISM \citep{decia2016ISMdepletion}, but is enhanced by $\sim0.4$ dex. However, when compared with the updated trends of the extended ISM sample \citep[see Appendix~B2 in][]{konstantopoulou2022ISMdepletion}, the agreement becomes satisfactory.
    \item S ($T_{\rm cond}$\,=\,672 K; moderately volatile): The [S/Zn] trend in our data is similar to the one observed in the ISM with a slight enhancement by $\sim$0.2 dex. However, when distant Ly-$\alpha$ absorbers are excluded (leaving only lines of sight in the Galaxy and the LMC), the agreement becomes satisfactory \citep[see Fig.~1 in][]{konstantopoulou2022ISMdepletion}. We highlight that despite being highly volatile, S may be depleted into dust grains to unknown extent, similarly to interstellar S in the Galaxy \citep{jenkins2009ISMdepletion} and towards the damped Ly-$\alpha$ absorbers \citep{decia2016ISMdepletion}.
    \item Mn ($T_{\rm cond}$\,=\,1123 K; moderately volatile): The patterns of [Mn/Zn] ratios match in our target sample and in the ISM. After accounting for the LTE underabundance of Mn in F and G stars within the solar neighbourhood \citep{battistini2015MnNucleosynthesis}, the slope of [Mn/Zn] vs [Fe/Zn] approaches unity \citep[see Table~3 in][]{decia2016ISMdepletion}.
    \item Cr ($T_{\rm cond}$\,=\,1291 K; moderately refractory): The [Cr/Zn] trends also match in our target sample and in the ISM. The Cr and Fe abundance ratios are identical within the error bars ([Cr/Zn]$\sim$1.04$\times$[Fe/Zn]) in the transition disc targets and in the ISM.
    \item Si ($T_{\rm cond}\,=\,1314$ K; moderately refractory): In evolved binaries, the slope of [Si/Zn] trend is --0.64, which is lower than the corresponding slope in the ISM. Since Si and Fe have similar condensation temperatures ($\Delta T_{\rm cond}\,=\,24$ K) and solar abundances ($\sim7.5$ dex), their absolute abundances $\log\varepsilon$(Si) and $\log\varepsilon$(Fe)\footnote{Absolute abundance of an element X is given by: $\log\varepsilon(X)\,=\,\log\frac{N(X)}{N(H)}\,=\,[{\rm X/H}] + \log\varepsilon_\odot$(X).} in transition disc targets show a 2:3 ratio \citep[Si and Fe abundances in the ISM show a 1:2 ratio;][]{decia2016ISMdepletion}.
    \item Mg ($T_{\rm cond}\,=\,1343$ K; moderately refractory): The [Mg/Zn] trend in our targets shows a steeper slope than in the ISM (--0.97 and --0.54, respectively). Given the close proximity of condensation temperatures ($\Delta T_{\rm cond}\,=\,5$ K) and solar abundances of Mg and Fe ($\sim7.5$ dex), their absolute abundances $\log\varepsilon$(Mg) and $\log\varepsilon$(Fe) in transition disc targets show 1:1 abundance ratio.
\end{enumerate}

To summarise, the observed abundance ratios in the transition disc post-AGB/post-RGB targets and in the ISM generally display similar trends for both volatile and refractory elements (O, S, Mn, and Cr). This may be attributed to the fact that all our sample targets are O-rich (C/O<1, see Table~\ref{tab:fnlabu_paper2}), similar to the O-rich ISM environments studied by \citet{decia2016ISMdepletion}. However, we detected lower slopes in the abundance trends of Si and Mg in the transition disc targets relative to the ISM.

As mentioned above, the [X/Zn] trends represent both the pure depletion and the nucleosynthetic over- or underabundance of an element X. To remove the nucleosynthetic effects, \citet{decia2016ISMdepletion} converted the Galactic abundance patterns from \citet{mcwilliam1997GalacticChemicalEvolution} to [Zn/Fe] scale and provided corrections for the slopes of [Cr/Zn], [Si/Zn], and [Mg/Zn]. Applying these corrections, we obtained [Si/Zn]$_{\rm corr}$ $\sim$ 0.76 $\times$ [Fe/Zn]$_{\rm corr}$, and [Cr/Zn]$_{\rm corr}$ $\sim$ [Mg/Zn]$_{\rm corr}$ $\sim$ 1.04 $\times$ [Fe/Zn]$_{\rm corr}$. Since the solar abundances of Cr and Fe differ by $\sim2$ dex, the slope of Cr trend points at independence of the depletion efficiency of an element from its absolute abundance in post-AGB/post-RGB binaries with transition discs. The corrected slopes of Si and Mg trends in post-AGB/post-RGB binaries with transition discs point to Si, Mg, and Fe being depleted in the stellar surface with a number abundance ratio of 0.76:1.04:1, respectively.

Previous studies of circumbinary discs around post-AGB/post-RGB binaries detected Mg-rich end members of olivines (forsterite) and pyroxenes (enstatite) and did not detect Fe-rich dust grains \citep{gielen2008SPITZERsurvey, gielen2009Depletion, gielen2010corrigendum, gielen2011silicates, hillen2015ACHerMinerals}. Despite this, our measurements of Fe depletion in transition disc targets suggest the existence of Fe grains, such as Fe alloys, Fe oxides, fayalite, or ferrosilite. This highlights the uniqueness of the depletion profile analysis in offering an independent insight into the dust composition in the transition discs around post-AGB/post-RGB binary stars. We note that precise and consistent modelling of infrared spectral features in transition disc targets is beyond the scope of this study, though combining our results with mid-infrared observations from current mid-infrared facilities like MIRI/JWST is a promising avenue for the future research.

\section{Conclusions}\label{sec:con_paper2}
In this study, we aimed to investigate the mechanisms responsible for depletion, properties of the depletion patterns, and correlations to other observational parameters in post-AGB/post-RGB binary stars with transition discs. In our analysis, we used high-resolution optical spectra from HERMES/Mercator and UVES/VLT.

Using E-iSpec, we performed a detailed chemical abundance analysis of six confirmed transition disc stars and six transition disc candidates in the Galaxy and in the LMC. In addition to the derived LTE abundances, we used \texttt{Balder} software to calculate the NLTE corrections for a representative subset of chemical elements from C to Fe. We found that these corrections significantly affect the surface C/O ratios of transition disc targets and modify the depletion patterns. The resulting NLTE-corrected depletion profiles are saturated, meaning that the current surface abundances of all transition disc targets trace the composition of the re-accreted gas, rather than the original photospheric material. Moreover, we confirmed that depletion efficiency in transition disc systems, as traced by [S/Ti] abundance ratio, is higher than in other post-AGB/post-RGB binary stars.

In addition, we explored correlations between the derived abundances and various observational parameters associated with the binary system (e.g., astrometric, photometric, orbital, pulsational). For transition disc targets, the detected correlations generally align with those reported in the literature for the broader sample of post-AGB/post-RGB binary stars. Notably, we identified a moderate correlation between stellar luminosity and the normalised [O/S] abundance ratio. This hints at the turn-off temperature $T_{\rm turn-off}$ in transition disc targets being typically lower for depletion profiles in post-AGB binaries than in post-RGB binaries.

We also investigated the links between chemical depletion in our target sample and other environments (around YSOs with transition discs and in the ISM). We confirmed that the homogeneously derived depletion strength ($\Delta_{\rm g/d,~pAGB/pRGB}\,=\,700-20000$) in post-AGB/post-RGB binaries with transition discs is significantly higher than the values observed in young stars with transition discs ($\Delta_{\rm g/d,~YSO} < 1\,000$). However, the depletion patterns in post-AGB/post-RGB binaries with transition discs, as traced by the [X/Zn] vs. [Zn/Fe] slopes, resemble those seen in the ISM for both volatile elements (O, S, Mn) and refractory elements (Cr, Fe). We also found that the refractory elements Si and Mg deviate from this trend, indirectly offering a rough estimate of the amount of unobservable Fe dust grains in the transition discs around post-AGB/post-RGB binaries. It is important to note that these findings are based on a limited sample size. Our ongoing analysis of optical spectra of post-AGB/post-RGB binaries in the Galaxy, SMC and LMC aims to expand the target sample to enable a more statistically robust investigation, allowing for further validation and confirmation of current results.

\section*{Acknowledgements}\label{sec:ack_paper2}
The spectroscopic results presented in this paper are based on observations made with the Mercator Telescope, operated on the island of La Palma by the Flemish Community, at the Spanish Observatorio del Roque de los Muchachos of the Instituto de Astrofisica de Canarias. This research is based on observations collected at the European Organisation for Astronomical Research in the Southern Hemisphere under ESO programmes 074.D-0619 and 092.D-0485. This research was supported by computational resources provided by the Australian Government through the National Computational Infrastructure (NCI) under the National Computational Merit Allocation Scheme and the ANU Merit Allocation Scheme (project y89).

MM1 acknowledges the International Macquarie Research Excellence Scholarship (iMQRES) program for the financial support during the research. MM1, DK, and MM2 acknowledge the ARC Centre of Excellence for All Sky Astrophysics in 3 Dimensions (ASTRO 3D), through project CE170100013. AMA acknowledges support from the Swedish Research Council (VR 2020-03940) and from the Crafoord Foundation via the Royal Swedish Academy of Sciences (CR 2024-0015).
\begin{savequote}[75mm]
\foreignlanguage{ukrainian}{
``Хто думає про науку, той любить її, а хто її любить, той ніколи не перестає вчитися, хоча б зовні він і здавався бездіяльним...''}
\qauthor{\foreignlanguage{ukrainian}{---Григорій Сковорода (1722-1794), \\український філософ, богослов, поет та педагог}}
``He who thinks about science loves it, and he who loves it never stops learning, even if outwardly he seems inactive...''
\qauthor{---Hryhorii Skovoroda (1722-1794), \\Ukrainian philosopher, theologist, poet, and teacher}
\end{savequote}

\chapter{Photospheric depletion in post-AGB/post-RGB binaries \texorpdfstring{\\with dust-poor discs}{}}\label{chp:pap3}
\graphicspath{{ch_paper3/figures_paper3/}} 

\clearpage

\textit{This chapter was prepared for publication as:}\\
\textbf{Abundance Analysis of Chemically Depleted Post-AGB/Post-RGB Binaries with Dust-Poor Discs}\\
Mohorian M., Kamath D., Menon M., Van Winckel H., Jian M., Amarsi A.~M.\\
\textit{To be submitted to PASA}\\
\\
\textit{In this chapter, we aim to investigate the evolutionary status of circumbinary discs around post-AGB/post-RGB stars using photospheric depletion as an indirect tracer. To achieve this, we analyse the spectra of 9 post-AGB/post-RGB binaries with dust-poor discs. This subset, which includes both mildly and strongly depleted systems, provides a critical opportunity to assess the evolutionary trajectory of dust-poor discs and their potential connection to the full disc and transition disc systems explored in previous chapters. By placing these findings within the broader context of disc evolution, this analysis provides new insights into the mechanisms driving photospheric chemical depletion in post-AGB and post-RGB binaries, thereby advancing our understanding of their complex evolutionary pathways.}\\
\\
\textbf{\textit{Author Contributions}}\\
M.\,Mohorian was primarily responsible for data reduction, analysis, and interpretation of the results presented in this chapter. D.\,Kamath provided feedback and engaged in regular discussions throughout each stage of the research process. M.\,Menon contributed to development of the \texttt{E-iSpec}. H.\,Van\,Winckel is the principal investigator of the proposals that resulted in the data used in this study. M.\,Jian and A.~M.\,Amarsi contributed to development of the NLTE analysis methodology. The chapter was written by M.\,Mohorian, with all co-authors providing feedback and comments.

\begin{tcolorbox}[colback=blue!10!white, boxrule=0mm]
    \section*{Original paper abstract}
    Post-AGB and post-RGB binaries, often surrounded by circumbinary discs of gas and dust, offer a critical window into the late stages of stellar and disc evolution. These systems are key to understanding how binary interactions shape disc structure, dynamics, and evolution. With an increasing number of identified post-AGB and post-RGB binaries, we are now uniquely positioned to investigate the mechanisms of disc formation and disc–binary interaction, exploring their implications for the photospheric chemical composition of these stars. In this study, we investigate the interactions in post-AGB/post-RGB binaries with dust-poor circumbinary discs, which exhibit depletion levels ranging from mild to strong. We analyse high-resolution optical spectra from HERMES/Mercator and UVES/VLT for 9 dust-poor disc targets in the Galaxy and the LMC. We use \texttt{E-iSpec} for a homogeneous derivation of atmospheric parameters and elemental abundances and apply \texttt{pySME} to calculate the NLTE corrections for key elements from carbon (C) to barium (Ba). Our results show that all targets exhibit saturated depletion profiles. We fitted these profiles using 2-piece linear functions with three free parameters: initial metallicity ([M/H]$_0$), turn-off temperature ($T_{\rm turn-off}$), and depletion scale ($\nabla_{\rm 100\,K}$). By analysing depletion trends in the dust-poor disc sample, we find that while the distributions of [M/H]$_0$ and $\nabla_{\rm 100\,K}$ are random, the $T_{\rm turn-off}$ distribution appears bimodal. To better understand this bimodality, we incorporate a combined sample of post-AGB/post-RGB binaries with full and transition discs from previous studies. Although our sample size is limited and may be subject to observational biases, trends suggest that dust-poor disc targets could be categorized into two subgroups: full-like dust-poor disc targets with $T_{\rm turn-off}\,>\,1\,100$\,K and transition-like dust-poor disc targets with $T_{\rm turn-off}\,<\,1\,100$\,K. The observed trends suggest that full and transition discs around post-AGB/post-RGB binaries may evolve into dust-poor discs over time, pointing at possibly different disc-formation mechanisms occurring in full and transition discs, which merits further confirmation with larger samples.
\end{tcolorbox}    

\section{Introduction}\label{sec:int_paper3}
Post-asymptotic giant branch (post-AGB) and post-red giant branch (post-RGB) binaries with circumbinary discs (CBDs) provide a unique window into disc-binary interactions and their influence on stellar surface chemistry. Toward the end of the AGB phase, low- and intermediate-mass stars ($0.8\,-\,8\,M_\odot$) lose most of their envelope mass through stellar winds \citep{vanwinckel2003Review, kobayashi2020OriginOfElements, kamath2023models}. In certain binary systems where the companion is typically a main-sequence star, with a mass distribution peaking at $M \sim 1.09\,M_\odot$ \citep{oomen2018OrbitalParameters}, mass loss—enhanced by poorly understood binary interactions—can prematurely terminate the AGB or RGB phase \citep{paczynski1971RLOF, vanwinckel2003Review}. This process leads to the formation of a post-AGB or post-RGB star and the surrounding CBD \citep{vanwinckel2003Review, kamath2016PostRGBDiscovery, vanwinckel2018Binaries}.

The presence of CBDs around post-AGB and post-RGB binaries has been observationally confirmed through the detection of dust excess, which manifests as a broad near-infrared (near-IR) excess in their spectral energy distributions (SEDs) \citep{deruyter2005discs, kamath2014SMC, kamath2015LMC, kamath2016PostRGBDiscovery, gezer2015WISERVTau, kluska2022GalacticBinaries}. Detailed imaging studies have revealed complex dust structures within these discs, including cavities, rings, and arc-like features, through interferometric \citep{kluska2019DiscSurvey, corporaal2023DiscParameters} and polarimetric techniques \citep{ertel2019Imaging, andrych2023Polarimetry, andrych2024IRAS08}. Additionally, the outer regions of CBDs exhibit signatures of dust crystallisation \citep{gielen2011silicates, hillen2015ACHerMinerals}, grain growth \citep{scicluna2020GrainGrowth}, and Keplerian rotation \citep{bujarrabal2015KeplerianRotation, gallardocava202389HerNebula}. 

\citet{kluska2022GalacticBinaries} classified CBDs into three types: i) full discs, where dust and gas are well mixed, ii) transition discs, which display inner dust gaps or cavities, and iii) dust-poor discs, where the SED lacks significant IR excess. \footnote{The absence of IR excess may also occur in gas-poor discs where dust has settled in the midplane rather than being genuinely dust-poor.}

The interactions between CBDs and their central binaries remain poorly understood \citep{heath2020DiscBinaryInteraction, penzlin2022DiscBinaryInteraction}. However, observational evidence reveals key signatures of these interactions, including jet formation and photospheric chemical depletion. Jets are linked to the interaction between the disc and the companion star \citep{bollen2022Jets, verhamme2024DiscWindModelling, deprins2024Jets}, while depletion results from the interaction between the disc and the post-AGB/post-RGB primary \citep{oomen2019depletion, oomen2020MESAdepletion}. 

In this study, we focus on photospheric chemical depletion, which occurs when gas in the CBD becomes fractionated from dust and is subsequently re-accreted onto the post-AGB/post-RGB star, altering its observed surface composition \citep{mosta2019ReaccretionInnerRim, munoz2019ReaccretionInnerRim, oomen2019depletion, oomen2018OrbitalParameters}. The efficiency of gas-dust fractionation depends on the condensation properties of the gas mixture, commonly described by 50\% condensation temperature $T_{\rm cond,\,50\%}$\footnote{At 50\% condensation temperature for a selected element, half of the element by abundance is in the gas phase, while another half is in the dust phase. The condensation temperatures are commonly calculated for solar mixture gas in chemical equilibrium at pressure $P\,=\,10^{-4}$\,bar.} \citep[$T_{\rm cond}$;][]{lodders2003CondensationTemperatures, wood2019CondensationTemperatures, agundez2020AGBDepletionModelling}. The fractionation process in the CBD separates gas containing elements with condensation temperatures $T_{\rm cond}\,\lesssim\,1\,250$\,K (volatile elements, including Na, S, Cu, and Zn) and dust containing elements with condensation temperatures $T_{\rm cond}\,\gtrsim\,1\,250$\,K (refractory elements, including Al, Si, Ti, Fe, and \textit{s}-process elements). Consequently, re-accreted volatile-rich matter dominates the original stellar surface composition, completely masking the nucleosynthetic products of AGB/RGB evolution \citep{deruyter2005discs, deruyter2006discs, oomen2018OrbitalParameters}. The resulting photospheric underabundance (depletion) of refractory elements in post-AGB/post-RGB binaries is particularly prominent in the relative [X/H] abundance scale and is commonly plotted against $T_{\rm cond}$ \citep[depletion profile;][]{deruyter2005discs, maas2005DiscPAGBs, oomen2019depletion}.

The photospheric depletion in post-AGB/post-RGB binaries is commonly explored using four key parameters \citep{vanwinckel2018Binaries, kluska2022GalacticBinaries}:
\begin{itemize}
    \item The turn-off temperature $T_{\rm turn-off}$, which marks the separation between weakly depleted and significantly depleted elements. In the previous studies, this parameter was estimated visually \citep{oomen2019depletion, kluska2022GalacticBinaries}.
    \item The initial metallicity [M/H]$_0$ ([S/H], [Zn/H], or their average, since Fe is depleted in post-AGB/post-RGB binaries), which traces the baseline composition of metals in the stellar surface. We note that volatile elements (including O, S, and Zn) may be mildly depleted \citep{mohorian2024EiSpec, mohorian2025TransitionDiscs}.
    \item The volatile-to-refractory abundance ratio ([Zn/Ti], [Zn/Fe], or [S/Ti]), which provides a scale for the efficiency of depletion \citep{waelkens1991depletion, oomen2019depletion, oomen2020MESAdepletion}. Based on [Zn/Ti] abundance ratios, post-AGB/post-RGB binaries are commonly categorised into three distinct groups: i) mildly depleted ([Zn/Ti]$\,\lesssim\,0.5$\,dex), ii) moderately depleted (0.5\,dex$\,\lesssim\,$[Zn/Ti]$\,\lesssim\,1.5$\,dex), and iii) strongly depleted ([Zn/Ti]$\,\gtrsim\,1.5$\,dex). We note that [Zn/Ti] abundance ratio depends on $T_{\rm turn-off}$ and [M/H]$_0$.
    \item The pattern of high-temperature end of [X/H] depletion profile: i) linear decline of [X/H] with increasing $T_{\rm cond}$ (`saturation') or ii) constant [X/H] with increasing $T_{\rm cond}$ (`plateau'). This difference between saturated and plateau profiles is hypothesised to stem from the timescale of dilution of re-accreted matter in stellar photosphere \citep{oomen2019depletion}.
\end{itemize}

CNO elements (C, nitrogen N, and oxygen O) are commonly omitted from the depletion analyses of post-AGB/post-RGB binaries, since the surface abundances of CNO elements are significantly altered by nucleosynthetic and mixing processes during the AGB/RGB evolution \citep{oomen2019depletion, oomen2020MESAdepletion, mohorian2024EiSpec, menon2024EvolvedBinaries}. However, current stellar evolutionary models predict that surface abundance of O may be unaffected by nucleosynthetic and mixing processes in RGB and low-mass ($M\,<\,2\,M_\odot$) AGB stars \citep{ventura2008aton3, karakas2014dawes, kobayashi2020OriginOfElements, kamath2023models}. We note that luminosity uncertainties of the majority of post-AGB/post-RGB targets do not allow to fully and definitively differentiate post-AGB and post-RGB binaries \citep{kluska2022GalacticBinaries, mohorian2024EiSpec, mohorian2025TransitionDiscs}. Hence, in this study, we exclude CNO elements from the depletion analysis of post-AGB/post-RGB binaries in the Galaxy and in the LMC.

The efficiency of photospheric depletion varies across the types of CBDs (i.e., full, transition, and dust-poor). Full disc targets show the mildest depletion ([Zn/Ti]\,$\lesssim$\,1\,dex), transition disc targets display the strongest depletion ([Zn/Ti]\,$\gtrsim$\,2\,dex), and dust-poor disc targets display a wide range of depletion efficiencies. We note that photospheric depletion is not limited to post-AGB and post-RGB binaries; a similar phenomenon was observed in young planet-hosting stars \citep[$\lambda$ B\"{o}o phenomenon;][]{venn1990lambdaBooStars, andrievsky2002lambdaBooAbundances, jura2015lambdaBoo, murphy2020lambdaBoo}. The depletion profiles in stars showing $\lambda$ B\"{o}o phenomenon suggest a possible connection between depletion mechanisms and the processes of planet formation \citep{kama2015DiscDepletionLinkinYSOs, jermyn2018Depletion}. Similarly, the depletion profiles in post-AGB/post-RGB binaries were theorised to stem from second-generation planet formation in their CBDs \citep{kluska2022GalacticBinaries, mohorian2024EiSpec}.

In this study, we investigate how photospheric depletion is connected to the evolution of CBDs around post-AGB/post-RGB binaries, providing a basis for understanding the nature of dust-poor discs. To achieve this, we analysed high-resolution optical spectra from HERMES/Mercator and UVES/VLT of 9 dust-poor disc targets in the Galaxy and in the LMC. The structure of paper is as follows: In Section~\ref{sec:sdo_paper3}, we describe our target sample, data, and observation details. In Section~\ref{sec:ana_paper3}, we outline the methodology and results of our chemical analysis of dust-poor disc targets. In Section~\ref{sec:dsc_paper3}, we discuss photospheric depletion in dust-poor disc targets and its connection to CBD evolution. Finally, in Section~\ref{sec:con_paper3}, we summarise the key findings of this study.

\section{Sample, data, and observations}\label{sec:sdo_paper3}
In this study, we focus on a subset of post-AGB/post-RGB binary stars with dust-poor discs. In Section~\ref{ssec:sdosam_paper3}, we present the target sample and explain the selection criteria. In Section~\ref{ssec:sdopht_paper3}, we present the photometric data used to derive luminosities of our target sample (using SED fitting and PLC relation; see Section~\ref{ssec:analum_paper3}). In Section~\ref{ssec:sdospc_paper3}, we briefly discuss the spectroscopic data used to derive precise atmospheric parameters and elemental abundances of the dust-poor disc targets (using \texttt{E-iSpec} and \texttt{pySME}; see Section~\ref{ssec:anaspc_paper3}).

\subsection{Sample selection}\label{ssec:sdosam_paper3}
The initial target sample consisted of post-AGB and post-RGB binary stars in the Galaxy \citep{szczerba2007TorunCatalogue, kluska2022GalacticBinaries} and the Magellanic Clouds \citep[from][]{kamath2014SMC, gielen2009Depletion, vanaarle2011PAGBsInLMC, kamath2015LMC}. From this initial sample, we selected dust-poor disc targets based on their IR magnitudes from the 2MASS Long Exposure (6X) Full Survey \citep{cutri20122MASS6X} and the AllWISE catalogue \citep{cutri2014AllWISE}, using the selection criteria $H-K < 0.37^m$ and $W_1 - W_3 < 2.3^m$ adopted from \citet{kluska2022GalacticBinaries}. We further refined our target selection by only retaining objects for which we had high-resolution optical spectra (see Section~\ref{ssec:anaspc_paper3}) and those with spectral types A to K. These spectral types allow for the optimal identification of \ion{Fe}{i} and \ion{Fe}{ii} spectral features. Additionally, we excluded one LMC object, MACHO 47.2496.8 (OGLE LMC-T2CEP-015), due to its observed \textit{s}-process enhancement \citep{menon2024EvolvedBinaries}. This chemical peculiarity suggests it follows a different evolutionary pathway, making it unsuitable for our study. As a result, our final target sample consists of 9 post-AGB/post-RGB binary stars -- 7 in the Galaxy and 2 in the LMC.

\begin{figure*}[ph!]
    \centering
    \includegraphics[width=.914\linewidth]{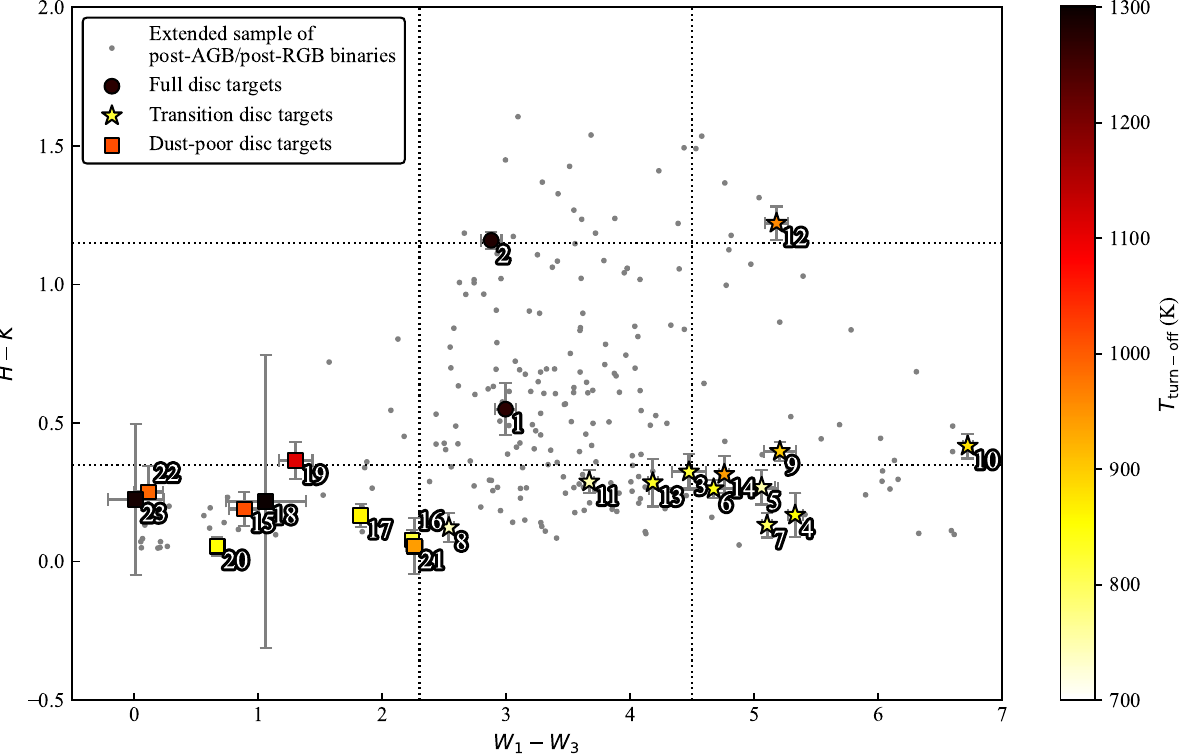}
    \includegraphics[width=.914\linewidth]{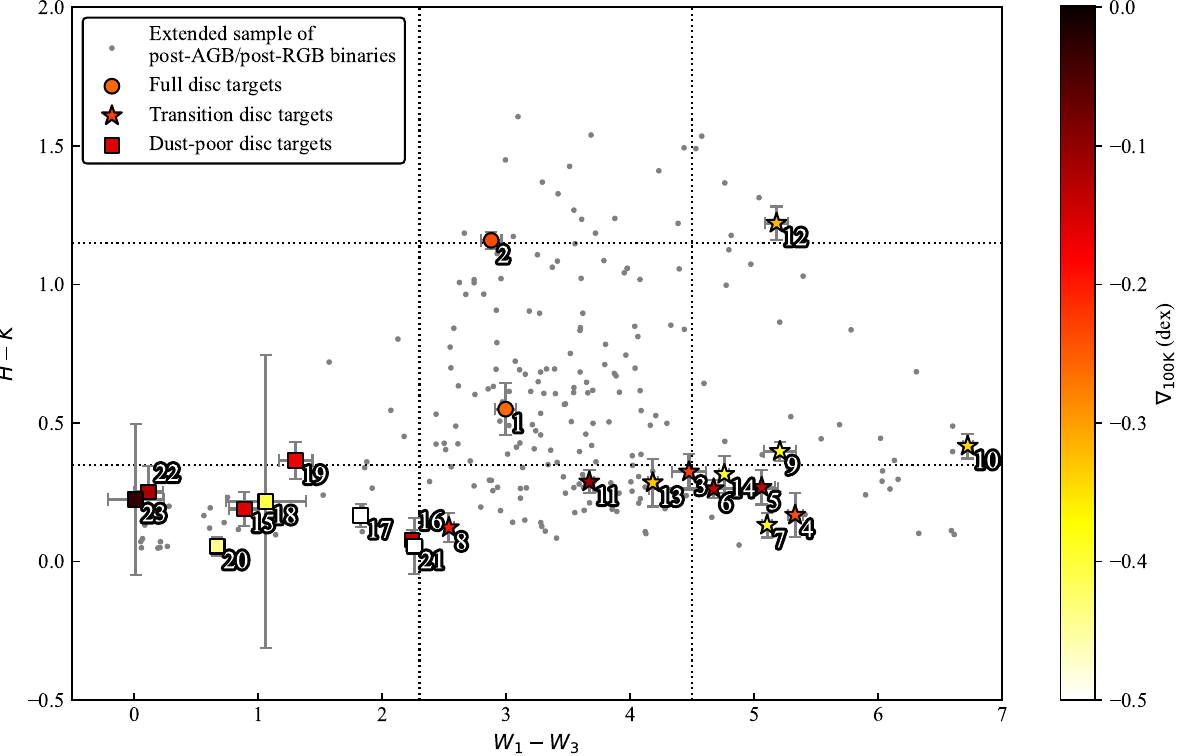}
    \caption[Updated IR colour-colour plot of post-AGB/post-RGB binary stars in the Galaxy and in the Magellanic Clouds]{Updated IR colour-colour plot of post-AGB/post-RGB binary stars in the Galaxy and in the Magellanic Clouds ($H$ and $K$ from 2MASS 6X; $W_1$ and $W_3$ from AllWISE). Grey dots represent the extended sample of disc targets in the Galaxy and the Magellanic Clouds \citep{szczerba2007TorunCatalogue, kamath2014SMC, kamath2015LMC, kluska2022GalacticBinaries}. Markers coloured by turn-off temperature $T_{\rm turn-off}$ (\textit{upper panel}) and depletion scale $\nabla_{\rm 100\,K}$ (\textit{lower panel}) represent the post-AGB/post-RGB binaries studied within this thesis (see Table~\ref{tab:respar_paper3}). Contours representing the demarcation between different disc categories are adopted from \citet{kluska2022GalacticBinaries}.}\label{fig:colplt_paper3}
\end{figure*}

In Table~\ref{tab:sample_paper3}, we present names, coordinates, and IR colours of our final sample of dust-poor disc targets. In Fig.~\ref{fig:colplt_paper3}, we display the IR colours of 9 dust-poor disc targets from this study (marked by squares). In Table~\ref{tab:litpar_paper3}, we provide relevant literature data for our dust-poor disc sample, including orbital and pulsational parameters, luminosities, and depletion parameters. Notably, all targets except BD+39\,4926 and BD+28\,772 are classified as RV\,Tauri variables. For all pulsating targets, we derived luminosities using period-luminosity-colour (PLC) relation (see Section~\ref{ssec:analum_paper3}).

\begin{sidewaystable}[ph!]
    \centering
    \footnotesize
    \caption[General details regarding dust-poor disc sample (names, coordinates, and photometric selection criteria; see Section~\ref{ssec:sdosam_paper3})]{General details regarding dust-poor disc sample (names, coordinates, and photometric selection criteria; see Section~\ref{ssec:sdosam_paper3}). $H$ and $K$ magnitudes originate from the 2MASS 6X catalogue, while the $W_1$ and $W_3$ magnitudes are sourced from the AllWISE catalogue (see Section~\ref{sec:sdo_paper3}).} \label{tab:sample_paper3}
    \begin{tabular}{|c|c|c|c|c|c|c|c|}
    \hline
        \multicolumn{3}{
        |c|}{\textbf{Names}} & \multicolumn{2}{|c|}{\textbf{Coordinates}} & \multicolumn{2}{|c|}{\textbf{Selection criteria}} \\ \hline
        \multirow{2}{*}{\textbf{Adopted}} & \multirow{2}{*}{\textbf{IRAS/OGLE}} & \multirow{2}{*}{\textbf{2MASS}} & \textbf{R.A.} & \textbf{Dec.} & \boldmath$H-K$ & \boldmath$W_1-W_3$ \\
        ~ & ~ & ~ & \textbf{(deg)} & \textbf{(deg)} & \textbf{(mag)} & \textbf{(mag)} \\\hline
        \multicolumn{7}{|c|}{\textit{Galactic dust-poor disc stars}} \\\hline
        SS Gem & 06054+2237 & J06083510+2237020 & 092.146283 & +22.617229 & 0.190$\,\pm\,$0.060 & 0.886$\,\pm\,$0.119 \\
        V382 Aur & 06338+5333 & J06375242+5331020 & 099.468426 & +53.517227 & 0.078$\,\pm\,$0.035 & 2.239$\,\pm\,$0.050 \\
        CC Lyr & - & J18335741+3138241 & 278.489225 & +31.640045 & 0.167$\,\pm\,$0.041 & 1.826$\,\pm\,$0.041 \\
        R Sct & 18448-0545 & J18472894-0542185 & 281.870622 & --05.705157 & 0.218$\,\pm\,$0.528 & 1.061$\,\pm\,$0.324 \\
        AU Vul & 20160+2734 & J20180588+2744035 & 304.524513 & +27.734322 & 0.364$\,\pm\,$0.066 & 1.301$\,\pm\,$0.136 \\
        BD+39 4926 & - & J22461123+4006262 & 341.546798 & +40.107304 & 0.056$\,\pm\,$0.034 & 0.668$\,\pm\,$0.040 \\ \hline
        \multicolumn{7}{|c|}{\textit{LMC dust-poor disc stars}} \\\hline
        J052204 & LMC-LPV-46487 & J05220425-6915206 & 80.517725 & --69.255730 & 0.056$\,\pm\,$0.100 & 2.257$\,\pm\,$0.068 \\
        J053254 & LMC-T2CEP-149 & J05325445-6935131 & 83.226904 & --69.586983 & 0.085$\,\pm\,$0.067 & 0.749$\,\pm\,$0.083$^a$ \\ \hline
        \multicolumn{7}{|c|}{\textit{Galactic dust-poor disc candidate}} \\\hline
        BD+28 772 & 05140+2851 & J05171546+2854180 & 79.314440 & +28.905008 & 0.225$\,\pm\,$0.272 & 0.012$\,\pm\,$0.227 \\ \hline
    \end{tabular}\\
    \textbf{Notes:} $^a$[3.6]-[8.0] from SAGE catalogue \citep{woods2011SAGE} was adopted as MIR colour for J053254. By scaling to WISE wavelengths ($\lambda_{W_1}\,=\,3.368\,\mu$m, $\lambda_{W_3}\,=\,12.082\,\mu$m), we estimate $W_1-W_3\,\sim\,1.483^m$ for this target, which is below the MIR cut-off magnitude of $2.3^m$ (see Section~\ref{ssec:sdosam_paper3}).
\end{sidewaystable}
\begin{sidewaystable}[ph!]
    \centering
    \footnotesize
    \caption{Parameters of dust-poor disc sample, including orbital, pulsational, and depletion parameters, as well as luminosity estimates adopted from the literature (see Section~\ref{sec:sdo_paper3}).} \label{tab:litpar_paper3}
    \begin{tabular}{|c|cc|cc|cc|ccc|}
    \hline
        & \multicolumn{2}{|c|}{\textbf{Orbital parameters}} & \multicolumn{2}{|c|}{\textbf{Pulsational parameters}} & \multicolumn{2}{|c|}{\textbf{Luminosity estimates}} & \multicolumn{3}{|c|}{\textbf{Depletion parameters}} \\
        \textbf{Name} & \boldmath$P_{\rm orb}$ & \boldmath$e$ & \boldmath$P_{\rm puls}$ & \textbf{RVb} & \boldmath$L_{\rm SED}$ & \boldmath$L_{\rm IR}/L_\ast$ & \textbf{[Zn/Ti]} & \boldmath$T_{\rm turn-off}$ & \textbf{Profile} \\
        ~ & \textbf{(d)} & ~ & \textbf{(d)} & ~ & \boldmath$(L_\odot)$ & ~ & \textbf{(dex)} & \textbf{(K)} & \textbf{pattern} \\\hline
        SS Gem & -- & -- & 44.525$^a$ & yes & 4540$\,\pm\,$1190 & 0.01 & 2.02$^g$ & 1100 & S \\
        V382 Aur & 597.4$\,\pm\,$0.2$^b$ & 0.30$\,\pm\,$0.02$^b$ & 29.06$^c$ & yes & 3540$\,\pm\,$810 & 0.02 & 0.87$^h$ & 800 & P \\
        CC Lyr & -- & -- & 23.634$^d$ & no & 650$\,\pm\,$210 & 0.03 & $\gtrsim$3$^i$ & 1000 & S \\
        R Sct & -- & -- & 70.8$^e$ & no & 2830$\,\pm\,$580 & 0.03 & 0.17$^j$ & -- & N \\
        AU Vul & -- & -- & 71.55$^a$ & no & 4220$\,\pm\,$1160 & 0.06 & -- & -- & -- \\
        BD+39 4926 & 871.7$\,\pm\,$0.4$^b$ & 0.024$\,\pm\,$0.006$^b$ & -- & no & 6120$\,\pm\,$1870 & 0.00 & 2.03$^k$ & 1000 & S \\ \hline
        J052204 & -- & -- & 30.172$^f$ & no & 4170$\,\pm\,$970 & -- & ... & -- & -- \\
        J053254 & -- & -- & 42.744$^f$ & no & 3210$\,\pm\,$1210 & -- & 0.6$^l$ & -- & -- \\ \hline
        BD+28 772 & -- & -- & -- & -- & -- & -- & -- & -- & -- \\ \hline
    \end{tabular}\\
    \textbf{Notes:} Pulsational and orbital parameters are adopted from the following studies: $^a$\cite{pawlak2019ASAS}, $^b$\cite{oomen2018OrbitalParameters}, $^c$\cite{hrivnak2008V382Aur}, $^d$\cite{zong2020LAMOST}, $^e$\cite{kalaee2019RSct}, $^f$\cite{soszynski2008OGLE}. Literature luminosity estimates are adopted from \citet{kluska2022GalacticBinaries} for Galactic binaries and from \citet{manick2018PLC}. [Zn/Ti] abundance ratios are adopted from the following studies: $^g$\cite{gonzalez1997CTOri}, $^h$\cite{hrivnak2008V382Aur}, $^i$\cite{maas2007t2cep}, $^j$\cite{giridhar2000RSct}, $^k$\cite{rao2012BD+394926}, $^l$\cite{gielen2009Depletion}. $T_{\rm turn-off}$ and profile patterns are adopted from \cite{oomen2019depletion}: `S' means `saturated', `P' means `plateau', `N' means `not depleted'.\\
\end{sidewaystable}

\subsection{Photometric data}\label{ssec:sdopht_paper3}
To construct SEDs, we compiled photometric magnitudes spanning the optical to far-IR wavelength range (see Table~\ref{tab:phomag_paper3}). This includes $UBVRI$ photometry in the Johnson-Cousins system \citep{johnson1953Filters, cousins1976Filters}; $B_T$ and $V_T$ bands from the Tycho-2 catalogue \citep{hog2000Tycho2}; $I$' band from the SDSS \citep{york2000SDSSphotometry}; $J$, $H$, and $K$ magnitudes from 2MASS \citep{skrutskie20062MASS}; mid-IR $W_1$ to $W_4$ bands from WISE \citep{wright2010WISE}, and far-IR fluxes from AKARI, IRAS, PACS, and SPIRE \citep{ishihara2010AKARI, neugebauer1984IRAS, poglitsch2010PACS, griffin2010SPIRE}. For BD+28\,772, UV magnitudes were adopted from the TD1 catalogue \citep[156.5-274 nm;][]{thompson1978TD1catalogue}.

\begin{sidewaystable}[ph!]
    \centering
    \footnotesize
    \caption[Photometric data of dust-poor disc sample (see Section~\ref{ssec:sdopht_paper3})]{Photometric data of dust-poor disc sample (see Section~\ref{ssec:sdopht_paper3}). The table includes units and central wavelengths (in $\mu$m) for each filter. This table is published in its entirety in the electronic edition of the paper. A portion is shown here for guidance regarding its form and content.}\label{tab:phomag_paper3}
    \begin{tabular}{|c|c|c|c|c|c|}\hline
        \textbf{Filter} & \textbf{TD1.1565 (II/59B/catalog)} & \textbf{TD1.1965 (II/59B/catalog)} & ... & \textbf{SPIRE.350 (SPIRE350)} & \textbf{SPIRE.500 (SPIRE500)} \\
        \textbf{Unit} & \textbf{cW/m2/nm} & \textbf{cW/m2/nm} & ~ & \textbf{Jy} & \textbf{Jy} \\
        \boldmath$\lambda$ \textbf{(\boldmath$\mu$m)} & \textbf{0.156} & \textbf{0.196} & ... & \textbf{348.438} & \textbf{500.412} \\ \hline
        SS Gem & -- & -- & ... & -- & -- \\
        V382 Aur & -- & -- & ... & 0.042$\pm$0.003 & 0.018$\pm$0.003 \\
        CC Lyr & -- & -- & ... & -- & -- \\
        R Sct & -- & -- & ... & -- & -- \\
        AU Vul & -- & -- & ... & -- & -- \\
        BD+39 4926 & -- & -- & ... & -- & -- \\
        ... & ... & ... & ... & ... & ... \\ \hline
    \end{tabular}
\end{sidewaystable}

\subsection{Spectroscopic data}\label{ssec:sdospc_paper3}
In this subsection, we present the high-resolution optical spectra used in this study, obtained with HERMES/Mercator and UVES/VLT. In Table~\ref{tab:obslog_paper3}, we present the observational log with measured radial velocities and periodicities in the spectra. To determine the radial velocities for all spectral visits of each dust-poor disc target, we used E-iSpec (see Section~\ref{ssec:anaspc_paper3}). We then analysed the resulting radial velocity curves to identify periodic variations. First, we applied the Lomb-Scargle periodogram to detect periodic signals \citep{lomb1976Periodogram, scargle1982Periodogram}. Then we refined derived periods using the Lafler-Kinman method to improve the precision of the derived periodicities \citep{lafler1965Periodogram}.

To derive precise atmospheric parameters and elemental abundances for our dust-poor disc targets, we selected spectra in the following way (see Table~\ref{tab:obslog_paper3}):
\begin{itemize}
    \item For targets observed with HERMES, we selected the optical visits with the highest S/N ratios for each target.
    \item For targets observed with UVES, we combined all available spectral visits.
\end{itemize}

\begin{figure*}[!ht]
    \centering
    \includegraphics[width=.99\linewidth]{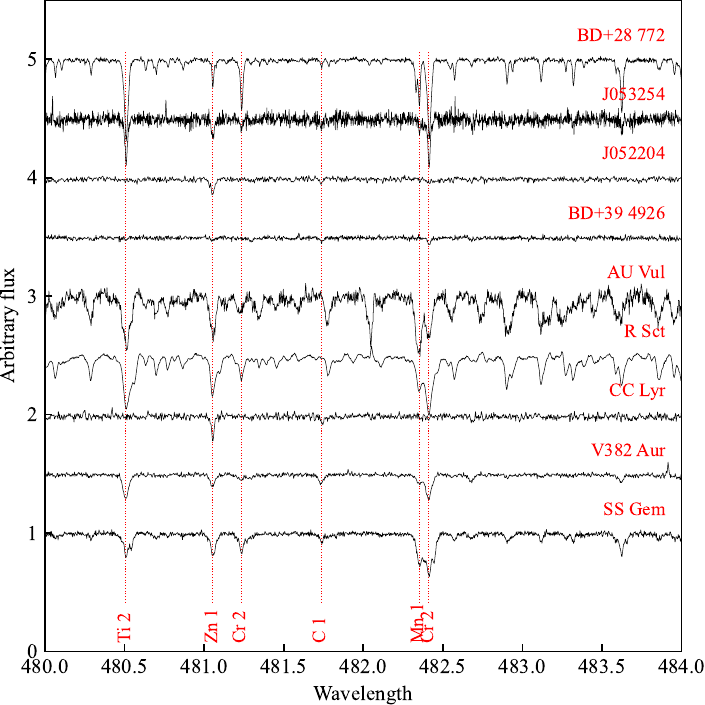}
    \caption[Spectra of all dust-poor disc targets across the region featuring lines from the volatile S and Zn]{Spectra of all dust-poor disc targets across the region featuring lines from the volatile S and Zn. Each spectrum is RV corrected, normalised, and offset in flux for clarity. Red dashed vertical lines mark the positions of spectral line peaks (for more details, see Section~\ref{ssec:anaspc_paper3}).}\label{fig:spcmon_paper3}
\end{figure*}

In Appendix~\ref{app:vis_paper3}, we provide details of all the spectral visits considered. In Fig.~\ref{fig:spcmon_paper3}, we display the sample spectra of dust-poor disc targets in the wavelength region near S and Zn lines to illustrate the quality of our obtained data. We note that [S/H] and [Zn/H] abundances are commonly used to represent the initial metallicity [M/H]$_0$ of post-AGB/post-RGB binaries, rather than [Fe/H] abundance, due to high efficiency of refractory depletion (see Section~\ref{sec:int_paper3}).

\begin{table}[ht]
    \centering
    \footnotesize
    \caption[Analysed spectral visits of dust-poor disc targets]{Analysed spectral visits of dust-poor disc targets. For more details on the criteria used for selecting spectral visits, see Section~\ref{ssec:sdospc_paper3}. For a complete observational summary of dust-poor disc targets, see Appendix~\ref{app:vis_paper3}.} \label{tab:obslog_paper3}
    \begin{tabular}{|c|c|c|c|c|c|c|}
    \hline
        \textbf{Name} & \textbf{Facility} & \textbf{ObsID} & \textbf{MJD} & \textbf{RV (km/s)} & \textbf{S/N} & \boldmath$P_{\rm RV}$ \textbf{(d)} \\ \hline
        SS Gem & H/M & \begin{tabular}{c} 273787,\\273788,\\273789 \end{tabular} & \begin{tabular}{c} 55232.98736,\\55232.99957,\\55233.01175 \end{tabular} & --15.8 & 45 & 44.66 \\ \hline
        V382 Aur & H/M & \begin{tabular}{c} 904077,\\904090 \end{tabular} & \begin{tabular}{c} 58450.08702,\\58450.20395 \end{tabular} & --84.6 & 65 & 599.1 \\ \hline
        CC Lyr & H/M & \begin{tabular}{c} 356745,\\356746,\\356747 \end{tabular} & \begin{tabular}{c} 55723.94027,\\55723.96863,\\55723.99005 \end{tabular} & --35.8 & 55 & 23.69 \\ \hline
        R Sct & H/M & 972206 & 59073.92861 & 36.2 & 35 & 71.54 \\ \hline
        AU Vul & H/M & \begin{tabular}{c} 596805,\\596806 \end{tabular} & \begin{tabular}{c} 56935.86608,\\56935.88750 \end{tabular} & --2.1 & 35 & 75.84 \\ \hline
        BD+39 4926 & H/M & \begin{tabular}{c} 244722,\\244723 \end{tabular} & \begin{tabular}{c} 55067.06677,\\55067.08689 \end{tabular} & --15.8 & 80 & 879.0 \\ \hline
        J052204 & U/V & \begin{tabular}{c} 456+472,\\739+776 \end{tabular} & \begin{tabular}{c} 58451.31574,\\58470.18473 \end{tabular} & 231.3 & 45 & -- \\ \hline
        J053254 & U/V & 494 & 53409.20586 & 268.8 & 40 & -- \\ \hline
        BD+28 772 & H/M & 427610 & 56210.06768 & --3.4 & 30 & 1596 \\ \hline
    \end{tabular}\\
    \textbf{Notes:} H/M denotes HERMES/Mercator, and U/V denotes UVES/VLT. RV is the radial velocity of the spectrum used in the analyses (for merged spectral visits, the range of RVs is <0.5 km/s). S/N is the average signal-to-noise ratio of the spectrum. $P_{\rm RV}$ is the period derived from RV curves, initially estimated using a Lomb-Scargle periodogram and refined with the Lafler-Kinman method.
\end{table}

\section{Derivation of luminosities, atmospheric parameters, and elemental abundances of dust-poor disc targets}\label{sec:ana_paper3}
In this section, we discuss the methodology and results of the photometric and spectral analysis of dust-poor disc targets. In Section~\ref{ssec:analum_paper3}, we outline the derivation of luminosities using SED fitting and PLC relation, based on the collected photometric data. In Section~\ref{ssec:anaspc_paper3}, we present the derivation of precise atmospheric parameters and elemental abundances using high-resolution optical spectra. In Section~\ref{ssec:ananlt_paper3}, we describe the calculation of NLTE corrections, which significantly impact the chemical analysis of post-AGB/post-RGB binaries. In Section~\ref{ssec:anares_paper3}, we present the depletion profiles of dust-poor (this study), transition \citep{mohorian2025TransitionDiscs}, and full disc targets \citep{mohorian2024EiSpec}.

\subsection{Luminosity estimation from SED fitting and PLC relation}\label{ssec:analum_paper3}
In this study, we derived the luminosities of our target sample using the following methods: i) SED fitting to calculate SED luminosities $L_{\rm SED}$ and ii) PLC relation to obtain PLC luminosities $L_{\rm PLC}$. In this subsection, we detail the procedures used for each method.

\begin{figure*}[ph!]
    \centering
    \includegraphics[trim=1cm 0 0.5cm 0, width=.49\linewidth]{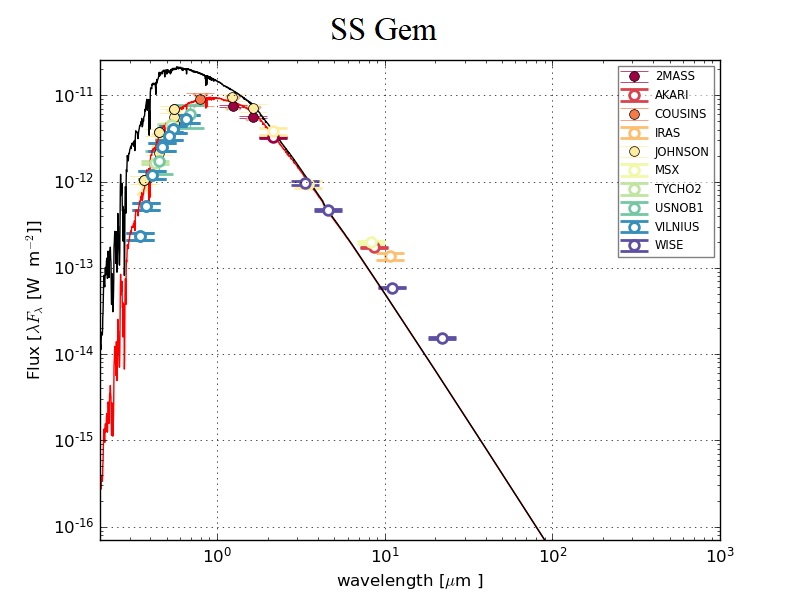}
    \includegraphics[trim=1cm 0 0.5cm 0, width=.49\linewidth]{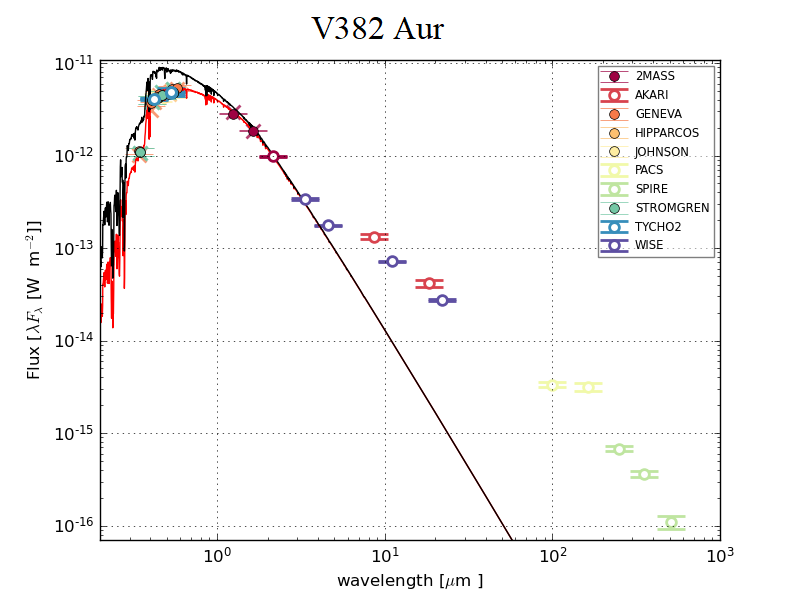}
    \includegraphics[trim=1cm 0 0.5cm 0, width=.49\linewidth]{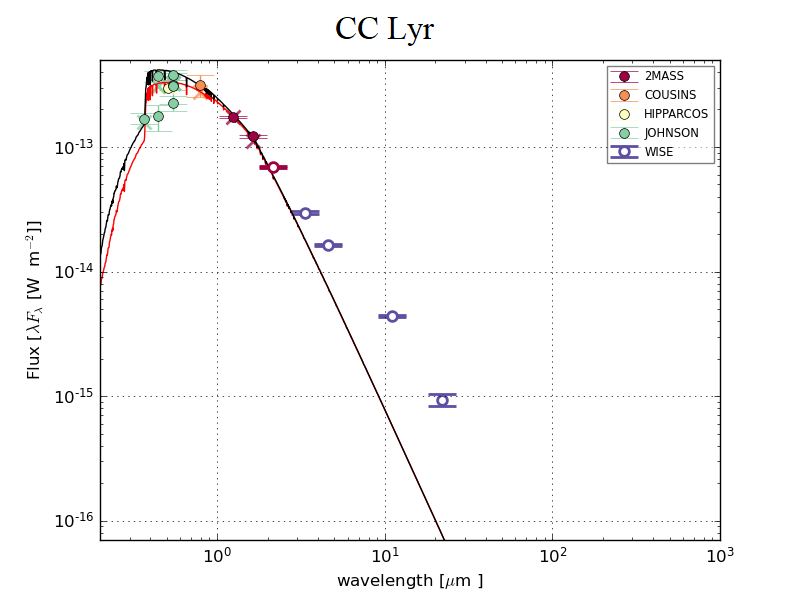}
    \includegraphics[trim=1cm 0 0.5cm 0, width=.49\linewidth]{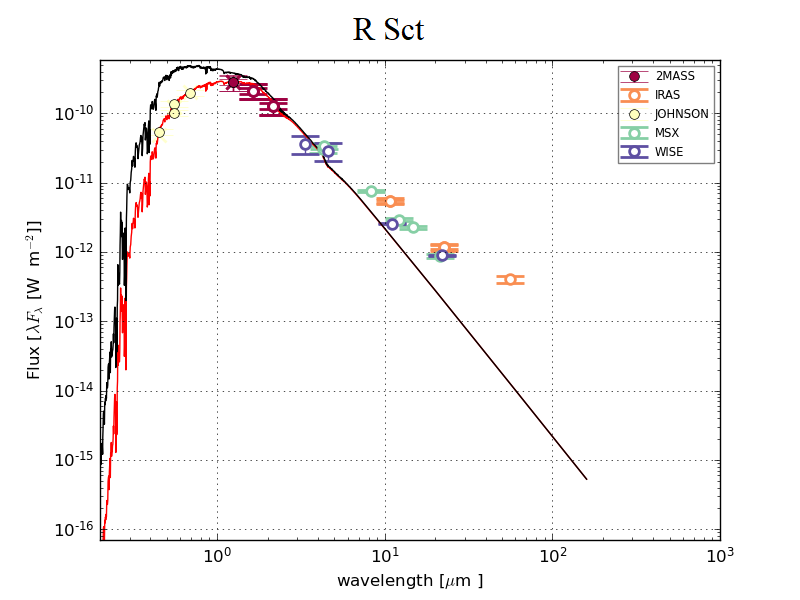}
    \includegraphics[trim=1cm 0 0.5cm 0, width=.49\linewidth]{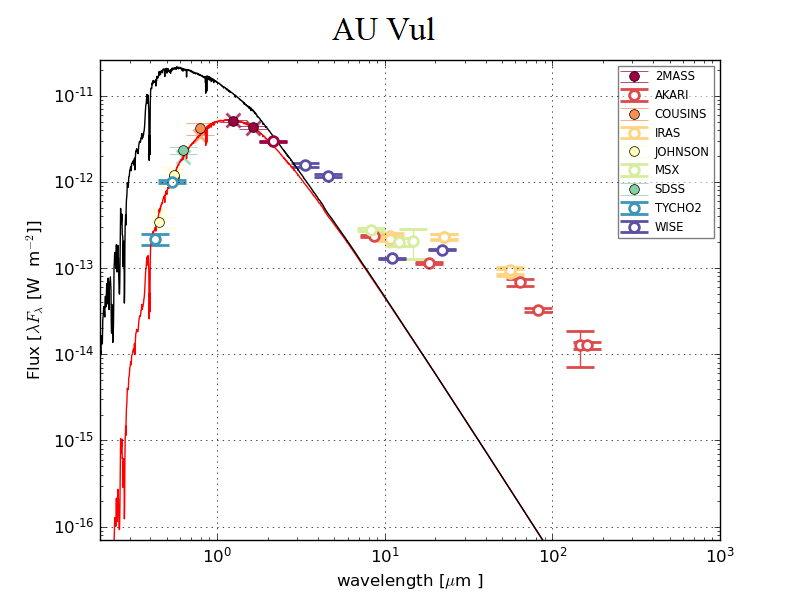}
    \includegraphics[trim=1cm 0 0.5cm 0, width=.49\linewidth]{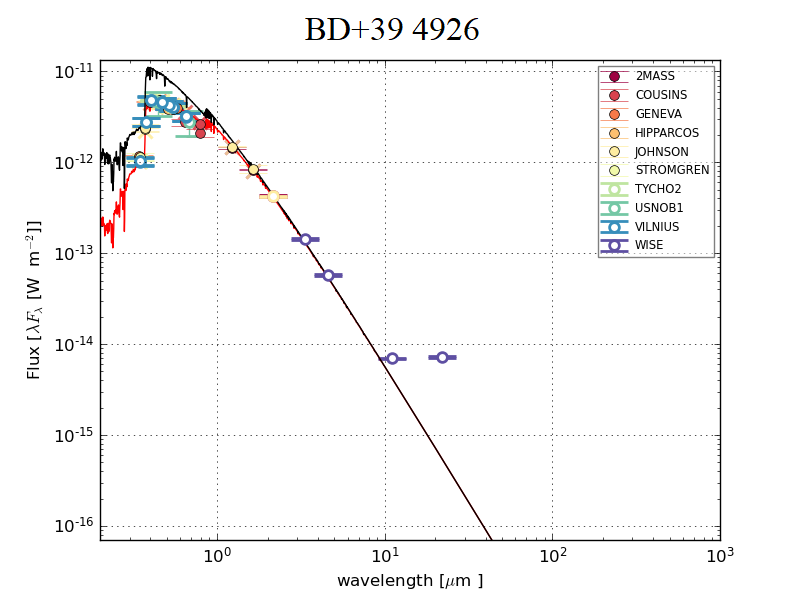}
    \caption[Spectral energy distribution plots of Galactic dust-poor disc stars]{Spectral energy distribution plots of Galactic dust-poor disc stars. The red solid curve corresponds to the reddened Kurucz model atmosphere, while the black solid curve represents Kurucz model atmosphere after de-reddening and scaling to the object. A legend within the plot clarifies the meaning of the symbols and colours used.}\label{figA:allSED1_paper3}
\end{figure*}

\begin{figure*}[ph!]
    \centering
    \includegraphics[trim=1cm 0 0.5cm 0, width=.49\linewidth]{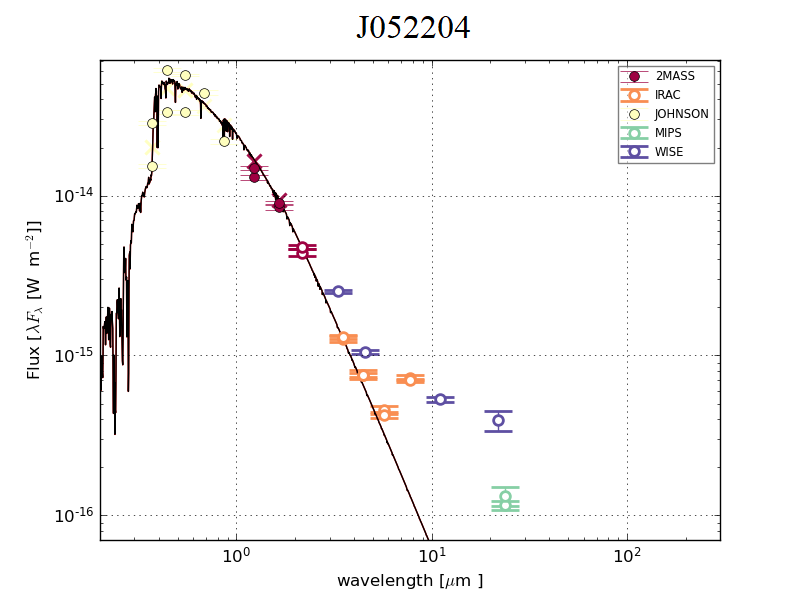}
    \includegraphics[trim=1cm 0 0.5cm 0, width=.49\linewidth]{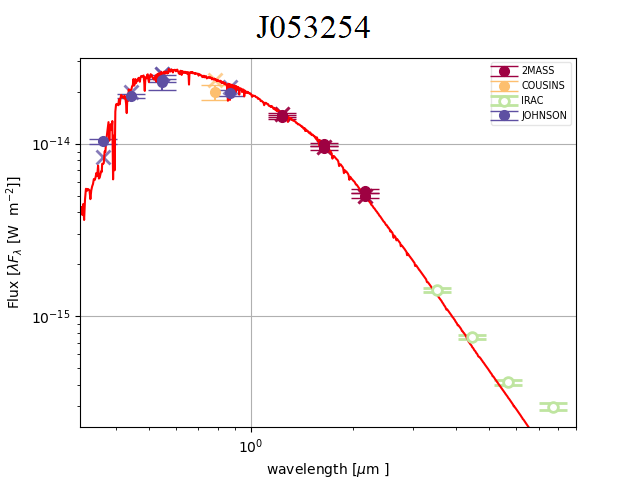}
    \includegraphics[trim=1cm 0 0.5cm 0, width=.44\linewidth]{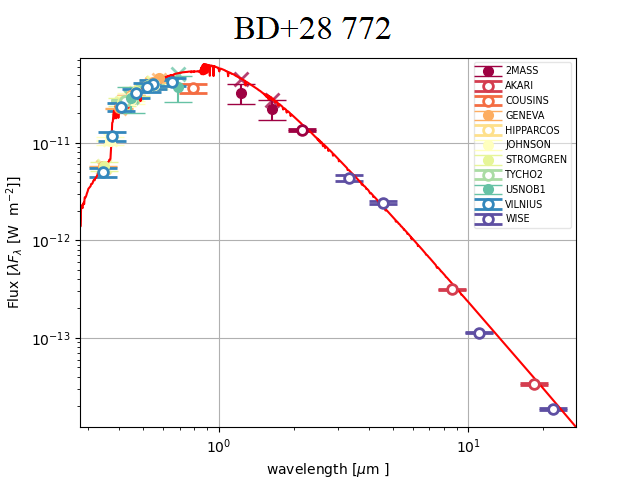}
    \caption[Spectral energy distribution plots of LMC dust-poor disc stars and Galactic dust-poor disc candidate]{Spectral energy distribution plots of LMC dust-poor disc stars and Galactic dust-poor disc candidate. The red solid curve corresponds to the reddened Kurucz model atmosphere, while the black solid curve represents Kurucz model atmosphere after de-reddening and scaling to the object. A legend within the plot clarifies the meaning of the symbols and colours used.}\label{figA:allSED2_paper3}
\end{figure*}

To calculate $L_{\rm SED}$, we followed the procedure outlined in \citet{mohorian2024EiSpec}. In brief, we fitted photometric data with spectra synthesised from Kurucz model atmospheres, integrating the bolometric IR luminosity and correcting for total reddening (interstellar and circumstellar). $\chi^2$ minimisation was performed until convergence of effective temperature $T_{\rm eff}$, surface gravity $\log g$, extinction (reddening) $E(B-V)$, and stellar angular size $\theta$. In our calculations, interstellar reddening follows the extinction law from \citet{cardelli1989SEDextinction} with $R_V\,=\,3.1^m$. We note that we used Bailer-Jones geometric distances ($z_{\rm BJ}$) with \textit{Gaia} EDR3 parallaxes \citep{bailerjones2021distances}, assuming isotropic radiation emission. Additionally, we did not explicitly account for pulsational variability, which resulted in higher $\chi^2$ values for pulsating variables. The uncertainties in $L_{\rm SED}$ are primarily caused by uncertainties in distance and photometric magnitudes. In Fig.~\ref{figA:allSED1_paper3} and \ref{figA:allSED2_paper3}, we show the SEDs of dust-poor disc targets.

To calculate $L_{\rm PLC}$, we applied the calibrated relation described in \citet{menon2024EvolvedBinaries}, given by
\begin{equation}
    M_{bol} = m_{cal}\cdot\log P_0 + c_{cal} - \mu + BC + 2.55\cdot(V-I)_0,
\end{equation}
where $M_{bol}$ is the absolute bolometric magnitude; $m_{cal}=-3.59^m$ and $c_{cal}=18.79^m$ are the calibrated slope and intercept of PLC relation, respectively; $P_0$ is the observed fundamental period of pulsation; $\mu=18.49$ is the distance modulus to the LMC (used to calibrate the relation); $BC$ is the bolometric correction based on the $T_{\rm eff}$ \citep{flower1996BoloCorr, torres2010BoloCorrErrata}; and $(V-I)_0$ is the intrinsic (de-reddened) colour based on the reddening value $E(B-V)$ from the SED fit. The uncertainties in $L_{\rm PLC}$ are primarily caused by uncertainties in $E(B-V)$.

\newpage

We note that we consider $L_{\rm PLC}$ to be more accurate than $L_{\rm SED}$. Therefore, for all pulsating variables in the dust-poor disc sample, we adopt $L_{\rm PLC}$ as the final luminosity for the subsequent analysis. For non-pulsating stars, BD+39\,4926 and BD+28\,772, we use $L_{\rm SED}$ as the final luminosity. In Table~\ref{tab:respar_paper3}, we provide the derived values and uncertainties for both $L_{\rm SED}$ and $L_{\rm PLC}$.

\begin{sidewaystable}[ph!]
    \centering
    \scriptsize
    \caption[Derived luminosities, atmospheric parameters, selected abundances and abundance ratios of dust-poor disc targets (see Sections~\ref{ssec:analum_paper3} and \ref{ssec:anaspc_paper3})]{Derived luminosities, atmospheric parameters, selected abundances and abundance ratios of dust-poor disc targets (see Sections~\ref{ssec:analum_paper3} and \ref{ssec:anaspc_paper3}). For a full list of [X/H] abundances, see Appendix~\ref{app:abu_paper3}.}\label{tab:respar_paper3}
    \begin{tabular}{|c|cc|cccc|cc|cc|}\hline
        \multirow{3}{*}{\textbf{Name}} & \multicolumn{2}{c|}{\textbf{Derived luminosities}} & \multicolumn{4}{c|}{\textbf{Atmospheric parameters}} & \multicolumn{2}{|c|}{\textbf{LTE ratios}} & \multicolumn{2}{|c|}{\textbf{NLTE ratios}} \\
        & \multirow{2}{*}{\boldmath$\log\dfrac{L_{\rm SED}}{L_\odot}$} & \multirow{2}{*}{\boldmath$\log\dfrac{L_{\rm PLC}}{L_\odot}$} & \boldmath$T_{\rm eff}$ & \boldmath$\log g$ & \textbf{[Fe/H]} & \boldmath$\xi_{\rm t}$ & \textbf{[Zn/Ti]} & \textbf{[Zn/Fe]} & \textbf{[S/Ti]} & \textbf{C/O} \\
        &&& \textbf{(K)} & \textbf{(dex)} & \textbf{(dex)} & \textbf{(km/s)} & \textbf{(dex)} & \textbf{(dex)} & \textbf{(dex)} & ~ \\ \hline
        SS Gem & 3.657$\,\pm\,$0.116 & $3.400_{-0.296}^{+0.099}$ & 6500$\,\pm\,$90 & 1.96$\,\pm\,$0.11 & --1.02$\,\pm\,$0.12 & 4.08$\,\pm\,$0.09 & 0.98$\,\pm\,$0.21 & 0.45$\,\pm\,$0.25 & 1.25$\,\pm\,$0.09 & 0.55$\,\pm\,$0.28 \\
        V382 Aur & 3.549$\,\pm\,$0.101 & $3.350_{-0.183}^{+0.113}$ & 6020$\,\pm\,$150 & 0.38$\,\pm\,$0.27 & --1.82$\,\pm\,$0.08 & 4.14$\,\pm\,$0.13 & 0.81$\,\pm\,$0.34 & 0.69$\,\pm\,$0.31 & 1.64$\,\pm\,$0.19 & 0.38$\,\pm\,$0.13 \\
        CC Lyr & 2.811$\,\pm\,$0.143 & $2.986_{-0.099}^{+0.211}$ & 6470$\,\pm\,$250 & 1.64$\,\pm\,$0.50 & --3.80$\,\pm\,$0.03 & 4.14$\,\pm\,$1.00 & -- & 2.95$\,\pm\,$0.44 & -- & 0.03$\,\pm\,$0.02 \\
        R Sct & 3.452$\,\pm\,$0.090 & $3.099_{-0.394}^{+0.113}$ & 5630$\,\pm\,$70 & 0.93$\,\pm\,$0.11 & --0.60$\,\pm\,$0.11 & 2.93$\,\pm\,$0.03 & 0.54$\,\pm\,$0.17 & 0.42$\,\pm\,$0.17 & 0.31$\,\pm\,$0.13 & 0.17$\,\pm\,$0.04 \\
        AU Vul & 3.626$\,\pm\,$0.123 & $3.816_{-0.141}^{+0.253}$ & 4740$\,\pm\,$40 & 0.00$\,\pm\,$0.09 & --0.95$\,\pm\,$0.08 & 4.32$\,\pm\,$0.04 & 0.40$\,\pm\,$0.27 & 0.08$\,\pm\,$0.36 & -- & 0.07$\,\pm\,$0.00 \\
        BD+39 4926 & 3.787$\,\pm\,$0.137 & -- & 7470$\,\pm\,$40 & 0.52$\,\pm\,$0.05 & --2.37$\,\pm\,$0.18 & 2.21$\,\pm\,$0.25 & 2.69$\,\pm\,$0.44 & 2.17$\,\pm\,$0.29 & 3.36$\,\pm\,$0.35 & 0.14$\,\pm\,$0.10 \\ \hline
        J052204 & 3.620$\,\pm\,$0.091 & $3.307_{-0.000}^{+0.225}$ & 7020$\,\pm\,$100 & 1.65$\,\pm\,$0.11 & --2.49$\,\pm\,$0.10 & 2.36$\,\pm\,$0.04 & 3.15$\,\pm\,$0.34 & 1.98$\,\pm\,$0.33 & -- & 0.54$\,\pm\,$0.21 \\
        J053254 & 3.507$\,\pm\,$0.139 & $3.843_{0.479}^{0.070}$ & 6040$\,\pm\,$90 & 0.47$\,\pm\,$0.13 & --1.66$\,\pm\,$0.09 & 2.36$\,\pm\,$0.03 & 0.87$\,\pm\,$0.27 & 0.45$\,\pm\,$0.23 & 1.31$\,\pm\,$0.15 & 0.81$\,\pm\,$0.21 \\ \hline
        BD+28 772 & 3.502$\,\pm\,$0.483 & -- & 6880$\,\pm\,$100 & 0.96$\,\pm\,$0.16 & --0.42$\,\pm\,$0.10 & 2.14$\,\pm\,$0.03 & --0.10$\,\pm\,$0.25 & --0.25$\,\pm\,$0.25 & 0.26$\,\pm\,$0.19 & 0.39$\,\pm\,$0.11 \\ \hline
    \end{tabular}\\
    \textbf{Note:} We adopt $L_{\rm PLC}$ for all targets except BD+39 4926 and BD+28 772, for which we adopt $L_{\rm SED}$.
\end{sidewaystable}

\subsection{Derivation of atmospheric parameters and elemental abundances using \texttt{E-iSpec}}\label{ssec:anaspc_paper3}
In this study, we derived atmospheric parameters and elemental abundances using \texttt{E-iSpec}, a semi-automated spectral analysis code \citep{mohorian2024EiSpec}, specifically tailored for evolved stars with complex atmospheres. \texttt{E-iSpec} is a modified version of the iSpec spectral analysis code \citep{blancocuaresma2014, blancocuaresma2019}. The analysis within \texttt{E-iSpec} is conducted using the LTE \texttt{Moog} transfer code \citep[operating with equivalent width method,][]{sneden2012Moog}, the VALD3 line list \citep{kupka2011vald}, solar abundances from \citet{asplund2021solar}, and 1D plane-parallel ATLAS9 model atmospheres \citep{castelli2003ATLAS9}. In Appendix~\ref{app:lst_paper3}, we present the final line list selected for precise derivation of atmospheric parameters and elemental abundances.

For atmospheric parameters, we adopt the original iSpec method for computing uncertainties. For elemental abundances, we compute the uncertainties using a quadrature sum of random and systematic components \citep[for more details, see][]{mohorian2025TransitionDiscs}. We note that the random component is set to 0.1 dex for elements with abundance derived from the single spectral line. Additionally, we assume that variations in metallicity affect the [X/Fe] abundance ratios, but not the [X/H] abundances.

In Table~\ref{tab:respar_paper3}, we present the derived atmospheric parameters and selected elemental abundance ratios of dust-poor disc targets ([Zn/Ti], [Zn/Fe], [S/Ti], and C/O). In Appendix~\ref{app:abu_paper3}, we present the derived LTE elemental abundances. In Fig.~\ref{fig:dpl3_paper3} and \ref{fig:dpl4_paper3}, we present the derived LTE abundances (marked as cyan circles).

\begin{figure}
    \centering
    \includegraphics[width=.41\linewidth]{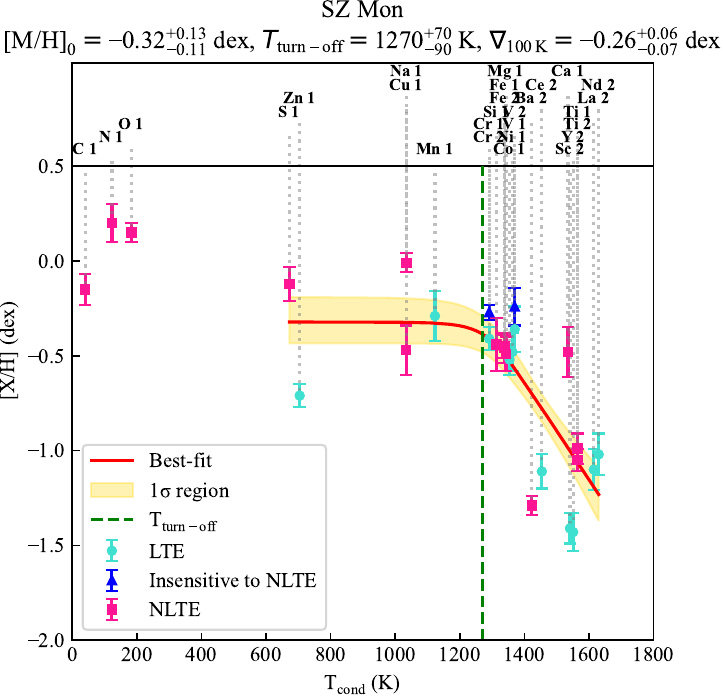}
    \includegraphics[width=.41\linewidth]{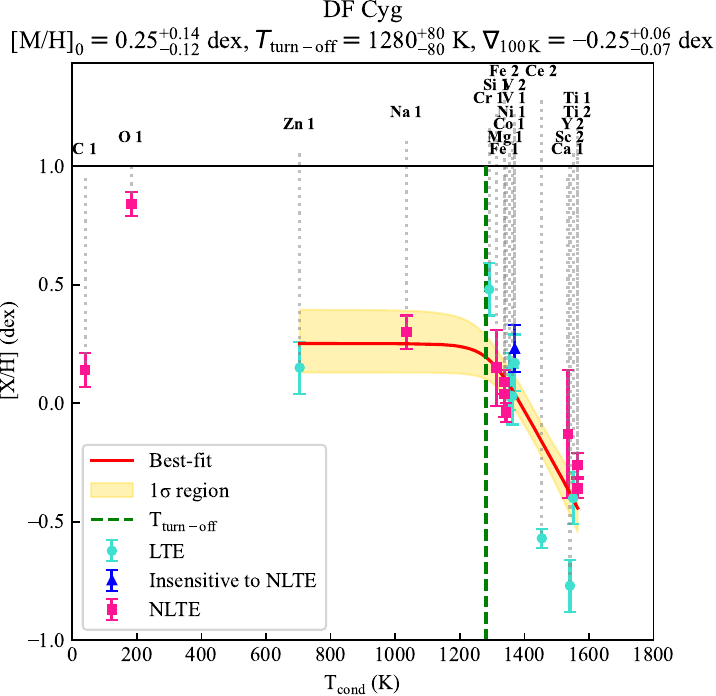}
    \includegraphics[width=.41\linewidth]{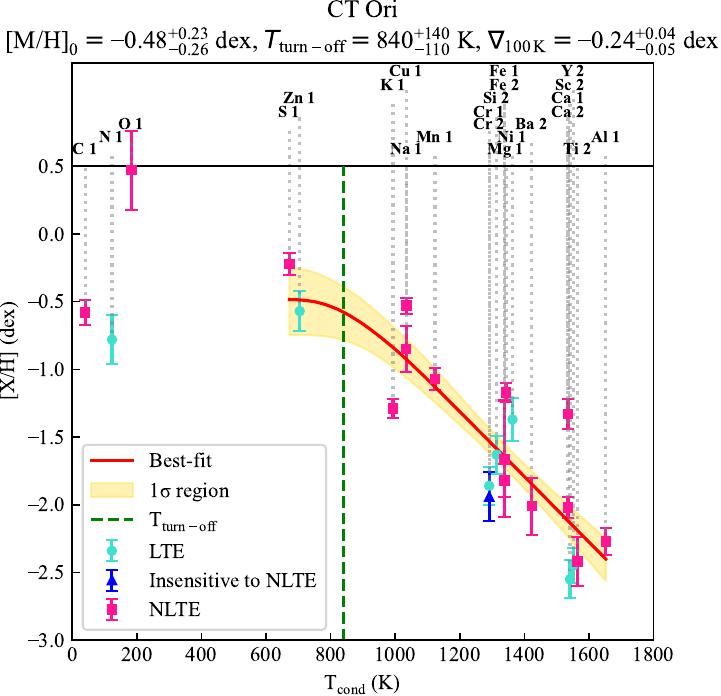}
    \includegraphics[width=.41\linewidth]{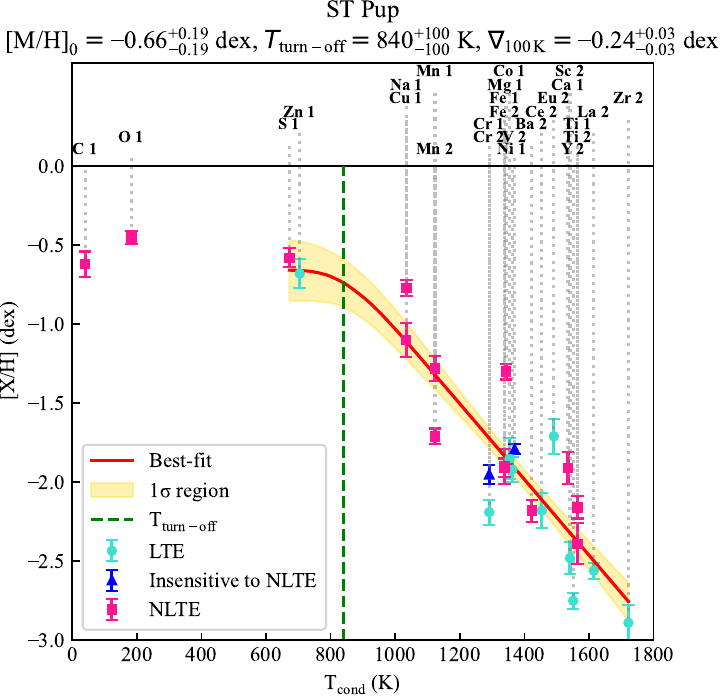}
    \includegraphics[width=.41\linewidth]{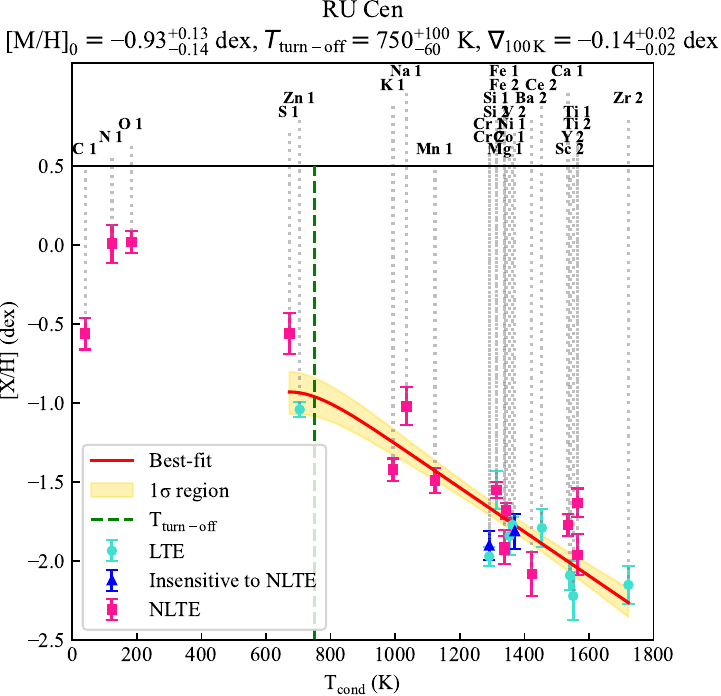}
    \includegraphics[width=.41\linewidth]{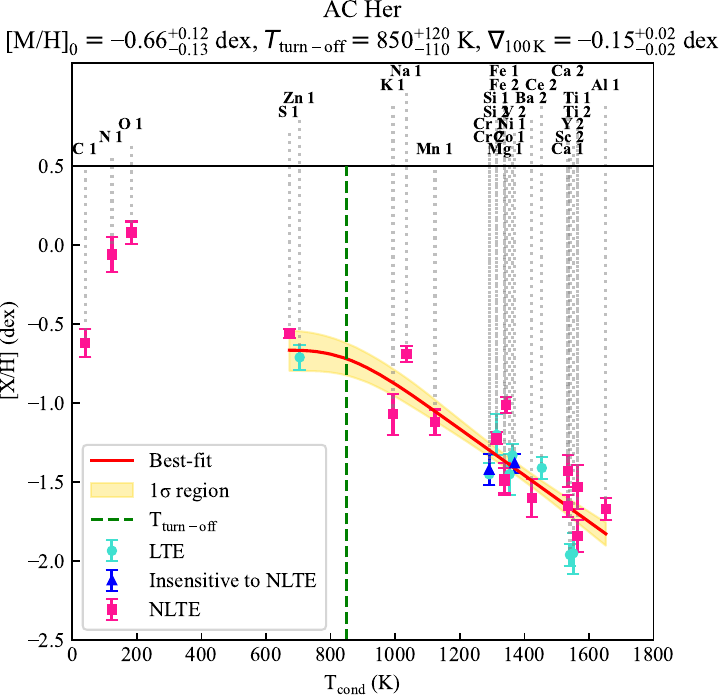}
    \caption[Elemental abundances of a subsample of post-AGB/post-RGB binaries with full and transition discs as functions of condensation temperature (part 1)]{Elemental abundances of a subsample of post-AGB/post-RGB binaries with full and transition discs as functions of condensation temperature \citep{lodders2003CondensationTemperatures, wood2019CondensationTemperatures}. The legend for the symbols and colours used is included within the plot. ``NLTE insensitive'' abundances are derived from spectral lines of \ion{V}{ii} and \ion{Cr}{ii}; for more details, see Section~\ref{ssec:anaspc_paper3}).}\label{fig:dpl1_paper3}
\end{figure}

\begin{figure}
    \centering
    \includegraphics[width=.41\linewidth]{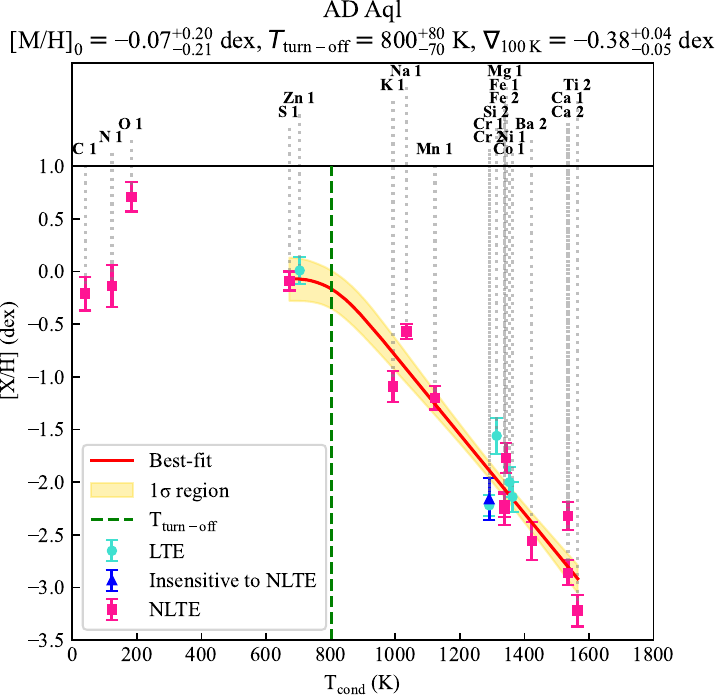}
    \includegraphics[width=.41\linewidth]{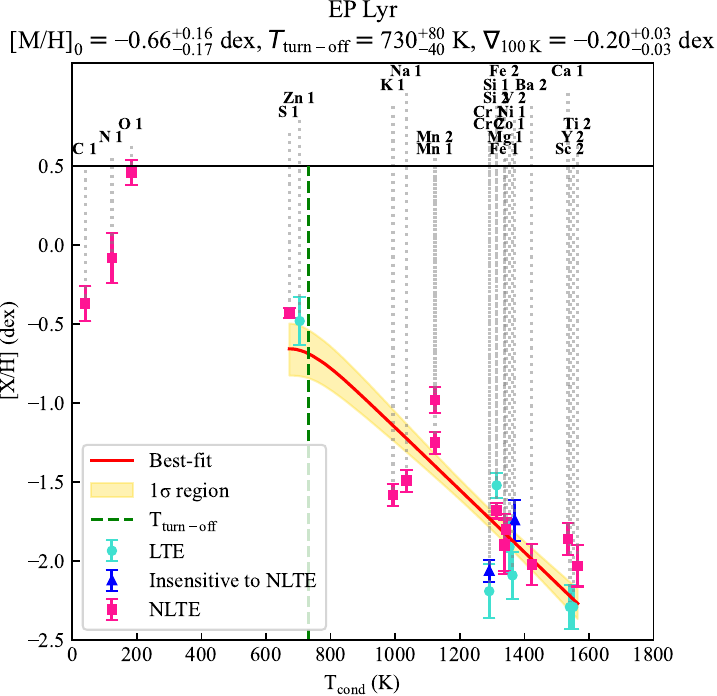}
    \includegraphics[width=.41\linewidth]{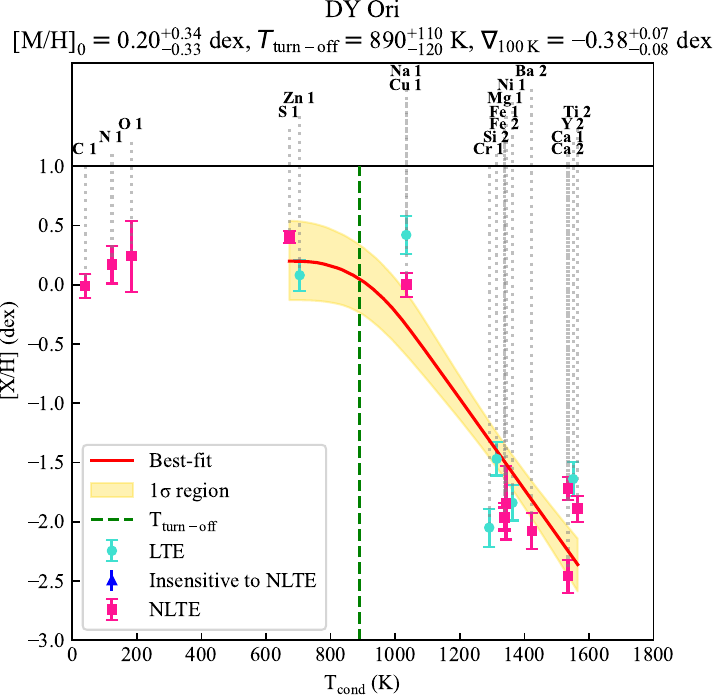}
    \includegraphics[width=.41\linewidth]{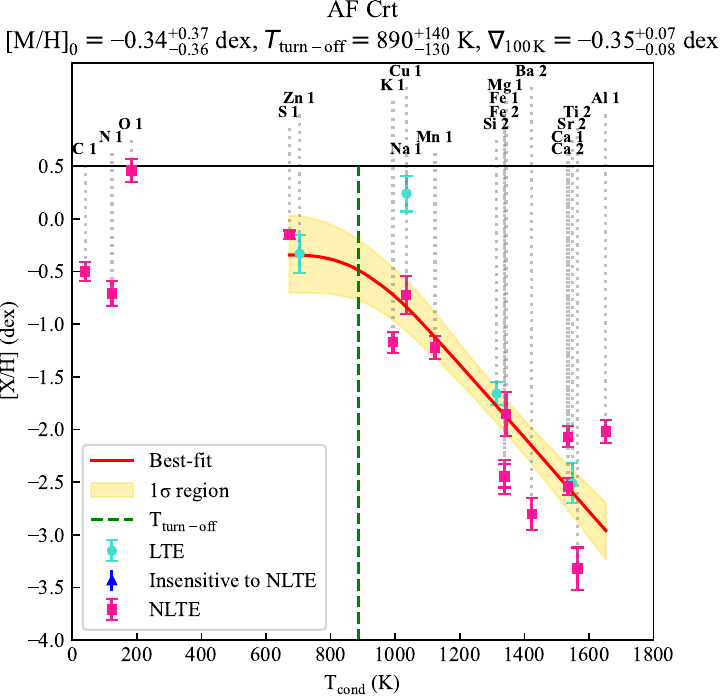}
    \includegraphics[width=.41\linewidth]{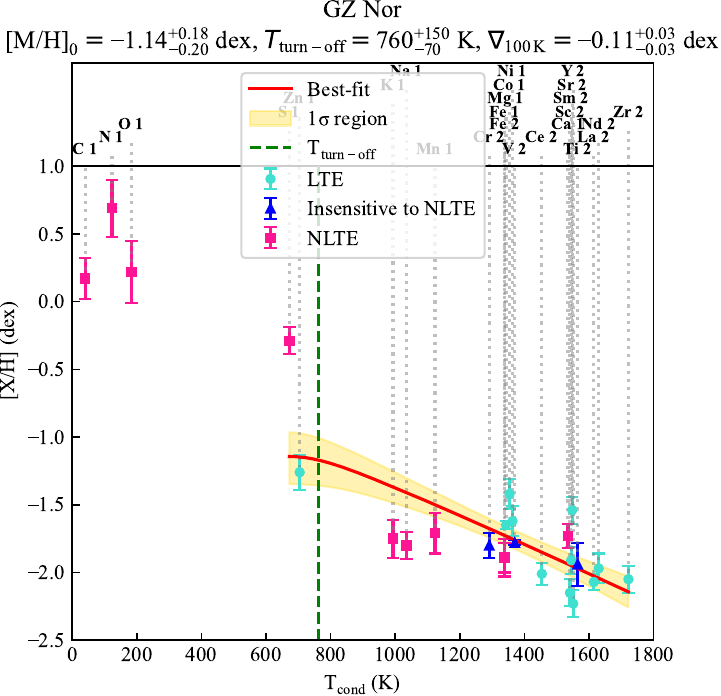}
    \includegraphics[width=.41\linewidth]{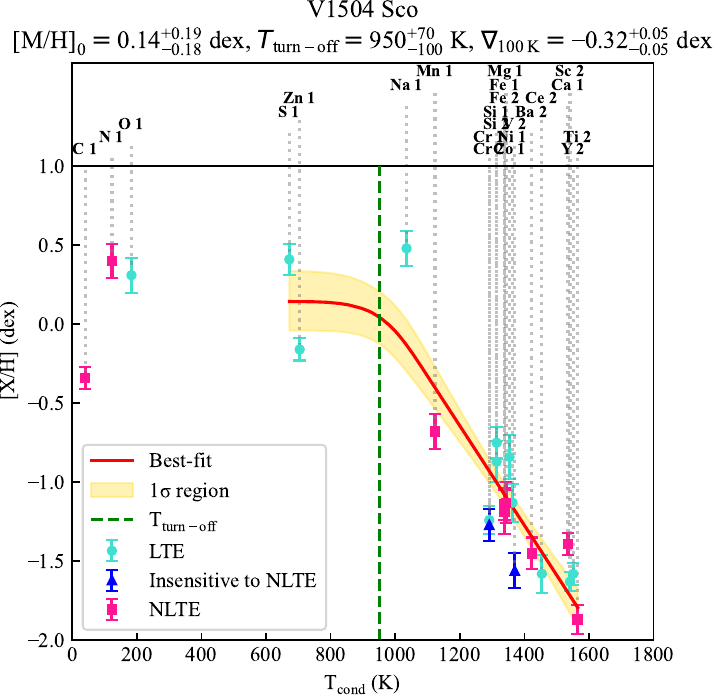}
    \caption[Elemental abundances of a subsample of post-AGB/post-RGB binaries with transition discs as functions of condensation temperature (part 2)]{Elemental abundances of a subsample of post-AGB/post-RGB binaries with transition discs as functions of condensation temperature \citep{lodders2003CondensationTemperatures, wood2019CondensationTemperatures}. The legend for the symbols and colours used is included within the plot. ``NLTE insensitive'' abundances are derived from spectral lines of \ion{V}{ii} and \ion{Cr}{ii}; for more details, see Section~\ref{ssec:anaspc_paper3}).}\label{fig:dpl2_paper3}
\end{figure}

\begin{figure}
    \centering
    \includegraphics[width=.41\linewidth]{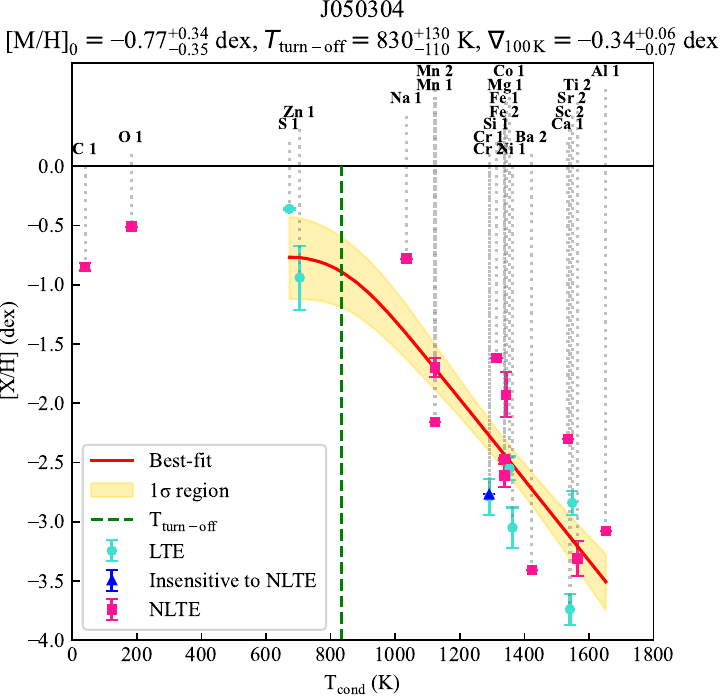}
    \includegraphics[width=.41\linewidth]{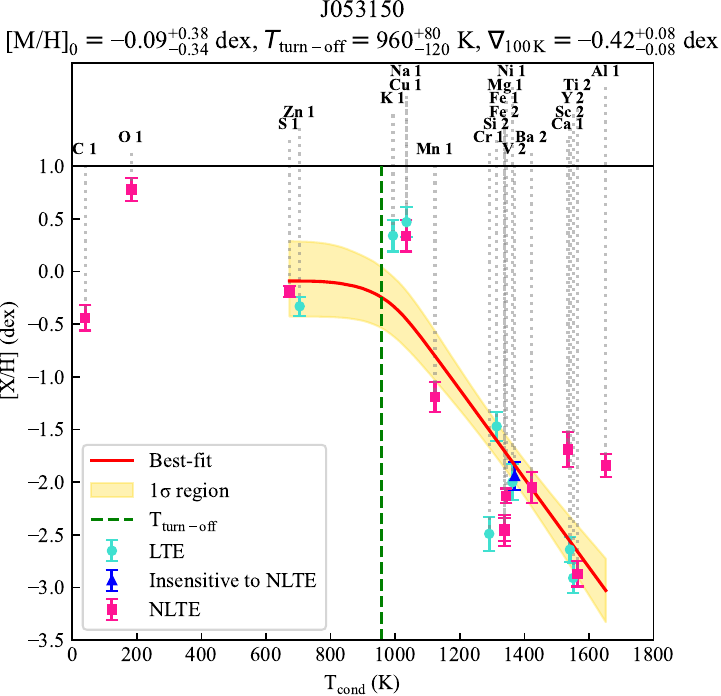}
    \includegraphics[width=.41\linewidth]{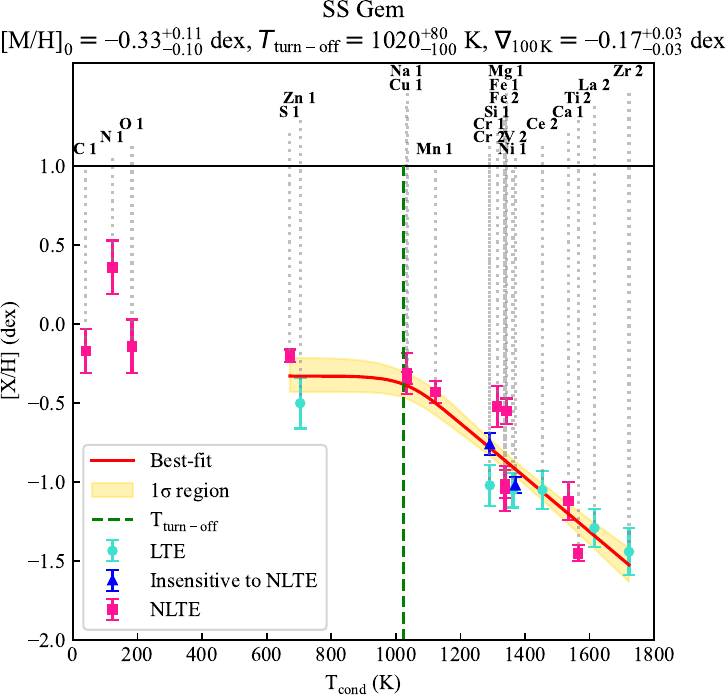}
    \includegraphics[width=.41\linewidth]{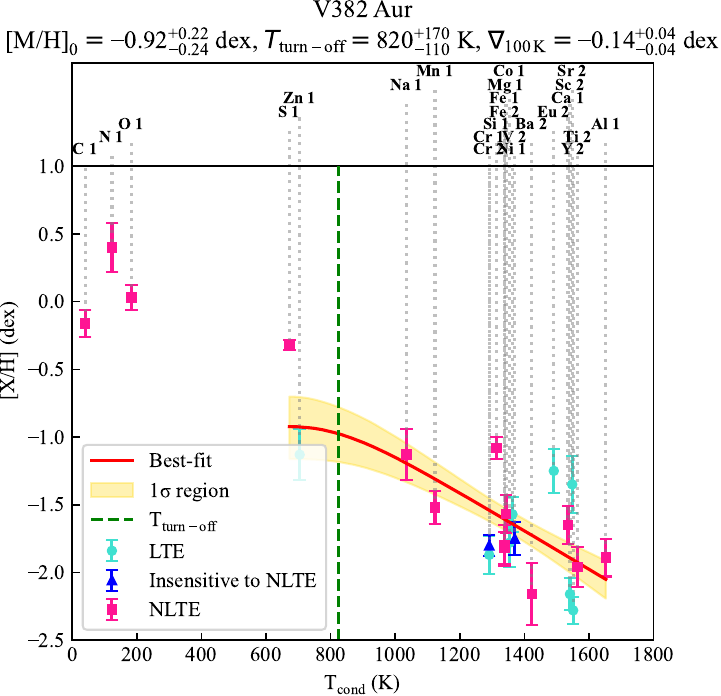}
    \includegraphics[width=.41\linewidth]{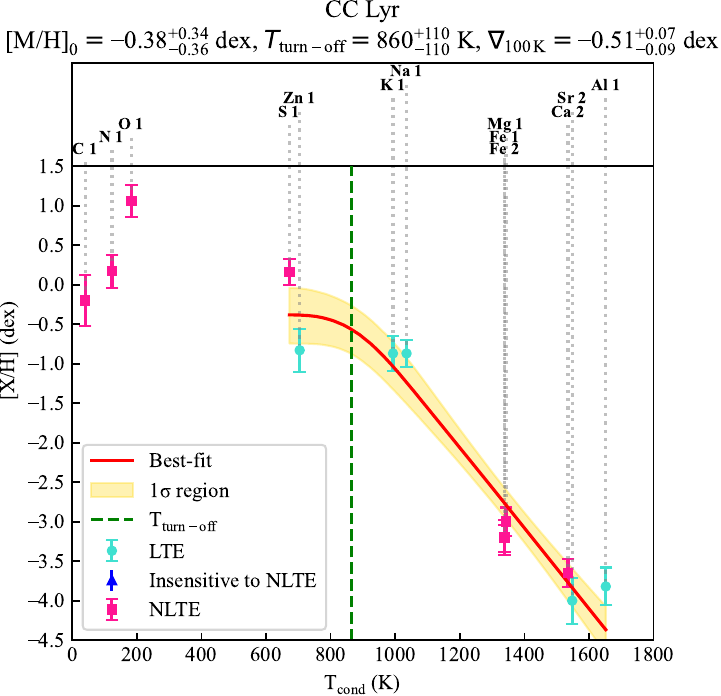}
    \includegraphics[width=.41\linewidth]{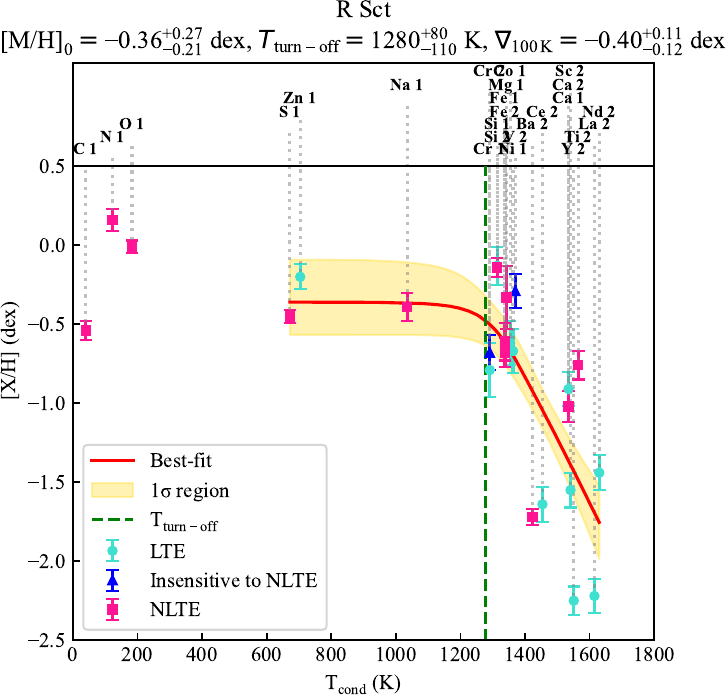}
    \caption[Elemental abundances of a subsample of post-AGB/post-RGB binaries with transition and dust-poor discs as functions of condensation temperature (part 3)]{Elemental abundances of a subsample of post-AGB/post-RGB binaries with transition and dust-poor discs as functions of condensation temperature \citep{lodders2003CondensationTemperatures, wood2019CondensationTemperatures}. The legend for the symbols and colours used is included within the plot. ``NLTE insensitive'' abundances are derived from spectral lines of \ion{V}{ii} and \ion{Cr}{ii}; for more details, see Section~\ref{ssec:anaspc_paper3}).}\label{fig:dpl3_paper3}
\end{figure}

\begin{figure}
    \centering
    \includegraphics[width=.41\linewidth]{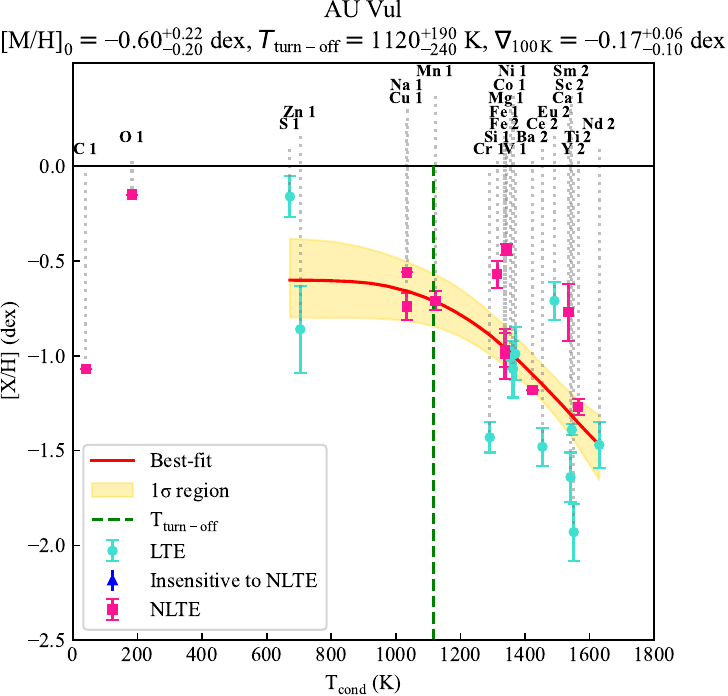}
    \includegraphics[width=.41\linewidth]{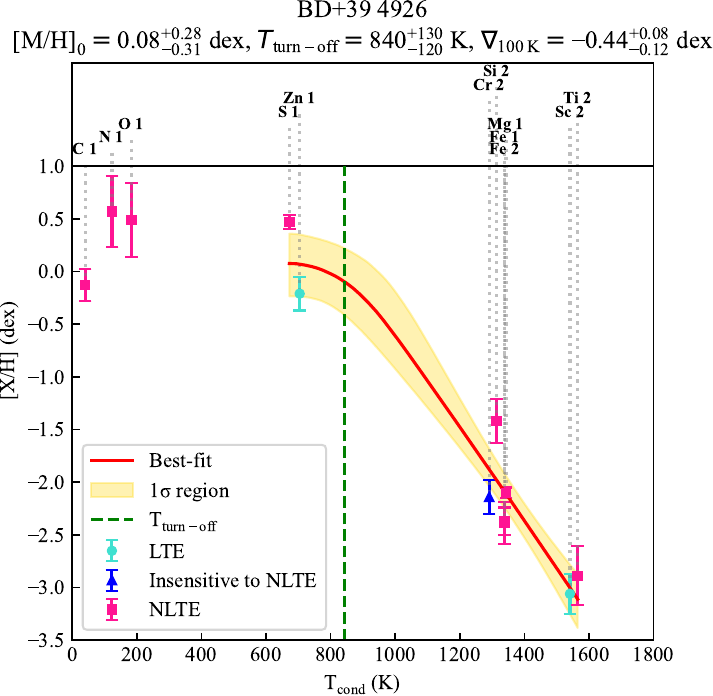}
    \includegraphics[width=.41\linewidth]{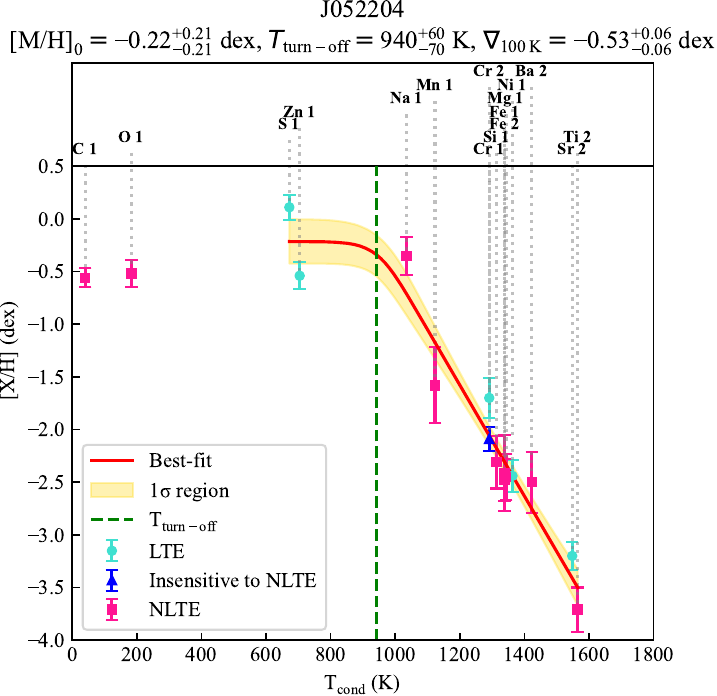}
    \includegraphics[width=.41\linewidth]{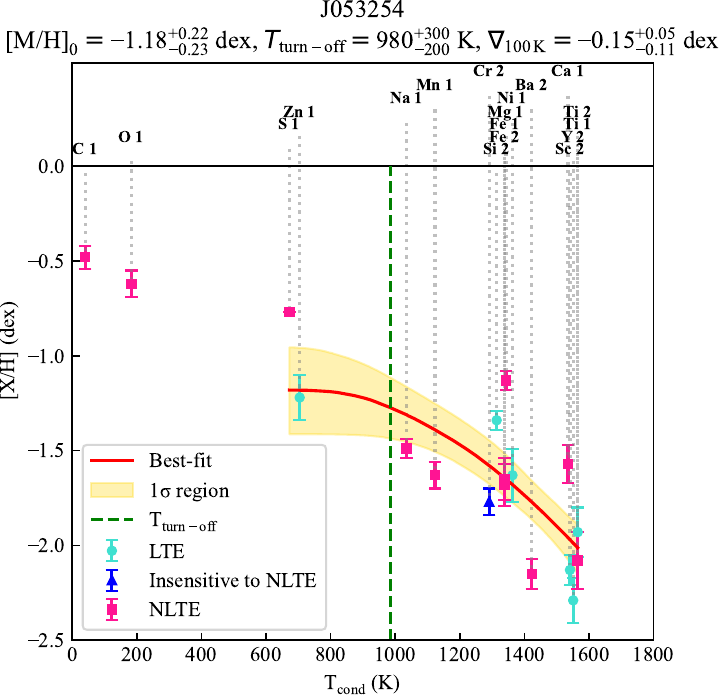}
    \includegraphics[width=.41\linewidth]{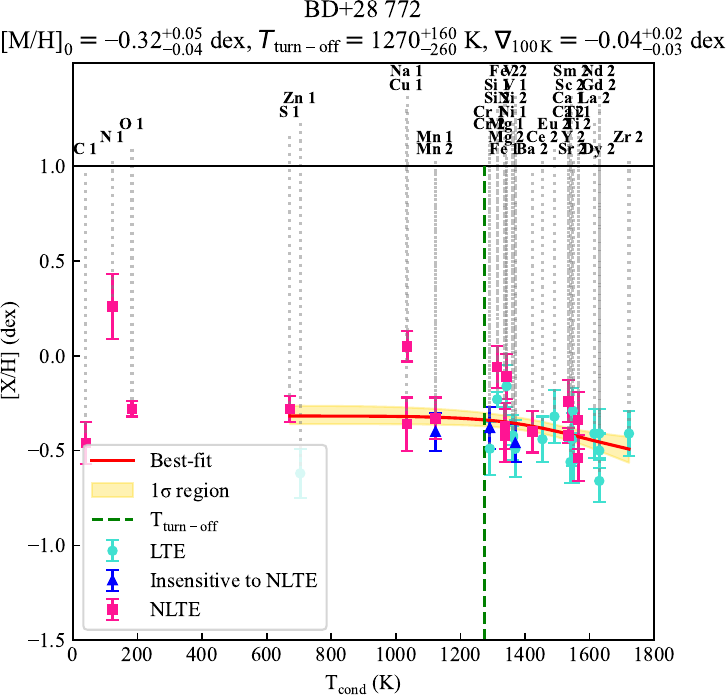}
    \caption[Elemental abundances of a subsample of post-AGB/post-RGB binaries with dust-poor discs as functions of condensation temperature (part 4)]{Elemental abundances of a subsample of post-AGB/post-RGB binaries with dust-poor discs as functions of condensation temperature \citep{lodders2003CondensationTemperatures, wood2019CondensationTemperatures}. The legend for the symbols and colours used is included within the plot. ``NLTE insensitive'' abundances are derived from spectral lines of \ion{V}{ii} and \ion{Cr}{ii}; for more details, see Section~\ref{ssec:anaspc_paper3}).}\label{fig:dpl4_paper3}
\end{figure}

\subsection{Calculation of NLTE corrections using \texttt{pySME}}\label{ssec:ananlt_paper3}
Departures from LTE generally (but not always) increase with higher $T_{\rm eff}$, lower $\log g$, and decreasing [Fe/H], as radiation becomes less balanced by thermal collisions at shorter wavelengths. This leads to over-ionisation in neutral species, including \ion{Fe}{i} and \ion{Ti}{i}, while their ionised counterparts, including \ion{Fe}{ii} and \ion{Ti}{ii}, remain largely unaffected.

In this study, we expanded our NLTE analysis beyond that of \citet{mohorian2025TransitionDiscs}. Previously, in \citet{mohorian2025TransitionDiscs} we used \texttt{Balder} to calculate NLTE corrections for individual spectral lines of C, N, O, Na, Mg, Al, Si, S, K, Ca, and Fe. In this work, we extended the range of elements by using \texttt{pySME} \citep{wehrhahn2023pySME} to calculate NLTE abundances for C, N, O, Na, Mg, Al, Si, S, K, Ca, Ti, Mn, Fe, Cu, and Ba. \texttt{pySME} uses pre-computed grids of NLTE departure coefficients $\beta_k = \frac{n_k^{\rm NLTE}}{n_k^{\rm LTE}}$, for each energy level $k$, with population $n_k$. Since the departure coefficients used in \texttt{pySME} were calculated for MARCS model atmospheres \citep{gustafsson2008MARCS} which are limited for hot metal-poor giants, we set $T_{\rm eff}\,=\,5\,750$\,K and/or $\log g\,=\,0.5$\,dex in NLTE calculations for all targets with higher $T_{\rm eff}$ and lower $\log g$, respectively. In Table~\ref{tab:nltmod_paper3}, we list the references for the model atoms used to calculate the departure coefficient grids. The NLTE corrections were determined as the differences between the NLTE and LTE abundances, as explained below.

To calculate NLTE abundances, we synthesised the LTE spectral lines in \texttt{pySME} using observed equivalent widths and derived LTE atmospheric parameters. Then, we fitted these lines with NLTE spectral lines. The differences between NLTE and LTE abundances derived from the \texttt{pySME} fitted lines provided the relative abundance correction $\Delta_i^{\rm diff.}=[{\rm X/H}]_i^{\rm NLTE} - [{\rm X/H}]_i^{\rm LTE}$ for each studied spectral line $i$. To calculate the uncertainty of NLTE abundance for an ionisation of an element, we added in quadratures the systematic uncertainty of LTE abundance (abundance deviations within the error bars of atmospheric parameters) and random uncertainty of individual NLTE abundances (0.1 dex for single line, standard deviation for more lines). 

In Appendix~\ref{app:abu_paper3}, we provide the derived NLTE elemental abundances of dust-poor disc targets. In Fig.~\ref{fig:dpl3_paper3} and \ref{fig:dpl4_paper3}, we present `NLTE-insensitive' abundances (\ion{V}{ii} and \ion{Cr}{ii}; navy triangles) and NLTE-corrected abundances (pink squares) of dust-poor disc targets.

\begin{table}[ht]
    \centering
    \footnotesize
    \caption{The references for model atoms used in this study (see Section~\ref{ssec:ananlt_paper3}).} \label{tab:nltmod_paper3}
    \begin{tabular}{|c|c|}
    \hline
        Element & Reference \\ \hline
        C & \cite{amarsi2020NLTEgalah} \\
        N & \cite{amarsi2020NLTEgalah} \\
        O & \cite{amarsi2020NLTEgalah} \\
        Na & \cite{amarsi2020NLTEgalah} \\
        Mg & \cite{amarsi2020NLTEgalah} \\
        Al & \cite{amarsi2020NLTEgalah} \\
        Si & \cite{amarsi2020NLTEgalah} \\
        S & Amarsi et al. (in prep.) \\
        K & \cite{amarsi2020NLTEgalah} \\
        Ca & \cite{amarsi2020NLTEgalah} \\
        Ti & \cite{mallinson2024NLTE} \\
        Mn & \cite{amarsi2020NLTEgalah} \\
        Fe & \cite{amarsi2022NLTEiron} \\
        Cu & Caliskan et al. (in prep.) \\
        Ba & \cite{amarsi2020NLTEgalah} \\ \hline
    \end{tabular}\\
\end{table}

\subsection{Quantification of depletion efficiency using three free parameters}\label{ssec:anares_paper3}
Depletion efficiency is primarily measured using volatile-to-refractory abundance ratios, including [Zn/Ti], [Zn/Fe], and [S/Ti] ratios (see Section~\ref{sec:int_paper3}). In Table~\ref{tab:respar_paper3}, we present [Zn/Ti], [Zn/Fe], and [S/Ti] ratios of dust-poor disc targets (see Section~\ref{ssec:anaspc_paper3}). We note that we use NLTE-corrected abundances of S, Ti, and Fe to derive these volatile-to-refractory abundance ratios (see Section~\ref{ssec:ananlt_paper3}). Although these abundance ratios allow for comparison with the literature, they do not fully determine which elements are significantly affected by depletion.

Another depletion parameter, $T_{\rm turn-off}$ was traditionally determined by visual inspection, making it subjective and difficult to apply consistently. In Table~\ref{tab:litpar_paper3}, we provide the literature $T_{\rm turn-off}$ values of dust-poor disc targets. Furthermore, the high-temperature pattern of the depletion profiles (distribution of [X/H] abundances for chemical elements with high condensation temperatures $T_{\rm cond}\,\geq\,T_{\rm turn-off}$) allows categorising post-AGB/post-RGB binaries into two types: `saturated' and `plateau' profiles (see Section~\ref{sec:int_paper3}). In Table~\ref{tab:litpar_paper3}, we list the reported patterns of depletion profiles from literature studies for dust-poor disc targets. Although this pattern classification is intuitive, it lacks a quantitative basis and does not allow direct comparisons between different systems.

In this study, we investigate the efficiency of the depletion in post-AGB/post-RGB binaries by exploring the shape of the depletion profiles (see Fig.~\ref{fig:dpl3_paper3} and \ref{fig:dpl4_paper3}). To achieve this goal, we introduce a new approach to quantify depletion efficiency by fitting depletion profiles with 2-piece linear functions, which have three free parameters:
\begin{itemize}
    \item The first piece describes the abundances of weakly depleted elements (initial metallicity [M/H]$_0$).
    \item The break represents the onset of significant depletion (turn-off temperature $T_{\rm turn-off}$).
    \item The second piece describes the abundance slope of significantly depleted elements in dex per 100\,K (depletion scale $\nabla_{\rm 100\,K}$).
\end{itemize}

Together, [M/H]$_0$, $T_{\rm turn-off}$, and $\nabla_{\rm 100\,K}$ enable describing the observed depletion profiles of post-AGB/post-RGB binaries in a complete and consistent way. The depletion trend was modelled using a two-piece linear function defined as:
\begin{equation}
    \text{[X/H]} = \text{[M/H]}_0
\end{equation}
for elements with $T_{\rm cond}<T_{\rm turn-off}$, and
\begin{equation}
    \text{[X/H]} = \text{[M/H]}_0+\nabla_{\rm 100\,K}\cdot\frac{T_{\rm cond}-T_{\rm turn-off}}{100}
\end{equation}
for elements with $T_{\rm cond}\geq T_{\rm turn-off}$.

The observed depletion profiles were fitted using a Bayesian approach implemented in PyMC5, using the No-U-Turn Sampler (NUTS) with 4 chains of 10\,000 tuning and 10\,000 sampling iterations each. The priors were placed on the model parameters as follows: $\text{[M/H]}_0 \sim \mathcal{N}(0,10)$, $\nabla_{\rm 100\,K} \sim \mathcal{N}(0,10)$, $T_{\rm turn-off} \sim \mathcal{U}(T_{\rm min}, \min(T_{\rm max}, 1500\,{\rm K}))$, where $\mathcal{N}(x,\sigma_x)$ is the normal distribution with mean $x$ and uncertainty $\sigma$ and $\mathcal{U}(x_{\rm min}, x_{\rm max})$ is the uniform distribution from minimum $x_{\rm min}$ to maximum $x_{\rm max}$. $T_{\rm min}$ and $T_{\rm max}$ are the minimum and maximum temperature in the individual observed depletion profile, respectively. The parameter uncertainties followed half-normal distributions ($\sigma_{param} \sim |\mathcal{N}(0,10)|$). These weakly informative priors are meant to reflect broad physical plausibility without imposing strong constraints. The elemental abundances themselves were used as observational data (not as parameters), with equal weighting across all fitted elements. Posterior predictions were generated by sampling the posterior distributions of the model parameters and credible intervals were derived from the $16^{\rm th}$ and $84^{\rm th}$ percentiles.

In Fig.~\ref{fig:dpl3_paper3} and \ref{fig:dpl4_paper3}, we present the depletion profile fits for dust-poor disc targets (red solid lines) with corresponding $1\sigma$-areas (yellow shaded regions). We note that CNO elements were excluded from the depletion fitting, as CNO surface abundances were significantly altered by nucleosynthetic and mixing processes during the AGB/RGB evolution, on scales comparable to and indiscernible from those of the depletion process.

We found that all derived NLTE depletion profiles of dust-poor disc targets show saturation (see Section~\ref{sec:int_paper3}). In Table~\ref{tab:alllum_paper3}, we list the final luminosities (derived in Section~\ref{ssec:analum_paper3}) and calculated depletion parameters of dust-poor disc targets. To further investigate depletion in dust-poor disc targets, we examined how the derived depletion parameters vary with final luminosity and effective temperature. We discovered a diversity of depletion parameters in dust-poor disc targets but did not detect any significant correlation or systematic trend (see Fig.~\ref{fig:dplteff_paper3} and \ref{fig:dpllum_paper3}).

Furthermore, we note that two Galactic dust-poor targets, CC Lyr (\#17) and BD+39 4926 (\#20) were previously categorised as extremely depleted targets due to extremely low metallicities \citep[{[Fe/H]}$\,\lesssim\,$--3\,dex;][]{kluska2022GalacticBinaries}, and their literature (LTE) depletion profiles displayed a strong decrease of [X/H] with $T_{\rm cond}$ \citep{maas2007t2cep, rao2012BD+394926}. In this study, we show that for CC Lyr (\#17) and BD+39\,4926 (\#20), depletion scales are not extreme within the post-AGB/post-RGB binary sample ($\nabla_{\rm 100\,K}\,=\,-0.51$ and --0.44\,dex; see Section~\ref{ssec:dscall_paper3}). We also note that dust-poor disc target BD+28\,772 (\#23) displays a rare depletion profile with unusually shallow depletion scale (for more details on this target, see Section~\ref{ssec:dscbd2_paper3}).

\section{Discussion}\label{sec:dsc_paper3}
In this section, we discuss the depletion of dust-poor disc targets in detail. In Section~\ref{ssec:dscall_paper3}, we analyse the depletion profiles of dust-poor disc targets in the broader context of post-AGB/post-RGB binaries. In Section~\ref{ssec:dscevo_paper3}, we discuss evolutionary links between the dust-poor, full, and transition discs around post-AGB/post-RGB binaries. Additionally, in Section~\ref{ssec:dscbd2_paper3}, we discuss the dust-poor disc target BD+28\,772 (\#23) with unusually shallow derived depletion scale $\nabla_{\rm 100\,K}$ (see Section~\ref{ssec:anares_paper3}).

\begin{table}[H]
    \centering
    \footnotesize
    \caption[Final luminosities and depletion parameters (see Section~\ref{sec:ana_paper3}) of the post-AGB/post-RGB binaries with dust-poor, full, and transition discs (see Section~\ref{sec:dsc_paper3})]{Final luminosities and depletion parameters (see Section~\ref{sec:ana_paper3}) of the post-AGB/post-RGB binaries with dust-poor, full, and transition discs (see Section~\ref{sec:dsc_paper3}).} \label{tab:alllum_paper3}
    \begin{tabular}{|c|c|c|c|c|c|}
    \hline
        \textbf{ID} & \textbf{Name} & \shortstack{\boldmath$L_{\rm final}$\\\boldmath$(L_\odot)$} & \shortstack{\textbf{[M/H]}\boldmath$_0$\\\textbf{(dex)}} & \shortstack{\boldmath$T_{\rm turn-off}$\\\textbf{(K)}} & \shortstack{\boldmath$\nabla_{\rm 100\,K}$\\\textbf{(dex)}} \\ \hline
        \multicolumn{6}{|c|}{\textit{Full disc targets \citep{mohorian2024EiSpec}}} \\ \hline
        1 & SZ Mon & 2.582$\,\pm\,$0.096 & $-0.32_{-0.11}^{+0.13}$ & $1270_{-90}^{+70}$ & $-0.25_{-0.07}^{+0.06}$ \\
        2 & DF Cyg & 2.818$\,\pm\,$0.069 & $+0.25_{-0.12}^{+0.14}$ & $1280_{-80}^{+80}$ & $-0.25_{-0.08}^{+0.06}$ \\ \hline
        \multicolumn{6}{|c|}{\textit{Transition disc targets \citep{mohorian2025TransitionDiscs}}} \\ \hline
        3 & CT Ori & 3.260$\,\pm\,$0.160 & $-0.49_{-0.25}^{+0.24}$ & $840_{-110}^{+140}$ & $-0.24_{-0.04}^{+0.04}$ \\
        4 & ST Pup & 2.964$\,\pm\,$0.170 & $-0.67_{-0.19}^{+0.19}$ & $840_{-100}^{+100}$ & $-0.24_{-0.03}^{+0.03}$ \\
        5 & RU Cen & 3.502$\,\pm\,$0.266 & $-0.93_{-0.14}^{+0.13}$ & $750_{-60}^{+100}$ & $-0.14_{-0.02}^{+0.02}$ \\
        6 & AC Her & 3.709$\,\pm\,$0.205 & $-0.67_{-0.13}^{+0.12}$ & $850_{-110}^{+120}$ & $-0.15_{-0.02}^{+0.02}$ \\
        7 & AD Aql & 2.836$\,\pm\,$0.290 & $-0.07_{-0.21}^{+0.20}$ & $800_{-80}^{+80}$ & $-0.38_{-0.05}^{+0.04}$ \\
        8 & EP Lyr & 3.960$\,\pm\,$0.232 & $-0.65_{-0.16}^{+0.16}$ & $730_{-40}^{+80}$ & $-0.20_{-0.03}^{+0.03}$ \\
        9 & DY Ori & 3.504$\,\pm\,$0.173 & $+0.20_{-0.33}^{+0.34}$ & $890_{-130}^{+120}$ & $-0.38_{-0.08}^{+0.07}$ \\
        10 & AF Crt & 2.510$\,\pm\,$0.170 & $-0.34_{-0.37}^{+0.36}$ & $890_{-130}^{+140}$ & $-0.35_{-0.08}^{+0.07}$ \\
        11 & GZ Nor & 2.708$\,\pm\,$0.288 & $-1.15_{-0.20}^{+0.18}$ & $760_{-70}^{+150}$ & $-0.11_{-0.03}^{+0.03}$ \\
        12 & V1504 Sco & 3.710$\,\pm\,$0.350 & $+0.14_{-0.17}^{+0.19}$ & $950_{-100}^{+70}$ & $-0.31_{-0.05}^{+0.05}$ \\
        13 & J050304 & 3.456$\,\pm\,$0.198 & $-0.77_{-0.34}^{+0.33}$ & $840_{-110}^{+120}$ & $-0.34_{-0.07}^{+0.06}$ \\
        14 & J053150 & 3.989$\,\pm\,$0.188 & $-0.08_{-0.34}^{+0.39}$ & $960_{-120}^{+80}$ & $-0.42_{-0.08}^{+0.08}$ \\ \hline
        \multicolumn{6}{|c|}{\textit{Dust-poor disc targets (this study)}} \\ \hline
        15 & SS Gem & 3.400$\,\pm\,$0.198 & $-0.33_{-0.10}^{+0.12}$ & $1030_{-100}^{+80}$ & $-0.17_{-0.03}^{+0.03}$ \\
        16 & V382 Aur & 3.350$\,\pm\,$0.148 & $-0.92_{-0.24}^{+0.22}$ & $820_{-110}^{+170}$ & $-0.14_{-0.04}^{+0.04}$ \\
        17 & CC Lyr & 2.986$\,\pm\,$0.155 & $-0.38_{-0.36}^{+0.34}$ & $860_{-100}^{+100}$ & $-0.51_{-0.08}^{+0.07}$ \\
        18 & R Sct & 3.099$\,\pm\,$0.254 & $-0.37_{-0.20}^{+0.26}$ & $1280_{-110}^{+80}$ & $-0.40_{-0.12}^{+0.11}$ \\
        19 & AU Vul & 3.816$\,\pm\,$0.197 & $-0.61_{-0.20}^{+0.22}$ & $1130_{-240}^{+180}$ & $-0.17_{-0.10}^{+0.06}$ \\
        20 & BD+39 4926 & 3.787$\,\pm\,$0.137 & $+0.08_{-0.31}^{+0.29}$ & $850_{-110}^{+130}$ & $-0.44_{-0.12}^{+0.08}$ \\
        21 & J052204 & 3.307$\,\pm\,$0.113 & $-0.22_{-0.21}^{+0.21}$ & $940_{-70}^{+60}$ & $-0.53_{-0.06}^{+0.06}$ \\
        22 & J053254 & 3.843$\,\pm\,$0.275 & $-1.17_{-0.24}^{+0.23}$ & $970_{-200}^{+300}$ & $-0.15_{-0.10}^{+0.05}$ \\
        23 & BD+28 772 & 3.502$\,\pm\,$0.483 & $-0.32_{-0.04}^{+0.05}$ & $1280_{-260}^{+160}$ & $-0.04_{-0.03}^{+0.02}$ \\ \hline
    \end{tabular}
\end{table}

\subsection{Comparing depletion profiles across dust-poor, full, and transition disc targets}\label{ssec:dscall_paper3}
To investigate the relationship between photospheric depletion and the evolutionary status of dust-poor CBDs around post-AGB/post-RGB binaries (see Section~\ref{ssec:dscevo_paper3}), we added to our sample of dust-poor disc targets the samples of full disc targets from \citet{mohorian2024EiSpec} and transition disc targets from \citet{mohorian2025TransitionDiscs}. This resulted in a combined sample of 23 post-AGB/post-RGB binaries containing 19 Galactic and 4 LMC targets. In Fig.~\ref{fig:colplt_paper3}, we present the IR colour-colour plot of combined post-AGB/post-RGB sample with full (marked as circles), transition (marked as stars), and dust-poor disc targets (marked as squares).

We adopted the final luminosities and LTE abundances of the full and transition disc targets reported in \citet{mohorian2024EiSpec} and \citet{mohorian2025TransitionDiscs}, which were also derived using \texttt{E-iSpec}. To ensure consistency in the NLTE depletion profiles across the combined sample, we calculated and applied NLTE corrections to the reported LTE abundances of the full and transition disc targets using the same approach as for the dust-poor disc targets (see Section~\ref{ssec:ananlt_paper3}). In Appendix~\ref{app:abu_paper3}, we provide the LTE and NLTE abundances of dust-poor, full, and transition disc targets. We note that for transition discs, the NLTE abundances derived in this study generally agree within the error bars (typically, $\sim$0.15 dex) with the NLTE abundances reported by \citet{mohorian2025TransitionDiscs}, with a median difference of 0.07\,dex and an average difference of 0.10\,dex. These differences are caused by the irregularity of the NLTE corrections at the boundaries of the MARCS grid ($T_{\rm eff}\,>$\,6\,000\,K, $\log g\,<\,1$\,dex, [Fe/H]\,$<$\,--1.5\,dex). We also note that within the post-AGB/post-RGB sample, the average NLTE corrections for volatile elements (such as C, N, O, and S) are $\approx\,-0.07$\,dex, whereas the average NLTE corrections for refractory elements (such as Al, Si, Ti, and Fe) are $\approx\,+0.08$\,dex. This confirms that although the NLTE corrections are crucial for precise abundance analysis of individual targets, these corrections only weakly affect the statistical distribution of the depletion parameters.

To investigate the depletion characteristics across different disc types, we analysed the depletion profiles of dust-poor, full, and transition disc targets. In Fig.~\ref{fig:dpl1_paper3}, \ref{fig:dpl2_paper3}, \ref{fig:dpl3_paper3}, and \ref{fig:dpl4_paper3}, we present the updated depletion profiles of dust-poor, full, and transition disc targets, along with the corresponding depletion fits (red solid lines) and their 1$\sigma$ confidence regions (yellow shaded areas). In Table~\ref{tab:alllum_paper3}, we list the final luminosities and derived depletion parameters of dust-poor, full, and transition disc targets.

\begin{figure}[ph!]
    \centering
    \includegraphics[trim=0 0.8cm 0 0.8cm, width=0.9\linewidth]{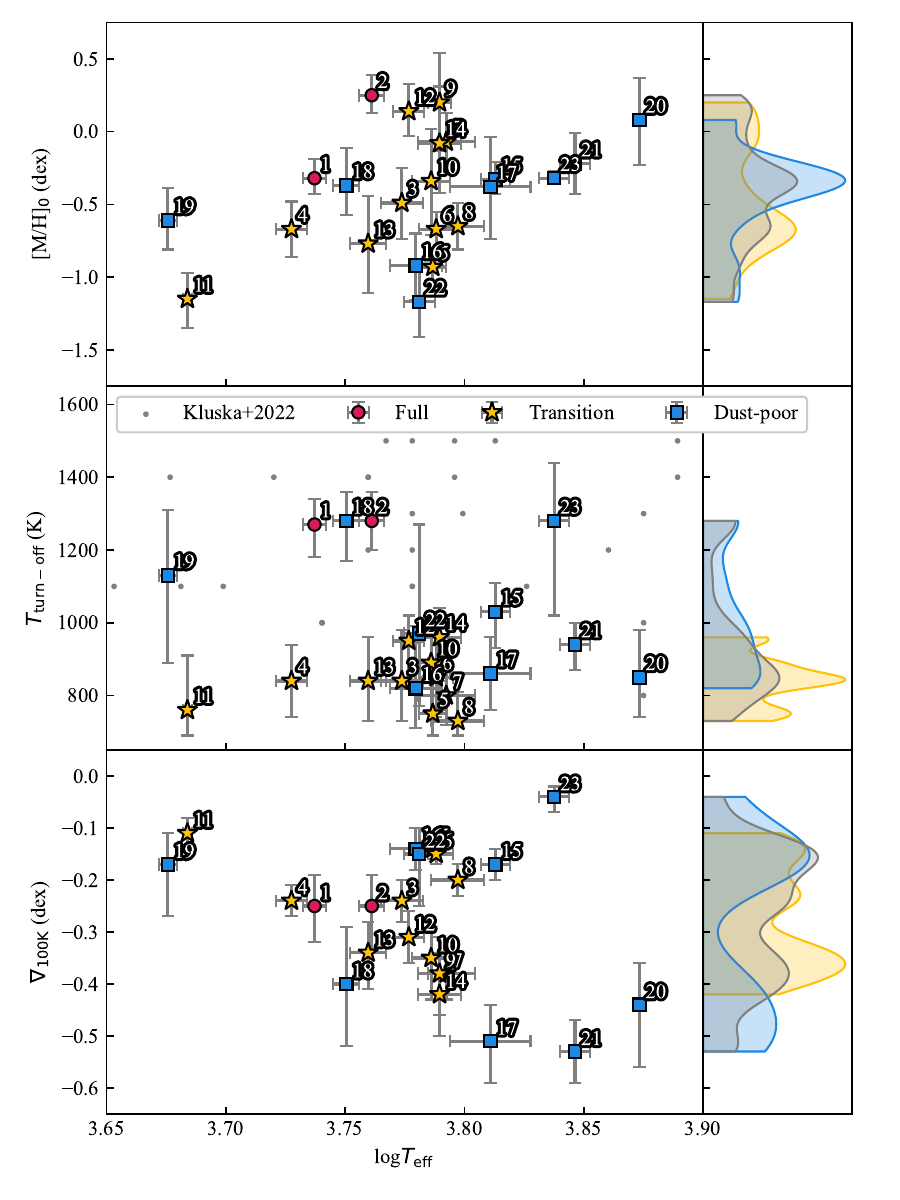}
    \caption[Distributions of depletion pattern parameters with effective temperature as a proxy for age on post-AGB/post-RGB track (\textit{upper panel}: initial metallicity {[M/H]}$_0$, \textit{middle panel}: $T_{\rm turn-off}$, \textit{lower panel}: depletion gradient $\nabla_{\rm100K}$)]{Distributions of depletion pattern parameters with effective temperature as a proxy for age on post-AGB/post-RGB track (\textit{upper panel}: initial metallicity [M/H]$_0$, \textit{middle panel}: $T_{\rm turn-off}$, \textit{lower panel}: depletion gradient $\nabla_{\rm100\,K}$). For target ID specification, see Table~\ref{tab:alllum_paper3}. Red circles represent full disc targets, yellow stars represent transition disc targets, blue squares represent dust-poor disc targets. Grey dots in $T_{\rm turn-off}$ subplot represent the Galactic full disc targets with visually estimated values of $T_{\rm turn-off}$ \citep{kluska2022GalacticBinaries}. Small right panels display the distributions of depletion parameters within the targets hosting transition discs (yellow), dust-poor discs (blue), and all types of discs (gray). For more details, see Section~\ref{sec:dsc_paper3}.}\label{fig:dplteff_paper3}
\end{figure}

\begin{figure}[ph!]
    \centering
    \includegraphics[trim=0 0.8cm 0 0.8cm, width=0.9\linewidth]{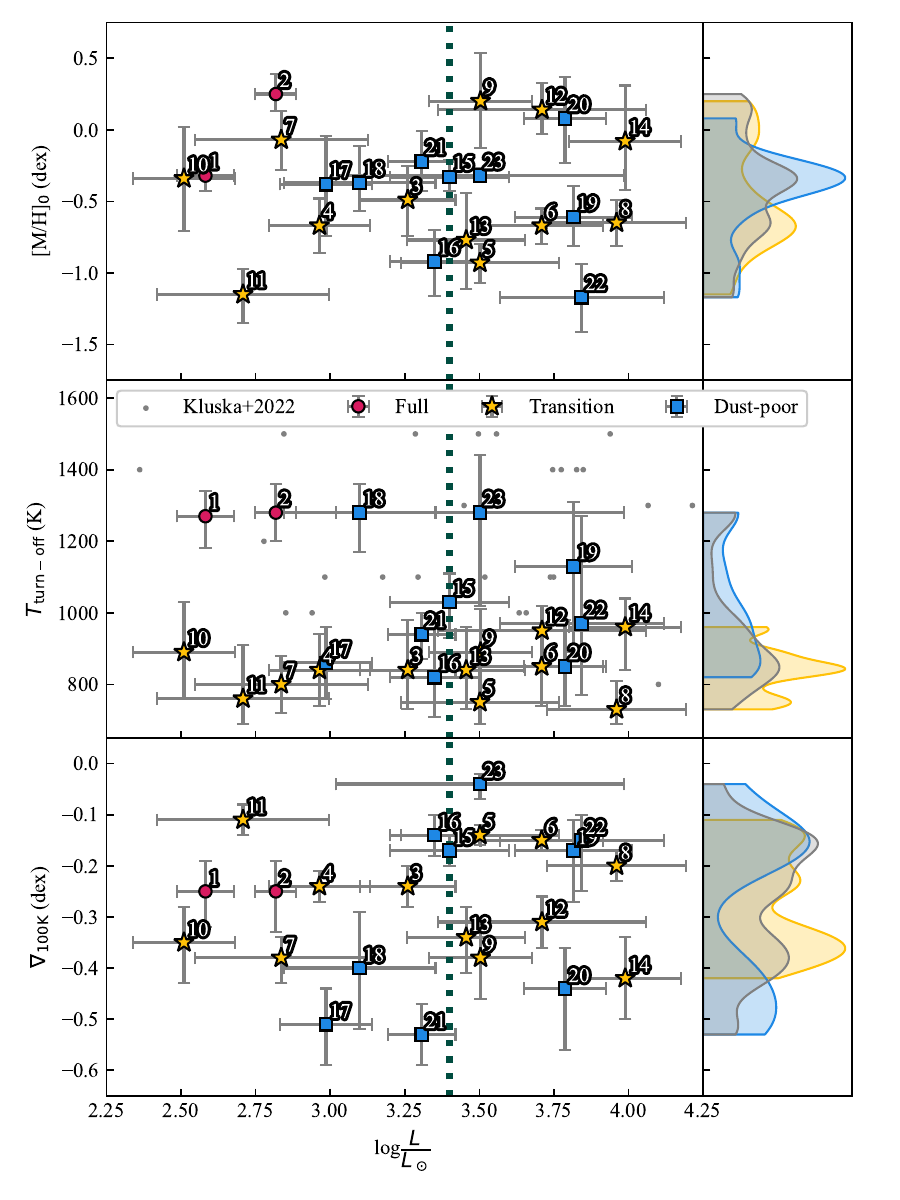}
    \caption[Distributions of depletion pattern parameters with adopted luminosity (\textit{upper panel}: initial metallicity {[M/H]}$_0$, \textit{middle panel}: $T_{\rm turn-off}$, \textit{lower panel}: depletion gradient $\nabla_{\rm100K}$)]{Distributions of depletion pattern parameters with adopted luminosity (\textit{upper panel}: initial metallicity [M/H]$_0$, \textit{middle panel}: $T_{\rm turn-off}$, \textit{lower panel}: depletion gradient $\nabla_{\rm100\,K}$). For target ID specification, see Table~\ref{tab:alllum_paper3}. Red circles represent full disc targets, yellow stars represent transition disc targets, blue squares represent dust-poor disc targets. Grey dots in middle panel represent the Galactic full disc targets with visually estimated $T_{\rm turn-off}$ values \citep{kluska2022GalacticBinaries}. Vertical dotted line represents rough demarcation between post-AGB and post-RGB binaries ($L\,\sim\,2\,500\,L_\odot$; $\log\frac{L}{L_\odot}\,\sim\,3.4$). Small right panels display the distributions of depletion parameters within the targets hosting transition discs (yellow), dust-poor discs (blue), and all types of discs (gray). For more details, see Section~\ref{sec:dsc_paper3}.}\label{fig:dpllum_paper3}
\end{figure}

To further examine trends in depletion efficiency, we also explored how the depletion parameters (initial metallicity [M/H]$_0$, turn-off temperature $T_{\rm turn-off}$, and depletion scale $\nabla_{\rm 100\,K}$) vary with stellar effective temperature and final luminosity (see Fig.~\ref{fig:dplteff_paper3} and \ref{fig:dpllum_paper3}). The effective temperature serves as a proxy for the evolutionary stage along the post-AGB/post-RGB track \citep{bertolami2016tracks, kamath2023models}, while final luminosity helps distinguish between more luminous post-AGB stars and less luminous post-RGB stars, with a rough demarcation at $L\,\sim\,2\,500\,L_\odot$ ($\log\frac{L}{L_\odot}\,\sim\,3.4$), depending on metallicity \citep{kamath2016PostRGBDiscovery}.

We note that our sample size remains limited and the observed trends may be influenced by small-number statistics and observational biases. A larger and more statistically robust sample will be necessary to confirm these findings and refine our understanding of depletion processes. The main findings of our comparison of the NLTE-corrected depletion profiles in the combined sample are summarised below:
\begin{itemize}
    \item In post-AGB and post-RGB binaries, the distributions of initial metallicity [M/H]$_0$ and depletion scale $\nabla_{\rm 100\,K}$ show considerable stochasticity within the sample, without a clear global trend. A small number of systems at the extremes of $\log T_{\rm eff}$ values stand out as notable outliers: GZ\,Nor (\#11), AU\,Vul (\#19), BD+39\,4926 (\#20), J052204 (\#21), and BD+28\,772 (\#23). Four of these targets belong to the dust-poor disc subsample (AU\,Vul, BD+39\,4926, J052204, and BD+28\,772) and are primarily responsible for the apparent bimodal distribution of the depletion scale within this subsample. The overall stochasticity of $\nabla_{\rm 100\,K}$ distribution indicates that current irradiation from the central binary may have minimal impact on the depletion process. Alternatively, this may indicate that CBDs evolve on short timescales, rapidly locking refractory elements into dust. Consequently, the initial chemical differences in the disc material, which arise from the differences in AGB/RGB nucleosynthesis, could be erased before the start of the re-accretion process. To understand the origins of the depletion parameter distribution, we need to develop a physical-chemical model of the CBDs, exploring the condensation, partial advection, and decoupling of the dust within these discs.
    \item In the post-RGB subsample, the $T_{\rm turn-off}$ values are different for the full and transition disc targets ($\sim\,1\,300$\,K and $\sim\,900$\,K, respectively). In the post-AGB subsample, the $T_{\rm turn-off}$ values of the transition disc targets concentrate at $\sim$900\,K (see Figures~\ref{fig:dplteff_paper3} and \ref{fig:dpllum_paper3}). This morphological difference between the full and transition disc targets suggests that different disc-formation mechanisms may be responsible for these disc types. Therefore, based on $T_{\rm turn-off}$, we categorised dust-poor disc targets into two subgroups: i) full-like disc targets with $T_{\rm turn-off}\,>\,1\,100$\,K (R Sct, \#18; AU Vul, \#19; and BD+28\,772, \#23) and ii) transition-like disc targets with $T_{\rm turn-off}\,<\,1\,100$\,K (SS Gem, \#15; V382 Aur, \#16; CC Lyr, \#17; BD+39\,4926, \#20; J052204, \#21; and J053254, \#22). In subsequent studies, we plan to confirm this connection between the $T_{\rm turn-off}$ parameter and the disc type by extending our homogeneous chemical study to full disc targets in the Galaxy and in the LMC.
    \item The depletion observed in post-AGB and post-RGB binaries is remarkably efficient with an average depletion scale $\nabla_{\rm 100\,K}\,\sim\,-0.3$\,dex \citep[for the reference, a typical average depletion scale observed in young and Sun-like planet-hosting stars is $\nabla_{\rm 100\,K}\,\sim\,-0.01$\,dex;][]{yun2024SolarDepletion}. This difference in depletion scales may arise due to a stronger fractionation of gas and dust in CBDs of post-AGB/post-RGB binaries, rather than in young planet-hosting stars \citep{mohorian2025TransitionDiscs}.
    \item Furthermore, the LMC targets -- J050304 (\#13), J053150 (\#14), J052204 (\#21), and J053254 (\#22) -- follow the distributions of the Galactic targets, with no significant deviations observed. This points to the minimal effect of initial metallicity on the efficiency of the depletion process.
\end{itemize}

Finally, we note that separation in $T_{\rm turn-off}$ values is the only significant variation detected in the distribution of depletion parameters across our combined sample of dust-poor, full, and transition disc targets. To gain deeper insight into photospheric depletion observed in post-AGB/post-RGB binaries, our subsequent work will focus on advancing the modelling of the depletion process through re-accretion, building on the initial works of \citet{oomen2019depletion, oomen2020MESAdepletion, martin2025ModellingDepletion}. This will involve incorporating more sophisticated disc dynamics, including photo-evaporation, torques, effects of central binary irradiation, and variations in the chemical composition of re-accreted material.


\subsection{Evolutionary scenarios for dust-poor disc hosts}\label{ssec:dscevo_paper3}
Based on our findings for depletion parameters (stochastic distribution of [M/H]$_0$ and $\nabla_{\rm 100\,K}$; bimodal distribution of $T_{\rm turn-off}$), the dust-poor disc sample could represent two distinct subgroups, which we classified as full-like and transition-like dust-poor discs (see Section~\ref{ssec:dscall_paper3}). We note that although our sample consists of just 23 post-AGB/post-RGB binaries, it offers valuable insights into the depletion process, especially considering the rarity of the post-AGB/post-RGB binaries, since this evolutionary stage lasts for $\lesssim\,0.1$\,Myr \citep{bertolami2016tracks}.

To interpret these subgroups in the context of disc evolution, we propose three possible scenarios for the nature of dust-poor discs in post-AGB/post-RGB binaries:
\begin{enumerate}
    \item \textit{Dust condensation processes were halted or did not operate long enough.} An example of such system may be AU Vul (\#19), which is the coolest (less evolved) target ($T_{\rm eff}\,=\,4740\,\pm\,40$\,K) within our combined sample of post-AGB/post-RGB binaries. Given the shallow $\nabla_{\rm 100\,K}$ value of AU Vul (\#19), it is possible that this target hosts a truly dust-poor disc. Alternatively, the shallow $\nabla_{\rm 100\,K}$ of AU Vul (\#19) could be caused by slower condensation processes in the CBD. The limited sample size of dust-poor disc targets, caused by the rarity of post-AGB/post-RGB binaries, presents a challenge in drawing definitive conclusions. In subsequent works, we will mitigate these constraints by conveying a systematic abundance analysis of full disc targets in the Galaxy and the Magellanic Clouds \citep{kluska2022GalacticBinaries, kamath2014SMC, kamath2015LMC}. This will offer a comprehensive understanding of cool ($T_{\rm eff}\sim\,4\,000$\,K) dust-poor disc targets within the whole post-AGB binary sample.
    \item \textit{The rates of disc re-accretion and decretion cleared most of the gas in the disc.} While the dust-poor disc sample does not appear systematically hotter (more evolved) than full or transition disc samples, 5 of 9 dust-poor disc targets -- SS Gem (\#15), CC Lyr (\#17), BD+39\,4926 (\#20), J052204 (\#21), and BD+28\,772 (\#23) -- have higher $T_{\rm eff}$ values ($T_{\rm eff}\,>\,6\,500$\,K) than full and transition disc targets. These hottest dust-poor disc targets may represent the final stages of CBD evolution for both full and transition discs. This scenario is plausible, as reported mass-loss rates of CBD material due to jets range from $10^{-8}$ to $10^{-6}\,M_\odot$/yr \citep{bollen2022Jets, verhamme2024DiscWindModelling}. Recently, \citep{deprins2024Jets} conducted the first MHD modelling of jets and found mass-loss rates ranging from $10^{-6}$ to $10^{-3}\,M_\odot$/yr. However, they noted that such high mass-loss rates are unrealistic for post-AGB/post-RGB binaries. Consequently, for the most conservative case, a binary system would lose $\sim\,10^{-3}\,M_\odot$ of CBD matter over the $\sim\,10^5$-year duration of the post-AGB phase \citep{bertolami2016tracks}. This mass is substantial compared to the total disc masses observed in post-AGB/post-RGB binaries, which are typically \citep[$\lesssim\,10^{-2}\,M_\odot$; see, e.g., Table 4.7 in][]{gallardocava2023thesis}. To confirm this scenario, we need to directly analyse the dust-poor discs with interferometric facilities like ALMA or VLTI.
    \item \textit{The discs become dust-poor due to planet formation processes.} In this scenario, the growth of pebbles/planetesimals creates pressure bumps and efficiently traps dust within the disc \citep{eriksson2020PebblesPlanetesimals, shibaike2023PebblesPlanetesimals, drazkowska2023DustPlanetFormation, price2025PebblesPlanetesimals}. Observational signatures of planet formation processes, including disc asymmetries, or planetary-induced spiral structures, could provide critical constraints on this hypothesis. To further investigate this possibility, we need to adapt the current planet formation models \citep[see, e.g.,][and references therein]{raymond2022PlanetFormationModels} of protoplanetary discs around young stars to the unique conditions reported for the CBDs around post-AGB/post-RGB binaries, including stellar irradiation, gas-to-dust ratios, distribution of gas and dust.
\end{enumerate}

We note that the proposed scenarios are not mutually exclusive and may not encompass the full diversity of dust-poor disc targets. Our ongoing observational and theoretical efforts will focus on identifying the evolution and mechanism(s) of photospheric depletion in post-AGB/post-RGB binaries and explaining why dust-poor and transition disc systems are relatively rare within the post-AGB/post-RGB sample.

\subsection{Case study: the least depleted dust-poor disc target BD+28 772}\label{ssec:dscbd2_paper3}
The dust-poor disc sample contains post-AGB/post-RGB binaries, which show depletion profiles with depletion scales $\nabla_{\rm 100\,K}$ ranging from $\sim\,0.1$ to $\sim\,0.5$\,dex per 100\,K. However, there is a unique dust-poor disc target, BD+28\,772, with an unusually shallow depletion scale ($\nabla_{\rm 100\,K}^{\rm BD+28\,772}\,=\,-0.04_{-0.03}^{+0.02}$\,dex; see Fig.\ref{fig:dpl4_paper3}). The derived luminosity ($L\,=\,3\,180_{-2\,130}^{+6\,480}\,L_\odot$) and effective temperature ($T_{\rm eff}\,=\,6\,880\,\pm\,100$\,K) of BD+28\,772 are typical within the post-AGB/post-RGB sample (see Table~\ref{tab:respar_paper3}). We note that the broad uncertainty range of final luminosity of BD+28\,772 ($L_{\rm SED}\,=\,3\,180_{-2\,140}^{+6\,460}$) does not allow to conclusively state whether BD+28\,772 is a post-AGB or post-RGB binary. Furthermore, the CNO abundances of BD+28\,772 are typical within the post-AGB/post-RGB sample ([C/H]\,=\,--0.46$\,\pm\,$0.11\,dex, [N/H]\,=\,+0.26$\,\pm\,$0.17\,dex, [O/H]\,=\,--0.28$\,\pm\,$0.04\,dex).

Among previously studied post-AGB/post-RGB binaries, there are 5 reported systems with depletion profiles similar to BD+28\,772: BZ Pyx, BT Lib, DS Aqr, V453 Oph, and V820 Cen \citep{gezer2015WISERVTau}. Out of these 5 systems, only BZ Pyx has a recently derived SED luminosity using \textit{Gaia} DR2 parallaxes \citep[$L_{\rm SED}^{\rm BZ Pyx}\,=\,380_{-70}^{+90}$;][]{oomen2019depletion}. Given the luminosities and depletion profiles of BD+28\,772 and BZ Pyx, we suggest that these targets (and potentially other `non-depleted' post-RGB binaries) could be post-RGB counterparts of `non-depleted' $s$-process enhanced post-AGB binaries \citep[e.g., J005107, MACHO 47.2496.8, and HD 158616][]{menon2024EvolvedBinaries}. In this case, the nitrogen enhancement by $\sim\,$0.6\,dex, noticeable in the depletion profile (see Fig.~\ref{fig:dpl4_paper3}), is likely a result of the first dredge-up and extra mixing during the RGB phase (see Fig.~\ref{fig:modszm_paper1} and \ref{fig:moddfc_paper1} in Section~\ref{sec:mod_paper1}). In subsequent works, we will investigate this hypothesis by simulating the CNO enrichment of BD+28\,772 with stellar evolution models \citep[e.g., ATON;][]{ventura2008aton3}.

\section{Conclusions}\label{sec:con_paper3}
The main aim of this study is to explore the connection between CBD evolution and photospheric depletion in post-AGB/post-RGB binaries with dust-poor discs. For this aim, we conducted a detailed chemical analysis using high-resolution optical spectra obtained with HERMES/Mercator and UVES/VLT. We used \texttt{E-iSpec} to derive the precise atmospheric parameters and elemental abundances of 9 post-AGB/post-RGB binaries with dust-poor discs in the Galaxy and in the LMC. Furthermore, we used \texttt{pySME} to calculate NLTE corrections for individual lines of a representative set of chemical elements from C to Ba. 

To consistently analyse the depletion profiles of dust-poor disc targets, we fitted them with 2-piece linear functions with 3 free parameters (initial metallicity [M/H]$_0$, turn-off temperature $T_{\rm turn-off}$, and depletion scale $\nabla_{\rm 100\,K}$). The analysis of NLTE-corrected depletion profiles of dust-poor disc targets revealed their saturation and bimodality of $T_{\rm turn-off}$ distribution.

To provide context for our results in dust-poor disc targets, we combined our sample with post-AGB/post-RGB binaries hosting full and transition discs. The distribution of [M/H]$_0$ and $\nabla_{\rm 100\,K}$ in the combined sample of post-AGB/post-RGB binaries exhibits continuity, with no evident systematic trends. In contrast, the distribution of $T_{\rm turn-off}$ in the combined sample of post-AGB/post-RGB binaries peaks around $\sim\,900$ and $\sim\,1\,300$\,K. Based on this feature of the $T_{\rm turn-off}$ distribution, we categorised dust-poor disc sample into two distinct subgroups: i) full-like disc targets and ii) transition-like disc targets. This categorisation implies that different disc-formation mechanisms may be responsible for the morphological difference of depletion profiles in full and transition disc targets. We note that our sample size remains limited and the observed trends may be influenced by small-number statistics and observational biases. The scope of this study will be expanded by our ongoing homogeneous abundance analysis of post-AGB/post-RGB binaries with full discs in the Galaxy and the LMC.

Additionally, we identified some peculiar targets within the dust-poor disc sample. One example is BD+28\,772 (\#23), which exhibits little to no chemical depletion, setting it apart from other post-AGB and post-RGB binaries in our combined sample of dust-poor, full, and transition disc targets. This distinct behaviour challenges current understanding of the depletion process and raises the question of whether BD+28\,772 (\#23) follows a unique evolutionary path or belongs to a previously unrecognised population of post-RGB binaries. To address this, we need to expand our homogeneous chemical study to include other reported `non-depleted' systems.

The diversity of depletion profiles in the dust-poor, full, and transition disc samples highlights the complexity of disc-binary interactions in post-AGB/post-RGB binary stars. To better understand the processes driving photospheric depletion in these systems, we are expanding both our sample and our methodological approach to enable determining whether depletion parameters can serve as indicators of planet formation processes in circumbinary discs around post-AGB/post-RGB binaries.

\section*{Acknowledgements}\label{sec:ack_paper3}
The spectroscopic results presented in this paper are based on observations made with the Mercator Telescope, operated on the island of La Palma by the Flemish Community, at the Spanish Observatorio del Roque de los Muchachos of the Instituto de Astrofisica de Canarias. This research is based on observations collected at the European Organisation for Astronomical Research in the Southern Hemisphere under ESO programmes 074.D-0619 and 092.D-0485. This research was supported by computational resources provided by the Australian Government through the National Computational Infrastructure (NCI) under the National Computational Merit Allocation Scheme and the ANU Merit Allocation Scheme (project y89).

MM1 acknowledges the International Macquarie Research Excellence Scholarship program (iMQRES) for the financial support during the research. MM1, DK, and HVW acknowledge the support from the Australian Research Council Discovery Project DP240101150. AMA acknowledges support from the Swedish Research Council (VR 2020-03940) and from the Crafoord Foundation via the Royal Swedish Academy of Sciences (CR 2024-0015). HVW acknowledges support from the Research Council, KU Leuven under grant number C14/17/082.

\begin{savequote}[75mm]
\foreignlanguage{ukrainian}{
``То трудно вірить, щоб погану одіж\\
Могла носить якась ідея гарна...''}
\qauthor{\foreignlanguage{ukrainian}{---Леся Українка, ``Руфін і Прісцілла'' (1911)}}
``It’s hardly true that such a great idea\\
Could wear the garments which are coarse or vile...''
\qauthor{---Lesia Ukrainka, ``Rufin and Priscilla'' (1911)}
\end{savequote}

\chapter{Conclusions \& Future prospects}\label{chp:dsc}
\graphicspath{{ch_discussion/figures/}} 

\clearpage

\textit{In this chapter, we summarise the main conclusions of this thesis (see Section~\ref{sec:dsccon}). Additionally, we discuss potential directions for future research, with a focus on binary evolution, chemical processes in circumbinary discs, and their implications for second-generation planet formation (see Section~\ref{sec:dscfut}).} 

\section{Conclusions}\label{sec:dsccon}
In this thesis, we explored the critical role of disc-binary interactions in shaping element and isotope production in post-AGB and post-RGB binaries hosting second-generation protoplanetary discs. We specifically focused on how the fractionation of volatile-rich gas and refractory-rich dust in circumbinary discs, followed by the re-accretion of dust-poor matter, leads to the characteristic photospheric chemical depletion observed in these systems.

To quantitatively assess the influence of disc-binary interactions on the surface composition of post-AGB/post-RGB binaries, we developed a consistent and homogeneous analysis framework E-iSpec, which is one of the key outcomes of this research. E-iSpec is a specialised spectral analysis tool designed for evolved stars with complex atmospheres, including post-AGB/post-RGB single and binary stars. While existing spectral analysis tools often struggle with dusty metal-poor giants, E-iSpec provides reliable methods for determining atmospheric parameters, elemental abundances, and, for the first time, isotopic ratios in the photospheres of these stars. Using E-iSpec, we analysed high-resolution optical and NIR spectra of a diverse sample of 23 post-AGB/post-RGB binaries hosting full, transition, and dust-poor discs (19 Galactic targets and 4 LMC targets), forming the foundation for our depletion analyses.

To assess the impact of disc-binary interactions on surface composition in Galactic post-RGB binaries, we conducted a detailed spectral analysis of SZ Mon and DF Cyg -- two targets with full circumbinary discs (discs, where the dust is distributed mostly similar to the gas). Our findings demonstrate that while the depletion patterns in both post-RGB and post-AGB binaries are largely similar, a key distinction emerges in the temperature onset of significant depletion: $T_{\rm turn-off}^{\rm post-AGB}\,\lesssim\,1\,100$ K compared to $T_{\rm turn-off}^{\rm post-RGB}\,\sim\,1\,300$ K. This difference hints at distinct efficiencies of disc-binary interaction processes in post-AGB and post-RGB binaries. Our analysis of the $^{12}$C/$^{13}$C isotopic ratio in SZ\,Mon and DF\,Cyg suggests a mild depletion of C. However, studying isotopic ratios in a larger sample of post-AGB/post-RGB binaries will be essential to confirm and strengthen this result.

Then, we performed a comprehensive spectral analysis of 12 Galactic and LMC post-AGB/post-RGB binaries with transition circumbinary discs (discs, which have dust inner cavities). Our findings highlight the critical importance of correcting for NLTE effects in post-AGB/post-RGB binaries ($\gtrsim$\,0.5\,dex for some elements). Notably, our study revealed that the updated NLTE depletion profiles for all studied binaries are saturated, indicating that the observed surface compositions reflect the chemical composition of re-accreted gas rather than the original stellar surface composition. Additionally, we detected a trend for turn-off temperatures $T_{\rm turn-off}$ in the transition disc sample: post-AGB binaries generally have lower $T_{\rm turn-off}$ values than post-RGB binaries. Furthermore, we compared chemical depletion in post-AGB/post-RGB binaries with transition discs, in YSOs with transition discs, and in the ISM. We showed that post-AGB/post-RGB depletion is on average $\sim\,10$ stronger than YSO depletion, but similar to ISM depletion.

Finally, we conducted a detailed spectral analysis of 9 Galactic and LMC post-AGB and post-RGB binaries with dust-poor circumbinary discs (discs, which display low to no IR excess in the SED). The resulting depletion profiles revealed a clear diversity in the dust-poor disc targets. To better understand this diversity, we expanded our sample to include full and transition disc systems from our previous spectral analysis studies, providing a broader comparative framework. We used pySME to consistently recalculate the NLTE corrections for this extended sample of post-AGB/post-RGB binaries. Using updated NLTE depletion profiles, we detected that based on $T_{\rm turn-off}$ values, dust-poor targets can be divided into two distinct subgroups (full-like and transition-like), pointing to multiple evolutionary pathways for circumbinary disc evolution. Furthermore, our findings reveal a striking bimodality of turn-off temperatures $T_{\rm turn-off}$ among post-RGB stars: $T_{\rm turn-off}\,\sim\,1300$\,K for targets hosting full discs and full-like dust-poor discs, $T_{\rm turn-off}\,\sim\,800$\,K for targets hosting transition discs and transition-like dust-poor discs. This highlights previously unrecognised bimodality in the depletion histories of post-RGB binaries. In addition, we reassessed the extreme chemical depletion reported for 2 post-AGB binaries with dust-poor discs (CC Lyr, \#17 and BD+39 4926, \#20) and showed that their depletion scale, $\nabla_{\rm 100\,K}\,\sim\,-0.5$ dex, does not reach the extreme levels previously suggested by their [Zn/Ti]\,$\sim$\,3 dex. Finally, we established fitting parameters [M/H]$_0$, $T_{\rm turn-off}$ and $\nabla_{\rm 100\,K}$ as consistent and independent variables for tracing the extent of photospheric chemical depletion in post-AGB/post-RGB binaries.

Overall, this research provides a comprehensive framework for investigating how disc-binary interactions govern photospheric chemical depletion in post-AGB/post-RGB systems. By homogeneously analysing high-resolution spectra of a diverse sample of post-AGB/post-RGB binaries, we established that the depletion profile in all studied stars is saturated, despite previous classifications. Our results demonstrate that while post-AGB and post-RGB binaries share similar depletion patterns, key differences emerge in $T_{\rm turn-off}$ values, with post-AGB stars displaying significant depletion at lower $T_{\rm turn-off}$ values than their post-RGB counterparts. The discovery of a bimodality in depletion histories among post-RGB binaries and the classification of dust-poor circumbinary discs into two distinct evolutionary pathways highlight the complexity of chemical processing in the circumbinary discs. Furthermore, we found no direct correlation between the newly defined depletion profile parameters ([M/H]$_0$, $T_{\rm turn-off}$, and $\nabla_{\rm 100\,K}$) and fundamental properties of the systems, reinforcing the notion that depletion operates independently of central binary and is governed primarily by disc evolution. These findings challenge previous assumptions and establish $T_{\rm turn-off}$ and $\nabla_{\rm 100\,K}$ as robust and independent tracers of chemical depletion in evolved binaries, providing a solid foundation for future modelling of disc-binary interactions in young and evolved stars (see Section~\ref{sec:dscfut}).

\section{Future prospects}\label{sec:dscfut}
In the coming decade, next-generation photometric and spectroscopic facilities will offer unprecedented opportunities to study post-AGB and post-RGB binaries with circumbinary discs. High-resolution spectrographs such as ANDES and MOSAIC on the ELT, along with Sloan Digital Sky Survey, 4MOST/VISTA, and MAVIS/VLT, will enable comprehensive and homogeneous chemical analyses of these systems in the Magellanic Clouds and fainter ($V > 14^m$) targets in the Galaxy, overcoming previous observational limitations.

Despite their rarity due to an evolutionary phase lasting only $\sim\,10^5$ years, post-AGB/post-RGB binaries exhibit remarkable chemical diversity. Our findings for transition and dust-poor disc systems highlight the need for systematic and consistent spectral analyses of full disc targets in the Galaxy and in the LMC, as well as post-AGB/post-RGB binaries in the SMC. This approach will enable a robust analysis of depletion parameters across the entire sample, building on the findings presented in this study. This will allow to robustly analyse the behaviour of the depletion parameters across the whole studied sample and to investigate the potential links between full, transition, and dust-poor discs, elaborating on the findings presented in this study.

Current and planned infrared facilities, including JWST ($H\,\lesssim\,30^m$) and METIS/ELT ($L'\,<\,21.8^m$), will provide essential spectroscopic data to investigate CNO isotopic ratios in stellar surfaces ($\sim$2\,$\mu$m), as well as carbonaceous and silicate emission features in circumbinary discs ($\sim$20\,$\mu$m). The CNO isotopic ratios will allow comparing the observed chemistry with theoretical predictions (using stellar evolutionary models, such as ATON), while the emission features will trace refractory-rich dust, which is trapped in the disc and is thus excluded from the re-accretion process that enriches the stellar surface. Combining depletion patterns and dust emission features will allow a more comprehensive investigation of the gas and dust distribution within circumbinary discs, refining our understanding of disc composition and evolution. In parallel, our research group is developing the theoretical framework for second-generation planet formation and using high-angular-resolution imaging to characterise the gas and dust properties of the circumbinary discs. Together, these efforts will deepen our understanding of the impact of disc-binary interactions on system evolution in post-AGB/post-RGB binaries.

A key priority for future studies is to understand why the distribution of depletion parameters remains highly stochastic and independent of key properties of post-AGB/post-RGB binaries, including surface temperature, luminosity, and pulsational and orbital parameters. Based on this independence, we suggest that the depletion patterns are mainly governed by processes occurring in the circumbinary discs. Testing this hypothesis will require expanding on the works of \citet{oomen2019depletion, oomen2020MESAdepletion, martin2025ModellingDepletion}, who used MESA code to model photospheric depletion through re-accretion. A crucial advancement in such modelling will involve incorporating more sophisticated disc dynamics (by including torques and photo-evaporation), stellar irradiation, and the diversity of depletion patterns of the re-accreted matter, as presented in this thesis.

Another critical avenue of future research is to quantify the extent of volatile depletion in post-AGB and post-RGB binary systems. Volatile elements (including O, S, and Zn) are weakly affected by dust condensation and often serve as tracers of initial stellar metallicity. The mild volatile depletion is further confirmed by $^{12}$C/$^{13}$C isotopic ratios observed in SZ\,Mon and DF\,Cyg. In this thesis, we assume that elements with condensation temperatures below $T_{\rm turn-off}$ represent the initial metallicity [M/H]$_0$, while elements with condensation temperatures above $T_{\rm turn-off}$ trace the depletion scale $\nabla_{\rm 100\,K}$. By comparing observed volatile abundances in post-AGB/post-RGB binaries with predictions from evolutionary models of AGB/RGB single stars (e.g., ATON, MESA, and FRUITY), we can gain insight into how depletion processes modify the abundance of volatile elements in post-AGB/post-RGB binaries.

This thesis has primarily advanced the observational characterisation of chemical depletion in post-AGB and post-RGB binaries. The next independent research steps emerging from this work are two-fold: i) completing the homogeneous abundance analysis of full-disc systems in the Galaxy and in the LMC, and ii) implementing more elaborate modelling of the depletion process, using the precise chemical constraints established in this study. In terms of elemental coverage, the current depletion profiles are largely limited by detectability, and there is little room for expanding the number of chemical elements in the depletion profiles. A possible exception may involve observing pulsating targets at cooler phases (at effective temperatures of 5\,000\,--\,5\,500\,K) using high-resolution spectrographs (e.g., UVES/VLT, HERMES/Mercator) to enhance the detectability of weaker atomic lines. These steps will help refine both the empirical foundation and the theoretical interpretation of the depletion phenomenon.

Although the post-AGB/post-RGB stage is relatively short, the systematic depletion of refractory elements makes post-AGB/post-RGB binaries valuable tracers of disc-mediated chemical processing. While individually rare, the fraction of LIM binary systems passing through the post-AGB/post-RGB stage may be high enough to influence Galactic chemical evolution. Excluding post-AGB/post-RGB binaries from the chemical evolution models risks overlooking processes that alter key abundance patterns of the gas phase, such as [X/Fe] abundance ratios and CNO isotopic ratios.

Finally, a major challenge lies in understanding the fundamental physical mechanisms driving disc-star and disc-binary interactions, which regulate photospheric depletion across various evolutionary stages, from young to evolved stars. Expanding our homogeneous chemical analysis to include refractory-depleted T~Tauri, Herbig Ae/Be, and Sun-like stars will establish a critical observational foundation for exploring the relationship between photospheric chemical depletion, disc evolution, and second-generation planet formation. This, in turn, will advance our understanding of the chemical enrichment of the Universe.

\appendix

\begin{savequote}[75mm]
\foreignlanguage{ukrainian}{``В житті все треба робити ритмічно, в хорошому темпі і ніколи не поспішати...''}
\qauthor{\foreignlanguage{ukrainian}{---Соломія Крушельницька (1872-1952), \\ українська оперна співачка, педагогиня}}
``In life, everything should be done rhythmically, at a good pace and never in a hurry...''
\qauthor{---Solomiya Krushelnytska (1872-1952), \\Ukrainian opera singer and teacher}
\end{savequote}

\chapter{Appendix to chapter 3}\label{chp:app3}

\clearpage

\section{Evolved candidates observed with APOGEE}\label{app:tar_paper1}
In this Appendix, we present the initial target sample for this study, which consisted of 36 post-RGB/post-AGB candidates (see Table~\ref{tabA:allapo_paper1}). The MW (Milky Way, Galactic) targets were chosen from \cite{gielen2011silicates, gezer2015WISERVTau, oomen2018OrbitalParameters, kamath2022GalacticSingles, kluska2022GalacticBinaries}, the SMC targets were selected from \cite{kamath2014SMC}, and the LMC targets were chosen from \cite{vanaarle2011PAGBsInLMC, woods2011SAGE, matsuura2014PAHsInPAGBs, kamath2015LMC}. The targets were selected by their IRAS, MSX and 2MASS colours such that they were feasible to be observed with the APOGEE survey (see Section~\ref{sssec:obsspcnir_paper1}) with an S/N greater than $\sim50$ needed for a precise chemical analysis. We filtered this initial sample using the following criteria:
\begin{enumerate}
    \item the final targets were previously observed with high-resolution optical spectrographs (UVES, HERMES). This allowed us to accurately determine atmospheric parameters and, if possible, the abundances of carbon, nitrogen, and oxygen (see Section~\ref{sec:san_paper1});
    \item the target's spectroscopically derived effective temperature should be below 5000 K, enabling the detection and study of CNO molecular bands (CO, CN).
\end{enumerate}

After performing the selection cuts, our final sample comprised two post-RGB binary stars: SZ~Mon and DF~Cyg (see Section~\ref{sec:tar_paper1}).

\begin{sidewaystable}[ph!]
    \centering
    \tiny
    \caption{Initial target sample containing all confirmed post-AGB/post-RGB stars which were observed with APOGEE.}\label{tabA:allapo_paper1}
    \begin{tabular}{|l|c|c|c|c|c|c|c|}
    \hline
        \textbf{Galaxy} & \textbf{2MASS name} & \textbf{IRAS/SAGE name}$^{a}$ & \textbf{Other Names} & \textbf{R.\,A.} & \textbf{Dec.} & \textbf{SED Type} & \textbf{Reference} \\
        ~ & ~ & ~ & ~ & \textbf{(deg)} & \textbf{(deg)} & ~ & ~ \\ \hline
        MW & J05075028+4824094 & IRAS 05040+4820 & BD+48 1220 & 076.959515 & 48.402634 & Shell & 1 \\
        ~ & J05365506+0854087 & IRAS 05341+0852 & -- & 084.229417 & 08.902422 & Shell & 1 \\
        ~ & J05405705+1014249 & IRAS 05381+1012 & BD+10 845 & 085.237725 & 10.240270 & Shell & 1 \\
        ~ & J06512784-0122158 & IRAS 06489-0118 & SZ Mon & 102.866018 & -01.371072 & Disc & 1 \\
        ~ & J06553181-0217283 & IRAS 06530-0213 & -- & 103.882582 & -02.291199 & Shell & 1 \\
        ~ & J15183614+0204162 & IRAS F15160+0215 & -- & 229.650616 & 02.071189 & Non-IR excess & 1 \\
        ~ & J15585827+2608046 & - & BD+26 2763 & 239.742810 & 26.134620 & Non-IR excess & 1 \\
        ~ & J17470327+2319454 & IRAS 17449+2320 & BD+23 3183 & 266.763660 & 23.329260 & Disc & 2 \\
        ~ & J19213906+3956080 & IRAS 19199+3950 & HP Lyr & 290.412790 & 39.935580 & Disc & 3 \\
        ~ & J19361378+0704184 & - & V870 Aql & 294.057430 & 07.071800 & Non-IR excess & 1 \\
        ~ & J19485394+4302145 & IRAS 19472+4254 & DF Cyg & 297.224760 & 43.037370 & Disc & 1 \\ \hline
        SMC & J00383006-7303340 & J003829.99-730334.1 & SSTISAGEMC J003830.04-730334.1 & 009.625263 & -73.059448 & Uncertain & 4 \\
        ~ & J00443128-7305496 & J004431.23-730549.3 & SSTISAGEMC J004431.30-730549.9 & 011.130125 & -73.097028 & Uncertain & 4 \\
        ~ & J00444111-7321361 & J004441.03-732136.0 & OGLE SMC-LPV-4910 & 011.171296 & -73.360039 & Shell & 4 \\
        ~ & J00445628-7322566 & J004456.21-732256.6 & SSTISAGEMC J004456.26-732256.8 & 011.234533 & -73.382408 & Uncertain & 4 \\
        ~ & J00494423-7252088 & J004944.15-725209.0 & SSTISAGEMC J004944.19-725208.9 & 012.434298 & -72.869118 & Uncertain & 4 \\
        ~ & J00510723-7341334 & J005107.19-734133.3 & SSTISAGEMC J005107.24-734133.3 & 012.780143 & -73.692612 & Disc & 4 \\
        ~ & J00522223-7335376 & J005222.19-733537.6 & SSTISAGEMC J005222.23-733537.6 & 013.092458 & -73.593778 & Uncertain & 4 \\
        ~ & J00530734-7344045 & J005307.35-734404.5 & SSTISAGEMC J005307.36-734404.5 & 013.280625 & -73.734583 & Uncertain & 4 \\
        ~ & J00551576-7125168 & J005515.71-712516.9 & SSTISAGEMC J005515.76-712516.9 & 013.815672 & -71.421349 & Uncertain & 4 \\
        ~ & J00590901-7106486 & J005908.99-710648.6 & SSTISAGEMC J005909.01-710648.5 & 014.787543 & -71.113503 & Disc & 4 \\ \hline
        LMC & J04524318-7047371 & J045243.16-704737.3 & SSTISAGEMC J045243.16-704737.3 & 073.179954 & -70.793655 & Disc & 5 \\
        ~ & J04562323-6927489 & J045623.21-692749.0 & SSTISAGEMC J045623.21-692749.0 & 074.096708 & -69.463611 & Disc & 5 \\
        ~ & J04565524-6827330 & J045655.23-682732.9 & SSTISAGEMC J045655.23-682732.9 & 074.230206 & -68.459175 & Disc & 5 \\
        ~ & J05022115-6913171 & J050221.17-691317.2 & SSTISAGEMC J050221.17-691317.2 & 075.588208 & -69.221444 & Shell & 5 \\
        ~ & J05145312-6917234 & J051453.10-691723.5 & SSTISAGEMC J051453.10-691723.5 & 078.721250 & -69.289861 & Uncertain & 5 \\
        ~ & J05172873-6942469 & J051728.71-694246.7 & - & 079.369745 & -69.713028 & Shell & 5 \\
        ~ & J05184886-7002469 & J051848.84-700247.0 & SSTISAGEMC J051848.84-700247.0 & 079.703619 & -70.046387 & Shell & 5 \\
        ~ & J05190686-6941539 & J051906.86-694153.9 & SSTISAGEMC J051906.86-694153.9 & 079.778608 & -69.698326 & Shell & 5 \\
        ~ & J05213559-6951572 & J052135.62-695157.1 & SSTISAGEMC J052135.62-695157.1 & 080.398297 & -69.865906 & Uncertain & 5 \\
        ~ & J05214797-7009569 & J052147.95-700957.0 & SSTISAGEMC J052147.95-700957.0 & 080.449915 & -70.165810 & Disc & 5 \\
        ~ & J05220425-6915206 & J052204.24-691520.7 & SSTISAGEMC J052204.24-691520.7 & 080.517725 & -69.255730 & Disc & 5 \\
        ~ & J05221852-6950134 & J052218.52-695013.3 & SSTISAGEMC J052218.52-695013.3 & 080.577187 & -69.837074 & Disc & 5 \\
        ~ & J05254820-6937002 & J052548.17-693700.1 & - & 081.450854 & -69.616737 & Uncertain & 5 \\
        ~ & J05345377-6908020 & J053453.75-690802.0 & SSTISAGEMC J053453.75-690802.0 & 083.724040 & -69.133910 & Disc & 6 \\
        ~ & J05455567-7057305 & J054555.68-705730.3 & SSTISAGEMC J054555.68-705730.3 & 086.481984 & -70.958481 & Disc & 5 \\ \hline
    \end{tabular}\\
    \textbf{Notes:} $^{a}$IRAS/SAGE names were adopted from \cite{kamath2014SMC} for SMC targets and from \cite{kamath2015LMC} for LMC targets. References for SED types: $^1$\cite{gezer2015WISERVTau}, $^2$Van Winckel (private communication), $^3$\cite{oomen2018OrbitalParameters}, $^4$\cite{kamath2014SMC}, $^5$\cite{kamath2015LMC}, $^6$\cite{vanaarle2011PAGBsInLMC}.
\end{sidewaystable}

\section{Additional data for SZ Mon and DF Cyg}\label{app:add_paper1}
In this Appendix, we provide the information about all optical and NIR observations of SZ~Mon and DF~Cyg.

In Tables~\ref{tabA:szmvis_paper1} and \ref{tabA:dfcvis_paper1}, we present a comprehensive summary of the optical and NIR observations conducted on two target stars: SZ~Mon and DF~Cyg, respectively. Table~\ref{tab:obslog_paper1} contains observational visits, which we selected for our chemical analysis.

\begin{table}[!ht]
    \centering
    \footnotesize
    \caption[Optical and near-infrared visits of SZ Mon]{All optical and near-infrared visits of SZ Mon. This table is published in its entirety in the electronic edition of the paper. A portion is shown here for guidance regarding its form and content.}\label{tabA:szmvis_paper1}
    \begin{tabular}{|c@{\hspace{0.2cm}}c@{\hspace{0.2cm}}c@{\hspace{0.2cm}}c@{\hspace{0.2cm}}c@{\hspace{0.2cm}}|} \hline
        \textbf{Visit \#} & \textbf{Obs. ID} & \textbf{BJD} & \textbf{Exp. (s)} & \textbf{RV (km/s)} \\ \hline
        \multicolumn{5}{|c|}{\textit{Mercator + HERMES}} \\ \hline
        SH\#1 & 00260298 & 2455160.7710 & 800 & --19.53$\pm$0.02 \\
        SH\#2 & 00272588 & 2455217.4926 & 1200 & 5.49$\pm$0.05 \\
        SH\#3 & 00313718 & 2455497.7400 & 850 & 50.83$\pm$0.30 \\
        SH\#4 & 00314466 & 2455507.6421 & 800 & 27.24$\pm$0.06 \\
        SH\#5 & 00326956 & 2455572.6357 & 1000 & 8.47$\pm$0.07 \\
        \multicolumn{5}{|c|}{\ldots} \\ \hline
    \end{tabular}
\end{table}

\begin{table}[!ht]
    \centering
    \footnotesize
    \caption[Optical and near-infrared visits of DF Cyg]{All optical and near-infrared visits of DF Cyg. This table is published in its entirety in the electronic edition of the paper. A portion is shown here for guidance regarding its form and content.}\label{tabA:dfcvis_paper1}
    \begin{tabular}{|c@{\hspace{0.2cm}}c@{\hspace{0.2cm}}c@{\hspace{0.2cm}}c@{\hspace{0.2cm}}c@{\hspace{0.2cm}}|} \hline
        \textbf{Visit \#} & \textbf{Obs. ID} & \textbf{BJD} & \textbf{Exp. (s)} & \textbf{RV (km/s)} \\ \hline
        \multicolumn{5}{|c|}{\textit{Mercator + HERMES}} \\ \hline
        DH\#1 & 00239921 & 2455010.5832 & 3149 & 2.04$\pm$0.47 \\
        DH\#2 & 00240154 & 2455013.6518 & 1800 & 6.03$\pm$0.31 \\
        DH\#3 & 00240155 & 2455013.6733 & 1800 & 5.85$\pm$0.32 \\
        DH\#4 & 00240156 & 2455013.6948 & 1800 & 5.79$\pm$0.30 \\
        DH\#5 & 00307940 & 2455475.4697 & 1200 & --53.72$\pm$0.17 \\
        \multicolumn{5}{|c|}{\ldots} \\ \hline
    \end{tabular}
\end{table}

In Table~\ref{tabA:optlst_paper1}, we provide the optical line lists for SZ~Mon and DF~Cyg. In Tables~\ref{tabA:nirlstszm_paper1} and \ref{tabA:nirlstdfc_paper1} (for SZ~Mon and DF~Cyg, respectively), we specify the spectral windows, which were sensitive to different elements and show the spectral features in these regions.

\begin{table}[!ht]
    \centering
    \footnotesize
    \caption[Optical line list for SZ Mon and DF Cyg]{Optical line list. This table is published in its entirety in the electronic edition of the paper. A portion is shown here for guidance regarding its form and content.}\label{tabA:optlst_paper1}
    \begin{tabular}{|r@{\hspace{0.25cm}}c@{\hspace{0.25cm}}c@{\hspace{0.25cm}}c@{\hspace{0.25cm}}|c@{\hspace{0.25cm}}c@{\hspace{0.25cm}}c@{\hspace{0.25cm}}|c@{\hspace{0.25cm}}c@{\hspace{0.25cm}}c@{\hspace{0.25cm}}|} \hline
        \multicolumn{4}{|c|}{\textbf{Atomic data}} & \multicolumn{3}{c}{\boldmath$W_{\lambda, {\rm SZ\,Mon}}$ \textbf{(m\AA)}} & \multicolumn{3}{|c|}{\boldmath$W_{\lambda, {\rm DF\,Cyg}}$ \textbf{(m\AA)}} \\
        \textbf{Element} & \boldmath$\lambda$ \textbf{(nm)} & \boldmath$\log gf$ & \boldmath$\chi$ \textbf{(eV)} & \textbf{SH\#73} & \textbf{SH\#29} & \textbf{SH\#47} & \textbf{DH\#83} & \textbf{DH\#26} & \textbf{DH\#54} \\ \hline
        \ion{C}{i} & 493.2049 & -1.658 & 7.685 & 93.9 & 96.6 & - & - & - & - \\
        \ion{C}{i} & 538.0325 & -1.616 & 7.685 & 86.3 & 95.1 & - & 106.0 & 112.6 & - \\
        \ion{C}{i} & 658.7610 & -1.003 & 8.537 & - & - & - & 80.3 & 93.4 & - \\
        \ion{C}{i} & 805.8612 & -1.275 & 8.851 & - & - & - & 43.6 & 58.4 & - \\ \hline
        \ion{N}{i} & 868.3403 & 0.105 & 10.330 & 50.9 & - & - & - & - & - \\ \hline
        \ion{O}{i} & 557.7339 & -8.204 & 1.967 & - & - & - & 49.3 & 49.8 & - \\
        \ion{O}{i} & 636.3776 & -10.258 & 0.020 & 47.2 & 65.2 & 73.5 & 66.4 & 58.6 & 37.2 \\ \hline
        \multicolumn{10}{|c|}{\ldots} \\ \hline
    \end{tabular}
\end{table}
\begin{table}[!ht]
    \centering
    \footnotesize
    \caption[Near-infrared line list for SZ\,Mon]{Near-infrared line list for SZ\,Mon. Iso code is isotopic code in the format ``$Z_10Z_2.A_10A_2$'', where $Z_1, Z_2$ are atomic numbers of the elements making up the molecule, and $A_1, A_2$ are the corresponding atomic masses. $W_{\lambda, {\rm theor}}$ are the equivalent widths calculated for metallicity-scaled solar abundances as a proxy of each line's impact on the overall profile. This table is published in its entirety in the electronic edition of the paper. A portion is shown here for guidance regarding its form and content.}\label{tabA:nirlstszm_paper1}
    \begin{tabular}{|c@{\hspace{0.15cm}}c@{\hspace{0.15cm}}c@{\hspace{0.15cm}}c@{\hspace{0.15cm}}c@{\hspace{0.15cm}}|} \hline
        \multicolumn{5}{|c|}{\textbf{SZ Mon}} \\
        \boldmath$\lambda$ \textbf{(nm)} & \textbf{Type} & \textbf{Species} & \textbf{Iso code} & \boldmath$W_{\lambda, {\rm theor}}$ \textbf{(m\AA)} \\ \hline
        \multicolumn{5}{|c|}{\textit{C}} \\ \hline
        \multicolumn{5}{|c|}{\textit{1688.9100-1689.1666}} \\ \hline
        1688.8645 & Mol & CO & 608.012016 & 13.63 \\
        1688.8988 & Mol & CO & 608.013016 & 6.21 \\
        1688.9209 & Mol & OH & 108.001018 & 1.27 \\
        1688.9371 & Mol & CO & 608.012017 & 30.77 \\
        1688.9390 & Mol & CO & 608.012016 & 2.52 \\
        1688.9473 & Ato & \ion{Fe}{i} & 26 & 13.2 \\
        \multicolumn{5}{|c|}{\ldots} \\ \hline
    \end{tabular}
\end{table}

\begin{table}[ht]
    \centering
    \footnotesize
    \caption[Near-infrared line list for DF\,Cyg]{Near-infrared line list for DF\,Cyg. Iso code is isotopic code in the format ``$Z_10Z_2.A_10A_2$'', where $Z_1, Z_2$ are atomic numbers of the elements making up the molecule, and $A_1, A_2$ are the corresponding atomic masses. $W_{\lambda, {\rm theor}}$ are the equivalent widths calculated for metallicity-scaled solar abundances as a proxy of each line's impact on the overall profile. This table is published in its entirety in the electronic edition of the paper. A portion is shown here for guidance regarding its form and content.}\label{tabA:nirlstdfc_paper1}
    \begin{tabular}{|c@{\hspace{0.15cm}}c@{\hspace{0.15cm}}c@{\hspace{0.15cm}}c@{\hspace{0.15cm}}c@{\hspace{0.15cm}}|} \hline
        \multicolumn{5}{|c|}{\textbf{DF Cyg}} \\
        \boldmath$\lambda$ \textbf{(nm)} & \textbf{Type} & \textbf{Species} & \textbf{Iso code} & \boldmath$W_{\lambda, {\rm theor}}$ \textbf{(m\AA)} \\ \hline
        \multicolumn{5}{|c|}{\textit{C}} \\ \hline
        \multicolumn{5}{|c|}{\textit{1557.6983-1558.6885}} \\ \hline
        1557.7381 & Mol & CO & 608.012016 & 2.70 \\
        1557.7448 & Mol & CO & 608.012016 & 2.69 \\
        1557.7506 & Mol & CN & 607.012015 & 4.84 \\
        1557.7526 & Ato & \ion{Fe}{i} & 26 & 1.33 \\
        1557.7581 & Mol & CO & 608.012016 & 2.71 \\
        1557.7783 & Mol & CO & 608.012016 & 2.67 \\
        \multicolumn{5}{|c|}{\ldots} \\ \hline
    \end{tabular}
\end{table}

\section{Comprehensive details on epoch selection}\label{app:epo_paper1}
In this Appendix, we extensively explain how we selected the optical (HERMES) and NIR (APOGEE) spectral visits, which we used for the analysis of chemical composition of two post-RGB binary targets, SZ~Mon and DF~Cyg (see Section~\ref{sec:san_paper1}).

First, we selected the APOGEE visits where we were able to detect prominent molecular bands in the spectra. This resulted in choosing a single NIR visit for each target (where we detected the molecular features of CO and CN): SA\#2 for SZ~Mon and DA\#1 for DF~Cyg.

However, to calculate the isotopic ratios using the APOGEE spectra, accurately derived atmospheric parameters are required (a model atmosphere). To derive these parameters, the ionisation analyses should be conducted, but the APOGEE spectra lacked the essential combination of spectral lines across different ionisation levels (e.g., Fe I and Fe II). Therefore, we used the HERMES spectra to determine atmospheric parameters and elemental abundances through excitation and ionisation analyses of atomic line transitions (see Section~\ref{sssec:stepar_paper1}). Naturally, we selected the HERMES visits with pulsation phases, which roughly matched those of SA\#2 and DA\#1 (out of 88 and 83 optical visits for SZ~Mon and DF~Cyg, respectively). Among these matching optical visits, we identified those visits, which exhibited the highest S/N. Specifically, for SZ~Mon, the chosen optical visit was identified as SH\#47, while for DF~Cyg, it was DH\#54.

Unfortunately, due to prominent blending in the above mentioned optical visits, the number of atomic spectral features we could analyse was significantly limited. On the contrary, at hotter phases the number of spectral features was smaller, hence the level of blending in the spectra was decreased. So, we included two additional optical visits for each target with the highest S/N values, which occurred at pulsation phases with higher temperatures. This approach enabled us to confirm the atmospheric parameters derived from SH\#47 and DH\#54, as well as to extend the number of studied chemical species. These visits are provided in Table~\ref{tab:obslog_paper1}: SH\#73 and SH\#29 for SZ~Mon, DH\#83 and DH\#26 for DF~Cyg.

Using the selected visits, we derived the atmospheric parameters, the chemical abundances, and the carbon isotopic ratios for SZ~Mon and DF~Cyg (see Section~\ref{sec:san_paper1}).

\section{E-iSpec testing highlights}\label{app:tst_paper1}
In this Appendix, we provide the information about the most important tests we performed for validating performance of E-iSpec: 
\begin{enumerate}
    \item derivation of atmospheric parameters of a diverse sample of post-AGB stars using EW method,
    \item calculation of elemental abundances of SZ~Mon and DF~Cyg using EW method (for optical visits) and SSF technique (for NIR visits) in all observations, which were selected for chemical analysis (see Section~\ref{sec:obs_paper1}),
    \item SSF of atomic (Fe) and molecular (C, N, and O) features in giant branch stars.
\end{enumerate}

We note that we use EW method for a few reasons. Firstly, the synthetic spectral fitting technique involves an additional free parameter (the broadening velocity), which makes this technique more demanding in our case (for our calculation of the abundance uncertainties, we also need to account for the abundance deviation caused by the uncertainty of the broadening parameter). In contrast, the equivalent width method is insensitive to the broadening velocity. Since there is one less parameter in the procedure, this method is more robust. Secondly, in iSpec, the equivalent width method (Moog) provides the similar results as the synthetic spectral fitting technique (Turbospectrum), meaning that atmospheric parameters and elemental abundances obtained with these two methods match within the uncertainty ranges \citep{blancocuaresma2014}. This interrelation of Moog and Turbospectrum was tested for a large set of Gaia FGKM Benchmark Stars covering a wide range in $T_{\rm eff}$ (3500 to 6600 K), $\log g$ (0.50 to 4.60 dex) and [Fe/H] (--2.70 to 0.30 dex): the resulting median difference in $T_{\rm eff}$ between the two methods was found to be 50 K, the median difference in $\log g$ was --0.02 dex, the median difference in [Fe/H] was 0.03 dex, and the median difference in $\xi_{\rm t}$ was --0.34 km/s \citep{blancocuaresma2019}. Additionally, carbon isotopic ratios $^{12}$C/$^{13}$C, which are the main focus of this study, are independent of the elemental abundances of other elements (except for indirect impacts like blends or continuum placement), hence it is also independent of the method we use to derive the elemental abundances of other elements.

In Table~\ref{tabA:tstsmp_paper1}, we show the results of Fe lines analysis using E-iSpec. The testing sample consisted of previously studied post-AGB stars in the SMC and the LMC \citep{desmedt2012j004441, desmedt2015LMC2sEnrichedPAGBs, kamath2017j005252, kamath2019depletionLMC}.

\begin{sidewaystable}[ph!]
    \centering
    \footnotesize
    \caption[Testing EW method performance for chemically peculiar evolved stars]{Testing EW method performance for chemically peculiar evolved stars. The atmospheric parameters are provided in the following format: literature/this study. The last line provides information on the ratio of automatically identified spectral lines out of those reported in the corresponding literature. The typical literature errors are: $\Delta T_{\rm eff}=250$ K, $\Delta\log g=0.5$ dex, $\Delta [$Fe/H$]$=0.5 dex, $\Delta\xi_{\rm t}=0.5$ km/s.}\label{tabA:tstsmp_paper1}
    \begin{tabular}{|r@{\hspace{0.15cm}}|c@{\hspace{0.25cm}}c@{\hspace{0.25cm}}c@{\hspace{0.25cm}}c@{\hspace{0.25cm}}c@{\hspace{0.25cm}}c@{\hspace{0.25cm}}|} \hline
        \textbf{Target} & \textbf{J051845}$^a$ & \textbf{J050356}$^a$ & \textbf{J050304}$^a$ & \textbf{J051848}$^b$ & \textbf{J004441}$^b$ & \textbf{J005252}$^c$ \\ \hline
        $T_{\rm eff}$ & 5000/5080$\pm$80 & 5500/5600$\pm$80 & 5750/5740$\pm$30 & 6000/5880$\pm$100 & 6250/6290$\pm$100 & 8250/8260$\pm$340 \\
        $\log g$ & 0.50/0.39$\pm$0.18 & 1.00/1.20$\pm$0.15 & 0.00/0.04$\pm$0.05 & 0.50/0.49$\pm$0.21 & 0.50/0.63$\pm$0.17 & 1.50/1.08$\pm$0.06 \\
        $[$Fe/H$]$ & --1.10/--1.09$\pm$0.17 & --0.60/--0.54$\pm$0.06 & --2.60/--2.56$\pm$0.15 & --1.00/--1.21$\pm$0.19 & --1.34/--1.33$\pm$0.14 & --1.18/--1.03$\pm$0.14 \\
        $\xi_{\rm t}$ & 2.50/2.83$\pm$0.07 & 2.00/2.91$\pm$0.08 & 2.20/2.32$\pm$0.06 & 2.80/2.63$\pm$0.04 & 3.50/2.80$\pm$0.05 & 2.00/1.62$\pm$0.16 \\ \hline
        \begin{tabular}{r}
            Identified\\
            lines (ratio)
        \end{tabular} & \begin{tabular}{c}
            151/155 \\
            (97.42\%)
        \end{tabular} & \begin{tabular}{c}
            55/65 \\
            (84.62\%)
        \end{tabular} & \begin{tabular}{c}
            70/73 \\
            (95.89\%)
        \end{tabular} & \begin{tabular}{c}
            147/163 \\
            (90.18\%)
        \end{tabular} & \begin{tabular}{c}
            72/97 \\
            (74.23\%)
        \end{tabular} & \begin{tabular}{c}
            128/131 \\
            (97.71\%)
        \end{tabular} \\ \hline
    \end{tabular}\\
    \textbf{Notes:} The target is $^a$depleted in refractory elements, $^b$strongly enhanced in \textit{s}-process elements, $^c$depleted in \textit{s}-process elements.
\end{sidewaystable}

In Table~\ref{tabA:tstabs_paper1}, we provide the chemical analysis of two evolved stars from \cite{masseron2019APOGEE+BACCHUS} with the lowest $T_{\rm eff}$, hence with the most significant spectral blending. Since \cite{masseron2019APOGEE+BACCHUS} used the same NIR line list, we were able to independently confirm the performance of our SSF technique with evolved stars.

\begin{table}[!ht]
    \centering
    \footnotesize
    \caption{Testing SSF technique performance for evolved stars with similar line lists \citep[M+19 =][]{masseron2019APOGEE+BACCHUS}.}\label{tabA:tstabs_paper1}
    \begin{tabular}{|l@{\hspace{0.25cm}}|c@{\hspace{0.25cm}}c@{\hspace{0.25cm}}c@{\hspace{0.25cm}}c@{\hspace{0.25cm}}|} \hline
        \textbf{2MASS name} & \multicolumn{2}{c}{\textbf{J16323061-1306301}} & \multicolumn{2}{c|}{\textbf{J19534103+1846056}} \\
        \textbf{Evol. stage} & \multicolumn{2}{c}{\textbf{RGB}} & \multicolumn{2}{c|}{\textbf{eAGB}} \\ \hline
        $T_{\rm eff}$ (K) & \multicolumn{2}{c}{4464} & \multicolumn{2}{c|}{4745} \\
        $\log g$ (dex) & \multicolumn{2}{c}{1.46} & \multicolumn{2}{c|}{1.86} \\
        $[$Fe/H$]$ (dex) & \multicolumn{2}{c}{--0.79} & \multicolumn{2}{c|}{--0.54} \\ \hline
        \textbf{Source} & \textbf{M+19} & \textbf{This study} & \textbf{M+19} & \textbf{This study} \\ \hline
        A(Fe) (dex) & 6.71$\pm$0.03 & 6.66$\pm$0.18 & 6.96$\pm$0.20 & 7.00$\pm$0.14 \\
        A(C) (dex) & 7.54$\pm$0.10 & 7.55$\pm$0.12 & 7.53$\pm$0.18 & 7.51$\pm$0.11 \\
        A(N) (dex) & 7.56$\pm$0.04 & 7.71$\pm$0.15 & 8.44$\pm$0.09 & 8.60$\pm$0.06 \\
        A(O) (dex) & 8.74$\pm$0.04 & 8.66$\pm$0.14 & 8.93$\pm$0.11 & 8.86$\pm$0.14 \\ \hline
    \end{tabular}
\end{table}

In Table~\ref{tabA:tststp_paper1}, we provide our results of chemical analysis performed on all selected observational visits of SZ~Mon and DF~Cyg to confirm the robustness of the final abundances. The atmospheric parameters of the infrared visits are bolded because they were fixed for these visits (see Section~\ref{ssec:saneis_paper1}).

\begin{table}[!ht]
    \centering
    \scriptsize
    \caption[Derived stellar parameters and abundances for optical and near-infrared visits of SZ Mon and DF Cyg]{Derived stellar parameters and abundances for optical and near-infrared visits of SZ Mon and DF Cyg. The abundances are provided in the form ``[X/H] (N)'', where N is number of spectral lines used to derive the abundance of the corresponding element.}\label{tabA:tststp_paper1}
    \begin{tabular}{|r|cccc|cccc|}\hline
        & \multicolumn{4}{|c|}{\textbf{SZ Mon}} & \multicolumn{4}{|c|}{\textbf{DF Cyg}} \\
        \textbf{Par$\backslash$Visit} & \textbf{00866282} & \textbf{00397261} & \textbf{00458609} & \textbf{APOGEE} & \textbf{00972481} & \textbf{00412205} & \textbf{00574546} & \textbf{APOGEE} \\ \hline
        $T_\textrm{eff}$ & 5460$\pm$60 & 5420$\pm$80 & 4520$\pm$40 & \textbf{4500} & 5770$\pm$70 & 5750$\pm$70 & 5210$\pm$120 & \textbf{4500} \\
        $\log g$ & 0.93$\pm$0.10 & 0.78$\pm$0.09 & 0.94$\pm$0.14 & \textbf{1.00} & 1.92$\pm$0.09 & 1.71$\pm$0.09 & 2.13$\pm$0.26 & \textbf{2.00} \\
        $[$Fe/H$]$ & --0.50$\pm$0.05 & --0.50$\pm$0.10 & --0.52$\pm$0.10 & \textbf{--0.50} & 0.05$\pm$0.05 & --0.04$\pm$0.08 & 0.01$\pm$0.02 & \textbf{0.00} \\
        $\xi_\textrm{t}$ & 4.37$\pm$0.08 & 4.50$\pm$0.13 & 4.56$\pm$0.05 & \textbf{4.50} & 3.97$\pm$0.03 & 4.66$\pm$0.05 & 5.81$\pm$0.20 & \textbf{4.00} \\ \hline
        \ion{C}{i} & --0.07 (2) & 0.00 (2) & -- & --0.19 (1) & 0.22 (3) & 0.26 (3) & -- & 0.27 (3) \\
        \ion{N}{i} & 0.46 (1) & -- & -- & 0.38 (15) & -- & -- & -- & 0.41 (7) \\
        \ion{O}{i} & 0.05 (1) & 0.16 (1) & 0.07 (1) & 0.16 (2) & 0.60 (1) & 0.50 (1) & -- & 0.50 (2) \\
        \ion{Na}{i} & 0.04 (2) & 0.07 (2) & 0.12 (2) & -- & 0.42 (2) & 0.33 (2) & 0.35 (1) & -- \\
        \ion{Mg}{i} & --0.37 (2) & -- & --0.38 (3) & -- & 0.00 (2) & --0.08 (2) & -- & -- \\
        \ion{Al}{i} & -- & -- & -- & --1.29 (2) & -- & -- & -- & --1.53 (3) \\
        \ion{Si}{i} & --0.38 (6) & --0.35 (4) & --0.35 (4) & --0.38 (4) & 0.23 (5) & 0.24 (3) & 0.15 (1) & -- \\
        \ion{S}{i} & 0.09 (3) & 0.20 (2) & -- & -- & -- & -- & -- & -- \\
        \ion{Ca}{i} & --0.66 (8) & --0.73 (7) & --0.80 (2) & -- & --0.21 (3) & --0.28 (3) & --0.26 (3) & -- \\
        \ion{Sc}{ii} & --1.49 (4) & --1.58 (3) & -- & -- & --0.80 (1) & --0.83 (1) & --0.70 (1) & -- \\
        \ion{Ti}{i} & --1.12 (2) & --1.19 (2) & -- & -- & --0.47 (1) & -- & -- & -- \\
        \ion{Ti}{ii} & --1.07 (6) & --1.05 (6) & -- & -- & --0.42 (2) & --0.39 (2) & -- & -- \\
        \ion{V}{i} & --0.43 (1) & --0.24 (1) & --0.34 (6) & -- & 0.29 (1) & 0.24 (1) & 0.28 (5) & -- \\
        \ion{V}{ii} & --0.34 (1) & --0.33 (1) & -- & -- & 0.24 (1) & -- & -- & -- \\
        \ion{Cr}{i} & --0.44 (2) & --0.53 (1) & --0.47 (6) & -- & 0.32 (2) & 0.30 (1) & 0.41 (1) & -- \\
        \ion{Cr}{ii} & --0.33 (3) & --0.42 (3) & -- & -- & -- & 0.36 (1) & -- & -- \\
        \ion{Mn}{i} & --0.32 (3) & --0.37 (3) & --0.41 (2) & --0.34 (1) & -- & --0.03 (1) & 0.05 (1) & -- \\
        \ion{Fe}{i} & --0.50 (29) & --0.49 (21) & --0.54 (36) & --0.46 (38) & 0.06 (24) & --0.05 (24) & 0.04 (8) & -- \\
        \ion{Fe}{ii} & --0.51 (6) & --0.52 (5) & --0.52 (4) & -- & 0.05 (3) & --0.05 (3) & 0.01 (2) & -- \\
        \ion{Co}{i} & --0.60 (3) & --0.55 (2) & --0.56 (6) & -- & 0.10 (4) & 0.05 (3) & 0.06 (3) & --0.25 (1) \\
        \ion{Ni}{i} & --0.52 (13) & --0.51 (8) & --0.53 (3) & --0.43 (2) & 0.07 (17) & --0.02 (17) & --0.04 (3) & -- \\
        \ion{Cu}{i} & --0.60 (1) & --0.7 (1) & -- & -- & -- & -- & -- & -- \\
        \ion{Zn}{i} & --0.73 (2) & --0.66 (2) & -- & -- & --0.14 (1) & -- & -- & -- \\
        \ion{Y}{ii} & --1.51 (3) & --1.48 (3) & --1.53 (2) & -- & --1.06 (1) & --1.11 (1) & --0.83 (1) & -- \\
        \ion{Ba}{ii} & --1.05 (1) & --1.14 (1) & -- & -- & -- & -- & -- & -- \\
        \ion{La}{ii} & --1.21 (1) & -- & -- & -- & -- & -- & -- & -- \\
        \ion{Ce}{ii} & --1.20 (2) & --1.11 (2) & -- & -- & --0.71 (2) & --0.71 (2) & -- & -- \\
        \ion{Nd}{ii} & --1.13 (1) & --1.29 (1) & -- & -- & -- & -- & -- & -- \\ \hline
    \end{tabular}
\end{table}
\begin{savequote}[75mm]
\foreignlanguage{ukrainian}{``Двоє дивляться вниз. Один бачить калюжу, другий -- зорі...''}
\qauthor{\foreignlanguage{ukrainian}{---Олександр Довженко (1894-1956), \\український письменник, кінорежисер, кінодраматург, художник, класик світового кінематографа}}
``Two people look down. One of them sees a puddle, while the other one sees the stars...''
\qauthor{---Oleksandr Dovzhenko (1894-1956), \\Ukrainian writer, film director, screenwriter, artist, classic of world cinema}
\end{savequote}

\chapter{Appendix to chapter 4}\label{chp:app4}

\clearpage

\section{Individual target details}\label{app:lit_paper2}
In this Appendix, we discuss the previous studies of our target sample, from which we adopted pulsational and orbital parameters, together with luminosity estimates (see Table~\ref{tab:litpar_paper2}). In Fig.~\ref{figA:allmap_paper2}, we show the astrometric distribution of transition disc targets.

\subsection{Transition disc stars}\label{ssec:tarhst_paper2}
In this subsection, we present the targets with circumbinary discs, for which the reported dust inner rims $R_{\rm in}$ are at least 2.5 times larger than the expected dust sublimation radius  $R_{\rm sub}$ \citep{corporaal2023DiscParameters}. The presence of large, dust-free cavities in the discs around CT Ori, ST Pup, RU Cen, AC Her, AD Aql, and EP Lyr indicates these systems have transition discs, similar to those found around YSOs.

\subsubsection{CT Ori (\#1)} 
CT Ori, characterised as an RV Tau variable (spectroscopic class B\footnote{RV Tau variables of spectroscopic classes A, B, and C are metal-rich, metal-poor with enhanced carbon, and metal-poor without carbon enhancement, respectively \citep{preston1963RVTauSpecGroups}.}), exhibits a fundamental\footnote{The fundamental pulsation period for Type II Cepheids is twice shorter than the double period which usually is the best-fit value for phase-folded light curves \citep{stobie1970PeriodInconsistency}.} pulsation period of 33.65 days with no RVb phenomenon (slow variation in mean magnitude with a long secondary period $P\sim600-2600$ days), as observed by \citet{kiss2007T2Cepheids}. The orbital parameters of CT Ori are not constrained yet. Based on SED fitting, the luminosity of CT Ori was estimated to be $L_{\rm SED}\,=\,15100L_\odot$ by \citet{oomen2019depletion}. \citet{kluska2022GalacticBinaries} obtained a disc-star luminosity ratio for CT Ori $L_{\rm IR}/L_\ast\,=\,0.55$. \citet{corporaal2023DiscParameters} confirmed the transition disc nature of CT Ori with $R_{\rm in}/R_{\rm sub}\sim4.5$. \citet{gonzalez1997CTOri} conducted a comprehensive abundance analysis of CT Ori and identified significant depletion with [Fe/H]\,=\,--2.0 dex and [Zn/Ti]\,=\,1.9 dex.

\subsubsection{ST Pup (\#2)} 
ST Pup is classified as a W Vir pulsating variable. \citet{walker2015STPup} has determined the fundamental pulsation period of ST Pup (18.73 days). Despite the absence of the RVb phenomenon, \citet{oomen2018OrbitalParameters} calculated the orbital parameters of ST Pup using data from the long-term radial-velocity monitoring campaign ($P_{\rm orb}\,=\,406\pm2$ d, $e\,=\,0.00+0.04$). \citet{oomen2019depletion} reported the luminosity estimate for ST Pup through SED fitting ($L_{\rm SED}\,=\,690L_\odot$). \citet{kluska2022GalacticBinaries} obtained a moderate disc-star luminosity ratio of $L_{\rm IR}/L_\ast\,=\,0.72$. According to \citet{corporaal2023DiscParameters}, the ratio of dust inner rim to dust sublimation radius for ST Pup is $R_{\rm in}/R_{\rm sub}\sim3$. \citet{gonzalez1996STPup} analysed the chemical composition of ST Pup, revealing significant depletion ([Fe/H]\,=\,--1.5 dex, [Zn/Ti]\,=\,2.1 dex).

\subsubsection{RU Cen (\#3)} 
RU Cen, classified as an RV Tau variable (spectroscopic class B), has a fundamental pulsation period of 32.37 days, as documented by \citet{bodi2019RVTauVars} (RVb phenomenon was not detected). By using radial velocities from various spectral observations, \citet{oomen2018OrbitalParameters} determined the orbital parameters of RU Cen to be $P_{\rm orb}\,=\,1489\pm10$ days and $e\,=\,0.62\pm0.07$. The luminosity of RU Cen was estimated to be $L_{\rm SED}\,=\,1100L_\odot$ based on SED fitting by \citet{oomen2019depletion}. \citet{kluska2022GalacticBinaries} provided a moderate disc-star luminosity ratio of $L_{\rm IR}/L_\ast\,=\,0.40$. \citet{corporaal2023DiscParameters} reported $R_{\rm in}/R_{\rm sub}\sim3.5$ for RU Cen. \citet{maas2002RUCenSXCen} conducted a detailed abundance analysis of RU Cen and reported significant depletion with [Fe/H]\,=\,--1.9 dex and [Zn/Ti]\,=\,1.0 dex.

\subsubsection{AC Her (\#4)} 
AC Her, an RV Tau pulsating variable of spectroscopic class B, has a fundamental pulsation period of 37.73 days as determined by \citet{giridhar1998RVTauVars}. There is no apparent RVb phenomenon in this star. Using radial velocities from long-term spectral observations, \citet{oomen2018OrbitalParameters} derived the orbital parameters of AC Her ($P_{\rm orb}\,=\,1189\pm1$ days and $e\,=\,0.0+0.05$). The luminosity of AC Her was estimated to be $L_{\rm SED}\,=\,2\,400\,L_\odot$ based on SED fitting by \citet{oomen2019depletion}. \citet{bollen2022Jets} modelled jets in this system and obtained an independent estimate of luminosity for this target $L_{\rm jet\,model}\,=\,3\,600\,L_\odot$. \citet{kluska2022GalacticBinaries} reported a low disc-star luminosity ratio $L_{\rm IR}/L_\ast\,=\,0.21$. In the study by \citet{corporaal2023DiscParameters}, the ratio of inner rim to sublimation radius for AC Her was found to be $R_{\rm in}/R_{\rm sub}\sim7.5$. \citet{giridhar1998RVTauVars} performed a detailed abundance analysis of AC Her and found moderate depletion with [Fe/H]\,=\,--1.4 dex and [Zn/Ti]\,=\,0.7 dex.

\subsubsection{AD Aql (\#5)} 
AD Aql, classified as an RV Tau variable (spectroscopic class B), has a fundamental pulsation period of 32.7 days, as documented by \citet{giridhar1998RVTauVars} (RVb phenomenon was not detected for this target). The orbital parameters of AD Aql are not studied yet. The luminosity of AD Aql was estimated to be $L_{\rm SED}\,=\,11\,500\,L_\odot$ based on SED fitting by \citet{oomen2019depletion}. \citet{kluska2022GalacticBinaries} obtained a rather moderate disc-star luminosity ratio of $L_{\rm IR}/L_\ast\,=\,0.51$. \citet{corporaal2023DiscParameters} reported $R_{\rm in}/R_{\rm sub}\sim6$ for AD Aql. \citet{giridhar1998RVTauVars} analysed the chemical composition of AD Aql and reported high depletion with [Fe/H]\,=\,--2.1 dex and [Zn/Ti]\,=\,2.5 dex.

\subsubsection{EP Lyr (\#6)} 
EP Lyr is an RV Tau pulsating variable of spectroscopic class B. This target possesses a fundamental pulsation period of 41.59 days \citep{bodi2019RVTauVars} with no evident long-period variation. \citet{oomen2018OrbitalParameters} employed radial velocities from long-term
radial-velocity monitoring campaign to derive the orbital parameters of EP Lyr ($P_{\rm orb}\,=\,1\,151\pm14$ days and $e\,=\,0.39\pm0.09$). \citet{oomen2019depletion} estimated the luminosity of EP Lyr through SED fitting ($L_{\rm SED}\,=\,5\,500\,L_\odot$). Furthermore, \citet{bollen2022Jets} modelled jets in this system and obtained an independent estimate of luminosity for EP Lyr $L_{\rm jet\,model}\,=\,7\,100\,L_\odot$. The disc-star luminosity ratio of $L_{\rm IR}/L_\ast\,=\,0.02$ \citep{kluska2022GalacticBinaries} is the lowest in the target sample. \citet{corporaal2023DiscParameters} derived the ratio of inner rim to sublimation radius $R_{\rm in}/R_{\rm sub}\sim3.5$ for EP Lyr. \citet{gonzalez1997EPLyrDYOriARPupRSgt} performed a detailed abundance analysis of EP Lyr, revealing moderate depletion with [Fe/H]\,=\,--1.8 dex and [Zn/Ti]\,=\,1.3 dex.

\subsection{Transition disc candidates}\label{ssec:tarcnd_paper2}
In this subsection, we present the targets classified by \citet{kluska2022GalacticBinaries} as category 2 ($W_1-W_3>4.5$) and category 3 ($2.3<W_1-W_3<4.5$, $H-K<0.3$). Even though category 3 of colour-colour plot is populated by both full disc and transition disc candidates \citep{kluska2022GalacticBinaries}, these subclasses may be distinguished by the depletion scale: [Zn/Ti]$_{\rm full-disc}\sim0$ dex, [Zn/Ti]$_{\rm transition-disc}>0.7$ dex.

\subsubsection{DY Ori (\#7)} 
DY Ori is an RV Tau pulsator of spectroscopic class B. ASAS \citep{pawlak2019ASAS} provides the fundamental pulsation period of DY Ori (30.155 days) with no obvious RVb phenomenon. However, \citet{oomen2018OrbitalParameters} used radial velocities of multiple spectral observations of DY Ori to derive the orbital parameters of this target ($P_{\rm orb}\,=\,1\,248\pm36$ d, $e\,=\,0.22\pm0.08$). \citet{oomen2019depletion} estimated the luminosity of DY Ori based on the SED fitting ($L_{\rm SED}\,=\,21\,500\,L_\odot$). \citet{kluska2022GalacticBinaries} reported a moderate disc-star luminosity ratio $L_{\rm IR}/L_\ast\,=\,0.55$. \citet{gonzalez1997EPLyrDYOriARPupRSgt} conducted a detailed abundance analysis of DY Ori and detected high depletion with [Fe/H]\,=\,--2.0 dex and [Zn/Ti]\,=\,2.1 dex.

\subsubsection{AF Crt (\#8)} 
AF Crt, a W Vir pulsator, has a double pulsation period of 31.5 days as determined by \citet{kiss2007T2Cepheids} without any apparent variation in mean magnitude on the order of hundreds of days. The orbital parameters of AF Crt are not constrained yet. The luminosity of AF Crt was estimated to be $L_{\rm SED}\,=\,280\,L_\odot$ based on SED fitting by \citet{oomen2019depletion}. \citet{kluska2022GalacticBinaries} derived a rather high disc-star luminosity ratio $L_{\rm IR}/L_\ast\,=\,1.83$. \citet{vanwinckel2012AFCrt} performed a detailed abundance analysis of AF Crt and reported high depletion of this target with [Fe/H]\,=\,--2.7 dex and [Zn/Sc]\,=\,3.4 dex (a slightly less refractory Sc was used for volatile-to-refractory ratio, since Ti abundance was not calculated).

\subsubsection{GZ Nor (\#9)} 
GZ Nor, classified as an RV Tau pulsator, demonstrates a fundamental pulsation period of 36.2 days \citep{gezer2019GKCarGZNor}. No evident RVb phenomenon was observed and no orbital parameters were derived for GZ Nor. \citet{oomen2019depletion} determined the luminosity of GZ Nor through SED fitting $L_{\rm SED}\,=\,1\,400\,L_\odot$. \citet{kluska2022GalacticBinaries} provided a mild disc-star luminosity ratio of $L_{\rm IR}/L_\ast\,=\,0.22$. \citet{gezer2019GKCarGZNor} studied the chemical composition of GZ Nor and detected moderate depletion with [Fe/H]\,=\,--2.1 dex and [Zn/Ti]\,=\,0.8 dex.

\subsubsection{V1504 Sco (\#10)} 
V1504 Sco is an RV Tau variable star of spectroscopic class B. \citet{kiss2007T2Cepheids} provided the fundamental pulsation period of this star (22.0 d) with a detected RVb phenomenon. Studying the mean magnitude variation, \citet{kiss2007T2Cepheids} derived the orbital period of V1504 Sco ($P_{\rm orb}\,=\,735\pm230$ d). \citet{oomen2019depletion} estimated the luminosity of V1504 Sco based on the SED fitting ($L_{\rm SED}\,=\,1\,100\,L_\odot$). \citet{kluska2022GalacticBinaries} reported an extreme disc-star luminosity ratio $L_{\rm IR}/L_\ast\,=\,4.69$. \citet{maas2005DiscPAGBs} conducted a detailed abundance analysis of V1504 Sco and detected high depletion with [Fe/H]\,=\,--1.0 dex and [Zn/Ti]\,=\,1.4 dex.

\subsubsection{LMC V0770 (\#11)} 
LMC V0770, an RV Tau pulsator, exhibits a fundamental pulsation period of 31.2 days, as determined by \citet{manick2018PLC}. No variation in the mean magnitude was detected and no orbital parameters were constrained for LMC V0770. \citet{oomen2019depletion} used SED fitting to calculate the luminosity of LMC V0770 ($L_{\rm SED}\,=\,3\,300\,L_\odot$). Moreover, \citet{manick2018PLC} used PLC relation to obtain an independent estimate of luminosity ($L_{\rm PLC}\,=\,2\,629\,L_\odot$). According to \citet{kluska2022GalacticBinaries}, the disc-star luminosity ratio for LMC V0770 is moderate $L_{\rm IR}/L_\ast\,=\,0.63$. \citet{kamath2019depletionLMC} analysed the chemical composition of LMC V0770 and reported a high depletion with [Fe/H]\,=\,--2.6 dex and [Zn/Ti]\,=\,2.3 dex.

\subsubsection{LMC V3156 (\#12)} 
LMC V3156, an RV Tau pulsator, has a fundamental pulsation period of 46.7 days, as provided by \citet{manick2018PLC}. There is no apparent RVb phenomenon in the light curves of this target and no known orbital parameters. The luminosity of LMC V3156 was estimated by \citet{manick2018PLC} using two methods: SED fitting and PLC relation ($L_{\rm SED}\,=\,5\,900\,L_\odot$, $L_{\rm PLC}=6\,989\,L_\odot$). \citet{vanaarle2011PAGBsInLMC} reported a disc-star luminosity ratio of $L_{\rm IR}/L_\ast\,=\,0.84$. \citet{reyniers2007LMC147} conducted a detailed abundance analysis of LMC V3156 and found significantly high depletion with [Fe/H]\,=\,--2.4 dex and [Zn/Ti]\,=\,2.5 dex.

\section{SED plots of the target sample}\label{app:sed_paper2}
\begin{figure}[!ht]
    \centering
    \includegraphics[width=.49\linewidth]{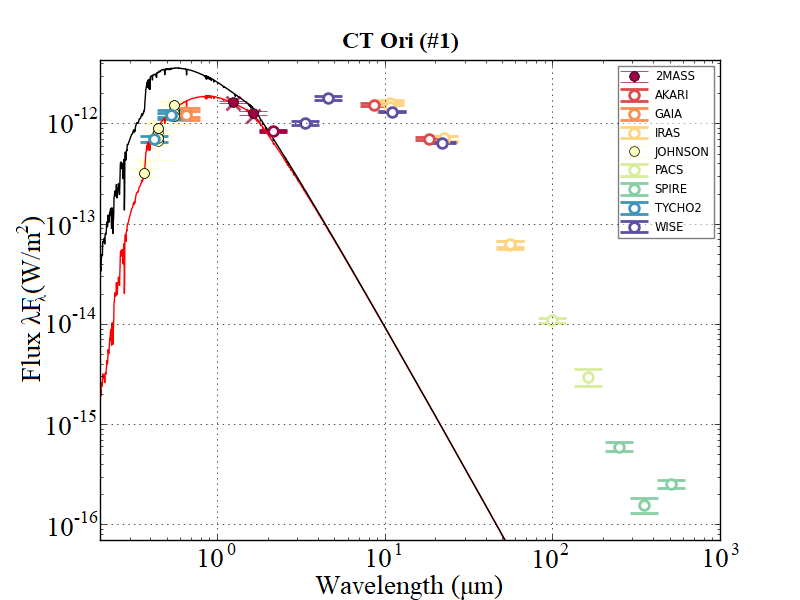}
    \includegraphics[width=.49\linewidth]{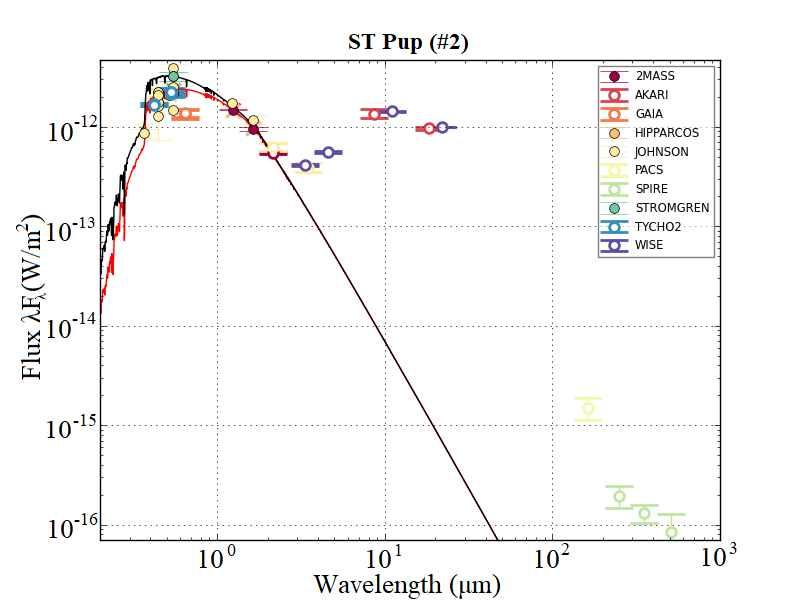}
    \includegraphics[width=.49\linewidth]{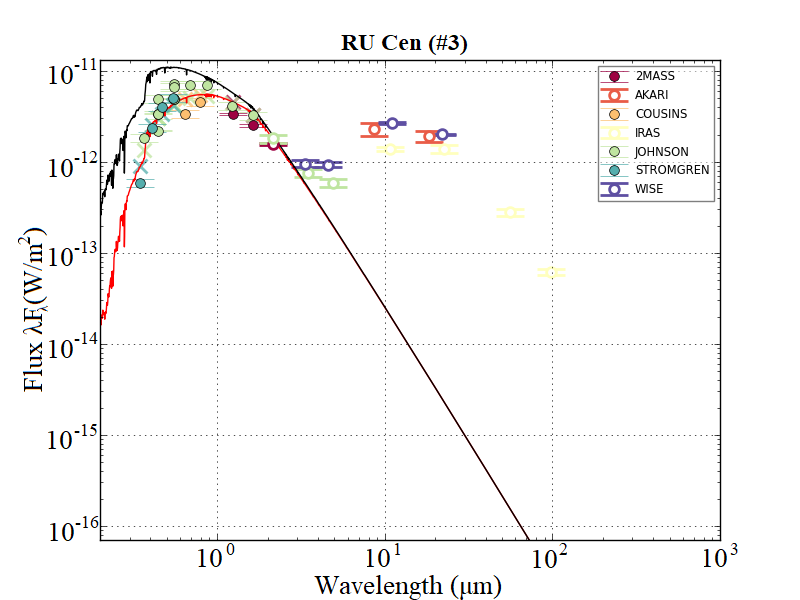}
    \includegraphics[width=.49\linewidth]{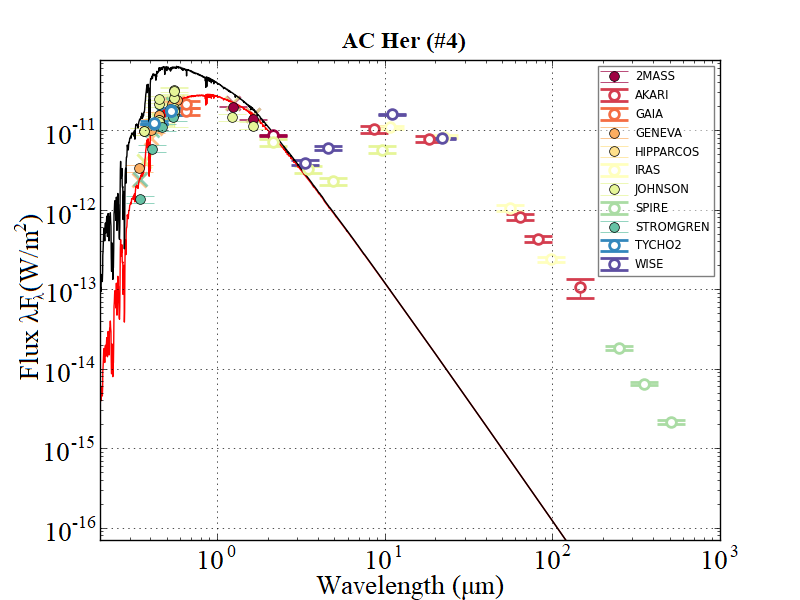}
    \includegraphics[width=.49\linewidth]{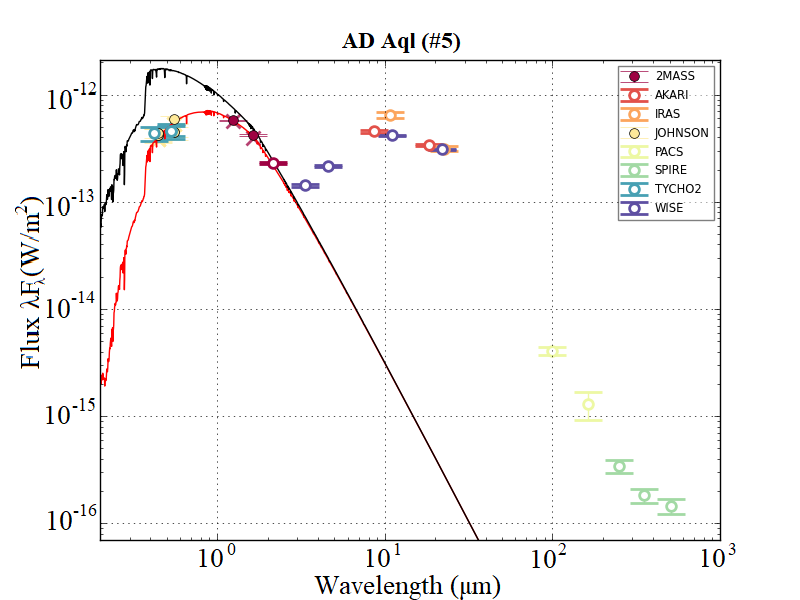}
    \includegraphics[width=.49\linewidth]{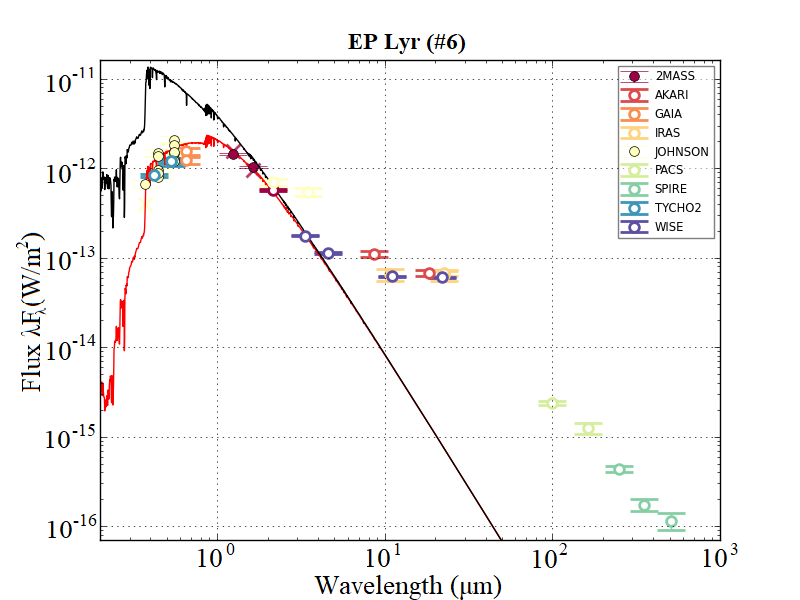}
    \caption[Spectral energy distribution of transition disc stars]{Spectral energy distribution of transition disc stars. The red solid line is the appropriate reddened Kurucz model atmosphere. The black solid line is the de-reddened model scaled to the object. We note that the photometric observations of our sample were obtained with different surveys at different time (pulsation phases). We also note that the IR dust excess in SED plots for AF Crt (\#8) and V1504 Sco (\#10) imply that we see these two targets edge-on. The legend for the symbols and colours used is included within the plot.}\label{figA:allSEDstr_paper2}
\end{figure}

\begin{figure}[!ht]
    \centering
    \includegraphics[width=.49\linewidth]{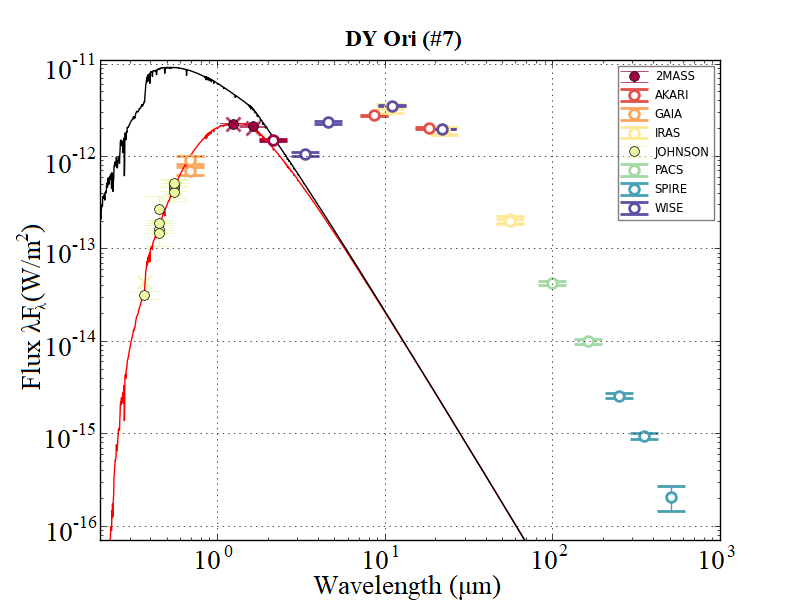}
    \includegraphics[width=.49\linewidth]{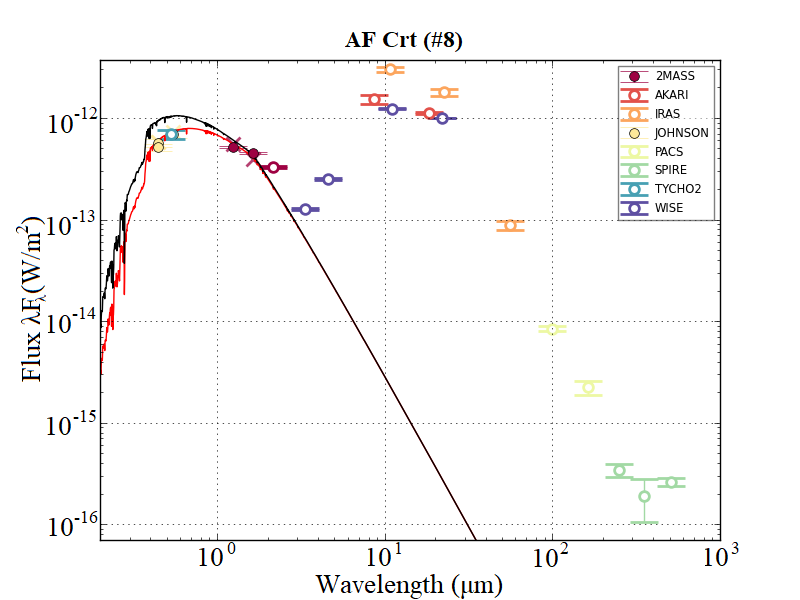}
    \includegraphics[width=.49\linewidth]{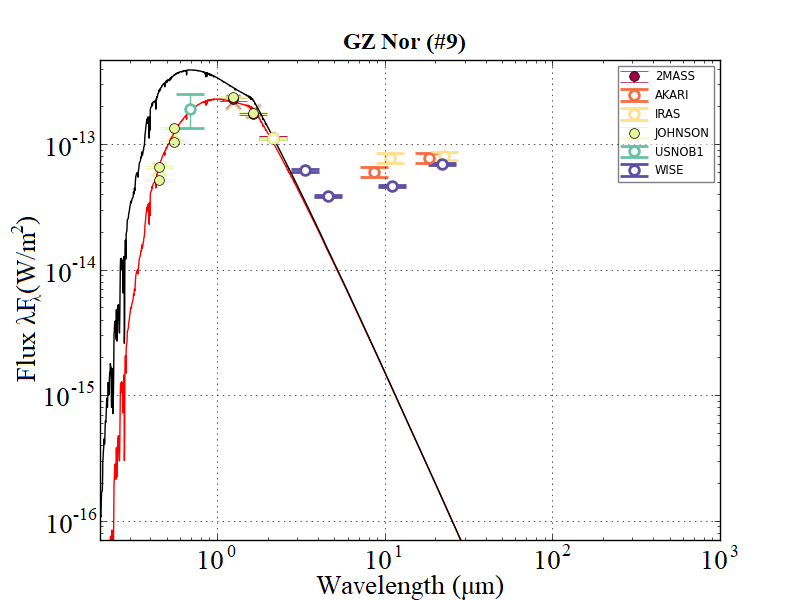}
    \includegraphics[width=.49\linewidth]{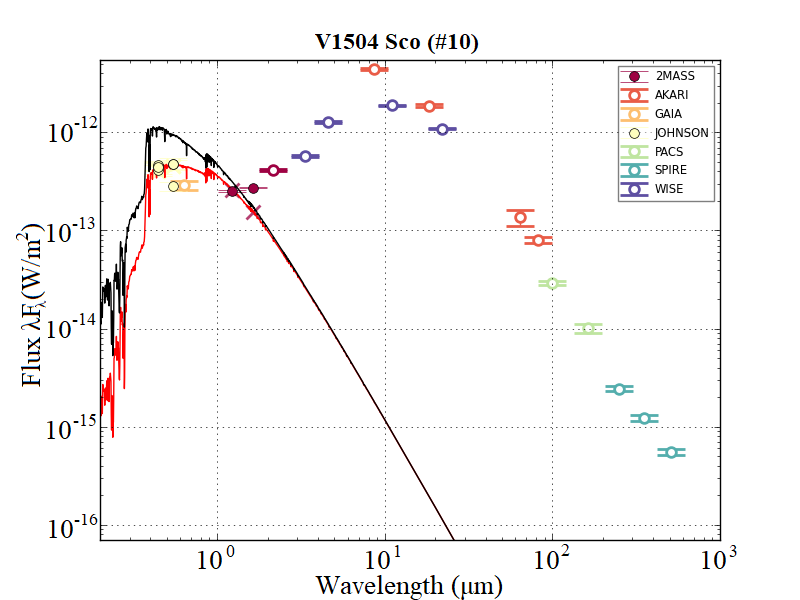}
    \includegraphics[width=.49\linewidth]{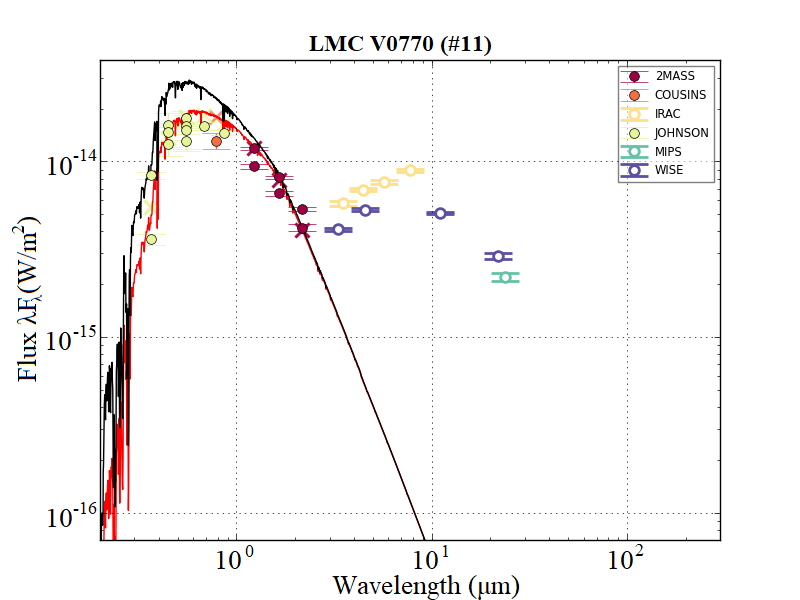}
    \includegraphics[width=.49\linewidth]{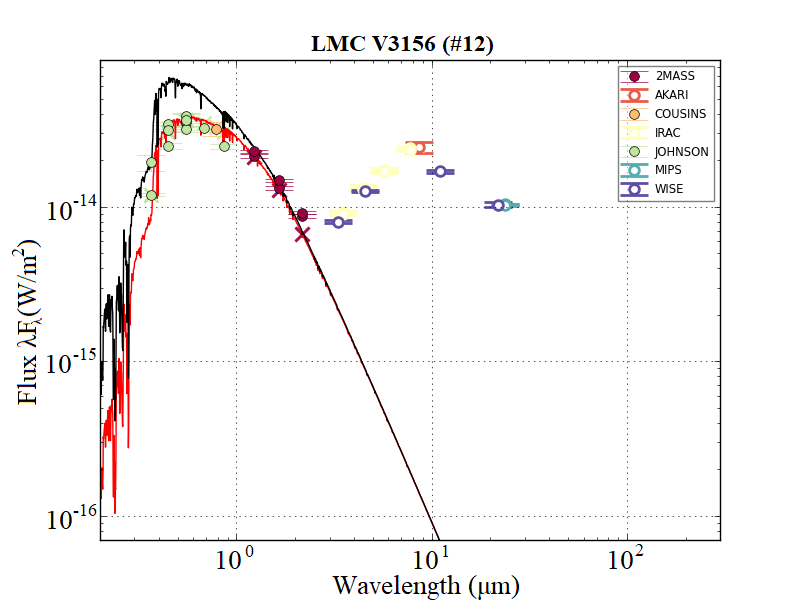}
    \caption[Spectral energy distribution of transition disc candidates]{Spectral energy distribution of transition disc candidates. The red solid line is the appropriate reddened Kurucz model atmosphere. The black solid line is the de-reddened model scaled to the object. We note that the photometric observations of our sample were obtained with different surveys at different time (pulsation phases). We also note that the IR dust excess in SED plots for AF Crt (\#8) and V1504 Sco (\#10) imply that we see these two targets edge-on. The legend for the symbols and colours used is included within the plot.}\label{figA:allSEDcnd_paper2}
\end{figure}

In Table~\ref{tabA:phomag_paper2}, we present the photometric data collected for SED plots of the target sample. In Fig.~\ref{figA:allSEDstr_paper2} and \ref{figA:allSEDcnd_paper2}, we provide our SED plots for all transition disc stars and candidates, respectively (see Section~\ref{ssec:dobpht_paper2}).

\begin{table}[!ht]
    \centering
    \footnotesize
    \caption[Photometric data for the target sample (see Section \ref{ssec:dobpht_paper2})]{Photometric data for the target sample (see Section \ref{ssec:dobpht_paper2}). For each filter we provide the units and the central wavelengths in $\mu$m. This table is published in its entirety in the electronic edition of the paper. A portion is shown here for guidance regarding its form and content.}\label{tabA:phomag_paper2}
    \begin{tabular}{|c|c|c|c|c|c|c|c|}\hline
        && \textbf{GENEVA.U} & \textbf{STROMGREN.U} & ... & \textbf{SPIRE.350} & \textbf{SPIRE.500} \\
        \textbf{ID} & \textbf{Name} & \textbf{mag} & \textbf{mag} && \textbf{Jy} & \textbf{Jy} \\
        && \textbf{0.342} & \textbf{0.346} & ... & \textbf{348.438} & \textbf{500.412} \\ \hline
        1 & CT Ori & -- & -- & ... & 0.018$\pm$0.003 & 0.042$\pm$0.004 \\
        2 & ST Pup & -- & -- & ... & 0.015$\pm$0.003 & 0.014$\pm$0.007 \\
        3 & RU Cen & -- & 12.105$\pm$99.999 & ... & -- & -- \\
        4 & AC Her & 9.419$\pm$0.116 & 11.203$\pm$99.999 & ... & 0.75$\pm$0.04 & 0.35$\pm$0.02 \\
        5 & AD Aql & -- & -- & ... & 0.021$\pm$0.003 & 0.024$\pm$0.004 \\
        6 & EP Lyr & -- & -- & ... & 0.020$\pm$0.003 & 0.019$\pm$0.004 \\
        ... & ... & ... & ... & ... & ... & ... \\ \hline
    \end{tabular}\\
    \textbf{Note:} We list uncertainties which are unavailable as `99.999'.
\end{table}

\section{Summary of optical spectral visits}\label{app:vis_paper2}
In Table~\ref{tabA:obslog_paper2}, we list all optical spectral visits considered in this work (HERMES and UVES; see Section~\ref{ssec:dobspc_paper2}) for our chemical analysis of the target sample (see Section~\ref{sec:san_paper2}).

\begin{table}[!ht]
    \centering
    \footnotesize
    \caption[All spectral visits of the target sample]{All spectral visits of the target sample. This table is published in its entirety in the electronic edition of the paper. A portion is shown here for guidance regarding its form and content.}\label{tabA:obslog_paper2}
    \begin{tabular}{|cc|cccc|}
    \hline
        \textbf{Visit} & \textbf{MJD} & \textbf{Phase} & \textbf{RV} & \textbf{eRV} & \textbf{SNR} \\ \hline
        \multicolumn{6}{|c|}{\textit{CT Ori (\#1) (127 visits)}} \\ \hline
        00273152 & 55221.91376 & 0.00 & 62.60 & 0.63 & 26.38 \\
        00314017 & 55500.22492 & 0.40 & 60.29 & 0.64 & 24.22 \\
        00314465 & 55507.12069 & 0.61 & 53.92 & 0.97 & 27.18 \\
        00325348 & 55553.04197 & 0.99 & 60.79 & 0.55 & 23.79 \\
        00327525 & 55576.97777 & 0.72 & 51.61 & 1.18 & 22.50 \\
        \multicolumn{6}{|c|}{\ldots} \\ \hline
    \end{tabular}
\end{table}

\section{Master line list of the target sample}\label{app:lst_paper2}
In Table~\ref{tabA:linlst_paper2}, we provide the combined optical line list, which we used to derive the atmospheric parameters and elemental abundances of all 12 transition disc targets (see Section~\ref{sec:san_paper2}).

\begin{sidewaystable}[ph!]
    \centering
    \tiny
    \caption[Combined optical line list of our sample]{Combined optical line list of our sample. This table is published in its entirety in the electronic edition of the paper. A portion is shown here for guidance regarding its form and content.}\label{tabA:linlst_paper2}
    \resizebox{0.99\textheight}{!}{
    \begin{tabular}{|c@{\hspace{0.1cm}}ccc|cccccccccccc|} \hline
        \multicolumn{4}{|c|}{\textbf{Atomic data}} & \multicolumn{12}{|c|}{\boldmath$W_\lambda$ \textbf{(m\AA)}} \\
        \textbf{Element} & \boldmath$\lambda$ & \boldmath$\log gf$ & \boldmath$\chi$ & \textbf{CT Ori} & \textbf{ST Pup} & \textbf{RU Cen} & \textbf{AC Her} & \textbf{AD Aql} & \textbf{EP Lyr} & \textbf{DY Ori} & \textbf{AF Crt} & \textbf{GZ Nor} & \textbf{V1504 Sco} & \textbf{LMC V0770} & \textbf{LMC V3156} \\
        ~ & \textbf{(nm)} & \textbf{(dex)} & \textbf{(eV)} & \textbf{(\#1)} & \textbf{(\#2)} & \textbf{(\#3)} & \textbf{(\#4)} & \textbf{(\#5)} & \textbf{(\#6)} & \textbf{(\#7)} & \textbf{(\#8)} & \textbf{(\#9)} & \textbf{(\#10)} & \textbf{(\#11)} & \textbf{(\#12)} \\ \hline
        \ion{C}{i} & 477.1730 & -1.866 & 7.488 & - & - & 82.6 & - & - & - & - & - & - & 64.6 & - & - \\
        \ion{C}{i} & 493.2049 & -1.658 & 7.685 & 123.6 & 56.7 & 89.8 & 62.8 & 109.0 & 68.0 & 86.5 & - & - & - & 72.3 & 66.3 \\
        \ion{C}{i} & 502.3841 & -2.210 & 7.946 & - & - & - & - & - & - & - & - & - & - & 26.4 & - \\
        \ion{C}{i} & 503.9057 & -1.790 & 7.946 & - & - & - & - & - & - & - & - & - & - & 63.8 & 33.7 \\
        \ion{C}{i} & 505.2144 & -1.303 & 7.685 & - & 94.3 & 123.0 & - & - & - & 134.1 & 124.5 & 92.0 & - & - & 90.4 \\
        \ion{C}{i} & 538.0325 & -1.616 & 7.685 & 120.0 & 52.8 & 98.6 & - & - & 68.3 & 83.8 & - & 57.6 & - & 81.4 & 57.0 \\
        \multicolumn{16}{|c|}{\ldots} \\ \hline
    \end{tabular}}
\end{sidewaystable}

\section{Individual depletion profiles}\label{app:dpl_paper2}
\begin{figure}[!ht]
    \centering
    \includegraphics[width=.95\linewidth]{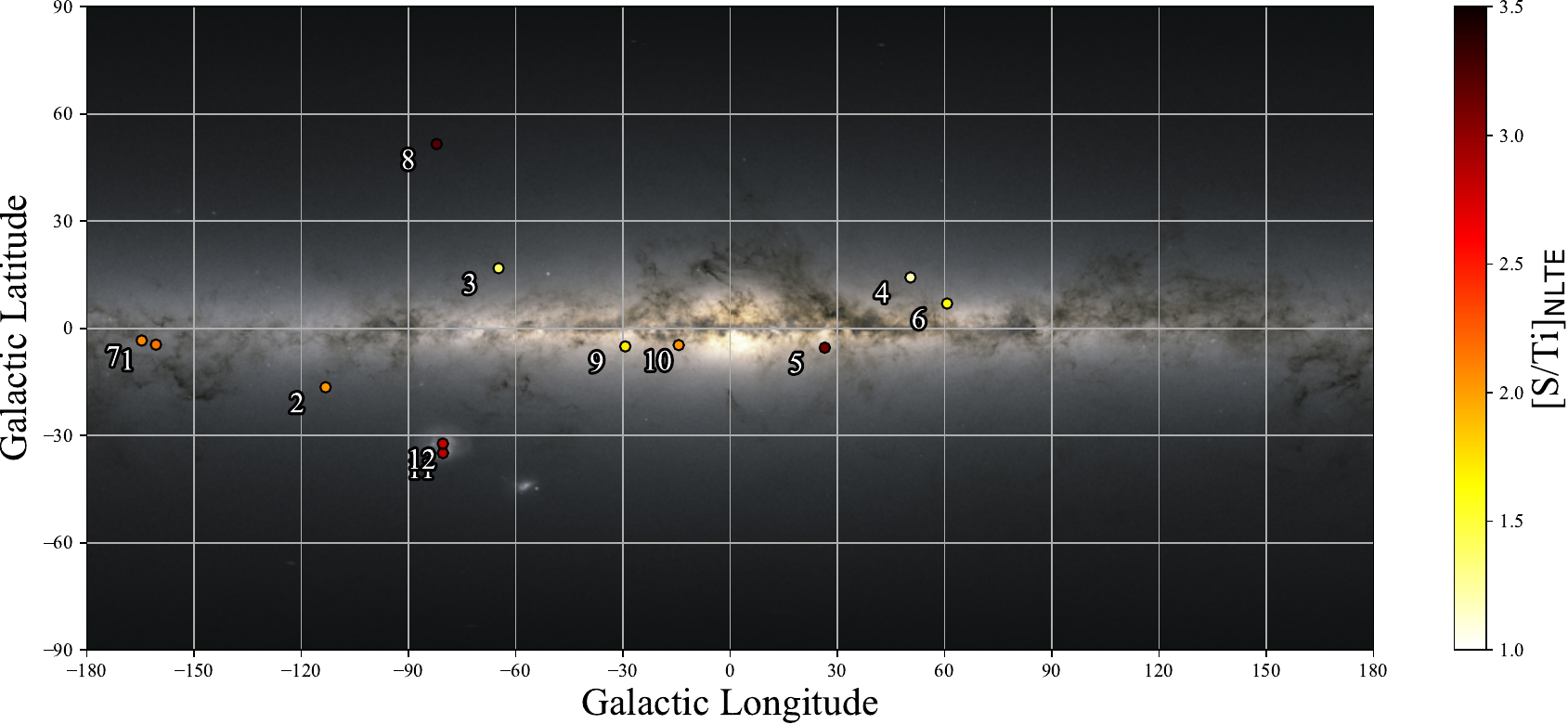}
    \caption[Spatial distribution of the transition disc targets]{Spatial distribution of the transition disc targets. The targets are labelled by their IDs and coloured by the [S/Ti]$_{\rm NLTE}$ values derived in this study (see Table~\ref{tab:fnlabu_paper2}).}\label{figA:allmap_paper2}
\end{figure}

In Table~\ref{tabA:fnlabu_paper2}, we provide all elemental abundances derived in this study (see Section~\ref{ssec:sanlte_paper2}). In Table \ref{tabA:fnlcor_paper2}, we provide the atmospheric parameters and differential NLTE corrections for transition disc targets (see Section~\ref{ssec:sannlte_paper2}). In Fig.~\ref{figA:allmap_paper2}, we present the spatial distribution of transition disc targets.

\begin{sidewaystable}[ph!]
    \centering
    \tiny
    \caption{1D LTE and 1D NLTE elemental abundances of transition disc targets in [X/H] scale.}\label{tabA:fnlabu_paper2}
    \resizebox{0.9\textheight}{!}{
    \begin{tabular}{|c|c|c|c|c|c|c|c|c|c|c|c|c|c|} \hline
        \multirow{2}{*}{\textbf{Ion}} & \multirow{2}{*}{\boldmath$T_{\rm cond}$} & \textbf{CT Ori} & \textbf{ST Pup} & \textbf{RU Cen} & \textbf{AC Her} & \textbf{AD Aql} & \textbf{EP Lyr} & \textbf{DY Ori} & \textbf{AF Crt} & \textbf{GZ Nor} & \textbf{V1504 Sco} & \textbf{LMC V0770} & \textbf{LMC V3156} \\
        && \textbf{(\#1)} & \textbf{(\#2)} & \textbf{(\#3)} & \textbf{(\#4)} & \textbf{(\#5)} & \textbf{(\#6)} & \textbf{(\#7)} & \textbf{(\#8)} & \textbf{(\#9)} & \textbf{(\#10)} & \textbf{(\#11)} & \textbf{(\#12)} \\ \hline
        \multicolumn{14}{|c|}{\textit{LTE}} \\ \hline
        \ion{C}{i} & 40 & --0.57$\pm$0.09 & --0.50$\pm$0.07 & --0.44$\pm$0.07 & --0.59$\pm$0.06 & --0.10$\pm$0.13 & --0.28$\pm$0.07 & 0.16$\pm$0.08 & --0.41$\pm$0.09 & 0.32$\pm$0.02 & --0.11$\pm$0.07 & --0.75$\pm$0.02 & --0.34$\pm$0.09 \\
        \ion{N}{i} & 123 & --0.78$\pm$0.18 & -- & 0.13$\pm$0.11 & --0.03$\pm$0.11 & --0.03$\pm$0.21 & 0.07$\pm$0.14 & 0.38$\pm$0.19 & --0.54$\pm$0.19 & 0.84$\pm$0.10 & 0.59$\pm$0.13 & -- & -- \\
        \ion{O}{i} & 183 & 0.03$\pm$0.09 & --0.50$\pm$0.05 & --0.01$\pm$0.12 & 0.01$\pm$0.08 & 0.75$\pm$0.17 & 0.14$\pm$0.14 & 0.32$\pm$0.17 & 0.43$\pm$0.17 & 0.50$\pm$0.06 & 0.31$\pm$0.11 & --0.40$\pm$0.10 & 0.78$\pm$0.14 \\
        \ion{Na}{i} & 1035 & --0.33$\pm$0.06 & --0.62$\pm$0.05 & --0.92$\pm$0.05 & --0.55$\pm$0.11 & --0.40$\pm$0.05 & --1.07$\pm$0.14 & 0.32$\pm$0.15 & 0.24$\pm$0.17 & --1.51$\pm$0.10 & 0.48$\pm$0.11 & --0.70$\pm$0.10 & 0.47$\pm$0.14 \\
        \ion{Mg}{i} & 1343 & --1.36$\pm$0.10 & --1.45$\pm$0.08 & --1.68$\pm$0.08 & --1.20$\pm$0.07 & --1.88$\pm$0.10 & --1.91$\pm$0.08 & --1.99$\pm$0.12 & --2.03$\pm$0.12 & --1.65$\pm$0.03 & --1.21$\pm$0.09 & --2.15$\pm$0.07 & --2.21$\pm$0.12 \\
        \ion{Al}{i} & 1652 & --2.40$\pm$0.16 & -- & -- & --1.86$\pm$0.13 & -- & -- & -- & --2.81$\pm$0.16 & -- & -- & --3.83$\pm$0.10 & --2.64$\pm$0.15 \\
        \ion{Si}{i} & 1314 & -- & -- & --1.48$\pm$0.10 & --1.18$\pm$0.03 & -- & --1.66$\pm$0.12 & -- & -- & -- & --0.75$\pm$0.10 & --1.99$\pm$0.10 & -- \\
        \ion{Si}{ii} & 1314 & --1.63$\pm$0.14 & -- & --1.55$\pm$0.12 & --1.20$\pm$0.13 & --1.56$\pm$0.17 & --1.52$\pm$0.08 & --1.47$\pm$0.14 & --1.66$\pm$0.11 & -- & --0.87$\pm$0.13 & -- & --1.47$\pm$0.14 \\
        \ion{S}{i} & 672 & --0.17$\pm$0.06 & --0.34$\pm$0.10 & --0.42$\pm$0.08 & --0.49$\pm$0.03 & 0.03$\pm$0.12 & --0.38$\pm$0.04 & 0.62$\pm$0.08 & --0.07$\pm$0.07 & --0.12$\pm$0.06 & 0.41$\pm$0.10 & --0.36$\pm$0.01 & 0.07$\pm$0.08 \\
        \ion{K}{i} & 993 & --0.47$\pm$0.15 & -- & --0.74$\pm$0.12 & --0.55$\pm$0.07 & --0.09$\pm$0.17 & --1.41$\pm$0.14 & -- & --0.81$\pm$0.15 & --1.31$\pm$0.10 & -- & -- & 0.34$\pm$0.15 \\
        \ion{Ca}{i} & 1535 & --1.64$\pm$0.11 & --2.04$\pm$0.06 & --1.86$\pm$0.05 & --1.45$\pm$0.07 & --2.43$\pm$0.08 & --2.02$\pm$0.11 & --1.76$\pm$0.14 & --2.22$\pm$0.11 & --1.92$\pm$0.03 & --1.41$\pm$0.06 & --2.56$\pm$0.10 & --1.79$\pm$0.12 \\
        \ion{Ca}{ii} & 1535 & --1.81$\pm$0.14 & -- & -- & --1.57$\pm$0.13 & --2.64$\pm$0.16 & -- & --2.04$\pm$0.13 & --2.23$\pm$0.15 & -- & -- & -- & -- \\
        \ion{Sc}{ii} & 1541 & --2.55$\pm$0.10 & --2.48$\pm$0.05 & --2.09$\pm$0.08 & --1.96$\pm$0.07 & -- & --2.29$\pm$0.11 & -- & -- & --2.15$\pm$0.10 & --1.63$\pm$0.06 & --3.74$\pm$0.09 & --2.64$\pm$0.10 \\
        \ion{Ti}{i} & 1565 & -- & --2.53$\pm$0.13 & --1.92$\pm$0.06 & --1.64$\pm$0.10 & -- & -- & -- & -- & -- & -- & -- & -- \\
        \ion{Ti}{ii} & 1565 & --2.48$\pm$0.11 & --2.44$\pm$0.04 & --1.99$\pm$0.07 & --1.85$\pm$0.07 & --3.32$\pm$0.12 & --2.20$\pm$0.10 & --1.77$\pm$0.14 & --3.45$\pm$0.15 & --1.94$\pm$0.03 & --1.81$\pm$0.07 & --3.40$\pm$0.06 & --2.95$\pm$0.09 \\
        \ion{V}{ii} & 1370 & -- & --1.79$\pm$0.02 & --1.81$\pm$0.11 & --1.38$\pm$0.06 & -- & --1.74$\pm$0.13 & -- & -- & --1.78$\pm$0.01 & --1.56$\pm$0.11 & -- & --1.94$\pm$0.13 \\
        \ion{Cr}{i} & 1291 & --1.86$\pm$0.13 & --2.19$\pm$0.07 & --1.97$\pm$0.06 & --1.45$\pm$0.07 & --2.22$\pm$0.10 & --2.19$\pm$0.17 & --2.05$\pm$0.16 & -- & -- & --1.24$\pm$0.08 & --2.79$\pm$0.09 & --2.49$\pm$0.14 \\
        \ion{Cr}{ii} & 1291 & --1.94$\pm$0.11 & --1.95$\pm$0.03 & --1.90$\pm$0.06 & --1.42$\pm$0.07 & --2.16$\pm$0.16 & --2.06$\pm$0.07 & -- & -- & --1.80$\pm$0.04 & --1.27$\pm$0.07 & --2.77$\pm$0.00 & -- \\
        \ion{Mn}{i} & 1123 & --1.62$\pm$0.13 & --1.70$\pm$0.06 & --1.82$\pm$0.05 & --1.23$\pm$0.07 & --1.58$\pm$0.16 & --1.39$\pm$0.14 & -- & --1.60$\pm$0.16 & --2.12$\pm$0.05 & --0.90$\pm$0.09 & --2.22$\pm$0.05 & --1.62$\pm$0.12 \\
        \ion{Mn}{ii} & 1123 & -- & --1.74$\pm$0.10 & -- & -- & -- & --1.35$\pm$0.13 & -- & -- & -- & -- & --2.20$\pm$0.10 & -- \\
        \ion{Fe}{i} & 1338 & --1.85$\pm$0.11 & --1.91$\pm$0.06 & --1.93$\pm$0.05 & --1.47$\pm$0.06 & --2.24$\pm$0.10 & --1.98$\pm$0.10 & --1.97$\pm$0.12 & --2.46$\pm$0.15 & --1.90$\pm$0.03 & --1.06$\pm$0.07 & --2.61$\pm$0.03 & --2.47$\pm$0.11 \\
        \ion{Fe}{ii} & 1338 & --1.89$\pm$0.07 & --1.92$\pm$0.03 & --1.95$\pm$0.07 & --1.48$\pm$0.07 & --2.21$\pm$0.14 & --2.02$\pm$0.08 & --1.98$\pm$0.08 & --2.49$\pm$0.11 & --1.89$\pm$0.04 & --1.03$\pm$0.07 & --2.50$\pm$0.02 & --2.48$\pm$0.09 \\
        \ion{Co}{i} & 1354 & -- & --1.85$\pm$0.13 & --1.84$\pm$0.12 & --1.45$\pm$0.13 & --2.00$\pm$0.14 & --1.91$\pm$0.16 & -- & -- & --1.42$\pm$0.11 & --0.84$\pm$0.14 & --2.55$\pm$0.10 & -- \\
        \ion{Ni}{i} & 1363 & --1.37$\pm$0.16 & --1.92$\pm$0.05 & --1.77$\pm$0.04 & --1.33$\pm$0.06 & --2.14$\pm$0.14 & --2.09$\pm$0.15 & --1.84$\pm$0.15 & -- & --1.62$\pm$0.06 & --1.13$\pm$0.12 & --3.05$\pm$0.12 & --2.00$\pm$0.17 \\
        \ion{Cu}{i} & 1034 & --1.10$\pm$0.17 & --1.46$\pm$0.09 & -- & -- & -- & -- & 0.42$\pm$0.16 & --1.13$\pm$0.18 & -- & -- & -- & --0.08$\pm$0.15 \\
        \ion{Zn}{i} & 704 & --0.57$\pm$0.12 & --0.68$\pm$0.06 & --1.04$\pm$0.04 & --0.71$\pm$0.06 & 0.01$\pm$0.11 & --0.48$\pm$0.12 & 0.08$\pm$0.13 & --0.33$\pm$0.13 & --1.26$\pm$0.08 & --0.16$\pm$0.07 & --0.94$\pm$0.19 & --0.33$\pm$0.09 \\
        \ion{Sr}{ii} & 1548 & -- & -- & -- & -- & -- & -- & -- & --2.51$\pm$0.19 & --1.54$\pm$0.10 & -- & --2.84$\pm$0.10 & -- \\
        \ion{Y}{ii} & 1551 & --2.45$\pm$0.13 & --2.75$\pm$0.05 & --2.22$\pm$0.11 & --1.95$\pm$0.10 & -- & --2.29$\pm$0.14 & --1.64$\pm$0.14 & -- & --2.23$\pm$0.10 & --1.58$\pm$0.07 & -- & --2.91$\pm$0.14 \\
        \ion{Zr}{ii} & 1722 & -- & --2.89$\pm$0.11 & --2.15$\pm$0.12 & -- & -- & -- & -- & -- & --2.05$\pm$0.10 & -- & -- & -- \\
        \ion{Ba}{ii} & 1423 & --1.92$\pm$0.15 & --2.15$\pm$0.12 & --2.01$\pm$0.08 & --1.52$\pm$0.09 & --2.53$\pm$0.16 & --2.14$\pm$0.16 & --1.97$\pm$0.13 & --2.92$\pm$0.17 & -- & --1.16$\pm$0.13 & --3.56$\pm$0.10 & --1.93$\pm$0.13 \\
        \ion{La}{ii} & 1615 & -- & --2.56$\pm$0.05 & -- & -- & -- & -- & -- & -- & --2.07$\pm$0.04 & -- & -- & -- \\
        \ion{Ce}{ii} & 1454 & -- & --2.18$\pm$0.11 & --1.79$\pm$0.12 & --1.41$\pm$0.07 & -- & -- & -- & -- & --2.01$\pm$0.04 & --1.58$\pm$0.12 & -- & -- \\
        \ion{Nd}{ii} & 1630 & -- & -- & -- & -- & -- & -- & -- & -- & --1.97$\pm$0.05 & -- & -- & -- \\
        \ion{Sm}{ii} & 1545 & -- & -- & -- & -- & -- & -- & -- & -- & --1.91$\pm$0.10 & -- & -- & -- \\
        \ion{Eu}{ii} & 1491 & -- & --1.71$\pm$0.11 & -- & -- & -- & -- & -- & -- & -- & -- & -- & -- \\ \hline
        \multicolumn{14}{|c|}{\textit{NLTE}} \\ \hline
        \ion{C}{i} & 40 & --0.70$\pm$0.09 & --0.61$\pm$0.07 & --0.57$\pm$0.07 & --0.69$\pm$0.06 & --0.24$\pm$0.13 & --0.41$\pm$0.07 & --0.07$\pm$0.08 & --0.55$\pm$0.09 & 0.09$\pm$0.04 & --0.26$\pm$0.06 & --0.87$\pm$0.01 & --0.52$\pm$0.09 \\
        \ion{N}{i} & 123 & --0.89$\pm$0.18 & -- & --0.01$\pm$0.10 & --0.16$\pm$0.10 & --0.17$\pm$0.21 & --0.12$\pm$0.15 & 0.08$\pm$0.19 & --0.80$\pm$0.19 & 0.70$\pm$0.10 & 0.39$\pm$0.13 & -- & -- \\
        \ion{O}{i} & 183 & --0.02$\pm$0.09 & --0.5$\pm$0.05 & --0.01$\pm$0.12 & 0.01$\pm$0.08 & 0.75$\pm$0.17 & 0.15$\pm$0.14 & 0.23$\pm$0.23 & 0.43$\pm$0.17 & --0.03$\pm$0.06 & 0.31$\pm$0.11 & --0.54$\pm$0.10 & 0.78$\pm$0.14 \\
        \ion{Na}{i} & 1035 & --0.49$\pm$0.06 & --0.74$\pm$0.05 & --1.03$\pm$0.06 & --0.70$\pm$0.11 & --0.55$\pm$0.05 & --1.45$\pm$0.14 & 0.07$\pm$0.12 & --0.53$\pm$0.17 & --1.69$\pm$0.10 & 0.15$\pm$0.11 & --0.82$\pm$0.10 & --0.35$\pm$0.14 \\
        \ion{Mg}{i} & 1343 & --1.33$\pm$0.11 & --1.40$\pm$0.08 & --1.75$\pm$0.04 & --1.20$\pm$0.09 & --1.86$\pm$0.08 & --1.94$\pm$0.07 & --2.04$\pm$0.12 & --2.07$\pm$0.09 & --1.58$\pm$0.05 & --1.29$\pm$0.06 & --2.14$\pm$0.05 & --2.18$\pm$0.12 \\
        \ion{Al}{i} & 1652 & --1.97$\pm$0.16 & -- & -- & --1.45$\pm$0.13 & -- & -- & -- & --2.21$\pm$0.16 & -- & -- & --3.17$\pm$0.10 & --2.06$\pm$0.15 \\
        \ion{Si}{i} & 1314 & -- & -- & --1.62$\pm$0.10 & --1.29$\pm$0.03 & -- & --1.79$\pm$0.12 & -- & -- & -- & --0.84$\pm$0.10 & --1.88$\pm$0.10 & -- \\
        \ion{S}{i} & 672 & --0.31$\pm$0.07 & --0.42$\pm$0.13 & --0.60$\pm$0.08 & --0.66$\pm$0.03 & --0.21$\pm$0.15 & --0.58$\pm$0.04 & 0.31$\pm$0.10 & --0.23$\pm$0.07 & --0.23$\pm$0.07 & 0.23$\pm$0.10 & --0.54$\pm$0.02 & --0.14$\pm$0.06 \\
        \ion{K}{i} & 993 & --0.99$\pm$0.15 & -- & --1.24$\pm$0.12 & --1.03$\pm$0.08 & --0.86$\pm$0.15 & --1.65$\pm$0.14 & -- & --1.1$\pm$0.15 & --1.55$\pm$0.10 & -- & -- & --0.44$\pm$0.15 \\
        \ion{Ca}{i} & 1535 & --1.55$\pm$0.12 & --1.87$\pm$0.05 & --1.75$\pm$0.05 & --1.39$\pm$0.07 & --2.30$\pm$0.08 & --1.92$\pm$0.11 & --1.68$\pm$0.14 & --2.05$\pm$0.11 & --1.79$\pm$0.03 & --1.36$\pm$0.07 & --2.34$\pm$0.10 & --1.67$\pm$0.13 \\
        \ion{Ca}{ii} & 1535 & --2.10$\pm$0.14 & -- & -- & --1.65$\pm$0.13 & --2.75$\pm$0.16 & -- & --2.29$\pm$0.12 & --2.38$\pm$0.15 & -- & -- & -- & -- \\
        \ion{Fe}{i} & 1338 & --1.70$\pm$0.11 & --1.74$\pm$0.06 & --1.76$\pm$0.05 & --1.32$\pm$0.06 & --2.04$\pm$0.10 & --1.81$\pm$0.10 & --1.81$\pm$0.12 & --2.19$\pm$0.15 & --1.79$\pm$0.03 & --1.01$\pm$0.07 & --2.37$\pm$0.03 & --2.14$\pm$0.11 \\
        \ion{Fe}{ii} & 1338 & --1.88$\pm$0.07 & --1.91$\pm$0.03 & --1.94$\pm$0.07 & --1.49$\pm$0.07 & --2.20$\pm$0.14 & --2.01$\pm$0.08 & --1.98$\pm$0.08 & --2.46$\pm$0.11 & --1.89$\pm$0.04 & --1.06$\pm$0.07 & --2.49$\pm$0.02 & --2.44$\pm$0.09 \\ \hline
    \end{tabular}}\\
    \textbf{Notes:} The condensation temperatures for C and N were adopted from \citet{lodders2003CondensationTemperatures}, the condensation temperatures for all other elements were adopted from \citet{wood2019CondensationTemperatures}. Uncertainties below 0.1\,dex are expected for elemental abundances derived from a small (2-4) number of lines displaying individual abundances, which are similar and stable within the uncertainty range of atmospheric parameters.
\end{sidewaystable}
\begin{sidewaystable}[ph!]
    \tiny
    \caption{Atmospheric parameters and differential NLTE corrections for studied elemental abundances of transition disc targets.}\label{tabA:fnlcor_paper2}
    \begin{tabular}{|c|c|c|c|c|c|c|c|c|c|c|c|c|}
    \hline
        \multirow{2}{*}{\textbf{Name}} & \textbf{CT Ori} & \textbf{ST Pup} & \textbf{RU Cen} & \textbf{AC Her} & \textbf{AD Aql} & \textbf{EP Lyr} & \textbf{DY Ori} & \textbf{AF Crt} & \textbf{GZ Nor} & \textbf{V1504 Sco} & \textbf{LMC V0770} & \textbf{LMC V3156} \\
        & \textbf{(\#1)} & \textbf{(\#2)} & \textbf{(\#3)} & \textbf{(\#4)} & \textbf{(\#5)} & \textbf{(\#6)} & \textbf{(\#7)} & \textbf{(\#8)} & \textbf{(\#9)} & \textbf{(\#10)} & \textbf{(\#11)} & \textbf{(\#12)} \\ \hline
        $T_{\rm eff}$ & 5940$\pm$120 & 5340$\pm$80 & 6120$\pm$80 & 6140$\pm$100 & 6200$\pm$170 & 6270$\pm$160 & 6160$\pm$70 & 6110$\pm$110 & 4830$\pm$20 & 5980$\pm$90 & 5750$\pm$100 & 6160$\pm$130 \\
        $\log g$ & 1.01$\pm$0.18 & 0.20$\pm$0.10 & 1.46$\pm$0.15 & 1.27$\pm$0.16 & 1.67$\pm$0.45 & 1.24$\pm$0.18 & 0.88$\pm$0.14 & 0.96$\pm$0.21 & 0.00$\pm$0.18 & 0.98$\pm$0.17 & 0.00$\pm$0.18 & 1.38$\pm$0.20 \\
        $[$Fe/H$]$ & --1.89$\pm$0.11 & --1.92$\pm$0.08 & --1.93$\pm$0.08 & --1.47$\pm$0.08 & --2.20$\pm$0.09 & --2.03$\pm$0.17 & --2.03$\pm$0.04 & --2.47$\pm$0.05 & --1.89$\pm$0.11 & --1.05$\pm$0.07 & --2.61$\pm$0.05 & --2.48$\pm$0.04 \\ \hline
        \ion{C}{i} & --0.13 & --0.11 & --0.13 & --0.10 & --0.14 & --0.13 & --0.23 & --0.14 & --0.23 & --0.15 & --0.12 & --0.18 \\
        \ion{N}{i} & --0.11 & -- & --0.14 & --0.13 & --0.14 & --0.19 & --0.30 & --0.26 & --0.14 & --0.20 & -- & -- \\
        \ion{O}{i} & --0.05 & 0.00 & 0.00 & 0.00 & 0.00 & 0.01 & --0.09 & 0.00 & --0.53 & 0.00 & --0.14 & 0.00 \\
        \ion{Na}{i} & --0.16 & --0.12 & --0.11 & --0.15 & --0.15 & --0.38 & --0.25 & --0.77 & --0.18 & --0.33 & --0.12 & --0.82 \\
        \ion{Mg}{i} & 0.03 & 0.05 & --0.07 & 0.00 & 0.02 & --0.03 & --0.05 & --0.04 & 0.07 & --0.08 & 0.01 & 0.03 \\
        \ion{Al}{i} & 0.43 & -- & -- & 0.41 & -- & -- & -- & 0.60 & -- & -- & 0.66 & 0.58 \\
        \ion{Si}{i} & -- & -- & --0.14 & --0.11 & -- & --0.13 & -- & -- & -- & --0.09 & 0.11 & -- \\
        \ion{S}{i} & --0.14 & --0.08 & --0.18 & --0.17 & --0.24 & --0.20 & --0.31 & --0.16 & --0.11 & --0.18 & --0.18 & --0.21 \\
        \ion{K}{i} & --0.52 & -- & --0.50 & --0.48 & --0.77 & --0.24 & -- & --0.29 & --0.24 & -- & -- & --0.78 \\
        \ion{Ca}{i} & 0.09 & 0.17 & 0.11 & 0.06 & 0.13 & 0.10 & 0.08 & 0.17 & 0.13 & 0.05 & 0.22 & 0.12 \\
        \ion{Ca}{ii} & --0.29 & -- & -- & --0.08 & --0.11 & -- & --0.25 & --0.15 & -- & -- & -- & -- \\
        \ion{Fe}{i} & 0.15 & 0.17 & 0.17 & 0.15 & 0.20 & 0.17 & 0.16 & 0.27 & 0.11 & 0.05 & 0.24 & 0.33 \\
        \ion{Fe}{ii} & 0.01 & 0.01 & 0.01 & --0.01 & 0.01 & 0.01 & 0.00 & 0.03 & 0.00 & --0.03 & 0.01 & 0.04 \\ \hline
    \end{tabular}
\end{sidewaystable}

\section{Individual correlation plots}\label{app:cor_paper2}
\begin{figure}[!ht]
    \centering
    \includegraphics[width=.49\linewidth]{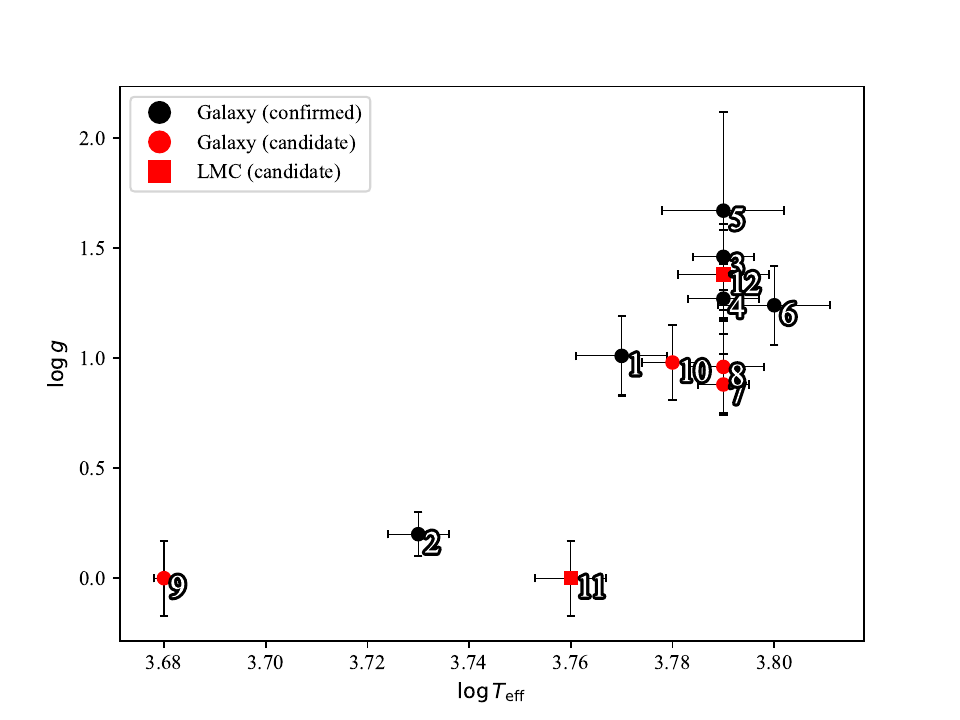}
    \includegraphics[width=.49\linewidth]{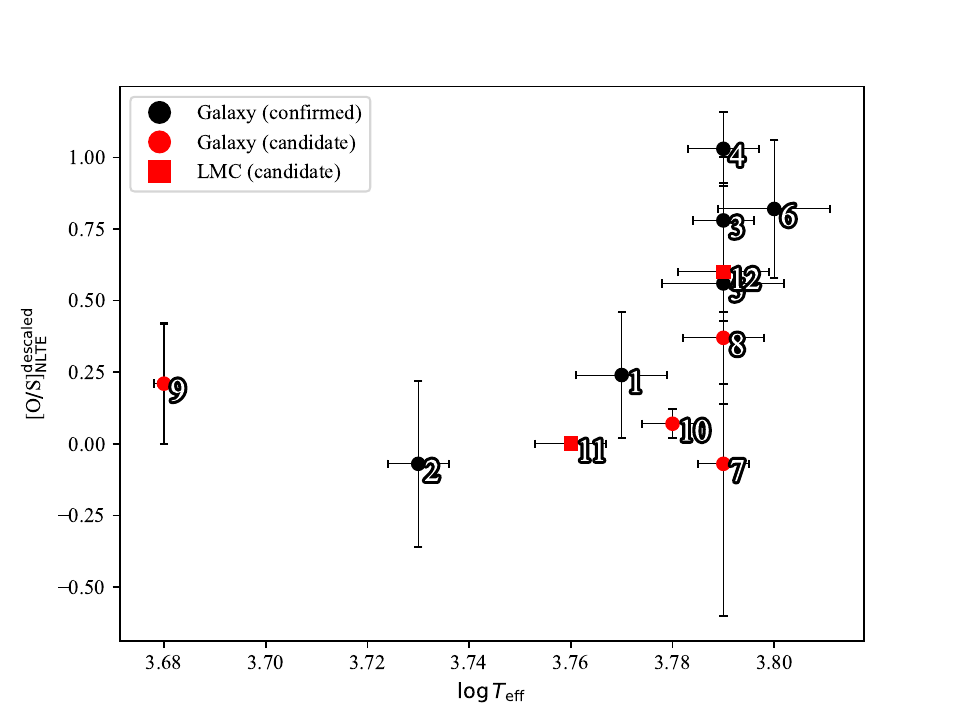}
    \includegraphics[width=.49\linewidth]{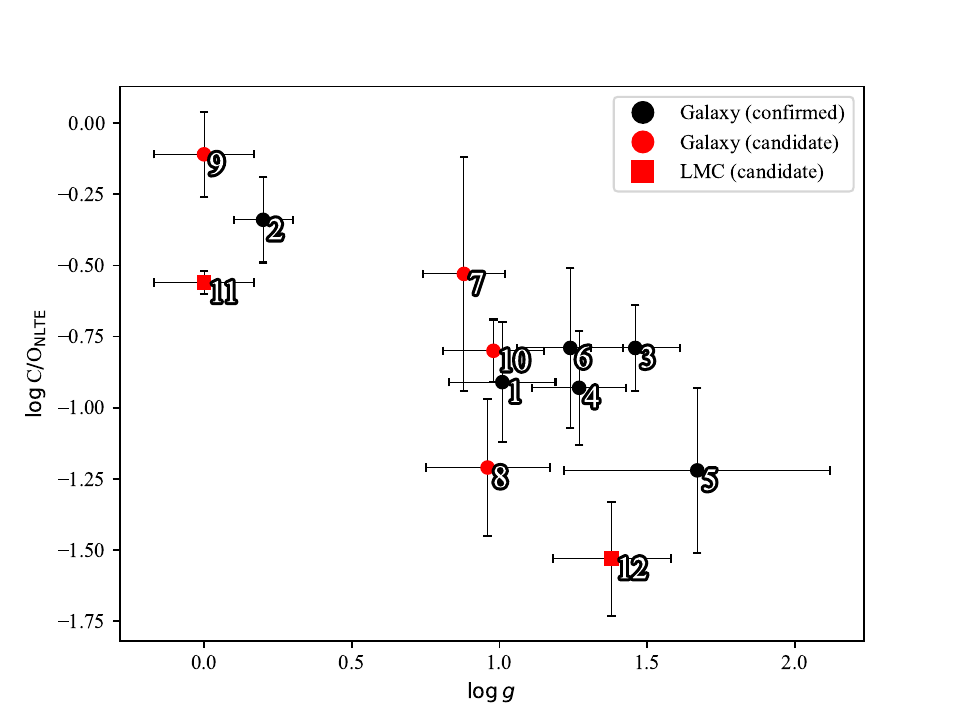}
    \includegraphics[width=.49\linewidth]{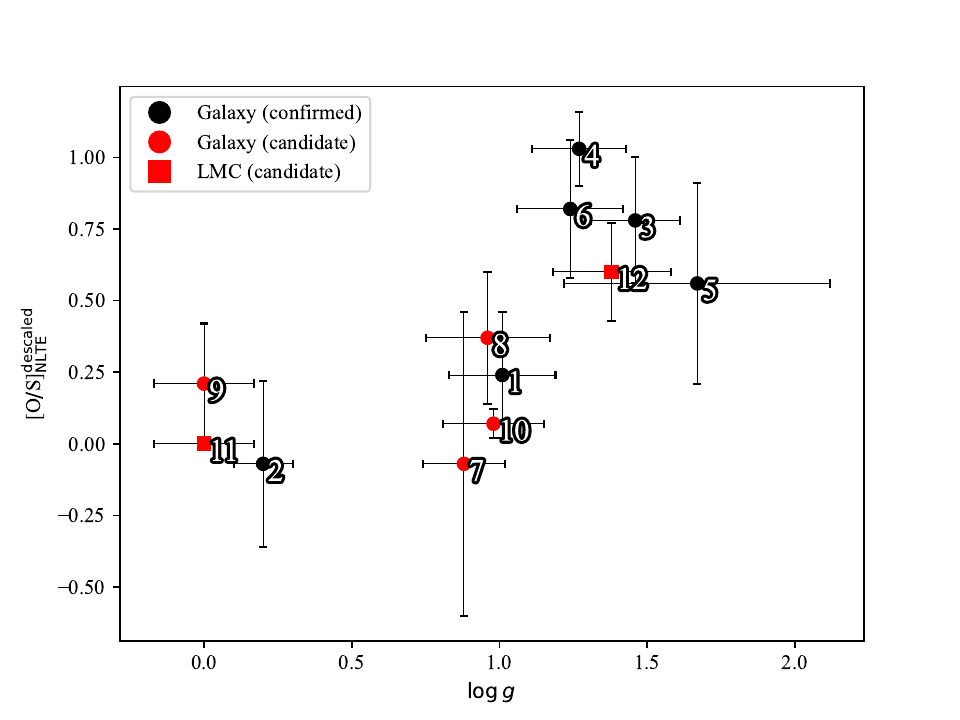}
    \caption[The strongest individual correlations in transition disc targets (see Section~\ref{ssec:dplcor_paper2})]{The strongest individual correlations in transition disc targets (see Section~\ref{ssec:dplcor_paper2}). The legend for the symbols and colours used is included within the plot. The targets are marked with their IDs.}\label{figA:allcor_paper2}
\end{figure}

In Fig.~\ref{figA:allcor_paper2}, we show the strong correlations and anti-correlations (with Spearman's coefficient $|\rho|\geq0.6$) between different parameters of the studied sample (see Section~\ref{ssec:dplcor_paper2}). We note that AF Crt (\#8) and V1504 Sco (\#10) were excluded from luminosity plots, since the luminosities of these edge-on targets are less reliable.

\begin{figure}[!ht]
    \centering
    \includegraphics[width=.49\linewidth]{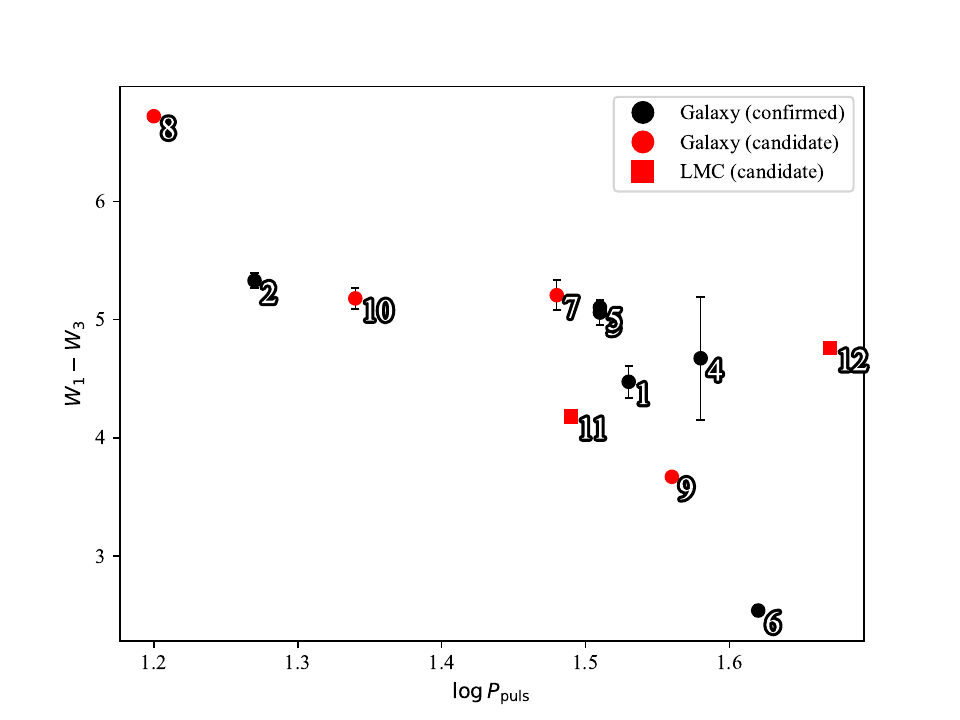}
    \includegraphics[width=.49\linewidth]{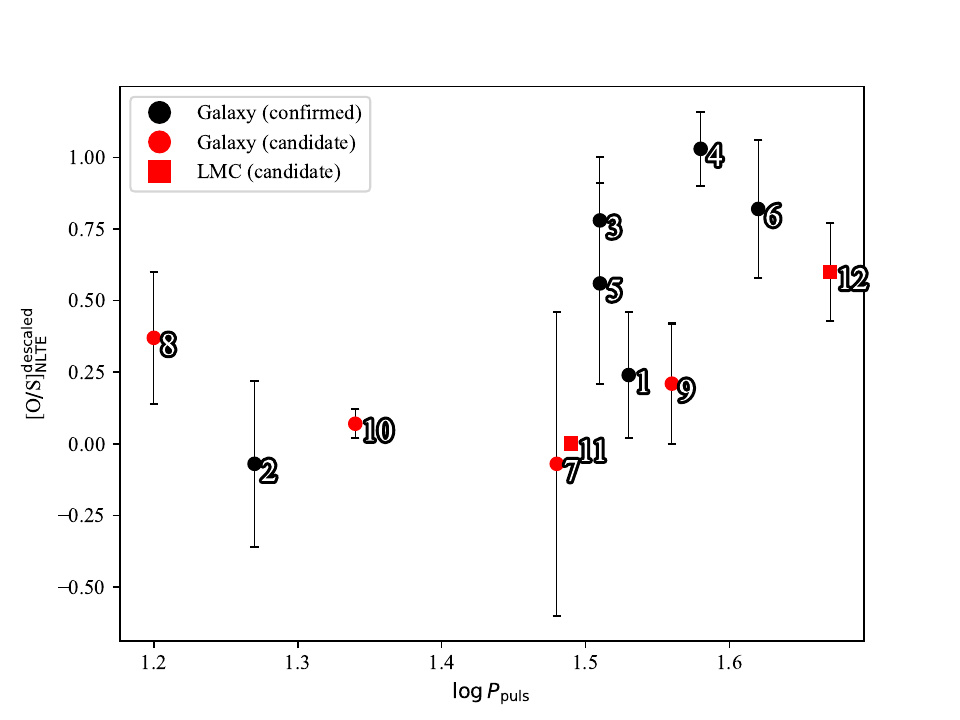}
    \includegraphics[width=.49\linewidth]{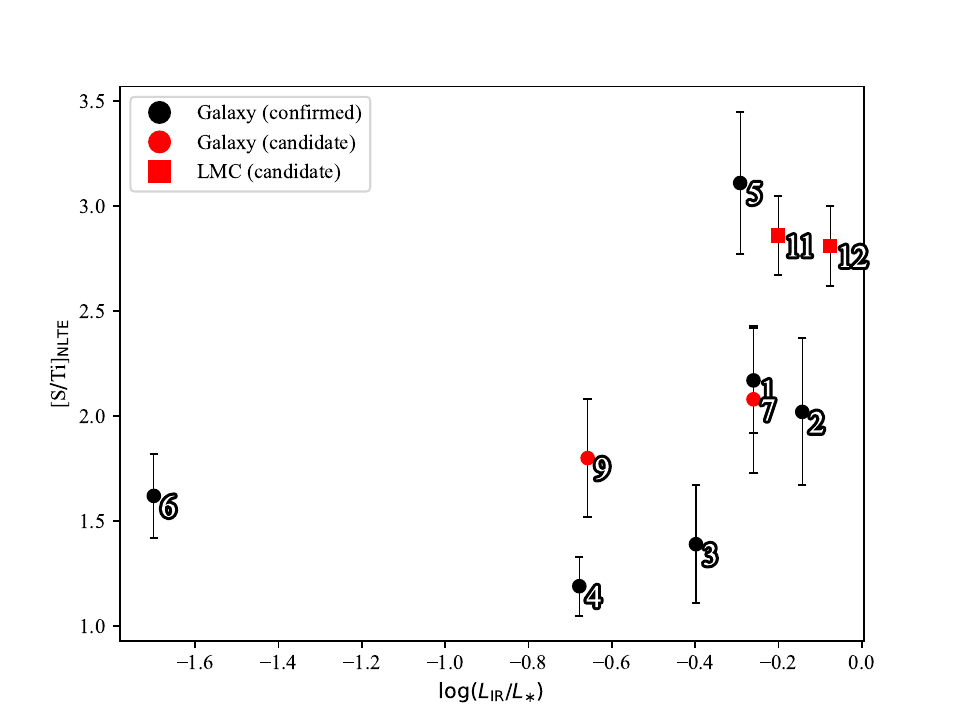}
    \includegraphics[width=.49\linewidth]{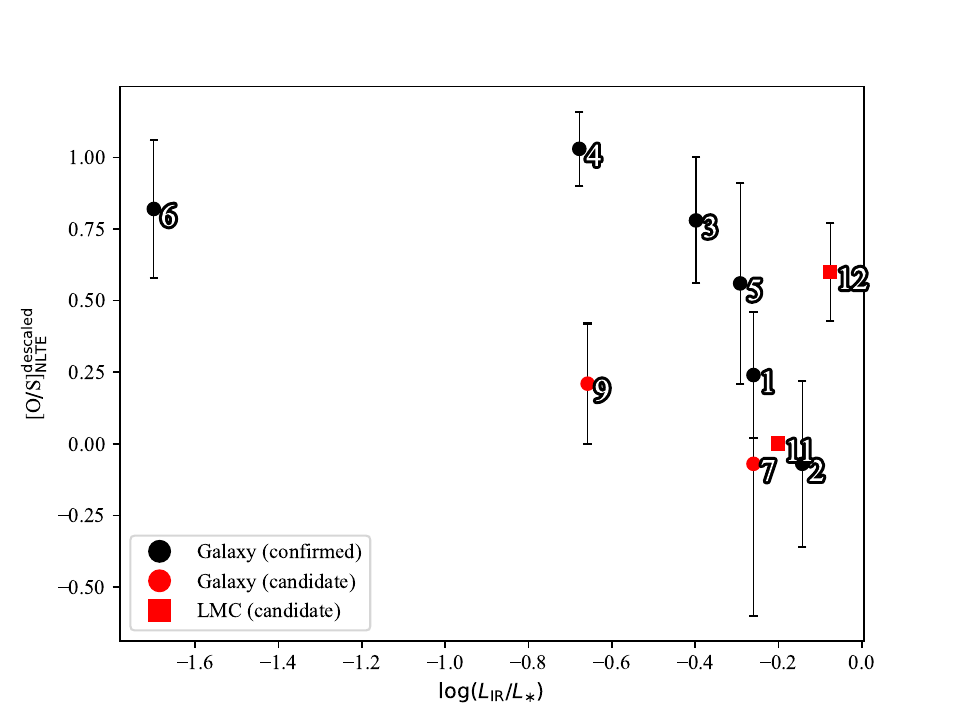}
    \includegraphics[width=.49\linewidth]{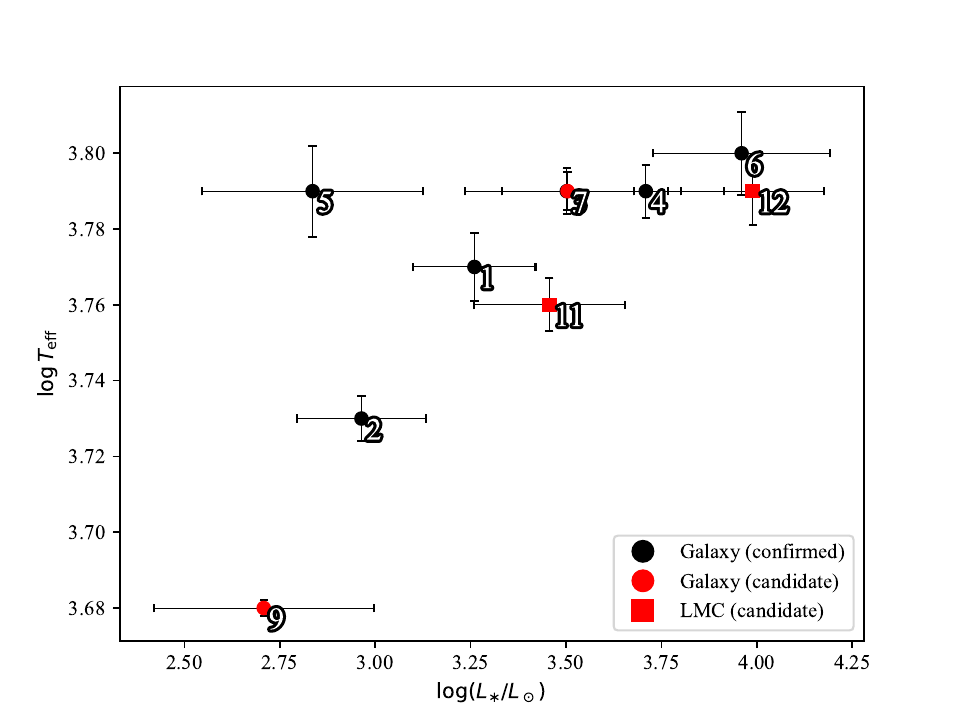}
    \includegraphics[width=.49\linewidth]{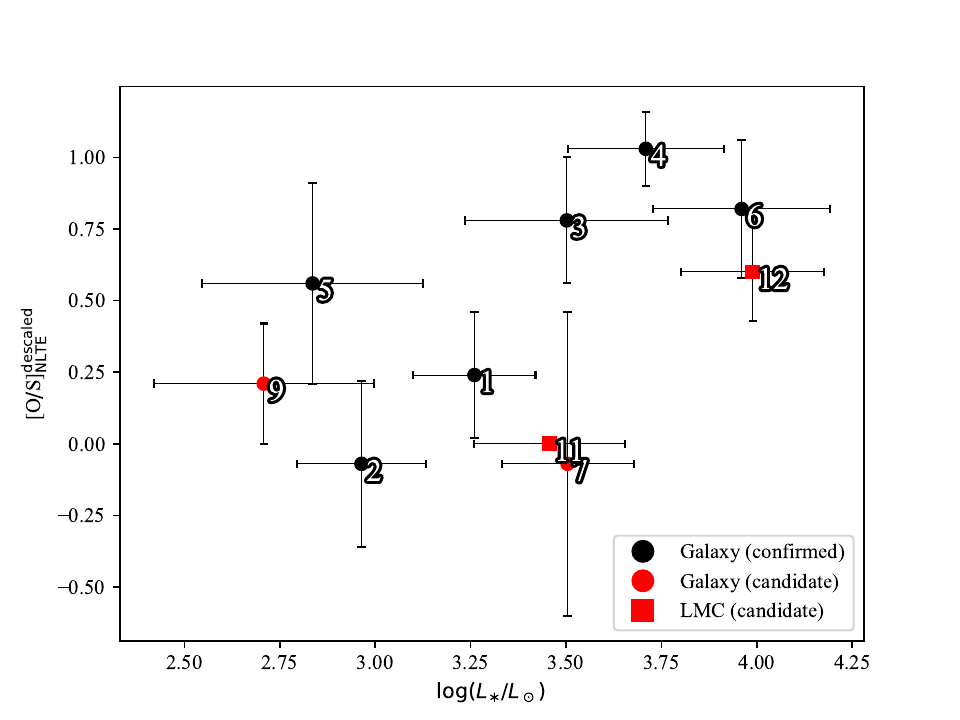}
    \caption[The strongest individual correlations in transition disc targets (see Section~\ref{ssec:dplcor_paper2})]{The strongest individual correlations in transition disc targets (see Section~\ref{ssec:dplcor_paper2}). The legend for the symbols and colours used is included within the plot. The targets are marked with their IDs.}\label{figA:allcor2_paper2}
\end{figure}
\begin{savequote}[75mm]
\foreignlanguage{ukrainian}{``Потреба любови все терпить, особливо, коли допомагає втішити когось при великих клопотах...''}
\qauthor{\foreignlanguage{ukrainian}{---Пилип Орлик (1672-1742), \\гетьман Війська Запорозького \\у вигнанні (1710—1742), поет, публіцист}}
``The need for love endures everything, especially when it helps to comfort someone in great trouble...''
\qauthor{---Pylyp Orlyk (1672-1742), \\Hetman of the Zaporozhian Host \\in exile (1710–1742), poet, publicist}
\end{savequote}

\chapter{Appendix to chapter 5}\label{chp:app5}

\clearpage



\section{Summary of optical spectral visits}\label{app:vis_paper3}
In this Appendix, we list all optical spectral visits considered in this study (see Table~\ref{tabA:obslog_paper3}) for the abundance analysis of post-AGB/post-RGB binaries with dust-poor discs (see Section~\ref{ssec:anaspc_paper3}).

\begin{table}[!ht]
    \centering
    \footnotesize
    \caption[Spectral visits of post-AGB/post-RGB binaries with dust-poor discs]{Spectral visits of post-AGB/post-RGB binaries with dust-poor discs (see Section~\ref{ssec:sdospc_paper3}). This table is published in its entirety in the electronic edition of the paper. A portion is shown here for guidance regarding its form and content.}\label{tabA:obslog_paper3}
    \begin{tabular}{|c|c|c|c|}
    \hline
        \textbf{ObsID} & \textbf{MJD} & \textbf{Phase} & \textbf{RV} \\ \hline
        \multicolumn{4}{|c|}{\textit{SS Gem (RV0=-8.4 km/s)}} \\ \hline
        253252 & 55124.23258 & 0.20 & 2.8 \\
        260288 & 55160.06403 & 0.00 & -13.6 \\
        272381 & 55214.99618 & 0.23 & -3.0 \\
        273153 & 55221.92900 & 0.39 & -1.1 \\
        273787 & 55232.98736 & 0.63 & -7.3 \\
        \multicolumn{4}{|c|}{\ldots} \\ \hline
    \end{tabular}\\
    \textbf{Notes:} RV0 is the radial velocity of the spectral visit with the best S/N ratio. RVrel are radial velocities relative to visit with the best S/N ratio.
\end{table}

\section{Master line list of the target sample}\label{app:lst_paper3}
In this Appendix, we show the combined optical line list (see Table~\ref{tabA:linlst_paper3}), which we used to derive the atmospheric parameters and elemental abundances of post-AGB/post-RGB binaries with dust-poor discs (see Section~\ref{ssec:anaspc_paper3}).

\begin{table}[!ht]
    \centering
    \footnotesize
    \caption[Optical line list of the whole studied sample of post-AGB/post-RGB binaries with full, transition, and dust-poor discs (see Section~\ref{sec:dsc_paper3})]{Optical line list of the whole studied sample of post-AGB/post-RGB binaries with full, transition, and dust-poor discs (see Section~\ref{sec:dsc_paper3}). This table is published in its entirety in the electronic edition of the paper. A portion is shown here for guidance regarding its form and content.}\label{tabA:linlst_paper3}
    \begin{tabular}{|c@{\hspace{0.1cm}}ccc|ccccc|} \hline
        \multicolumn{4}{|c|}{\textbf{Atomic data}} & \multicolumn{5}{|c|}{\boldmath$W_\lambda$ \textbf{(m\AA)}} \\
        \textbf{Element} & \boldmath$\lambda$ & \boldmath$\log gf$ & \boldmath$\chi$ & \textbf{SZ Mon} & \textbf{DF Cyg} & \textbf{\ldots} & \textbf{J053254} & \textbf{BD+28 772} \\
        ~ & \textbf{(nm)} & \textbf{(dex)} & \textbf{(eV)} & \textbf{(\#1)} & \textbf{(\#2)} & \textbf{\ldots} & \textbf{(\#22)} & \textbf{(\#23)} \\ \hline
        \ion{C}{i} & 402.9414 & -2.135 & 7.488 & -- & -- & \ldots & -- & -- \\ 
        \ion{C}{i} & 422.8326 & -1.914 & 7.685 & -- & -- & \ldots & -- & -- \\
        \ion{C}{i} & 426.9019 & -1.637 & 7.685 & -- & -- & \ldots & -- & -- \\
        \ion{C}{i} & 437.1367 & -1.962 & 7.685 & -- & -- & \ldots & -- & -- \\
        \ion{C}{i} & 477.0021 & -2.437 & 7.483 & -- & -- & \ldots & -- & 31.90 \\
        \multicolumn{9}{|c|}{\ldots} \\ \hline
    \end{tabular}
\end{table}

\section{Individual abundances of dust-poor disc targets}\label{app:abu_paper3}
In this Appendix, we present the derived LTE and NLTE elemental abundances of dust-poor disc targets (see Tables~\ref{tabA:reslte_paper3} and \ref{tabA:resnlt_paper3}, respectively). For transition discs, the NLTE abundances obtained in this study are generally consistent within the error bars with those reported by \citet{mohorian2025TransitionDiscs}, showing a median difference of 0.07\,dex and an average difference of 0.10\,dex. These discrepancies likely arise from the interpolation of NLTE corrections on an irregular grid. For more details on the abundance analysis, see Section~\ref{ssec:anaspc_paper3}.

\begin{sidewaystable}[ph!]
    \centering
    \footnotesize
    \caption[LTE {[X/H]} abundances of the whole studied sample of post-AGB/post-RGB binaries with full, transition, and dust-poor discs (see Section~\ref{sec:dsc_paper3})]{LTE [X/H] abundances of the whole studied sample of post-AGB/post-RGB binaries with full, transition, and dust-poor discs (see Section~\ref{sec:dsc_paper3}).}\label{tabA:reslte_paper3} 
    \begin{tabular}{|c|c|ccccc|}
    \hline
        \textbf{Adopted} & \boldmath$T_{\rm cond}$ & \textbf{SZ Mon} & \textbf{DF Cyg} & \textbf{\ldots} & \textbf{J053254} & \textbf{BD+28\,772} \\
        \textbf{name} & \textbf{(K)} & \textbf{(\#1)} & \textbf{(\#2)} & \textbf{\ldots} & \textbf{(\#22)} & \textbf{(\#23)} \\ \hline
        \multicolumn{7}{|c|}{\textit{LTE}} \\ \hline
        \ion{C}{i} & 40 & --0.09$\pm$0.08 & 0.27$\pm$0.09 & \ldots & --0.42$\pm$0.06 & --0.39$\pm$0.09 \\
        \ion{N}{i} & 123 & 0.44$\pm$0.14 & -- & \ldots & -- & 0.68$\pm$0.08 \\
        \ion{O}{i} & 183 & 0.15$\pm$0.11 & 0.73$\pm$0.12 & \ldots & --0.63$\pm$0.12 & --0.11$\pm$0.05 \\
        \ldots & \ldots & \ldots & \ldots & \ldots & \ldots & \ldots\\ \hline
    \end{tabular}\\
    \textbf{Note:} The condensation temperatures for C and N were adopted from \cite{lodders2003CondensationTemperatures}, the condensation temperatures for all other elements were adopted from \cite{wood2019CondensationTemperatures}.
\end{sidewaystable}

\begin{sidewaystable}[ph!]
    \centering
    \footnotesize
    \caption[NLTE {[X/H]} abundances of the whole studied sample of post-AGB/post-RGB binaries with full, transition, and dust-poor discs (see Section~\ref{sec:dsc_paper3})]{NLTE [X/H] abundances of the whole studied sample of post-AGB/post-RGB binaries with full, transition, and dust-poor discs (see Section~\ref{sec:dsc_paper3}).}\label{tabA:resnlt_paper3} 
    \begin{tabular}{|c|c|ccccc|}
    \hline
        \textbf{Adopted} & \boldmath$T_{\rm cond}$ & \textbf{SZ Mon} & \textbf{DF Cyg} & \textbf{\ldots} & \textbf{J053254} & \textbf{BD+28\,772} \\
        \textbf{name} & \textbf{(K)} & \textbf{(\#1)} & \textbf{(\#2)} & \textbf{\ldots} & \textbf{(\#22)} & \textbf{(\#23)} \\ \hline
        \multicolumn{7}{|c|}{\textit{NLTE}} \\ \hline
        \ion{C}{i} & 40 & --0.15$\pm$0.08 & 0.14$\pm$0.07 & \ldots & --0.48$\pm$0.06 & --0.46$\pm$0.11 \\
        \ion{N}{i} & 123 & 0.20$\pm$0.14 & -- & \ldots & -- & 0.26$\pm$0.17 \\
        \ion{O}{i} & 183 & 0.15$\pm$0.11 & 0.84$\pm$0.05 & \ldots & --0.62$\pm$0.12 & --0.28$\pm$0.04 \\
        \ldots & \ldots & \ldots & \ldots & \ldots & \ldots & \ldots\\ \hline
    \end{tabular}\\
    \textbf{Note:} The condensation temperatures for C and N were adopted from \cite{lodders2003CondensationTemperatures}, the condensation temperatures for all other elements were adopted from \cite{wood2019CondensationTemperatures}.
\end{sidewaystable}
 
\backmatter

\bibliography{references.bib}

\end{document}